\documentclass[titlepage]{tcibook}
\usepackage{fancyhea}
\usepackage{work}
\usepackage{bm}       
\usepackage{graphicx}
\usepackage{hyperref}      
\usepackage[usenames,dvipsnames]{color} 
\usepackage{amssymb}
\usepackage{amsmath}
\usepackage{chngcntr}
\usepackage{booktabs}
\usepackage{multirow}
\usepackage{multicol}
\counterwithout{figure}{chapter}
\usepackage{titlesec}

\usepackage[utf8]{inputenc}

\newcommand{\nc}{\newcommand}  

\def\beq{\begin{equation}}
\def\eeq#1{\label{#1}\end{equation}}
\def\eeqn{\end{equation}}

\newenvironment{Eqnarray}%
   {\arraycolsep 0.14em\begin{eqnarray}}{\end{eqnarray}}
\def\beqa{\begin{Eqnarray}}
\def\eeqa#1{\label{#1}\end{Eqnarray}}
\def\eeqan{\end{Eqnarray}}

\nc{\ra}{\rightarrow}  
\nc{\slsh}{\slash\hspace*{-0.22cm}}
\def\Re{{\cal R \mskip-4mu \lower.1ex \hbox{\it e}\,}}
\def\Im{{\cal I \mskip-5mu \lower.1ex \hbox{\it m}\,}}

\nc{\vev}[1]{ \left\langle {#1} \right\rangle }
\nc{\bra}[1]{ \langle {#1} | }
\nc{\ket}[1]{ | {#1} \rangle }
\nc{\fb}{\,{\rm fb}^{-1}}
\nc{\ev}{{\rm eV}}
\nc{\kev}{{\rm keV}}
\nc{\Mev}{{\rm MeV}}
\nc{\gev}{{\rm GeV}}
\nc{\tev}{{\rm TeV}}
\nc{\mev}{{\rm MeV}}

\def\del{\partial}
\def\Dslash{\not{\hbox{\kern-4pt $D$}}}
\def\dslash{\not{\hbox{\kern-2pt $\del$}}}
\def\pslash{\not{\hbox{\kern-2pt $p$}}}
\def\ETmiss{ \not{\hbox{\kern-4pt $E$}}_T }

\def\One{\ensuremath{\bf 1}\xspace}

\def\msb{{\bar{\ssstyle M \kern -1pt S}}}

\newcommand{\cmbexp}{{CMB-S4}}
\newcommand{\cobe}{{\sl COBE}}
\newcommand{\wmap}{{\sl WMAP}}
\newcommand{\planck}{{\sl Planck}}

\newcommand{\neff}{\ensuremath{N_\mathrm{eff}}}
\newcommand{\fsky}{\ensuremath{f_\mathrm{sky}}}

\definecolor{orange}{rgb}{1,0.3,0}

\def\Neff{N_{\rm eff}}
\def\Nf{N_{\rm eff}}
\def\gs{g_{\star}}
\def\Mpl{M_{\rm P}}
\def\lsim{\raise-.75ex\hbox{$\buildrel<\over\sim$}}

\DeclareUrlCommand\email{\urlstyle{rm}}

\begin{document}


\def\bibname{References}

\bibliographystyle{utphys}  

\raggedbottom

\pagenumbering{roman}

\parindent=0pt
\parskip=8pt
\setlength{\evensidemargin}{0pt}
\setlength{\oddsidemargin}{0pt}
\setlength{\marginparsep}{0.0in}
\setlength{\marginparwidth}{0.0in}
\marginparpush=0pt


\renewcommand{\chapname}{chap:intro_}
\renewcommand{\chapterdir}{.}
\renewcommand{\arraystretch}{1.25}
\addtolength{\arraycolsep}{-3pt}

\pagenumbering{roman} 
\chapter*{CMB-S4 Science Book\\ First Edition}
\vskip -9.5pt
\hbox to\headwidth{%
       \leaders\hrule height1.5pt\hfil}
\vskip-6.5pt
\hbox to\headwidth{%
       \leaders\hrule height3.5pt\hfil}

  \begin{center}
   {\Large\bf
      CMB-S4 Collaboration\\
      \bigskip
      August 1, 2016
   }
%
%
 \end{center}
\eject

\setcounter{page}{1}

\begin{center}
  {\Large \bf Executive Summary}
\end{center}

The next generation ``Stage-4" ground-based cosmic microwave background (CMB) experiment, CMB-S4, consisting of dedicated telescopes equipped with highly sensitive superconducting cameras operating at the South Pole, the high Chilean Atacama plateau, and possibly northern hemisphere sites, will provide a dramatic leap forward in our understanding of the fundamental nature of space and time and the evolution of the Universe. CMB-S4 will be designed to cross critical thresholds in testing inflation, determining the number and masses of the neutrinos, constraining possible new light relic particles, providing precise constraints on the nature of dark energy, and testing general relativity on large scales. 

CMB-S4 is intended to be the definitive ground-based CMB project. It will deliver a highly constraining data set with which any model for the origin of the primordial fluctuations---be it inflation or an alternative theory---and their evolution to the structure seen in the Universe today must be consistent.   
While we have learned a great deal from CMB measurements, including discoveries that have pointed the way to new physics, we have only begun to tap the information encoded in CMB polarization, CMB lensing and other secondary effects.  The discovery space from these and other yet to be imagined effects will be  maximized by designing CMB-S4 to produce high-fidelity maps, which will also ensure enormous legacy value for CMB-S4.

CMB-S4 is the logical successor to the Stage-3 CMB projects which will operate over the next few years. For maximum impact, CMB-S4 should be implemented on a schedule that allows a transition from Stage~3 to Stage~4 that is as seamless and as timely as possible, preserving the expertise in the community and ensuring a continued stream of CMB science results. This timing is also necessary to ensure the optimum synergistic enhancement of the science return from contemporaneous optical surveys (e.g., LSST, DESI, Euclid and WFIRST).   Information learned from the ongoing Stage-3 experiments can be easily incorporated into CMB-S4 with little or no impact on its design. In particular, additional information on the properties of Galactic foregrounds would inform  the detailed distribution of detectors among frequency bands in CMB-S4.  
The sensitivity and fidelity of the multiple band foreground measurements needed to realize the goals of CMB-S4 will be provided by CMB-S4 itself, at frequencies just below and above those of the main CMB channels. 

This timeline is possible because CMB-S4 will use proven existing technology that has been developed and demonstrated by the CMB experimental groups over the last decade. There are, to be sure, considerable technical challenges presented by the required scaling-up of the instrumentation and by the scope and complexity of the data analysis and interpretation.  CMB-S4 will require: scaled-up superconducting detector arrays with well-understood and robust material properties and processing techniques; high-throughput mm-wave telescopes and optics with unprecedented precision and rejection of systematic contamination; full internal characterization of astronomical foreground emission; large cosmological simulations and theoretical modeling with accuracies yet to be achieved; and computational methods for extracting minute correlations in massive, multi-frequency data sets contaminated by noise and a host of known and unknown signals.

CMB-S4 is well aligned with the plan put forth by the Particle Physics Project Prioritization Panel (P5) report {\it Building for Discovery: Strategic Plan for U.S. Particle Physics in the Global Context}, which recommends ``Support CMB experiments as part of the core particle physics program. The multidisciplinary nature of the science warrants continued multiagency support.''  The P5 justification for this recommendation emphasized, ``These measurements are of central significance to particle physics. Particle physics groups at the DOE laboratories have unique capabilities, e.g., in sensor technology and production of large sensor arrays that are essential to future CMB experiments as the technological sophistication and scale of the experiments expands. The participation of particle physicists in cases in which they contribute unique expertise is warranted. For these reasons, substantially increased particle physics funding of CMB research and projects is appropriate in the context of continued multiagency partnerships.''  The overall project/activity plan of the P5 report has the CMB-S4 project proceeding under all funding scenarios considered.  CMB-S4 is also endorsed as one of only three priority research initiatives in 
the NRC report {\it A Strategic Vision for NSF Investments in Antarctic and Southern Ocean Research}. 

CMB measurements are also being proposed with instrumentation on satellites and high-altitude balloons. Compared to ground-based observations, these measurements have the advantage of not having to observe through Earth's absorbing and emitting atmosphere, which causes additional measurement noise (balloon-based measurements suffer only small residual noise).  
To minimize the additive noise, ground-based CMB measurements are pursued at wavelengths within broad atmospheric windows bordered by strong atmospheric absorption lines. Within these windows, the chief source of noise impacting the measurements is emission from temporally and spatially varying water vapor, often referred to as sky noise. This sky noise makes precision measurements on the largest angular scales exceedingly difficult. These are also the angular scales at which the CMB signal is most heavily contaminated by astronomical foreground emission.
Compared to satellite based experiments, however, ground-based measurements have the advantage of deploying large aperture telescopes (diameters of several meters), which are needed to achieve the high angular resolution as dictated by the CMB-S4 science goals. While a large telescope could in principle be launched into orbit at an exorbitant cost, the ground-based measurements have been proven to work well at these angular scales,  are far more economical, and are able to take advantage of the latest technological advances.  

There is therefore a natural synergy for future satellite-based and ground-based measurements. CMB-S4 will make the definitive CMB measurements at angular scales from tens of degrees to arc minutes at wavelengths which straddle the peak of the CMB emission. On the largest angular scales, corresponding at multipoles $\ell < 20$, the definitive measurements will need to be done by a satellite covering a range of wavelengths.  The complementarity of ground and satellite measurements will be particularly important in the search for the signature of inflationary gravitational waves in the CMB polarization, so called B mode polarization, as there should be distinct features at degree angular scales imprinted when the CMB decoupled and at very low multipoles when the universe reionized. 

This Science Book sets the scientific goals for CMB-S4 and the measurements required to achieve them. It thereby provides the basis for proceeding with the detailed experimental design. We now provide summaries of the primary science drivers, observables, and analysis/computing issues, each of which is developed in depth in a dedicated chapter of the book.

\subsection*{Inflation: investigating the origin of primordial perturbations and the beginning of time}

Inflation, a period of accelerated expansion of the early Universe, is the leading paradigm for explaining the origin of the primordial density perturbations that grew into the CMB anisotropies and eventually into the stars and galaxies we see around us. In addition to primordial density perturbations, the rapid expansion creates primordial gravitational waves that imprint a characteristic polarization pattern onto the CMB. If our Universe is described by a typical model of inflation that naturally explains the statistical properties of the density perturbations, CMB-S4 will detect this signature of inflation. A detection of this particular polarization pattern would open a completely new window onto the physics of the early Universe and provide us with an additional relic left over from the hot big bang. This relic would constitute our most direct probe of the very early Universe and transform our understanding of several aspects of fundamental physics. Because the polarization pattern is due to quantum fluctuations in the gravitational field during inflation, it would provide insights into the quantum nature of gravity. The strength of the signal, encoded in the tensor-to-scalar ratio $r$, would provide a direct measurement of the expansion rate of the Universe during inflation. A detection with CMB-S4 would point to inflationary physics near the energy scale associated with grand unified theories and would provide additional evidence in favor of the idea of the unification of forces. Knowledge of the scale of inflation would also have broad implications for many other aspects of fundamental physics, including ubiquitous ingredients of string theory like axions and moduli. 
 
Even an upper limit of $r<0.002$ at $95\%$ CL achievable by CMB-S4, over an order of magnitude stronger than current limits, would significantly advance our understanding of inflation. It would rule out the most popular and most widely studied classes of models and dramatically impact how we think about the theory. To some, the remaining class of models would be contrived enough to give up on inflation altogether. Furthermore, CMB-S4 is in a unique position to probe the statistical properties of primordial density perturbations through measurements of primary anisotropies in the temperature and polarization of the CMB with unprecedented precision, providing us with invaluable information about the early Universe.

\subsection*{Neutrinos: setting the neutrino mass scale and testing the 3-neutrino paradigm}

Neutrinos are the least explored corner of the Standard Model of particle physics.  The 2015 Nobel Prize recognized the discovery of neutrino oscillations, which shows that they have mass. However, the overall scale of the masses and the full suite of mixing parameters are still not measured.  Cosmology offers a unique view of neutrinos; they were produced in large numbers in the high temperatures of the early universe and left a distinctive imprint in the cosmic microwave background and on the large-scale structure of the universe. Therefore, CMB-S4 and large-scale structure surveys together will have the power to detect properties of neutrinos that supplement those probed by large terrestrial experiments such as short- and long-baseline as well as neutrino-less double beta decay experiments.

Specifically, while long baseline experiments are sensitive to the differences in the masses of the different types of neutrinos, CMB-S4  will probe the sum of all the neutrino masses. The current lower limit on the sum of neutrino masses imposed by oscillation experiments is \mbox{$\sum m_\nu=58\,\rm{meV}$}. CMB-S4, in conjunction with upcoming baryon acoustic oscillation surveys, will measure this sum with high significance. Once determined, the sum of neutrino masses will inform the prospects for future neutrino-less double beta decay experiments that aim to determine whether neutrinos are their own anti-particle. Furthermore, an upper limit below \mbox{$\sum m_\nu=105\,\rm{meV}$} would disfavor the inverted mass hierarchy. Finally, CMB-S4 is particularly sensitive to the possible existence of additional neutrinos that interact even more weakly than the neutrinos in the Standard Model. These so-called {\it sterile} neutrinos are also being vigorously pursued with short baseline experiments around the world. So the combination of CMB-S4, large scale structure surveys, and terrestrial probes adds up to a comprehensive assault on the three-neutrino paradigm.

\subsection*{Light Relics: searching for new light particles}

New light particles appear in many attempts to understand both the observed laws of physics and extensions to higher energies.  These light particles are often deeply tied to the underlying symmetries of nature and can play crucial roles in understanding some of the great outstanding problems in physics.  In most cases, these particles interact too weakly to be produced at an appreciable level in Earth-based experiments, making them experimentally elusive.  At the very high temperatures believed to be present in the early Universe, however, even extremely weakly coupled particles can be produced prolifically and can reach thermal equilibrium with the Standard Model particles. Light particles (masses less than $0.1$ eV) produced at early times survive until the time when the CMB is emitted and direct observations become possible.  Neutrinos are one example of such a relic found in the Standard Model.  Extensions of the Standard Model also include a wide variety of possible light relics including axions, sterile neutrinos, hidden photons, and gravitinos.  As a result, the search for light relics from the early Universe with CMB-S4 can shed light on some of the most important questions in fundamental physics, complementing existing collider searches and efforts to detect these light particles in the lab.  
 
Light relics contribute to the total energy density in radiation in the Universe during the radiation era and significantly alter the appearance of the CMB at small angular scales (high multipole number $\ell$). The energy density in radiation controls both the expansion rate of the Universe at that time and the fluctuations in the gravitational potential in which the baryons and photons evolve.  Through these effects, CMB-S4 can provide an exquisite measurement of the total energy density in light weakly-coupled particles, often parametrized by the quantity $\Neff$.  Any additional light particle that decoupled from thermal equilibrium with the Standard Model produces a change to the density equivalent to $\Delta \Neff \geq 0.027$ per effective degree of freedom of the particle.  This is a relatively large contribution to the radiation density that arises from the democratic population of all species during thermal equilibrium.  Conservative configurations of CMB-S4 can reach $\sigma(\Neff) \sim 0.02-0.03$, which will test the minimal contribution of any light relic with spin at $2\sigma$ and at 1$\sigma$ for any particle with zero spin.  $\Neff$ is a unique measurement to cosmology, and it is likely that these thresholds can only be reached by observing the CMB with the angular resolution and sensitivity attainable by a CMB-S4 experiment.

\subsection*{Dark Matter: searching for heavy WIMPS and extremely light axions}

Dark matter is required to explain a host of cosmological observations such as the velocities of galaxies in galaxy clusters, galaxy rotation curves, strong and weak lensing measurements, and the acoustic peak structure of the CMB. While most of these observations could be explained by non-luminous baryonic matter, the CMB provides overwhelming evidence that $85\%$ of the matter in the Universe is non-baryonic, presumably a new particle never observed in terrestrial experiments. Because dark matter has only been observed through its gravitational effects, its microscopic properties remain a mystery. Identifying its nature and its connection to the rest of physics is one of the prime challenges of high energy physics.

Weakly interacting massive particles (WIMPs) are one well-motivated candidate that naturally appears in many extensions of the standard model. A host of experiments are hoping to detect them: deep underground ton-scale detectors, gamma-ray observatories, and the Large Hadron Collider. The CMB provides a complementary probe through annihilation of dark matter into Standard Model particles.   In the WIMP paradigm, the processes that allow dark matter to be created often allow the particles to annihilate with one another. The rate for this process governs how many of the particles remain today and, for a given WIMP mass, is well-constrained by the known dark matter abundance. The same annihilation process injects a small amount of energy into the CMB, slightly distorting its anisotropy power spectrum. CMB-S4 will probe dark matter masses a few times larger than those probed by current CMB experiments.

Dark matter need not be heavy or thermally produced. Axions provide one compelling example that appears in many extensions of the standard model and is often invoked as solution to some of the most challenging problems in particle physics. Although axions are often extremely light, they can naturally furnish some or all of the dark matter non-thermally. Their effects on the expansion rate of the Universe, on the clustering, and on the local composition of the Universe through quantum fluctuations in the axion field all lead to subtle modifications of the CMB and lensing power spectra. CMB-S4 will improve current limits by as much as an order of magnitude and for some range of masses would be sensitive to axions contributing as little as $1\%$ to the energy density of dark matter.  As for WIMPs, there is an active program of direct and indirect experimental searches for axions that will complement the CMB, and the interplay can reveal important insights into both axions and cosmology.

\subsection*{Dark Energy: measuring cosmic acceleration and testing general relativity}

The discovery almost 20 years ago that the expansion of the universe is accelerating presented a profound challenge to our laws of physics, one that we have yet to conquer. Our current framework can explain these observations only by invoking a new substance with bizarre properties ({\it dark energy}) or by changing the century-old, well-tested theory of general relativity invented by Einstein. The current epoch of acceleration is much later than the epoch from which the photons in the CMB originate, and the behavior of dark energy or modifications of gravity do not significantly influence the properties of the primordial CMB. However, during their long journey to our telescopes, CMB photons occasionally interact with the intervening matter and can have their trajectories and their energies slightly distorted. These distortions---gravitational lensing by intervening mass and energy gain by scattering off hot electrons---are small, but powerful experiments currently online have already detected them, and CMB-S4 will exploit them to the fullest extent, enabling us to learn about the mechanism driving the current epoch of acceleration.

The canonical model is that acceleration is driven by a cosmological constant. Although theoretically implausible, this model does satisfy current constraints, so a simple target for CMB-S4 is to test the many predictions this model makes at late times. Using the gravitational lensing of the CMB, the abundance of galaxy clusters, and cosmic velocities, CMB-S4 will measure both the expansion rate $H$ and the amount of clustering, quantified by the parameter $\sigma_8$, as a function of time. The constraints from CMB-S4 alone will be at the sub-percent level on each and, when combined with other experiments, will reach below a tenth of a percent, particularly when the power of CMB-S4 is also harnessed to calibrate these other probes. These constraints will be among the most powerful tests of the cosmological constant; more crucially, this simultaneous sensitivity to expansion and growth will allow us to distinguish the dark energy paradigm from a failure of general relativity. Models for acceleration in this latter class abound, and CMB-S4 will constrain the parameters of these as well.

\subsection*{CMB lensing: mapping all the mass in the Universe}

The distribution of matter in the Universe contains a wealth of information about the primordial density perturbations and the forces that have shaped our cosmological evolution. Mapping this distribution is one of the central goals of modern cosmology. Gravitational lensing provides a unique method to map the matter between us and distant light sources, and lensing of the CMB, the most distant light source available, allows us to map the matter between us and the surface of last scattering.

Gravitational lensing of the CMB can be measured because the statistical properties of the primordial CMB are exquisitely well-known. As CMB photons travel to Earth from the last scattering surface, they are deflected by intervening matter which distorts the observed pattern of CMB anisotropies and modifies their statistical properties. These distortions can be used to create a map of the gravitational potential that altered the photons' paths. The gravitational potential encodes information about the formation of structure in the Universe and, indirectly, cosmological parameters like the sum of the neutrino masses.  CMB-S4 is expected to produce high-fidelity maps over large fractions of the sky, improving on the signal-to-noise of  the \planck\ lensing maps by more than an order of magnitude.  These maps will inform many of the science targets discussed throughout the book and can also be used to calibrate and enhance results of upcoming galaxy redshift surveys or any other maps of the matter distribution.  Unfortunately, lensing also obscures our view of the CMB.  By measuring and removing the effects of lensing from the CMB maps, we sharpen our view of primordial gravitational waves and our understanding of the very early Universe more generally.

\subsection*{Data Analysis, Simulations \& Forecasting}

Extracting science from a CMB dataset is a complex, iterative process requiring expertise in both physical and computational sciences. An integral part of the analysis process is played by high-fidelity simulations of the millimeter-wave sky and the experiment's response to the various sources of emission. Fast-turnaround versions of these sky and instrument simulations play a key role at the instrument design stage, allowing exploration of instrument configuration parameter space and projections for science yield. In all three of these areas (analysis, simulations, forecasting), the large leap in detector count and complexity of CMB-S4 over fielded experiments presents challenges to current methods. Some of these challenges are purely computational---for example, performing full time-ordered-data simulations for CMB-S4 will require computing resources and distributed computing tools significantly beyond what was required for \planck. Other challenges are algorithmic, including finding the optimal way to separate the CMB signal of interest from foregrounds and how to optimally combine data from different experimental platforms. To meet these challenges, we will bring the full intellectual and technical resources of the CMB community to bear, in an effort analogous to the unified effort among hardware groups to build the CMB-S4 instrument. A wide cross-section of the CMB theory, phenomenology, and analysis communities has already come together to produce the forecasts shown elsewhere in this document, including detailed code comparisons and agreement on unified frameworks for forecasting.

\eject

\begin{center}
  {\Large \bf Preface}
\end{center}
\bigskip

This Science Book is the product of a large, global community of scientists who are united in support of proceeding with CMB-S4, which will make key advances in our understanding of the fundamental nature of space and time and the evolution of the Universe. 
The CMB-S4 concept was conceived during the 2013 Snowmass Cosmic Frontier planning exercise. Through the Snowmass process including two meetings and numerous telecons, the CMB experimental groups and the broader cosmology community came together to produce two influential CMB planning papers, endorsed by over 90 scientists, that outlined the science case as well as the CMB-S4 instrumental concept \cite{Abazajian:2013vfg,Abazajian:2013oma}.  It became clear that an enormous increase in the scale of ground-based CMB experiments would be needed to achieve the exciting scientific goals, necessitating a phase change in the ground-based CMB experimental program. To realize CMB-S4, a partnership of the university-based CMB groups, the broader cosmology community and the national laboratories would be needed.

Based on the Snowmass papers and with additional information from the CMB experimental groups, the 2014 report of the Particle Physics Project Prioritization Panel (P5) included CMB-S4 in their recommended program.  After the P5 report was released, the CMB community began a series of semi-annual workshops to advance CMB-S4. The first of these was a dedicated session at the  {\it Cosmology with the CMB 
and its Polarization} workshop at the University of Minnesota January 14-16, 2015 attended by over 90 scientists.
Discussions focused on the unique and vital role of the future ground-based CMB program and its synergy with a possible future satellite mission. It was decided that the community would draft a detailed CMB-S4 Science Book.  The second and third workshops, {\it Cosmology with CMB-S4} held at the University of Michigan September 21-22, 2015 with over 100 participants,
and {\it Cosmology with CMB-S4} held at the Lawrence Berkeley National Laboratory March 7-9, 2016 with over 160 participants,
were dedicated to developing the Science Book.  Working groups for each CMB-S4 science thrust were responsible for preparing and leading dedicated sessions at the workshop and for drafting the corresponding chapters of this book. In addition, a small writing group was responsible for integrating the Science Book. Through the workshops, numerous teleconferences, postings on the CMB-S4 wiki, contributions to the github Science Book repository, and feedback on drafts,  over 200 scientists have contributed to this first edition of the CMB-S4 Science Book.

\clearpage
\newcounter{affilcount}

\newcommand{\affil}[1]{\refstepcounter{affilcount}\label{#1}}

\def\Irvinetext{University of California, Irvine}
\affil{Irvine}
\def\Illtext{University of Illinois, Urbana-Champaign}
\affil{Ill}
\def\Penntext{University of Pennsylvania}
\affil{Penn}
\def\Slactext{SLAC National Accelerator Laboratory}
\affil{Slac}
\def\Princetontext{Princeton University}
\affil{Princeton}
\def\Hopkinstext{Johns Hopkins University}
\affil{Hopkins}
\def\Stanfordtext{Stanford University}
\affil{Stanford}
\def\Oxfordtext{University of Oxford, United Kingdom}
\affil{Oxford}
\def\Fermitext{Fermilab National Accelerator Laboratory}
\affil{Fermi}
\def\Chicagotext{University of Chicago}
\affil{Chicago}
\def\Sandiegotext{University of California, San Diego}
\affil{Sandiego}
\def\Nisttext{National Institute of Standards and Technology}
\affil{Nist}
\def\Sissatext{Scuola Internazionale Superiore di Studi Avanzati, Trieste, Italy}
\affil{Sissa}
\def\Berkeleytext{University of California, Berkeley}
\affil{Berkeley}
\def\Apctext{APC - Universit{\'e} Paris Diderot/USPC, CNRS/IN2P3, France}
\affil{Apc}
\def\Cambridgetext{University of Cambridge, United Kingdom}
\affil{Cambridge}
\def\Iaptext{Institut d'Astrophysique de Paris, Sorbonne Universit{\'e}s-UPMC / Centre National de la Recherche Scientifique}
\affil{Iap}
\def\Anltext{Argonne National Laboratory}
\affil{Anl}
\def\Cincytext{University of Cincinnati}
\affil{Cincy}
\def\Citatext{Canadian Institute For Theoretical Astrophysics, Canada}
\affil{Cita}
\def\Lbnltext{Lawrence Berkeley National Laboratory}
\affil{Lbnl}
\def\Manchestertext{University of Manchester, United Kingdom}
\affil{Manchester}
\def\Perimetertext{Perimeter Institute, Canada}
\affil{Perimeter}
\def\Harvardtext{Harvard University}
\affil{Harvard}
\def\Sapienzatext{Sapienza - Universit{\`a} di Roma, Italy}
\affil{Sapienza}
\def\Dartmouthtext{Dartmouth College}
\affil{Dartmouth}
\def\Kwazulutext{University of KwaZulu-Natal, South Africa}
\affil{Kwazulu}
\def\Novatext{Villanova University}
\affil{Nova}
\def\Cornelltext{Cornell University}
\affil{Cornell}
\def\Jpltext{Jet Propulsion Laboratory}
\affil{Jpl}
\def\Caltechtext{California Institute of Technology}
\affil{Caltech}
\def\Arcetritext{Istituto Nazionale di Astrofisica - Osservatorio Astrofisico di Arcetri, Italy}
\affil{Arcetri}
\def\Mcgilltext{McGill University, Canada}
\affil{Mcgill}
\def\Ilptext{Institut Lagrange Paris - Universit{\'e} Pierre et Marie Curie /  Centre National de la Recherche Scientifique, France}
\affil{Ilp}
\def\Ccatext{Center for Computational Astrophysics}
\affil{Cca}
\def\Muddtext{Harvey Mudd College}
\affil{Mudd}
\def\Icetext{Institut de Ci{\'e}ncies de l'Espai, Spain}
\affil{Ice}
\def\Stockholmtext{Stockholm University, Sweden}
\affil{Stockholm}
\def\Iastext{Institute for Advanced Study}
\affil{Ias}
\def\Haverfordtext{Haverford College}
\affil{Haverford}
\def\Michigantext{University of Michigan}
\affil{Michigan}
\def\Ubctext{University of British Columbia, Canada}
\affil{Ubc}
\def\Bouldertext{University of Colorado Boulder}
\affil{Boulder}
\def\Minnesotatext{University of Minnesota}
\affil{Minnesota}
\def\Laltext{Laboratoire de l'Acc{\'e}l{\'e}rateur Lin{\'e}aire, France}
\affil{Lal}
\def\Columbiatext{Columbia University}
\affil{Columbia}
\def\Osutext{The Ohio State University}
\affil{Osu}
\def\Torontotext{University of Toronto, Canada}
\affil{Toronto}
\def\Fsutext{Florida State University}
\affil{Fsu}
\def\Davistext{University of California, Davis}
\affil{Davis}
\def\Pitttext{University of Pittsburgh}
\affil{Pitt}
\def\Ferraratext{Universit{\`a} di Ferrara, Italy}
\affil{Ferrara}
\def\Sussextext{University of Sussex, United Kingdom}
\affil{Sussex}
\def\Stonytext{Stony Brook University}
\affil{Stony}
\def\Madisontext{University of Wisconsin - Madison}
\affil{Madison}
\def\Santabarbaratext{University of California, Santa Barbara}
\affil{Santabarbara}
\def\Lpsctext{Laboratoire de Physique Subatomique \& Cosmologie - Institut National de Physique Nucl{\'e}aire et de Physique des Particules (Centre National de la Recherche Scientifique) / Universit{\'e} Grenoble Alpes}
\affil{Lpsc}
\def\Kingstext{King's College London, United Kingdom}
\affil{Kings}
\def\Asutext{Arizona State University}
\affil{Asu}
\def\Ceatext{Commissariat {\`a} l'Energie Atomique - Saclay, France}
\affil{Cea}
\def\Usctext{University of Southern California}
\affil{Usc}
\def\Bohrtext{Niels Bohr Institute, Denmark}
\affil{Bohr}
\def\Orsaytext{Institut d'Astrophysique Spatiale - Orsay, France}
\affil{Orsay}
\def\Sfutext{Simon Fraser University, Canada}
\affil{Sfu}
\def\Oslotext{University of Oslo, Norway}
\affil{Oslo}
\def\Melbournetext{University of Melbourne, Australia}
\affil{Melbourne}
\def\Casetext{Case Western Reserve University}
\affil{Case}
\def\Pennstatetext{The Pennsylvania State University}
\affil{Pennstate}
\def\Brookhaventext{Brookhaven National Laboratory}
\affil{Brookhaven}
\def\Swarthmoretext{Swarthmore College}
\affil{Swarthmore}
\def\Hpdtext{High Precision Devices, Boulder, Colorado}
\affil{Hpd}
\def\Umasstext{University of Massachusetts, Amherst}
\affil{Umass}
\def\Iucaatext{Inter-University Centre for Astronomy and Astrophysics, India}
\affil{Iucaa}
\def\Saotext{Smithsonian Astrophysical Observatory}
\affil{Sao}
\def\Mittext{Massachusetts Institute of Technology}
\affil{Mit}
\def\Mpatext{Max-Planck-Institut f{\"u}r Astrophysik, Germany}
\affil{Mpa}
\def\Browntext{Brown University}
\affil{Brown}
\def\Syracusetext{Syracuse University}
\affil{Syracuse}
\def\Gsfctext{National Aeronautics and Space Administration Goddard Space Flight Center}
\affil{Gsfc}

\begin{center}
  {\Large \bf List of Endorsers}
\end{center}
\bigskip


The following people have endorsed the science case for CMB-S4 as presented here and many of them have contributed directly to the writing of this document:

\begin{raggedright}

Kevork N.~Abazajian\textsuperscript{\ref{Irvine}},
Peter Adshead\textsuperscript{\ref{Ill}},
James Aguirre\textsuperscript{\ref{Penn}},
Zeeshan Ahmed\textsuperscript{\ref{Slac}},
Simone Aiola\textsuperscript{\ref{Princeton}},
Yacine Ali-Haimoud\textsuperscript{\ref{Hopkins}},
Steven W.~Allen\textsuperscript{\ref{Stanford},\ref{Slac}},
David Alonso\textsuperscript{\ref{Oxford}},
Adam Anderson\textsuperscript{\ref{Fermi}},
James Annis\textsuperscript{\ref{Fermi}},
John W.~Appel\textsuperscript{\ref{Hopkins}}, 
Douglas E.~Applegate\textsuperscript{\ref{Chicago}}, 
Kam S.~Arnold\textsuperscript{\ref{Sandiego}}, 
Jason E.~Austermann\textsuperscript{\ref{Nist}},
Carlo Baccigalupi\textsuperscript{\ref{Sissa}}, 
Darcy Barron\textsuperscript{\ref{Berkeley}}, 
James G.~Bartlett\textsuperscript{\ref{Apc}}, 
Ritoban Basu Thakur\textsuperscript{\ref{Chicago}},
Nicholas Battaglia\textsuperscript{\ref{Princeton}},
Daniel Baumann\textsuperscript{\ref{Cambridge}},
Karim Benabed\textsuperscript{\ref{Iap}}, 
Amy N.~Bender\textsuperscript{\ref{Anl}}, 
Charles L.~Bennett\textsuperscript{\ref{Hopkins}},
Bradford A.~Benson\textsuperscript{\ref{Fermi}},
Colin A.~Bischoff\textsuperscript{\ref{Cincy}},
Lindsey Bleem\textsuperscript{\ref{Anl}}, 
J.~Richard Bond\textsuperscript{\ref{Cita}}, 
Julian Borrill\textsuperscript{\ref{Lbnl},\ref{Berkeley}}, 
François R.~Bouchet\textsuperscript{\ref{Iap}}, 
Michael L.~Brown\textsuperscript{\ref{Manchester}},
Christopher Brust\textsuperscript{\ref{Perimeter}}, 
Victor Buza\textsuperscript{\ref{Harvard}}, 
Karen Byrum\textsuperscript{\ref{Anl}}, 
Giovanni Cabass\textsuperscript{\ref{Sapienza}}, 
Erminia Calabrese\textsuperscript{\ref{Oxford}}, 
Robert Caldwell\textsuperscript{\ref{Dartmouth}}, 
John E.~Carlstrom\textsuperscript{\ref{Chicago},\ref{Anl}}, 
Anthony Challinor\textsuperscript{\ref{Cambridge}}, 
Clarence L.~Chang\textsuperscript{\ref{Anl}}, 
Hsin C.~Chiang\textsuperscript{\ref{Kwazulu}},
David T.~Chuss\textsuperscript{\ref{Nova}},
Asantha Cooray\textsuperscript{\ref{Irvine}}, 
Nicholas F. Cothard\textsuperscript{\ref{Cornell}}, 
Thomas M.~Crawford\textsuperscript{\ref{Chicago}}, 
Brendan Crill\textsuperscript{\ref{Jpl}},
Abigail Crites\textsuperscript{\ref{Caltech}},
Francis-Yan Cyr-Racine\textsuperscript{\ref{Harvard}},
Francesco de Bernardis\textsuperscript{\ref{Cornell}}, 
Paolo de Bernardis\textsuperscript{\ref{Sapienza}},
Tijmen de Haan\textsuperscript{\ref{Berkeley}},
Jacques Delabrouille\textsuperscript{\ref{Apc}}, 
Marcel Demarteau\textsuperscript{\ref{Anl}},
Mark Devlin\textsuperscript{\ref{Penn}}, 
Sperello di Serego Alighieri\textsuperscript{\ref{Arcetri}}, 
Eleonora di Valentino\textsuperscript{\ref{Iap}},
Clive Dickinson\textsuperscript{\ref{Manchester}},
Matt Dobbs\textsuperscript{\ref{Mcgill}},
Scott Dodelson\textsuperscript{\ref{Fermi}}, 
Olivier Dore\textsuperscript{\ref{Jpl}},
Joanna Dunkley\textsuperscript{\ref{Princeton}},
Cora Dvorkin\textsuperscript{\ref{Harvard}},
Josquin Errard\textsuperscript{\ref{Ilp}},
Thomas Essinger-Hileman\textsuperscript{\ref{Hopkins}},
Giulio Fabbian\textsuperscript{\ref{Sissa}},
Stephen Feeney\textsuperscript{\ref{Cca}},
Simone Ferraro\textsuperscript{\ref{Berkeley}}, 
Jeffrey P.~Filippini\textsuperscript{\ref{Ill}},
Raphael Flauger\textsuperscript{\ref{Sandiego}}, 
Aurelien A.~Fraisse\textsuperscript{\ref{Princeton}},
George M.~Fuller\textsuperscript{\ref{Sandiego}},
Patricio A.~Gallardo\textsuperscript{\ref{Cornell}},
Silvia Galli\textsuperscript{\ref{Iap}},
Jason Gallicchio\textsuperscript{\ref{Mudd}},
Ken Ganga\textsuperscript{\ref{Apc}},
Enrique Gaztanaga\textsuperscript{\ref{Ice}},
Martina Gerbino\textsuperscript{\ref{Stockholm}},
Mandeep S.~S.~Gill\textsuperscript{\ref{Stanford}},
Yannick Giraud-Héraud\textsuperscript{\ref{Apc}},
Vera Gluscevic\textsuperscript{\ref{Ias}},
Sunil Golwala\textsuperscript{\ref{Caltech}},
Krzysztof M.~Gorski\textsuperscript{\ref{Jpl}},
Daniel Green\textsuperscript{\ref{Berkeley}},
Daniel Grin\textsuperscript{\ref{Haverford}},
Evan Grohs\textsuperscript{\ref{Michigan}},
Riccardo Gualtieri\textsuperscript{\ref{Ill}}, 
Jon E.~Gudmundsson\textsuperscript{\ref{Stockholm}},
Grantland Hall\textsuperscript{\ref{Harvard}},
Mark Halpern\textsuperscript{\ref{Ubc}},
Nils W.~Halverson\textsuperscript{\ref{Boulder}},
Shaul Hanany\textsuperscript{\ref{Minnesota}},
Shawn Henderson\textsuperscript{\ref{Cornell}},
Jason W.~Henning\textsuperscript{\ref{Chicago}},
Sophie Henrot-Versille\textsuperscript{\ref{Lal}},
Sergi R.~Hildebrandt\textsuperscript{\ref{Caltech}},
J. Colin Hill\textsuperscript{\ref{Columbia}},
Christopher M.~Hirata\textsuperscript{\ref{Osu}},
Eric Hivon\textsuperscript{\ref{Iap}},
Ren\'{e}e Hlo\v{z}ek\textsuperscript{\ref{Toronto}},
Gilbert Holder\textsuperscript{\ref{Ill}},
William Holzapfel\textsuperscript{\ref{Berkeley}},
Wayne Hu\textsuperscript{\ref{Chicago}},
Johannes Hubmayr\textsuperscript{\ref{Nist}},
Kevin M.~Huffenberger\textsuperscript{\ref{Fsu}},
Kent Irwin\textsuperscript{\ref{Stanford},\ref{Slac}},
Bradley R.~Johnson\textsuperscript{\ref{Columbia}}, 
William C.~Jones\textsuperscript{\ref{Princeton}},
Marc Kamionkowski\textsuperscript{\ref{Hopkins}},
Brian Keating\textsuperscript{\ref{Sandiego}},
Sarah Kernasovskiy\textsuperscript{\ref{Stanford}},
Reijo Keskitalo\textsuperscript{\ref{Lbnl},\ref{Berkeley}},
Theodore Kisner\textsuperscript{\ref{Lbnl},\ref{Berkeley}},
Lloyd Knox\textsuperscript{\ref{Davis}},
Brian J.~Koopman\textsuperscript{\ref{Cornell}},
Arthur Kosowsky\textsuperscript{\ref{Pitt}},
John Kovac\textsuperscript{\ref{Harvard}},
Ely D.~Kovetz\textsuperscript{\ref{Hopkins}},
Nicoletta Krachmalnicoff\textsuperscript{\ref{Sissa}},
Chao-Lin Kuo\textsuperscript{\ref{Stanford},\ref{Slac}},
Akito Kusaka\textsuperscript{\ref{Berkeley}},
Nicole A.~Larsen\textsuperscript{\ref{Chicago}},
Massimiliano Lattanzi\textsuperscript{\ref{Ferrara}},
Charles R.~Lawrence\textsuperscript{\ref{Jpl}},
Maude Le Jeune\textsuperscript{\ref{Apc}},
Adrian T.~Lee\textsuperscript{\ref{Berkeley},\ref{Lbnl}},
Antony Lewis\textsuperscript{\ref{Sussex}},
Marc Lilley\textsuperscript{\ref{Iap}},
Thibaut Louis\textsuperscript{\ref{Iap}},
Marilena Loverde\textsuperscript{\ref{Stony}},
Amy Lowitz\textsuperscript{\ref{Madison}},
Philip M.~Lubin\textsuperscript{\ref{Santabarbara}},
Juan J.~F.~Macias-Perez\textsuperscript{\ref{Lpsc}},
Mathew S. Madhavacheril\textsuperscript{\ref{Princeton}},
Adam Mantz\textsuperscript{\ref{Stanford}},
David J.~E.~Marsh\textsuperscript{\ref{Kings}},
Silvia Masi\textsuperscript{\ref{Sapienza}},
Philip Mauskopf\textsuperscript{\ref{Asu}},
Jeffrey McMahon\textsuperscript{\ref{Michigan}},
Pieter Daniel Meerburg\textsuperscript{\ref{Cita}},
Alessandro Melchiorri\textsuperscript{\ref{Sapienza}},
Jean-Baptiste Melin\textsuperscript{\ref{Cea}},
Stephan Meyer\textsuperscript{\ref{Chicago}},
Joel Meyers\textsuperscript{\ref{Cita}},
Amber D.~Miller\textsuperscript{\ref{Usc}},
Laura M.~Mocanu\textsuperscript{\ref{Chicago}},
Lorenzo Moncelsi\textsuperscript{\ref{Caltech}},
Julian B.~Munoz\textsuperscript{\ref{Hopkins}},
Andrew Nadolski\textsuperscript{\ref{Ill}},
Toshiya Namikawa\textsuperscript{\ref{Stanford}},
Pavel Naselsky\textsuperscript{\ref{Bohr}},
Paolo Natoli\textsuperscript{\ref{Ferrara}},
Ho Nam Nguyen\textsuperscript{\ref{Stony}},
Michael D.~Niemack\textsuperscript{\ref{Cornell}},
Stephen Padin\textsuperscript{\ref{Chicago},\ref{Anl}},
Luca Pagano\textsuperscript{\ref{Orsay}},
Lyman Page\textsuperscript{\ref{Princeton}},
Robert Bruce Partridge\textsuperscript{\ref{Haverford}},
Guillaume Patanchon\textsuperscript{\ref{Apc}},
Timothy J.~Pearson\textsuperscript{\ref{Caltech}},
Marco Peloso\textsuperscript{\ref{Minnesota}},
Julien Peloton\textsuperscript{\ref{Sussex}},
Olivier Perdereau\textsuperscript{\ref{Lal}},
Laurence Perotto\textsuperscript{\ref{Lpsc}},
Francesco Piacentini\textsuperscript{\ref{Sapienza}},
Michel Piat\textsuperscript{\ref{Apc}},
Levon Pogosian\textsuperscript{\ref{Sfu}},
Clement Pryke\textsuperscript{\ref{Minnesota}},
Benjamin Racine\textsuperscript{\ref{Oslo}},
Srinivasan Raghunathan\textsuperscript{\ref{Melbourne}},
Alexandra Rahlin\textsuperscript{\ref{Fermi}},
Marco Raveri\textsuperscript{\ref{Chicago}},
Christian L.~Reichardt\textsuperscript{\ref{Melbourne}},
Mathieu Remazeilles\textsuperscript{\ref{Manchester}},
Graca Rocha\textsuperscript{\ref{Jpl}},
Natalie A.~Roe\textsuperscript{\ref{Lbnl}},
Aditya Rotti\textsuperscript{\ref{Fsu}},
John Ruhl\textsuperscript{\ref{Case}},
Laura Salvati\textsuperscript{\ref{Sapienza}},
Emmanuel Schaan\textsuperscript{\ref{Princeton}},
Marcel M.~Schmittfull\textsuperscript{\ref{Berkeley}},
Douglas Scott\textsuperscript{\ref{Ubc}},
Neelima Sehgal\textsuperscript{\ref{Stony}},
Sarah Shandera\textsuperscript{\ref{Pennstate}},
Christopher Sheehy\textsuperscript{\ref{Brookhaven}},
Blake D. Sherwin\textsuperscript{\ref{Berkeley}},
Erik Shirokoff\textsuperscript{\ref{Chicago}},
Eva Silverstein\textsuperscript{\ref{Stanford}},
Sara M.~Simon\textsuperscript{\ref{Michigan}},
Tristan L.~Smith\textsuperscript{\ref{Swarthmore}},
Michael Snow\textsuperscript{\ref{Hpd}},
Lorenzo Sorbo\textsuperscript{\ref{Umass}},
Tarun Souradeep\textsuperscript{\ref{Iucaa}},
Suzanne T.~Staggs\textsuperscript{\ref{Princeton}},
Antony A.~Stark\textsuperscript{\ref{Sao}},
Glenn D.~Starkman\textsuperscript{\ref{Case}},
George F.~Stein\textsuperscript{\ref{Cita}},
Jason R.~Stevens\textsuperscript{\ref{Cornell}},
Radek Stompor\textsuperscript{\ref{Apc}},
Kyle T.~Story\textsuperscript{\ref{Stanford}},
Chris Stoughton\textsuperscript{\ref{Fermi}},
Meng Su\textsuperscript{\ref{Mit}},
Rashid Sunyaev\textsuperscript{\ref{Mpa}},
Aritoki Suzuki\textsuperscript{\ref{Berkeley}},
Grant P.~Teply\textsuperscript{\ref{Sandiego}},
Peter Timbie\textsuperscript{\ref{Madison}},
Jesse I.~Treu\textsuperscript{\ref{Princeton}},
Matthieu Tristram\textsuperscript{\ref{Lal}},
Gregory Tucker\textsuperscript{\ref{Brown}},
Sunny Vagnozzi\textsuperscript{\ref{Stockholm}},
Alexander van Engelen\textsuperscript{\ref{Cita}},
Eve M.~Vavagiakis\textsuperscript{\ref{Cornell}},
Joaquin D.~Vieira\textsuperscript{\ref{Ill}},
Abigail G.~Vieregg\textsuperscript{\ref{Chicago}}, 
Sebastian von Hausegger\textsuperscript{\ref{Bohr}},
Benjamin Wallisch\textsuperscript{\ref{Cambridge}},
Benjamin D.~Wandelt\textsuperscript{\ref{Iap}},
Scott Watson\textsuperscript{\ref{Syracuse}},
Nathan Whitehorn\textsuperscript{\ref{Berkeley}},
Edward J.~Wollack\textsuperscript{\ref{Gsfc}},
W.~L.~Kimmy Wu\textsuperscript{\ref{Berkeley}},
Zhilei Xu\textsuperscript{\ref{Hopkins}},
Ki Won Yoon\textsuperscript{\ref{Slac}},
Matias Zaldarriaga\textsuperscript{\ref{Ias}}.

\clearpage

\begin{multicols}{2}

\scriptsize

\parskip=4pt

\noindent
\textsuperscript{\ref{Irvine}}\Irvinetext

\noindent
\textsuperscript{\ref{Ill}}\Illtext

\noindent
\textsuperscript{\ref{Penn}}\Penntext

\noindent
\textsuperscript{\ref{Slac}}\Slactext

\noindent
\textsuperscript{\ref{Princeton}}\Princetontext

\noindent
\textsuperscript{\ref{Hopkins}}\Hopkinstext

\noindent
\textsuperscript{\ref{Stanford}}\Stanfordtext

\noindent
\textsuperscript{\ref{Oxford}}\Oxfordtext

\noindent
\textsuperscript{\ref{Fermi}}\Fermitext

\noindent
\textsuperscript{\ref{Chicago}}\Chicagotext

\noindent
\textsuperscript{\ref{Sandiego}}\Sandiegotext

\noindent
\textsuperscript{\ref{Nist}}\Nisttext

\noindent
\textsuperscript{\ref{Sissa}}\Sissatext

\noindent
\textsuperscript{\ref{Berkeley}}\Berkeleytext

\noindent
\textsuperscript{\ref{Apc}}\Apctext

\noindent
\textsuperscript{\ref{Cambridge}}\Cambridgetext

\noindent
\textsuperscript{\ref{Iap}}\Iaptext

\noindent
\textsuperscript{\ref{Anl}}\Anltext

\noindent
\textsuperscript{\ref{Cincy}}\Cincytext

\noindent
\textsuperscript{\ref{Cita}}\Citatext

\noindent
\textsuperscript{\ref{Lbnl}}\Lbnltext

\noindent
\textsuperscript{\ref{Manchester}}\Manchestertext

\noindent
\textsuperscript{\ref{Perimeter}}\Perimetertext

\noindent
\textsuperscript{\ref{Harvard}}\Harvardtext

\noindent
\textsuperscript{\ref{Sapienza}}\Sapienzatext

\noindent
\textsuperscript{\ref{Dartmouth}}\Dartmouthtext

\noindent
\textsuperscript{\ref{Kwazulu}}\Kwazulutext

\noindent
\textsuperscript{\ref{Nova}}\Novatext

\noindent
\textsuperscript{\ref{Cornell}}\Cornelltext

\noindent
\textsuperscript{\ref{Jpl}}\Jpltext

\noindent
\textsuperscript{\ref{Caltech}}\Caltechtext

\noindent
\textsuperscript{\ref{Arcetri}}\Arcetritext

\noindent
\textsuperscript{\ref{Mcgill}}\Mcgilltext

\noindent
\textsuperscript{\ref{Ilp}}\Ilptext

\noindent
\textsuperscript{\ref{Cca}}\Ccatext

\noindent
\textsuperscript{\ref{Mudd}}\Muddtext

\noindent
\textsuperscript{\ref{Ice}}\Icetext

\noindent
\textsuperscript{\ref{Stockholm}}\Stockholmtext

\noindent
\textsuperscript{\ref{Ias}}\Iastext

\noindent
\textsuperscript{\ref{Haverford}}\Haverfordtext

\noindent
\textsuperscript{\ref{Michigan}}\Michigantext

\noindent
\textsuperscript{\ref{Ubc}}\Ubctext

\noindent
\textsuperscript{\ref{Boulder}}\Bouldertext

\noindent
\textsuperscript{\ref{Minnesota}}\Minnesotatext

\noindent
\textsuperscript{\ref{Lal}}\Laltext

\noindent
\textsuperscript{\ref{Columbia}}\Columbiatext

\noindent
\textsuperscript{\ref{Osu}}\Osutext

\noindent
\textsuperscript{\ref{Toronto}}\Torontotext

\noindent
\textsuperscript{\ref{Fsu}}\Fsutext

\noindent
\textsuperscript{\ref{Davis}}\Davistext

\noindent
\textsuperscript{\ref{Pitt}}\Pitttext

\noindent
\textsuperscript{\ref{Ferrara}}\Ferraratext

\noindent
\textsuperscript{\ref{Sussex}}\Sussextext

\noindent
\textsuperscript{\ref{Stony}}\Stonytext

\noindent
\textsuperscript{\ref{Madison}}\Madisontext

\noindent
\textsuperscript{\ref{Santabarbara}}\Santabarbaratext

\noindent
\textsuperscript{\ref{Lpsc}}\Lpsctext

\noindent
\textsuperscript{\ref{Kings}}\Kingstext

\noindent
\textsuperscript{\ref{Asu}}\Asutext

\noindent
\textsuperscript{\ref{Cea}}\Ceatext

\noindent
\textsuperscript{\ref{Usc}}\Usctext

\noindent
\textsuperscript{\ref{Bohr}}\Bohrtext

\noindent
\textsuperscript{\ref{Orsay}}\Orsaytext

\noindent
\textsuperscript{\ref{Sfu}}\Sfutext

\noindent
\textsuperscript{\ref{Oslo}}\Oslotext

\noindent
\textsuperscript{\ref{Melbourne}}\Melbournetext

\noindent
\textsuperscript{\ref{Case}}\Casetext

\noindent
\textsuperscript{\ref{Pennstate}}\Pennstatetext

\noindent
\textsuperscript{\ref{Brookhaven}}\Brookhaventext

\noindent
\textsuperscript{\ref{Swarthmore}}\Swarthmoretext

\noindent
\textsuperscript{\ref{Hpd}}\Hpdtext

\noindent
\textsuperscript{\ref{Umass}}\Umasstext

\noindent
\textsuperscript{\ref{Iucaa}}\Iucaatext

\noindent
\textsuperscript{\ref{Sao}}\Saotext

\noindent
\textsuperscript{\ref{Mit}}\Mittext

\noindent
\textsuperscript{\ref{Mpa}}\Mpatext

\noindent
\textsuperscript{\ref{Brown}}\Browntext

\noindent
\textsuperscript{\ref{Syracuse}}\Syracusetext

\noindent
\textsuperscript{\ref{Gsfc}}\Gsfctext

\normalsize

\end{multicols}

\parskip=8pt

\end{raggedright}

\clearpage

\tableofcontents


\def\as#1{[{\bf AS:} {\it #1}] }


\eject
\pagenumbering{arabic} 
\setcounter{page}{1}
 
\chapter{Exhortations}
\label{chap:intro}


\bigskip

Fourteen billion years ago, in the first fraction of a second of our Universe's existence, the most extreme high-energy physics experiment took place. The ability to use the cosmic microwave background (CMB) to investigate this fantastic event, at energy scales as much as a trillion times higher than can be obtained at CERN, is at the very core of our quest to understand the fundamental nature of space and time and the physics that drive the evolution of the Universe. 

The CMB allows direct tests of models of the quantum mechanical origin of all we see in the Universe. Subtle correlations in its anisotropy imparted by the interplay of gravitational and quantum physics at high energies contain information on the unification of gravity and quantum physics. Separately, correlations induced on the background at later times encode details about the distribution of all the mass, ordinary and dark, in the Universe, as well as the properties of the neutrinos, including the number of neutrino species and types, and their still unknown masses. 

The purpose of this book is to set the scientific goals to be addressed by the next generation ground-based cosmic microwave background experiment, CMB-S4, consisting of dedicated telescopes at the South Pole, the high Chilean Atacama plateau and possibly a northern hemisphere site, all equipped with new superconducting cameras. CMB-S4 
is envisioned to be the definitive CMB experiment. It will enable a dramatic leap forward in cosmological studies by crossing critical thresholds in testing inflation, in the number and masses of the neutrinos, in finding possible new light relics, in  constraining the nature of dark energy, and in testing general relativity on large scales.

We begin this chapter with a brief history and the current status of CMB measurements and cosmological results.  This is followed by a general overview of how the CMB-S4 science goals, as outlined in the executive summary, lead to general aspects of the instrument design. Based on these considerations, we present a strawman configuration that serves as an initial jumping-off point for exploring instrument configuration parameter space in the following science chapters.  Lastly, this chapter provides a brief overview of the path from the ongoing Stage-3 experiments to realizing CMB-S4. 

\section{Brief History and Current Status of CMB measurements}
\label{sec:background}

Since the discovery of the cosmic microwave background (CMB) 50 years ago, CMB measurements have led to spectacular insights into the fundamental workings of space and time, from the quantum mechanical origin of the Universe at extremely high energies through the growth of structure and the emergence of the dark energy that now dominates the energy density of the Universe. Studies of the CMB connect physics at the smallest scales and highest energies with the largest scales in the Universe, roughly 58 orders of magnitude in length scale. They connect physics at the earliest times to the structure that surrounds us now, over 52 magnitudes in time scale. 

The deep connections of CMB studies and particle physics predate the discovery of the background, going back to the 1940s when Alpher and Gamow were considering a hot, dense, early Universe as a possible site for nucleosynthesis. To produce the
amount of helium observed in the local Universe, they concluded there
had to be about $10^{10}$\ thermal photons for every nucleon. Alpher and Herman subsequently predicted that this background of photons would persist to the present day as a thermal bath at a few degrees Kelvin.

The continuing, remarkably successful, story of CMB studies is one driven by the close interplay of theory and phenomenology with increasingly sensitive and sophisticated experiments. The high degree of isotropy of the CMB across the sky, to nearly a part of one in a hundred thousand, led to the theory of inflation and cold dark matter in the 1980's. It was not until 1992 that instruments aboard the \cobe\ satellite led to the discovery of the anisotropy, and pinned the level of anisotropy for the following higher angular resolution measurements to characterize. In 2006 the \cobe\ measurements of the background anisotropy and its black-body spectrum were recognized with the second Nobel Prize for CMB research; the first was awarded in 1978 to Penzias and Wilson for the discovery of the CMB.  In the decade after the \cobe\ results, measurements with ground and balloon-based instruments revealed the acoustic peaks in the CMB angular power spectrum, which showed that the Universe was geometrically flat in accordance with predictions of inflation and provided strong support for contemporary claims for an accelerating Universe based on observations of type Ia supernovae (SNe), which were recognized with the 2011 Nobel Prize in physics. These early anisotropy measurements also provided an estimate of the universal baryon density and found it to be in excellent agreement with the level estimated at $t \sim 1$ second by big bang nucleosynthesis (BBN) calculations constrained to match the observed elemental abundances. The CMB measurements also clearly showed that dark matter was non-baryonic. The polarization anisotropy was discovered ten years after \cobe\ at the level predicted from temperature anisotropy measurements. The now standard $\Lambda$CDM cosmological model was firmly established.

\begin{figure}[t]
\centering \includegraphics[width=0.9\textwidth]{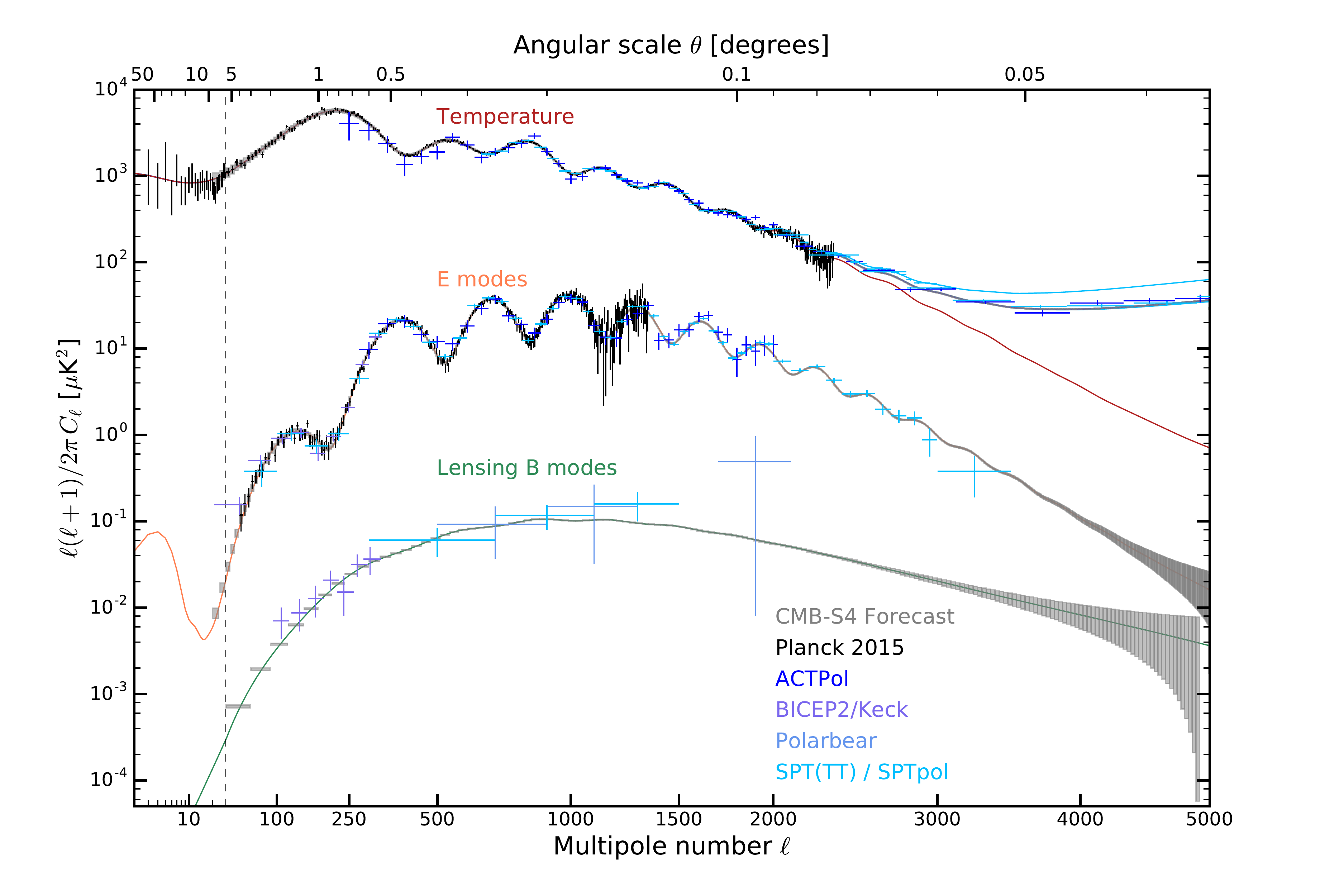}
\caption{Current measurements of the angular power spectrum of the CMB temperature and polarization anisotropy. The horizontal axis is scaled logarithmically in multipole $\ell$ left of the vertical dashed line ($\ell < 30$) and as $\ell^{0.6}$ at higher multipole.  Best-fit models of residual foregrounds plus primary CMB anisotropy power for TT datasets are also plotted. To illustrate the expected improvements with CMB-S4, the projections for a strawman instrumental configuration are shown in grey (binned with $\Delta\ell = 5$ for TT and EE spectra and $\Delta\ell = 30$ for BB) for a $\Lambda$CDM with $r =0$ cosmological model.}
\label{fig:CurrentCMB}
\end{figure}

Two CMB satellites have mapped the entire sky since \cobe, first \wmap\ with moderate angular resolution (as fine as 12 arcminutes), followed by \planck\ with resolution as fine as 5 arcminutes. Higher-resolution maps of smaller regions of the sky have been provided by ground-based experiments, most notably by the 10m South Pole Telescope (SPT) and the 6m Actacama Cosmology Telescope (ACT). The primary CMB temperature anisotropy is now well characterized through the damping tail, i.e., to multipoles $\ell \sim 3000$, and secondary anisotropies have been measured to multipoles up to  $10,000$.  Figure~\ref{fig:CurrentCMB} shows the current state of the temperature and polarization anisotropy measurements and the expected improvements with CMB-S4.

The $\Lambda$CDM model continues to hold up stunningly well, even as the precision of the CMB determined parameters has increased substantially. Inflationary constraints include limits on curvature constrained to be less than 3\% of the energy density,
departures from Gaussianity are bounded at the level of 1 part in $10^4$, 
and 
the predicted small departure from pure scale invariance of the primordial fluctuations is detected at 5-sigma confidence. Also of interest to particle physics, the effective number of light relativistic species (i.e., neutrinos and any yet identified ``dark radiation") is shown to be within 10\% of $\neff = 3.046$, the number predicted by BBN.  The sum of the masses of the neutrinos is found to be less than 0.6 eV. Dark matter is shown to be non-baryonic at $> 40$ sigma. Early dark energy models are highly constrained as are models of decaying dark matter. 

There remains much science to extract from the CMB, including: 1) using CMB B-mode polarization to search for primordial gravitational waves to constrain the energy scale of inflation and to test alternative models, and to provide insights into quantum gravity; 2) obtaining sufficiently accurate and precise determinations of the effective number of light relativistic species (dark radiation) to search for new light relics, and to allow independent and rigorous tests of BBN and our understanding of the evolution of the Universe at $t = 1$\ sec; 3) a detection of the sum of the neutrino masses, even at the minimum mass allowed by oscillation experiments and in the normal hierarchy; 4) using secondary CMB anisotropy measurements to provide precision tests of dark energy through its impact on the growth of structure; and 5) testing general relativity and constraining alternate theories of gravity on large scales.

The best cosmological constraints come from analyzing the combination of primary and secondary CMB anisotropy measurements with other cosmological probes, such as baryon acoustic oscillations (BAO) and redshift space distortions, weak lensing, galaxy and galaxy cluster surveys, Lyman-alpha forest measurements, local determinations of the Hubble constant, observations of type Ia SNe, and others. The CMB primary anisotropy measurements provide highly complementary data for the combined analysis;  by providing a precision measurement of the Universe at $z = 1100$, the CMB data leads to tight predictions for measurements of the late time Universe for any adopted cosmological model---measurements of the Hubble constant, the BAO scale, and the normalization of the present day matter fluctuation spectrum being excellent examples. Secondary CMB measurements provide late-time probes directly from the CMB measurement, e.g., CMB lensing, the SZ effects and SZ cluster catalogs, which will provide critical constraints on the standard cosmological models and extensions to it. The cosmological reach of future cosmological surveys at all wavelengths will be greatly extended by their joint analyses with secondary CMB anisotropy measurements.

\section{CMB-S4 Design Considerations}

The CMB-S4 science goals, as outlined in the executive summary, and detailed in the following chapters, lead to several general aspects of the instrument design. We briefly summarize the general design considerations below.

\subsection{Raw sensitivity considerations and detector count}

\begin{figure}[t]
\centering \includegraphics[width=0.6\textwidth]{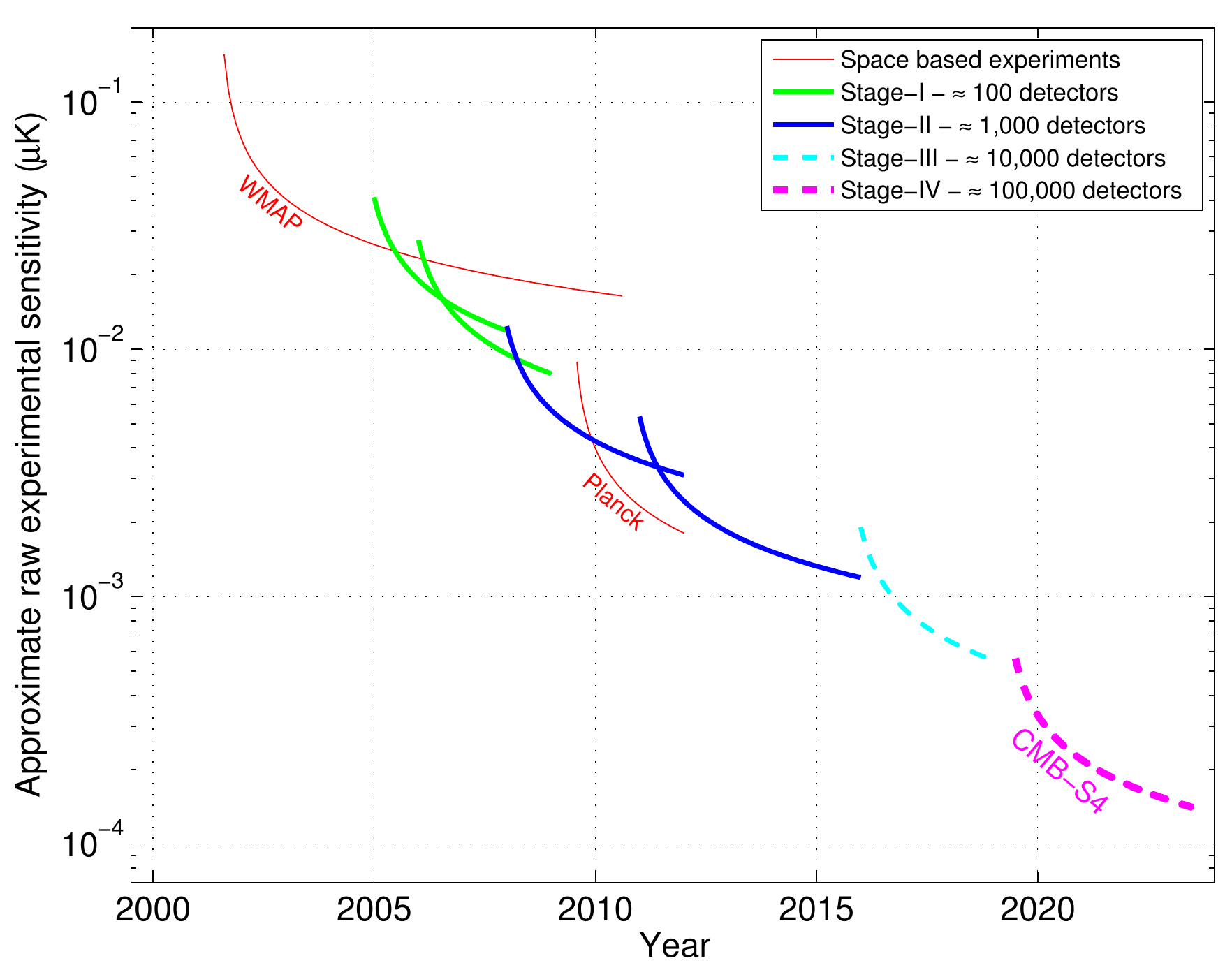}
\caption{Plot illustrating the evolution of the raw sensitivity of CMB
  experiments, which scales as the total number of
  bolometers. Ground-based CMB experiments are classified into Stages
  with Stage II experiments having $O$(1000) detectors, Stage III
  experiments having $O$(10,000) detectors, and a Stage IV experiment
  (such as \cmbexp) having $O$(100,000) detectors. Figure from Snowmass  CF5
  Neutrino planning document.}
\label{fig:expt_progress-intro}
\end{figure}

The sensitivity of CMB measurements has increased enormously since Penzias and Wilson's discovery in 1965, following a Moore's Law like scaling, doubling every roughly 2.3 years. Fig.~\ref{fig:expt_progress-intro} shows the sensitivity of recent experiments, expectations for upcoming Stage-3 experiments, characterized by order 10,000 detectors on the sky, and the projection for a Stage 4 experiment with order 100,000 detectors. To obtain many of the CMB-S4 science goals requires of order $1~\mu$K arcminute sensitivity over roughly half of the sky, which for a four-year survey requires of order 500,000 CMB-sensitive detectors. 

To maintain the Moore's Law-like scaling requires a major leap forward, a phase change in the mode of operation of the ground based CMB program.  Two constraints drive the change:  1) CMB detectors are background-limited, so more pixels are needed on the sky to increase sensitivity; and 2) the pixel count for existing CMB telescopes are nearing saturation.  Even using multichroic pixels and wide field of view optics, these CMB telescopes are expected to field only tens of thousands of polarization detectors, far fewer than needed to meet the CMB-S4 science goals. 

CMB-S4 thus requires multiple telescopes, each with a maximally outfitted focal plane of pixels utilizing superconducting, background limited, CMB detectors. To achieve the large sky coverage and to take advantage of the best atmospheric conditions, the South Pole and the Chilean Atacama sites are baselined, with the possibility of adding a new northern site to increase sky coverage to the entire sky not contaminated by prohibitively strong Galactic emission.

\subsection{Degree angular scale (low $\ell$) sensitivity}

At the largest angular scales (low $\ell$)---the angular scales that must be measured well to pursue inflationary 
B modes---the CMB polarization anisotropy is highly contaminated by foregrounds. Galactic synchrotron dominates at low frequencies and galactic dust at high frequencies, as recently shown by the \planck\ and \planck/BICEP/KECK polarization results. Multi-band polarization measurements are required to distinguish the primordial polarized signals from the foregrounds. 

Adding to the complexity of low multipole CMB observations is the need to reject the considerable atmospheric noise contributions over the large scans needed to extract the low 
$\ell$ polarization. While the spatial and temporal fluctuations of the atmosphere are not expected to be polarized, any mismatches in the polarized beams or detector gains will lead to systematic contamination of the measured polarization by the much stronger unpolarized signal, referred to as T-P leakage. These issues can be mitigated by including additional modulations into the instrument design, such as bore-sight rotation or modulation of the entire optics with a polarization modulation scheme in front of the telescope. Implementing such modulations is easier for small telescopes, although they could in principle be implemented on large telescopes as well. The cost of a small aperture telescope is dominated by the detectors, making it feasible to deploy multiple telescopes each optimized for a single band, or perhaps multiple bands within the relatively narrow atmosphere windows.

It is therefore an attractive option for CMB-S4 to include dedicated small aperture telescopes for pursuing low-$\ell$ polarization. The default plan for CMB-S4 is to target the recombination bump, with E-mode and B-mode polarization down to $\ell \sim 20$. If Stage 3 experiments demonstrate that it is feasible to target the reionization bump at $\ell < 20$ from the ground, those techniques may be incorporated to extend the reach of CMB-S4. More likely, however, this is the $\ell$ range for which CMB-S4 will be designed to be complementary to balloon-based and satellite based measurements. 

\subsection{Subdegree angular scale (high $\ell$) sensitivity}

At the highest angular resolution (high $\ell$)---the angular scales needed for de-lensing the inflationary B modes, constraining \neff\ and $\Sigma m_\nu$,  investigating dark energy and performing gravity tests with secondary CMB  anisotropy---the CMB polarization anisotropy is much less affected by both foregrounds and atmospheric noise. In fact, it should be possible to measure the primary CMB anisotropy in E-mode polarization to multipoles significantly higher than is possible in TT, thereby extending the lever arm to measure the spectral index and running of the primordial scalar (density) fluctuations. CMB lensing benefits from $\ell_{max}$ of order 5000 and secondary CMB measurements are greatly improved with $\ell_{max}$ of order 10,000 and higher, requiring large-aperture telescopes with diameters of several meters. Owing to the steep scaling of telescope cost with aperture diameter, it is likely not cost-effective to consider separate large aperture telescopes each optimized for a single frequency band. 

CMB-S4 is therefore envisioned to include dedicated large-aperture, wide-field-of-view telescopes equipped with multiple band detector arrays.

\section{A strawman instrument configuration}
\label{sec:strawman}

The rough conceptual design outlined above clearly needs to be refined.  The first priorities are to determine the specific measurements needed to meet the requirements for each of the science goals---the purpose of this Science Book---and then to translate them into instrumentation design specifications. We need to determine:  the required resolution and sensitivity; the number of bands to mitigate foreground contamination, which is likely to be function of angular scale; the required sky coverage; the beam specifications; the scanning strategy and instrument stability; etc. 

Determining these specifications requires simulations, informed by the best available data and phenomenological models.  Only when we have these specifications in hand can we design the instrument and answer such basic questions as the number and sizes of the telescopes.  This will be, of course, an iterative process, involving detailed simulations and cost considerations. At this time the Science Book is a working document with this first edition focused primarily on defining the possible reach in the  key science areas, along with the simulations needed to refine the science case and set the specifications of the needed measurements. This will set the stage for defining the instrument.

On the other hand, we need a jumping-off point for exploring instrument configuration parameter space.  
Simple, back-of-the-envelope calculations make it clear that achieving the science goals outlined above requires a raw sensitivity equivalent to roughly 500,000 detectors operating for four years, though we may find that certain science goals push us to yet greater detector count. This order-of-magnitude level of sensitivity is appropriate for both measuring the tensor-to-scalar ratio $r$ and to the ``non-$r$'' science goals, but the other specifications for the instrument and survey (resolution, sky coverage, band placement) potentially pull in different directions for these two sets of goals. For this reason, we choose as a baseline for parameter forecasts two separate instrument configurations, one which we will optimize for $r$ constraints and one for non-$r$ science, with the detector effort split evenly between the two configurations. If the optimization exercise tells us that the two configurations are similar enough, then the two surveys can be re-merged. 

For the ``$r$'' survey, the strawman configuration consists of an array of small-aperture ($\sim 1$m) telescopes and a separate large-aperture telescope to measure and remove the lensing contamination on the patch of sky targeted by the small-aperture array. The $10^6$ detector years (250,000 detectors operating for four years) is split between the small- and large-aperture efforts in a way that optimizes the combination of noise and lensing residuals. The known foregrounds at 100 to 150~GHz, synchrotron and thermal dust, require the small-aperture effort to be split into at least three bands, but to guard against potential foreground complexity any realistic configuration would have many more. There are four accessible atmospheric windows in the frequency range at which the CMB peaks, centered at roughly 35, 90, 150, and 250~GHz. In the strawman configuration considered here, each of these windows is split into two bands. The total detector effort for the small-aperture telescopes is split between the eight bands to optimize the combination of noise and foreground residuals. The parameter space that can then be explored to discover what is necessary to reach the target sensitivity to $r$ include fraction of sky covered, band placement, and total detector count.

For the ``non-$r$'' survey, the strawman configuration consists of an array of medium-to-large-aperture telescopes,  with the full $10^6$ detector years (250,000 detectors operating for four years) dedicated to a small number of frequency bands near the peak of the CMB. The key instrumental parameters to investigate for the neutrino, light-relic, and dark energy science goals are angular resolution, sky coverage, and total detector effort. 

\section{The Road from Stage 3 to Stage 4}
\label{sec:context}

\begin{figure}[t]
\centering \includegraphics[width=0.8\textwidth]{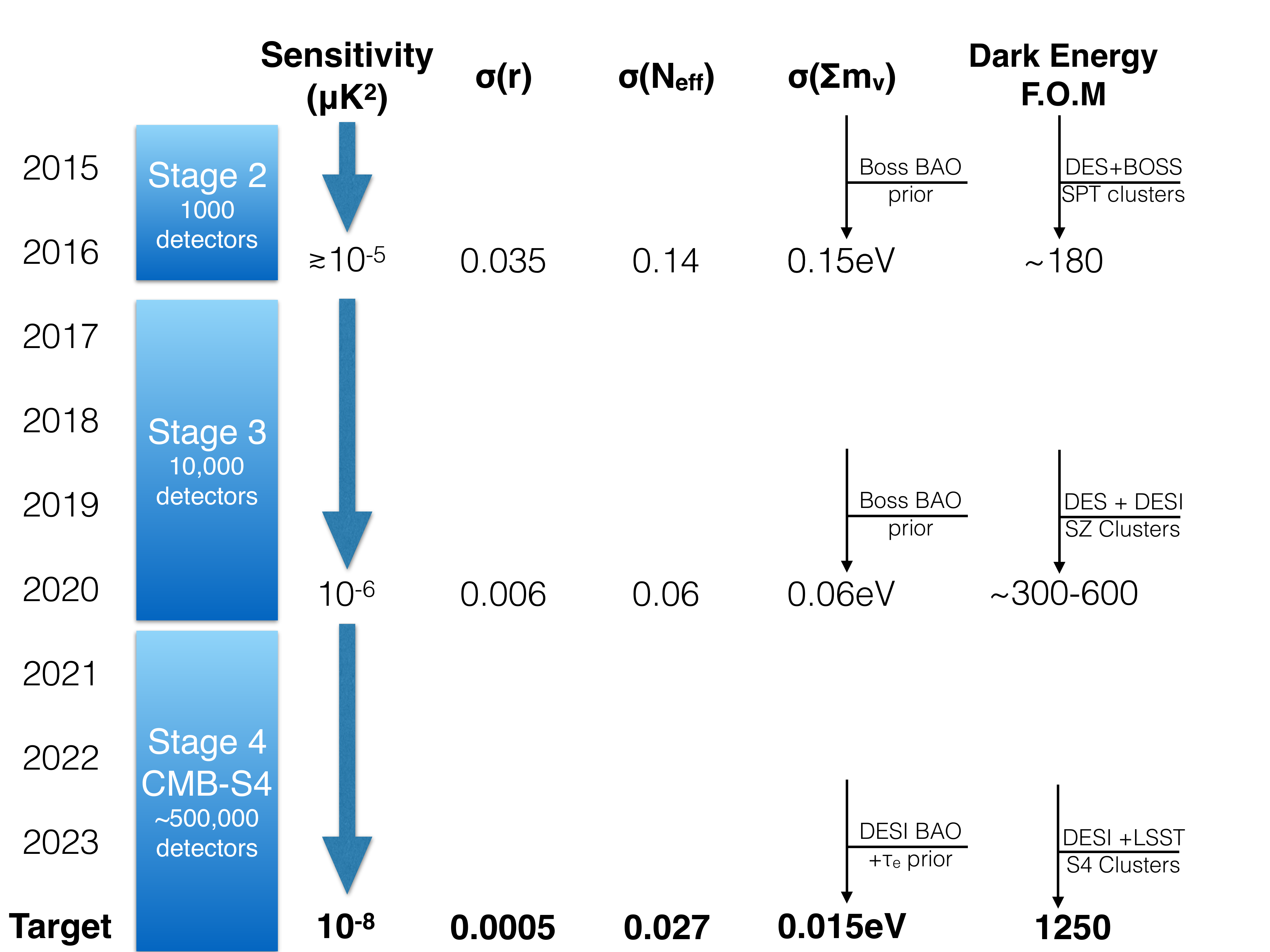}
\vskip 10pt \caption{Schematic timeline showing the expected increase in sensitivity ($\mu$K$^2$) and the corresponding improvement for a few of the key cosmological parameters for Stage-3, along with the threshold-crossing aspirational goals targeted for CMB-S4.}
\label{fig:science_timeline-intro}
\end{figure}

The Stage-2 and Stage-3 experiments are logical technical and scientific stepping stones to CMB-S4. Ongoing R\&D directed toward achieving the scaling up required to CMB-S4 is being pursued at several universities and national labs. 
Figure~\ref{fig:science_timeline-intro} shows the timeline of the expected increase in sensitivity and the corresponding improvement for a few of the key cosmological parameters for Stage-3, along with the threshold-crossing aspirational goals  targeted for CMB-S4.

\begin{figure}[ht]
\centering \includegraphics[trim=1in 0in 1.2in 0in, clip, width=0.8\textwidth,]{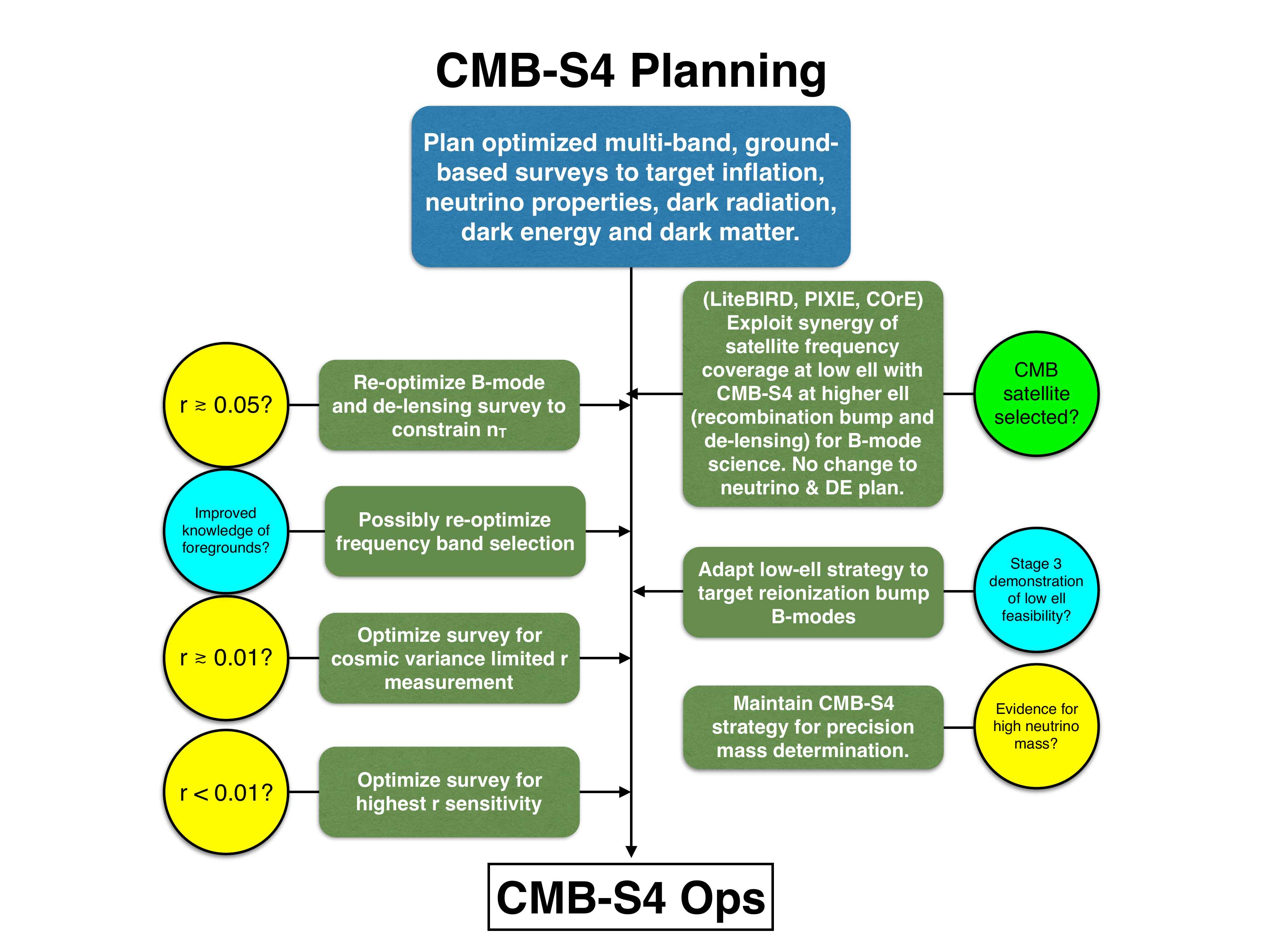}
\caption{Schematic chart showing how scientific findings (yellow circles), technical advances (blue circles) and satellite decisions by various agencies (green circles) would affect the science goals, the survey strategy and possibly the design of CMB-S4 (green boxes)}
\label{fig:flowchart}
\end{figure}

Finally, in Fig.~\ref{fig:flowchart} we show how the scientific findings (yellow circles), the technical advances (blue circles) and satellite selections (green circles) would affect the science goals, survey strategy and possibly the design of CMB-S4.

\chapter{Inflation}
\label{sec:inflation}


\bigskip

\begin{quotation}

\end{quotation}
\section{Introduction}

Proposed at first to solve certain conceptual puzzles of the hot big bang
model, inflation was soon recognized to be a highly compelling mechanism for
the generation of primordial perturbations
(e.g.\ \cite{Abbott:1986kb,Linde:2005ht,Lemoine:2008zz,EllisWands}).
The simplest models of inflation predict that our Hubble patch should have
almost zero mean curvature, and highly Gaussian, adiabatic primordial density perturbations with a nearly (but not quite)
scale-invariant power spectrum.
These generic features correspond closely with cosmological observations---to 
date, we have fairly tight upper limits on mean curvature, primordial
non-Gaussianity, and the amplitudes of any non-adiabatic (isocurvature)
contributions to the perturbations. We also have a detection of a small
departure from scale invariance, consistent with the expectations of simple
inflationary models.

These successes, and the difficulties in the construction of alternative
scenarios that are also consistent with the data, have led most cosmologists to
see inflation as our best bet for the creation of the primordial perturbations,
while acknowledging that many questions remain. If inflation did occur, how did
it occur? Was there a single effective field dominating the dynamics of both
the background expansion and the perturbations, or were multiple fields
involved? How did it begin? Are ground-state fluctuations truly the source of density
perturbations? What is the connection of inflation to the rest of physics?
Are there observations that could falsify inflation?

CMB-S4 will provide answers to some of these questions by opening a new window on the study of inflation, and on the generation of primordial perturbations in general. Thus far, observed anisotropies can all be interpreted as resulting from density perturbations.  With CMB-S4 we have an opportunity to investigate a spectacular prediction of the inflationary paradigm: primordial gravitational waves. The sensitivity of CMB measurements to gravitational waves or tensor perturbations arises from the generation of polarization in the CMB: scalar perturbations produce only curl-free E-mode polarization to first order, while tensor perturbations produce divergence-free B-mode polarization as well. Thus a measurement of B-mode polarization in the CMB (with the standard caveats relating to foregrounds and gravitational lensing) is a direct measurement of the amplitude of tensor perturbations.

Of course we do not know that inflation is correct---and the value of this new window is much more general than the inflationary paradigm itself.  Nevertheless, we can use the inflationary picture as a concrete framework for exploring the potential impact of these measurements. The tensor sector offers a more direct probe of the dynamics of the inflationary expansion because tensor perturbations are an inevitable consequence of the degrees of freedom of the spacetime metric obeying the uncertainty principle. In other words, the existence of an inflationary epoch in the Universe's past directly implies the existence of a background of tensor perturbations. Furthermore, the amplitude of the tensor perturbations depends only on the rate of expansion during inflation. In contrast, the amplitude of the scalar perturbations depends on both the amplitude and slope of the effective potential of the field responsible for inflation, and more generally on the sound speed of the inflaton field as well.

In addition to probing the origin of all structure in the Universe, opening the tensor sector would give us access to physics at energy scales more than $10^{11}$ times higher than those probed at the LHC. A detection with CMB S4 would reveal the energy scale of inflation to be near $10^{16}$ GeV. If the tensor perturbations are detectable, we are already probing physics at these energies via the scalar perturbations, but we cannot know this until the tensor perturbations are actually detected. 

We currently only have upper limits on the amplitude of tensor perturbations---limits that are only marginally stronger than those that can be inferred from measuring temperature anisotropies \cite{Ade:2015lrj,Array:2015xqh}. To detect tensor perturbations we need to dramatically improve measurements of CMB polarization. In the tensor sector, CMB-S4 will improve current constraints by over an order of magnitude. This is especially interesting because it allows this next-generation instrument to reach theoretically well-motivated thresholds for the tensor-to-scalar ratio $r$ (the ratio of power in tensor modes to power in scalar modes), which consequently serves as the primary inflationary science driver for the design. 

It is worth pointing out explicitly that these tensor perturbations are by definition gravitational waves. With the recent LIGO detections \cite{Abbott:2016blz,Abbott:2016nmj}, we have entered the era of gravitational wave astronomy, and with CMB-S4 we will be targeting the ultra long-wavelength end of nature's gravitational wave spectrum. We expect the background of inflation-produced gravitational waves to be nearly scale-invariant, and that waves with frequencies $\sim 10^{15}$ times higher than those detectable with CMB-S4 may one day be detectable with a space-based observatory \cite{Caligiuri:2014sla}, greatly enhancing the value of any CMB-S4 detection. 

Inflation predicts B-mode fluctuations sourced by primordial gravitational waves. But more generally, the B-mode signal carries information about both the spectrum of primordial perturbations in the tensor (and vector) components of the metric and any physics that affected the evolution of those modes once they re-entered the horizon. Furthermore, the inflationary sector is not isolated from the rest of particle physics at high energies. In the context of specific proposals for physics beyond the standard model, including dark matter models, a detection of B modes can have consequences for predictions for the post-inflation spectrum of particles and their thermodynamics. These models may also provide observables other than the amplitude of B modes that constrain inflationary physics. The rich interplay of inflation models and other physics beyond the standard model is discussed in detail later in this chapter and in Chapters 4 and 5.

{\it A detection of primordial gravitational waves would open a completely new window on the physical processes of the early Universe and reveal a new length scale of particle physics, far below those accessible with terrestrial particle colliders. }

If the overall amplitude of the B-mode signal is large enough to be detected at high significance by the CMB-S4 instrument, we will be able characterize the statistics of the tensor perturbations. Investigating the scale-dependence of the amplitude of tensor fluctuations and their Gaussianity will allow us to determine if the signal is consistent with the amplification of quantum vacuum fluctuations of the metric during inflation. If CMB-S4 measurements are consistent with a nearly scale-invariant and a weakly non-Gaussian spectrum, a detection would:
\begin{itemize}
 \item identify the energy scale of inflation;
 \item provide strong evidence that gravity is quantized, at least at the linear level;
 \item provide strong evidence that the complete theory of quantum gravity must accommodate a Planckian field range for the inflaton.
\end{itemize}

Departures from a nearly scale-invariant, Gaussian spectrum would reveal new physics beyond the simplest inflationary models. There are currently a few models with significantly different signatures generated by a richer inflationary or post-inflationary sector, and these predictions would be tested. However, given the lack of observational constraints on physics at such high energy scales there is also enormous discovery potential. Polarization data will provide new consistency checks on the current dominant theoretical framework, including model-independent constraints on the graviton mass and constraints on alternatives to inflation.

Even in the absence of a detection CMB-S4 would provide invaluable information about the physics of the early Universe. Many of the most appealing inflation models have simple monomial potentials. Current observations already put considerable pressure on those (e.g. the $m^2\phi^2$ potential), but do not probe the entire interesting parameter space. CMB-S4 should be designed to comprehensively rule out or detect the remaining monomial models. However, to be more than an incremental improvement it should also push for the first time into a completely new regime, reaching well into the only other remaining viable class of models that can naturally explain the observed value of the spectral index. Purely in terms of dynamics or scale, there is not a sharp threshold on what may ultimately be possible or interesting within this class. For example, models with a field range or characteristic mass scale of $0.1 M_p$ extend down to a tensor-to-scalar ratio below $r=10^{-4}$. However, Lagrangians of special interest in this class (eg, Starobinsky and Higgs inflation) predict a significantly larger amplitude of primordial gravitational waves, $r\sim0.003$. Thus, an upper limit of $r<0.002$ at $95\%$ CL would rule them out. The remaining allowed space would be restricted enough to dramatically change how we think about inflation, and perhaps force us to rethink the paradigm altogether. Certainly some would consider all natural models to be ruled out. This is where we believe CMB-S4 should aim. Section~\ref{sec:upperLimits} discusses the details of this argument, and the spectral index and $r$ values for various models are illustrated in Figure~\ref{fig:nsrp01}.

In the next section, Section~\ref{sec:basics}, we provide a basic introduction to the inflationary paradigm in its simplest form. Section~\ref{sec:needs} discusses the forecasted sensitivity of CMB-S4 designed to reach the theoretically motivated goal of  tensor-to-scalar ratio $r<0.002$ at $95\%$ CL and some of the experimental, observational, and analysis challenges involved in achieving this sensitivity. In Section~\ref{sec:detection} we review in detail what a detection of primordial gravitational waves would mean and what follow-up measurements could be done to further characterize any signal. Section~\ref{sec:upperLimits} explains in detail the implications of a robust upper limit of $r<0.002$ at $95\%$ CL. Sections \ref{sec:scalar} through \ref{sec:PMF} describe the significant gains CMB-S4 will allow in constraining other aspects of the primordial Universe, both standard and more speculative. These include characterizing the scalar power spectrum; constraining spatial curvature, scalar non-Gaussianity, and isocurvature modes; further probing CMB ``anomalies''; and constraining cosmic strings. We summarize in Section~\ref{sec:summary}.

\section{Basics of cosmological inflation}
\label{sec:basics}

In this section, we introduce the essential concepts of cosmological inflation. We do this in two stages, first giving a broad outline of the important concepts (closely following \cite{Carlstrom:2015cck}) and then proceeding to a more technical definition.  Rather than surveying the extensive historical literature, we refer the reader to several review articles and books for further information
\cite{LythRiotto,LiddleLyth,Mukhanov:2005sc,Baumann:2009ds,Linde:2005dd,EllisWands}.

\subsection{Inflation basics I: A heuristic picture}

\begin{figure}[ht]
\begin{center}
\,\raisebox{4.5mm}{\includegraphics[width=0.2688\textwidth]{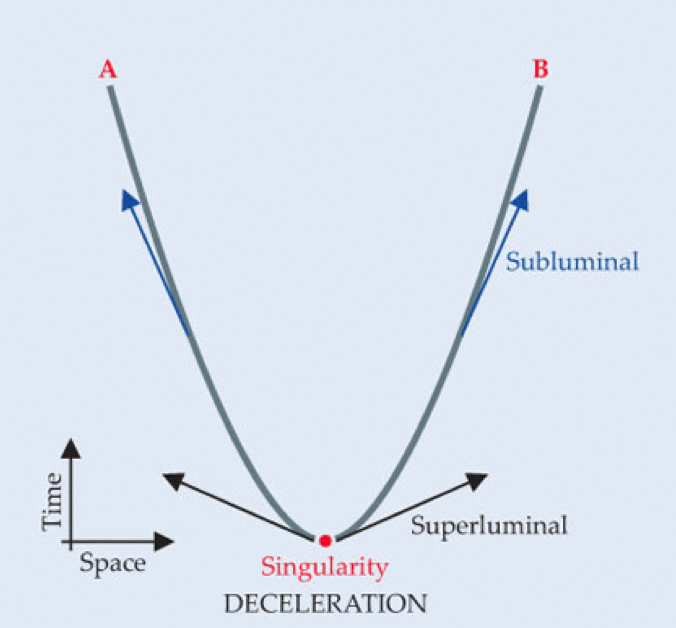}
\includegraphics[width=0.3098\textwidth]{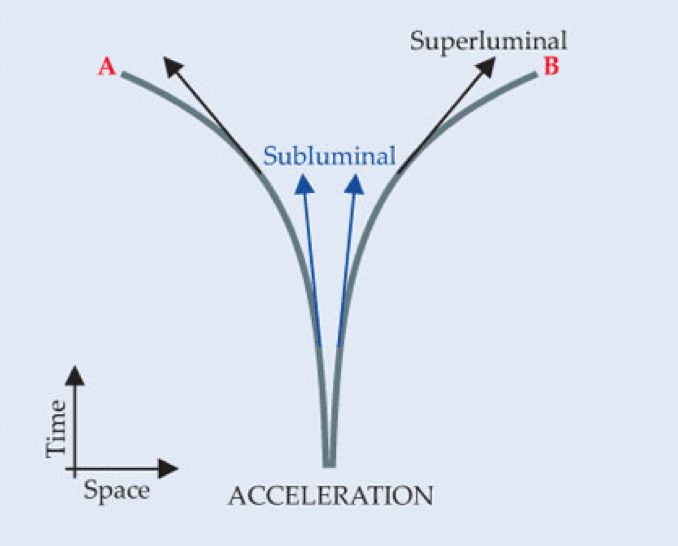}}
\includegraphics[width=0.349\textwidth]{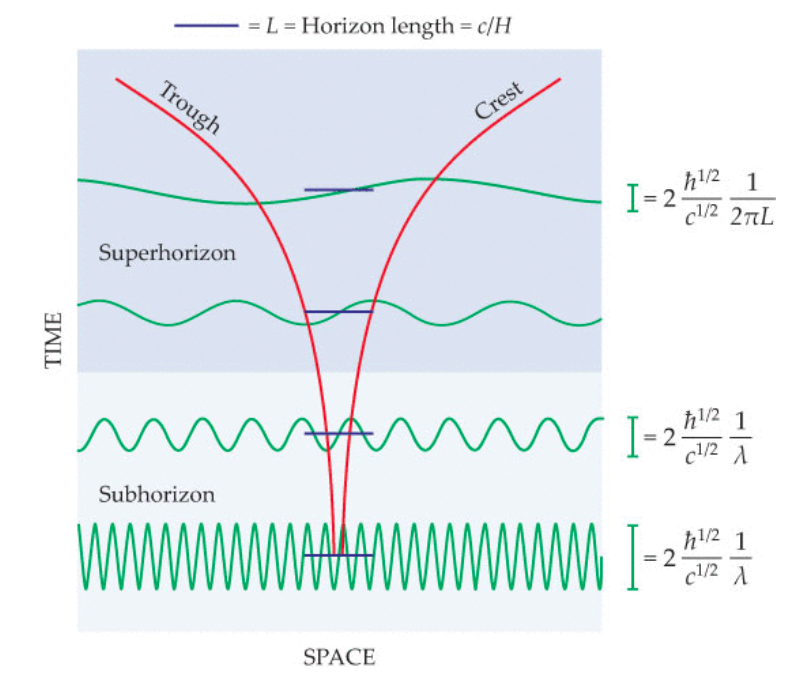}
\end{center}
\caption{{\bf Left and center panels}: In an expanding universe, the distance between two separated points increases over time, simply due to the expansion of the space between them. The two panels here show the spacetime trajectories of two comoving points, A and B. For the decelerating expansion illustrated in the left panel, the separation rate is greater in the past and even exceeds the speed of light at sufficiently early times. Thus A and B go from being out of causal contact---unable to influence each other---to being in causal contact. In an accelerating universe, the separation rate is smaller in the past. The two points go from being in causal contact to being out of causal contact. In the inflationary universe scenario, an early epoch of acceleration---the inflationary era---smoothly maps onto a long period of deceleration. Thus two points can go from being in causal contact to out of causal contact and, much later, back into causal contact. \newline
\mbox{\bf Right panel}: Fluctuations in the value of the inflaton field, which is responsible for the accelerating expansion of the cosmos, evolve differently, depending on whether their wavelength $\lambda$ is less than or greater than the horizon length $L = c/H$. When $\lambda \ll L$, the uncertainty principle limits how smooth the field can be. As a result, the amplitude of the fluctuation is inversely proportional to $\lambda$ and thus decreases as the Universe expands (and the influence of the uncertainty principle is reflected by the appearance of Planck's constant $\hbar$ in the expression for the amplitude). As $\lambda$ becomes larger than the horizon, the crests and troughs of the wave cease to be in causal contact, so the amplitude stops evolving. For superhorizon evolution, the asymptotic value of the amplitude corresponds to replacing the wavelength in the subhorizon case with $2\pi L$. Eventually, cosmic expansion stretches the fluctuations to astrophysically large length scales. Elsewhere in this document $\hbar$ and $c$ are set equal to unity. 
}
\label{fig:PTfigs}
\end{figure}

Inflation is, by definition, a period of accelerating expansion. As illustrated in Fig.~\ref{fig:PTfigs}, an accelerating universe has a causal structure very different from that of a decelerating universe. In a decelerating universe, a pair of separated comoving particles evolves from being causally disconnected---in which case the particles, unable to influence each other, are said to be ``superhorizon''---to being causally connected, or ``subhorizon.'' In an accelerating universe, the opposite occurs. In the inflationary scenario, the Universe undergoes an accelerating stage, which is followed by a long period of deceleration.

In view of the early period of accelerating expansion, two separated regions in the Universe that are now causally disconnected could have been able to interact with each other during the inflationary epoch (which of course was one of the original motivations for inflation). Causally connected perturbations in those two regions---for example, an underdensity in one and an overdensity in the other---could thus have been created at very early times. Quantum mechanics provides a mechanism for generating such perturbations, and in fact makes them unavoidable. Quantum mechanical fluctuations initially created with sub-nuclear wavelengths are stretched by the cosmic expansion to millimeter length scales within a tiny fraction of a second; at present these fluctuations are on astrophysical scales. Thus observations of cosmic structure give us an opportunity to probe physics on extremely small length scales.

Accelerating expansion requires the Universe to have an energy density that dilutes relatively slowly with expansion. In inflationary models, such an energy density is usually obtained via the introduction of a new field $\phi$, called the inflaton field with Lagrangian density, in the simplest cases, given by
\begin{equation}
{\cal L} = - \frac{1}{2} \partial_\mu \phi \partial^\mu \phi - V(\phi)
\end{equation}
where $V(\phi)$ is a potential energy density. 

A generic inflaton field configuration will not lead to inflation. But if there is a large enough patch of space in which $\phi$ takes values for which the potential is sufficiently flat, $\phi$ will rapidly evolve to satisfy the ``slow-roll condition,'' namely $\frac{1}{2} \left(d\phi/dt\right)^2 \ll V(\phi)$. When both the spatial and temporal derivatives of the inflaton field are small, $V(\phi)$ is nearly constant in time and makes the dominant contribution to the energy density, $\rho$. Under such conditions, and given the Friedmann equation $\dot a/a \propto \sqrt{\rho}$, the patch inflates. In the limit that the energy density is completely constant in time, the scale factor grows as $e^{H t}$, and points separated by more than $c/H$ are causally disconnected.

A standard assumption in the calculation of inflationary perturbation spectra is that the field is as smooth as it possibly can be, and still be consistent with the uncertainty principle. As Fig.~\ref{fig:PTfigs} shows, these fluctuations will be stretched to astrophysically large length scales by cosmic expansion. In an inflationary scenario quantum fluctuations provide the initial seeds of all structure in the Universe. 

As $\phi$ rolls toward the potential minimum, $V(\phi)$ eventually becomes smaller than $\frac{1}{2}(d\phi/dt)^2$. At this point the slow-roll condition is no longer met, and inflation ends. Decays of the inflaton to other particles---irrelevant during inflation because the decay products were quickly diluted by expansion---then become important. The remaining energy in the $\phi$ field converts to a thermal bath of the particles of the standard model, and perhaps other particles as well---beginning what is usually called the ``hot big bang'' model.

The small but nonzero spatial fluctuations in $\phi$ cause inflation to end at different times in different locations. In a sense, we can see these spatial variations in the end of the inflationary epoch when we observe anisotropies in the CMB.  In those regions where inflation ends relatively early, the mass density is lower due to the extra expansion that the region has undergone since the end of inflation. Thus the slightly different expansion histories of different locations result in density differences; those small density perturbations eventually grow under the influence of gravity to create all the structures we observe in the Universe today.

The spacetime metric itself presumably also obeys the uncertainty principle. As a result, we expect gravitational waves to be produced during inflation as well. Just as with fluctuations of the inflaton field, they have their amplitude set to a value proportional to the Hubble parameter $H$ during inflation. Detecting the influence of that gravitational-wave background on the CMB would allow cosmologists to infer $H$ and hence the energy scale of the inflationary potential. Observations of density perturbations, by contrast, provide a relatively indirect look at the inflationary era. As emphasized already, CMB-S4 is poised to detect, or place interesting upper limits on, the amplitude and spectrum of inflation-produced gravitational waves via their signature in B-mode polarization.

\subsection{Inflation basics II: Quantifying the predictions}

The overall evolution of the Universe is well modeled by a Friedmann-Lema\^{\i}tre-Robertson-Walker line element
\begin{equation}
ds^2=-dt^2+a^2(t)\left[\frac{dr^2}{1-kr^2}+r^2d\Omega^2\right]\,,
\end{equation}
where $k=0$ for a flat spatial geometry, $k=\pm1$ allows for spatial curvature, and the time evolution is specified by the scale factor, $a(t)$. The Hubble parameter, $H=\dot{a}/a$, gives the rate of expansion of the Universe. A period of inflation will drive the spatial curvature close to zero, in good agreement with current observations. We will assume spatial flatness and set $k=0$ for most considerations, but see Section~\ref{sec:curvature} for a discussion of CMB-S4 constraints on curvature. 

Since the period of cosmic inflation has to end, there must exist a clock, or scalar degree of freedom. According to the uncertainty principle this clock must fluctuate, generating density perturbations that are adiabatic. In the most economic scenarios, these density perturbations are the seeds that grow into the anisotropies observed in the CMB and the stars and galaxies around us. Other degrees of freedom could, of course, also be present during this phase and might even be responsible for the generation of the density perturbations we observe. In this section, for simplicity, we restrict the analysis to the case of fluctuations of the clock field only, propagating with sound speed $c_s=1$.

For these early times, the ADM formalism \cite{Arnowitt:1962hi}
provides a convenient parametrization of the line element in the presence of perturbations
\begin{eqnarray}
\label{eq:metric}
ds^2&=&-N^2dt^2 +h_{ij}(dx^i+N^idt)(dx^j+N^jdt)\,\nonumber\\
h_{ij}&=&a^2(t)[e^{2\zeta}\delta_{ij}+\gamma_{ij}]\,.
\end{eqnarray}

The equations of motion for $N$ (the lapse) and $N^i$ (the shift) are the Hamiltonian and momentum constraints, while $\zeta$ 
and $\gamma_{ij}$ contain the dynamical scalar and tensor degrees of freedom. In scenarios with matter sources other than a scalar field there may also be vector perturbations. These rapidly decay and can be neglected unless they are actively sourced in the post-inflationary Universe, e.g. by cosmic strings.

Because the equations of motion are invariant under translations, and the perturbations small enough to work in perturbation theory, it is convenient to work with the Fourier transforms
\begin{equation}
\zeta(t,\mathbf{x})=\int \frac{d^3 k}{(2\pi)^3}\zeta(t,\mathbf{k})e^{i \mathbf{k}\cdot\mathbf{x}}+{\rm h.c.}\qquad{\rm and}\qquad\gamma_{ij}(t,\mathbf{x})=\sum\limits_\sigma\int\frac{d^3k}{(2\pi)^3}\gamma_\sigma(t,\mathbf{k})e_{ij}(\mathbf{k},\sigma)e^{i \mathbf{k}\cdot\mathbf{x}}+{\rm h.c.}\,,
\end{equation}
where $e_{ij}(\mathbf{k},\sigma)$ is the transverse-traceless polarization tensor for the graviton, $\sigma$ labels the polarization states of the gravitational waves,
and `h.c.' stands for the Hermitian conjugate. The solutions oscillate when the modes are deep inside the horizon, $k\gg aH$. By definition, the modes exit the horizon when $k=aH$ and in single-field models approach a constant outside the horizon when $k\ll aH$.

The statistical properties of the scalar and tensor fluctuations, $\zeta$ and $\gamma_\sigma$, at times sufficiently late so that they have frozen out, provide the link between the primordial era and the observed CMB today as well as other probes of the structure of the late Universe. For a universe that is statistically homogeneous and isotropic and in which the primordial fluctuations are Gaussian, the information about the statistical properties is contained in the 2-point correlation functions
\begin{eqnarray}
\langle\zeta(\mathbf{k})\zeta(\mathbf{k}^{\prime})\rangle&=&(2\pi)^3\delta^3(\mathbf{k}+\mathbf{k}^{\prime})\frac{2\pi^2}{k^3}\mathcal{P}_{\zeta}(k),\nonumber\\
\langle\gamma_\sigma(\mathbf{k})\gamma_{\sigma^{\prime}}(\mathbf{k}^{\prime})\rangle&=&(2\pi)^3\delta_{\sigma\sigma^{\prime}}\delta^3(\mathbf{k}+\mathbf{k}^{\prime})\frac{2\pi^2}{k^3}\frac{1}{2}\mathcal{P}_{\rm t}(k),\nonumber\\
\end{eqnarray}
where the factor of $1/2$ in the second to last line accounts for the fact that the measured power includes contributions from each of the two graviton polarizations. In single-field slow-roll inflation, the gauge-invariant combination of metric and scalar field fluctuations that is conserved outside the horizon has the power spectrum
\begin{equation}
\label{eq:inf_Pzeta}
\mathcal{P}_{\zeta}(k)=\frac{1}{2\epsilon M_{\rm P}^2}\left.\left(\frac{H}{2\pi}\right)^2\right|_{k=aH},
\end{equation}
where $\epsilon=-\dot{H}/H^2$ is the first slow-roll parameter, and $M_{\rm P}=1/\sqrt{8\pi G}$ is the reduced Planck mass. As indicated, the Hubble parameter and $\epsilon$ are to be evaluated at horizon exit, when the wavenumber $k$ is equal to the inverse comoving Hubble radius. In the absence of additional sources, the tensor power spectrum generated by inflation is
\begin{equation}
\label{eq:inf_Pt}
\mathcal{P}_{\rm t}(k)=\frac{8}{M_{\rm P}^2}\left.\left(\frac{H}{2\pi}\right)^2\right|_{k=aH}.
\end{equation}

It is convenient to introduce the logarithmic derivatives of these power spectra 
\begin{equation}\label{eq:specind}
n_{\rm s}(k)-1\equiv\frac{d\ln \mathcal{P}_{\zeta}}{d\ln k}\qquad{\rm and}\qquad n_{\rm t}(k)\equiv \frac{d\ln \mathcal{P}_{\rm t}}{d\ln k}\,.
\end{equation}
If the Hubble rate and slow-roll parameter only weakly depend on time as in slow-roll inflation, these will be $n_{\rm s}(k)\simeq 1$ and $n_{\rm t}(k)\simeq 0$, nearly independent of scale, and can be expanded around a pivot scale $k_\ast$ accessible by the CMB
\begin{equation}
n_{\rm s}(k)-1=n_{\rm s}-1+\left.\frac{dn_{\rm s}(k)}{d\ln k}\right|_{k_\ast}\ln(k/k_\ast)+\dots \qquad{\rm and}\qquad n_{\rm t}(k)=n_{\rm t}+\left.\frac{dn_{\rm t}(k)}{d\ln k}\right|_{k_\ast}\ln(k/k_\ast)+\dots \,.
\end{equation}
In this approximation, the power spectra are
\begin{eqnarray}\label{eq:power_spectra_power_law}
\mathcal{P}_{\zeta}(k)&=& A_{\rm s}\left(\frac{k}{k_\ast}\right)^{n_{\rm s}-1+\frac{1}{2}\left.\frac{dn_{\rm s}}{d\ln k}\right|_{k=k_\ast}\ln(k/k_\ast)+\dots}\,,\nonumber\\
\mathcal{P}_{\rm t}(k)&=& A_{\rm t} \left(\frac{k}{k_\ast}\right)^{n_{\rm t}+\frac{1}{2}\left.\frac{dn_{\rm t}}{d\ln k}\right|_{k=k_\ast}\ln(k/k_\ast)+\dots}\,,
\end{eqnarray}
where $A_{\rm s}$, $A_{\rm t}$ are the scalar and tensor amplitudes, and $n_{\rm s}$ and $n_{\rm t}$, are the scalar and tensor spectral indices, respectively, both at the pivot scale. 
The tensor-to-scalar ratio, $r$, is the relative power in the two types of fluctuations at a chosen pivot scale $k_\ast$ (e.g. values of 0.002 and
0.05\,Mpc$^{-1}$ have been used in previous studies)
\begin{equation}
r=\frac{A_{\rm t}}{A_{\rm s}}\,.
\end{equation}

The power spectra of $\zeta$ and $\gamma_\sigma$ are time-independent as long as the modes are outside the horizon, and only begin to evolve once the modes of interest re-enter the horizon at late times. In particular, they set the initial conditions for the system of equations governing the time evolution of the Universe from a temperature of around $10^9$ K (when electrons and positrons have just annihilated) to the present. To exhibit the link between the primordial perturbations and late time observables explicitly, note that in a spatially flat universe, the contributions of primordial scalar perturbations to the angular power spectra of temperature or E-mode anisotropies are given by
\begin{equation}
C^{({\rm s,t})}_{XX,\ell}=\int \frac{dk}{k}\mathcal{P}_\zeta(k)\left|\int\limits_0^{\tau_0} d\tau S_X^{({\rm s,t})}(k,\tau)j_\ell(k(\tau_0-\tau))\right|^2\,,
\label{eq:clscalar}
\end{equation}
where $j_\ell$ is a spherical Bessel function that encodes the (spatially flat) geometry of the Universe and $S_X^{({\rm s,t})}(k,\tau)$ with $X=T,E$ are source functions for scalar and tensor modes that encode the evolution of the modes in the hot big bang Universe.
At linear order, scalar perturbations only contribute to the angular power spectra of temperature and E-mode polarization and the cross-spectrum of temperature and E-mode polarization, while the tensor perturbations in addition generate B-mode polarization. The primordial contribution of the tensor perturbations to the angular power spectrum of B modes is 
\begin{equation}
C^{({\rm t})}_{BB,\ell}=\int \frac{dk}{k}\mathcal{P}_{\rm t}(k)\left|\int\limits_0^{\tau_0} d\tau S_B^{({\rm t})}(k,\tau)j_\ell(k(\tau_0-\tau))\right|^2\,.
\label{eq:clbb}
\end{equation}
where $S_B^{({\rm t})}(k,\tau)$ is the appropriate source function. 

\begin{figure}[h]
\begin{center}
\includegraphics[width=4in]{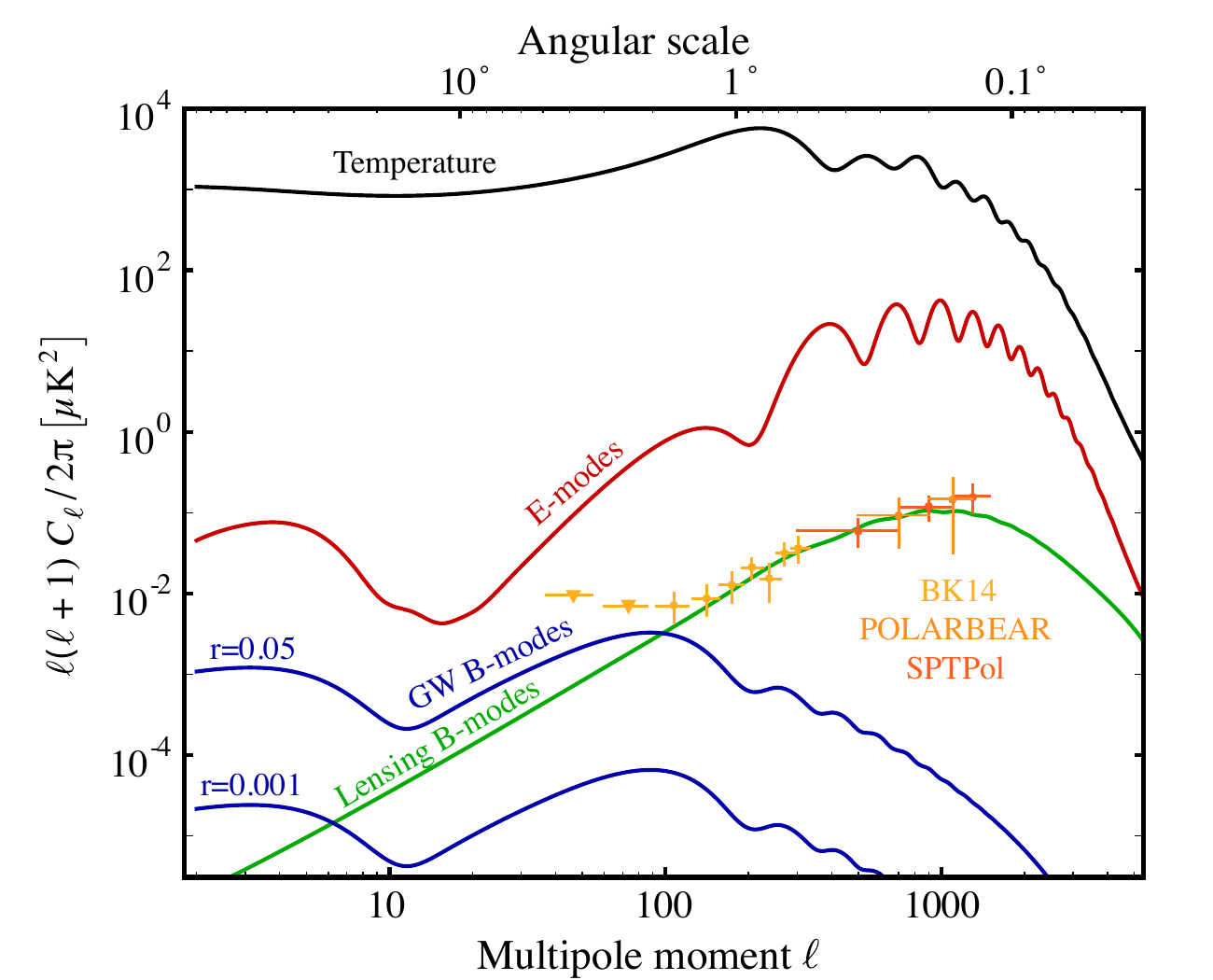}
\end{center}
\caption{Theoretical predictions for the temperature (black), 
E-mode (red), and tensor B-mode (blue) power spectra. Primordial 
B-mode spectra are shown for two representative values of the tensor-to-scalar
ratio: $r=0.001$ and $r=0.05.$ 
The contribution to tensor B modes from scattering at recombination peaks at $\ell \sim 80$
and from reionization at $\ell < 10$.
Also shown are expected values for the contribution to B modes from gravitationally lensed E modes (green).
Current measurements of the B-mode spectrum are shown for {BICEP}2/{\em Keck Array} (light orange), POLARBEAR (orange), and SPTPol (dark orange). 
The lensing contribution to the B-mode spectrum can be partially removed by measuring the 
E and exploiting the non-Gaussian statistics of the lensing.
}
\label{fig:clall}
\end{figure}

The results of calculations using Eqs.~(\ref{eq:clscalar}) and (\ref{eq:clbb}), performed with the Code for Anisotropies in the Microwave Background (CAMB, \cite{Lewis:1999bs}), are shown in Fig.~\ref{fig:clall}. Results for the temperature and E-mode spectra are given by the black and red lines, respectively, while the result for the tensor B-mode spectrum is given by the blue lines for two possible values of $r$. Also shown are predictions for the B-mode spectrum generated by gravitational lensing of E modes (green line). B-mode signal shape and detection prospects, as well as contamination mitigation, will be discussed further in the next section.  Finally, the data points in Fig.~\ref{fig:clall} show current constraints on the B-mode power from lensing, residual foreground, and potentially primordial gravitational waves from { BICEP}2/{\it Keck Array} \cite{Ade:2015fwj}, POLARBEAR~\cite{Ade:2014afa}, and SPTpol \cite{Keisler:2015hfa}.

At present, bounds on the tensor-to-scalar ratio from the temperature and E-mode anisotropies are comparable to those from B-mode observations \cite{Planck:2013jfk,Ade:2015lrj}. The temperature constraints are now cosmic variance-limited, and the E-mode constraints are approaching that level.  However, there is no such limit on the B modes, so that improvements (and a potential detection) with CMB-S4 will rely on measurements of $C_\ell^{BB}$---most likely targeting the degree-scale ``recombination'' feature in the primordial B-mode spectrum. 

\section{Sensitivity forecasts for $r$}
\label{sec:needs}

Achieving the CMB-S4 target sensitivity of $\sigma(r) \sim 10^{-3}$ will require exquisite measurements of the B-mode power spectrum. 
It is expected that CMB-S4 will target the degree-scale recombination feature rather than the tens-of-degree-scale reionization feature (see Fig.~\ref{fig:clall}), because these largest scales are difficult to access from the ground due to atmosphere and sidelobe pickup (though some Stage-3 ground-based experiments are attempting this measurement, notably CLASS \cite{Essinger-Hileman:2014pja}).

\begin{figure}[h]
\begin{center}
\includegraphics[width=0.6\textwidth]{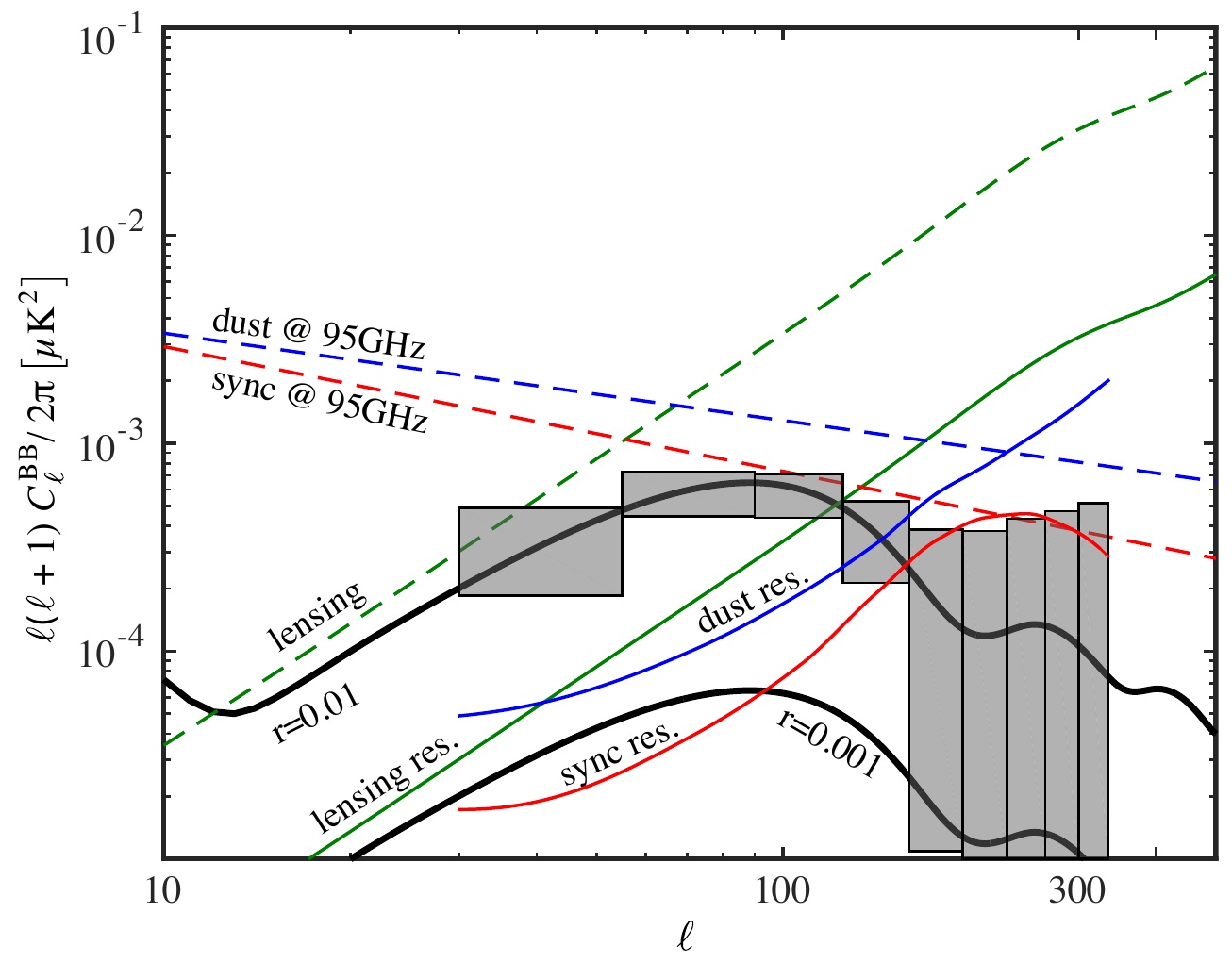}
\end{center}
\caption{
Bin-by-bin forecasted tensor constraints for r=0.01, $f_\mathrm{sky} = 0.03$, 
and the default detector effort ($10^{6}$ detector years).
The boxes denote the forecasted CMB-S4 erorr bars.  
Primordial B-mode spectra are shown for two representative values of 
the tensor-to-scalar ratio: r=0.001 and r=0.01. The dashed green line shows the
$\Lambda$CDM expectation for the B modes induced by gravitational lensing of E modes, 
with the solid line showing the residual lensing power after delensing. The
dashed blue and red lines show the dust and synchrotron (current upper limit) model 
assumed in the forecasting, at the foreground minimum of 95\,GHz. The levels of dust
and synchrotron are equal to the ones reported in \cite{Array:2015xqh}. The 
contribution of dust and synchrotron to the vertical error bars are shown in 
solid blue and red lines. Since these are calculated from a multi-frequency 
optimization, the ``effective frequency'' at which these foreground residuals are 
defined varies with each bin, allowing the residual lines to go above the input 
foreground model lines which are defined at a fixed frequency of 95\,GHz. 
Furthermore, due to the low frequency channels having larger 
beam sizes than the higher frequency ones, in the higher bins, the primordial CMB 
component will be constrained at a higher effective frequency. Defining the 
foreground residuals at these effective frequencies will yield a higher amplitude
for the dust residual, and a lower amplitude for the synchrotron residual, resulting
in the respective shapes of the solid blue and red lines.}
\label{fig_clBBr01}
\end{figure}

As can be seen from Fig.~\ref{fig:clall}, the first requirement for this level of sensitivity to $r$ is a substantial leap forward in raw instrument sensitivity. 
For ground-based bolometric detectors, which are individually limited in sensitivity by the random arrival of background photons, this means a large increase in detector count. 
The forecasts in this section use a baseline of 250,000 detectors operating for four years (or $10^6$ detector years), dedicated solely to maximizing sensitivity to $r$. 
It will be necessary to split this total effort among many electromagnetic frequencies, to separate the CMB from polarized Galactic foregrounds. 
The forecasts here assume eight frequency bands, ranging from 30 to 270\,GHz.
Contamination from gravitationally lensed E modes must also be mitigated.
While a precise prediction for the cosmological mean of the lensing B-mode power spectrum can be made and subtracted from the observed spectrum, there will be a sample variance residual between this prediction and the real lensing B modes on a particular patch of sky.
To suppress this sample variance, it will be necessary to ``delens'' the B-mode maps with a prediction for the lensing signal from that particular patch of sky, constructed from the E-mode map and some tracer of the lensing potential (see Section~\ref{delens} for details).
Forecasts in this section assign part of the total detector count to a dedicated delensing effort, assumed to be a large-aperture ($\ge 6$-meter) telescope at a single frequency
(see below for a discussion of assumptions about aperture size).
Finally, from the relative amplitudes of the temperature, E-mode, and B-mode power spectra, it is clear that instrumental systematics that mix temperature or E-mode power into B modes must also be controlled to an extremely low level. 
To account for real-world inefficiencies, including non-ideal detector performance and yield, observing efficiency, bad weather, data filtering, and cuts, the forecasts in this section use scaled versions of achieved power-spectrum covariance matrices from the { BICEP}2/{\em Keck Array} experiments.
This conservative assumption accounts for many difficult to quantify factors that result in worse constraints on r than a naive, raw-sensitivity calculation would imply.
Further details of the forecasting methodology, including assumptions regarding foreground properties and delensing efficiency, can be found in Section~\ref{sec_specforecast}.
Fig.~\ref{fig_clBBr01} shows some of the inputs to and assumptions of the forecasting code, including foregrounds, B-mode spectrum error bars (including sample variance on the $r=0.01$ spectrum for 3\% of the sky and noise variance using the scaled noise covariance for the default detector count), and delensing efficiency (the value shown is appropriate for 3\% of the sky, $r=0.01$, and the default detector count---see below for details).

\begin{figure}[th]
\begin{center}
\includegraphics[width=0.49\textwidth]{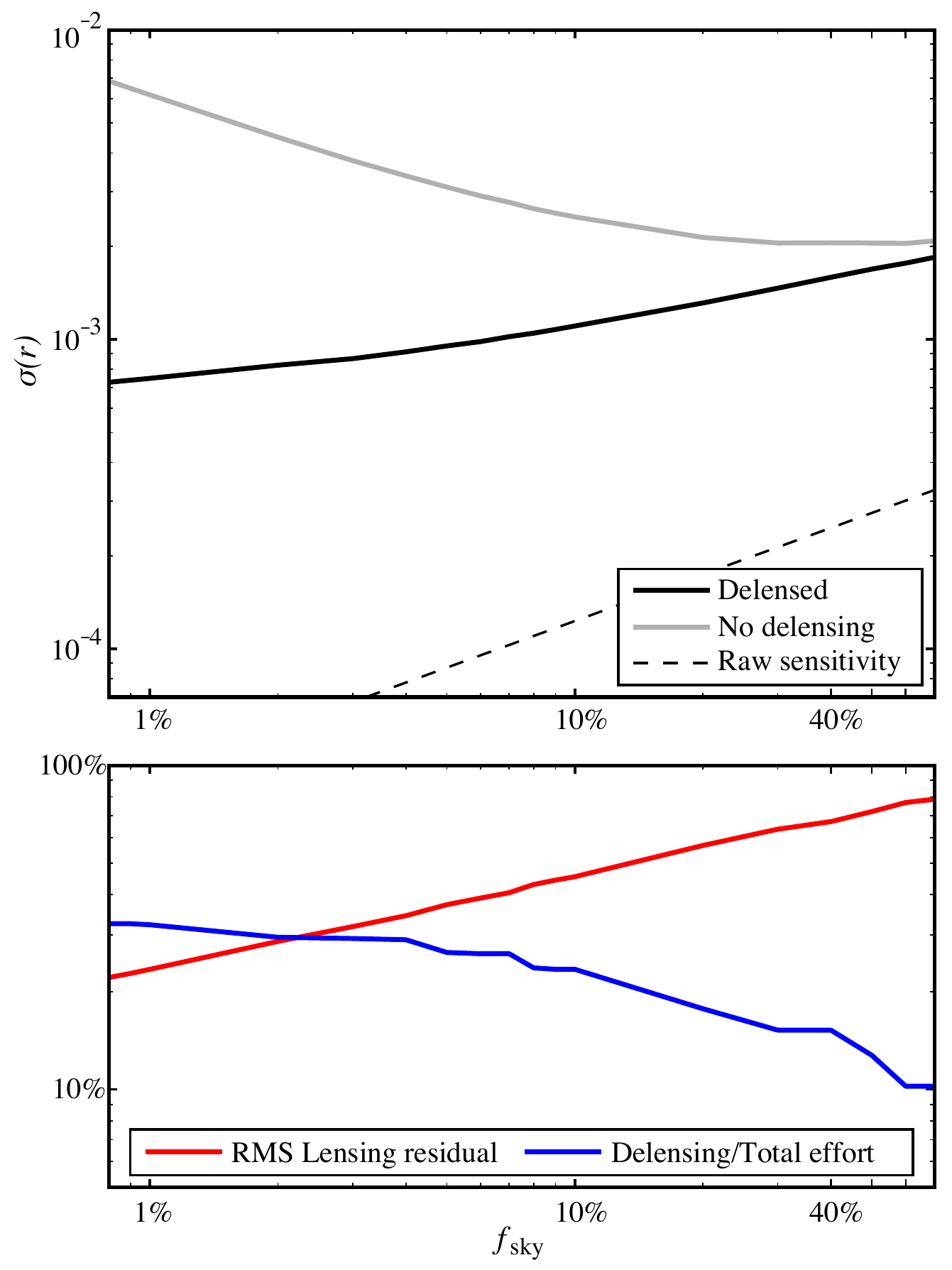}
\includegraphics[width=0.49\textwidth]{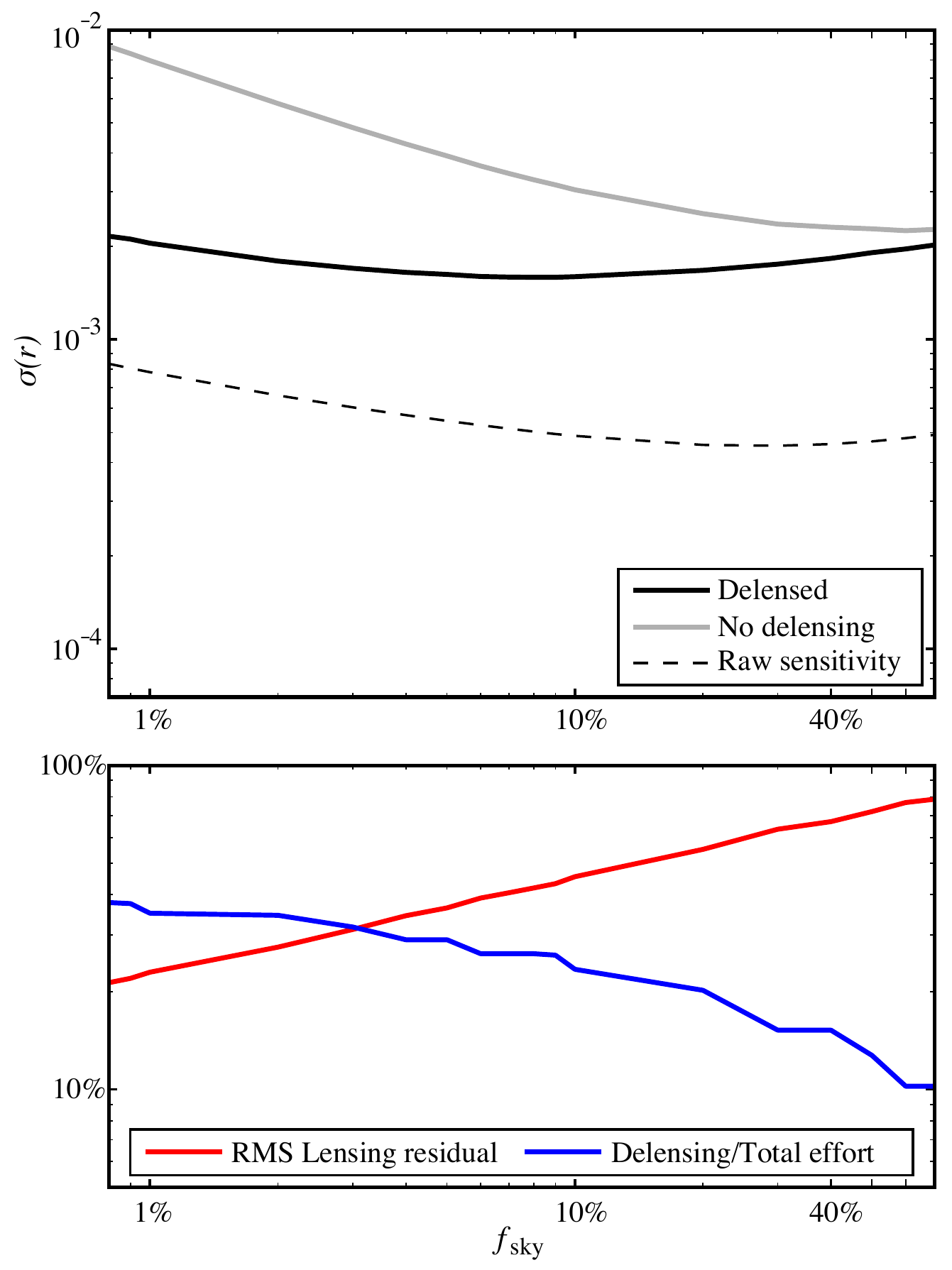}
\end{center}
\caption{(Top panels)
Uncertainty forecasts on $r$, as a function of $f_\mathrm{sky}$, for an
$10^6$ detector-years of effort (150\,GHz equivalent), assuming $r=0$ (left panel) 
and $r=0.01$ (right panel). The forecasting procedure is specifically targeted 
towards optimizing tensor-to-scalar parameter constraints in the presence of 
Galactic foregrounds and gravitational lensing of the CMB. The optimization 
assumes an amount of achieved delensing that varies with $f_\mathrm{sky}$ and a
level of dust decorrelation. In addition to the ``delensed'' case, two more cases are
 included to quantify the importance of delensing (``no delensing'' case), and 
foregrounds + residual lensing (``raw sensitivity'' case), towards achieving the 
desired r constrains.
(Bottom panels)
For the ``delensed'' case, an rms lensing residual and the fraction of the total 
deep survey effort devoted towards delensing are included as a function of 
$f_\mathrm{sky}$. For a detailed description of the forecasting framework and the 
assumptions made, see Section~\ref{sec_specforecast}}
\label{fig_rforecast1}
\end{figure}

The trade-off between raw sensitivity, ability to remove foregrounds, and ability to delens results in a complicated optimization problem with respect to sky coverage.
Fig.~\ref{fig_rforecast1} shows the $r$ sensitivity forecast for CMB-S4 as a function of the observed sky fraction for the case that we only have an upper limit (i.e.\ assuming $r=0$, left) or for the case of a detection (here assuming $r=0.01$, right).
Focusing on the $r=0$ case, we see that an effort devoted to an initial detection of $r$ will benefit from a deep survey that targets a small sky area.
In fact, there is no minimum in Fig.~\ref{fig_rforecast1}. Taken at face value, this optimization drives us to as small a sky fraction as possible. 
Several real-world constraints caution against this extreme interpretation, the most important of which is the level to which we will rely on delensing at the smallest sky fractions.
For example, as shown in the bottom-left panel of Fig.~\ref{fig_rforecast1}, achieving the forecasted sensitivity to $r$ for a survey targeting 1\% of the sky will require an 80\% reduction in the map rms level of the CMB lensing B modes. 
While this is achievable from a sensitivity standpoint (see Section~\ref{delens}), systematics and foregrounds will need to be carefully considered.
There are other real-world concerns not completely captured in the forecasting code that would work in the other direction, steepening the optimization curve at high sky fractions.
For now we assume identical foreground behavior in all parts of the sky (equivalent to the measured behavior in the { BICEP}2/{\em Keck Array} region), while in fact the average amplitude---and possibly the complexity---of foregrounds will increase as larger sky fractions are targeted \cite{Adam:2014bub,Aghanim:2016cps}.
This effect might appear to increase our preference for a small area survey that focuses on the cleanest regions. However, we need to be cautious until we know more about foregrounds at these sensitivity levels.
Likewise, our ability to identify and address instrumental systematics is often limited by the noise level of the maps, so deeper maps can serve as a guard against instrumental problems.
Balancing the forecasting results with these real-world concerns, for subsequent plots we have chosen 3\% as the default sky fraction for CMB-S4 $r$ constraints (assuming a true value of $r=0$).

\begin{figure}[!ht]
\begin{center}
\includegraphics[width=0.49\textwidth]{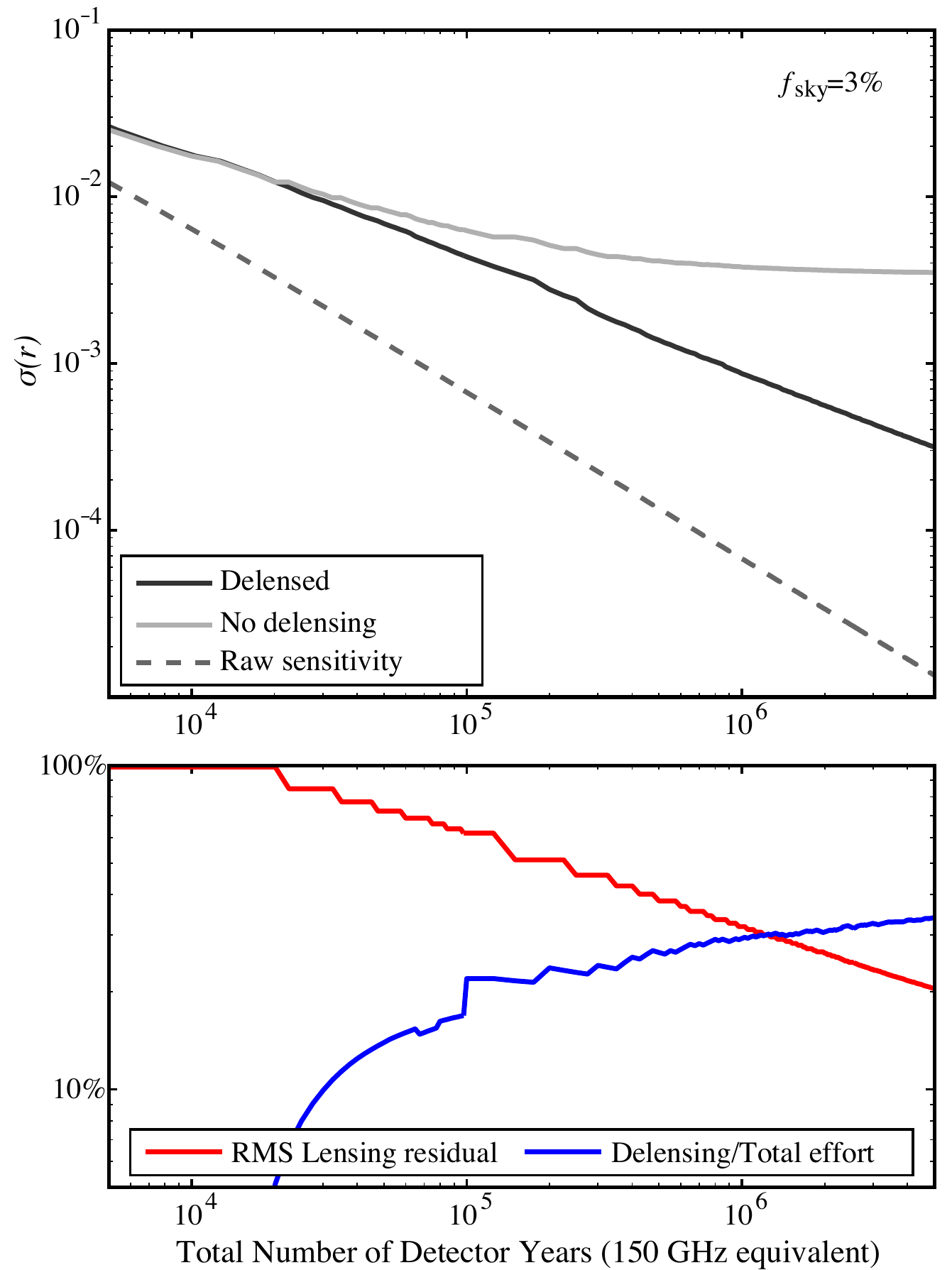}
\includegraphics[width=0.49\textwidth]{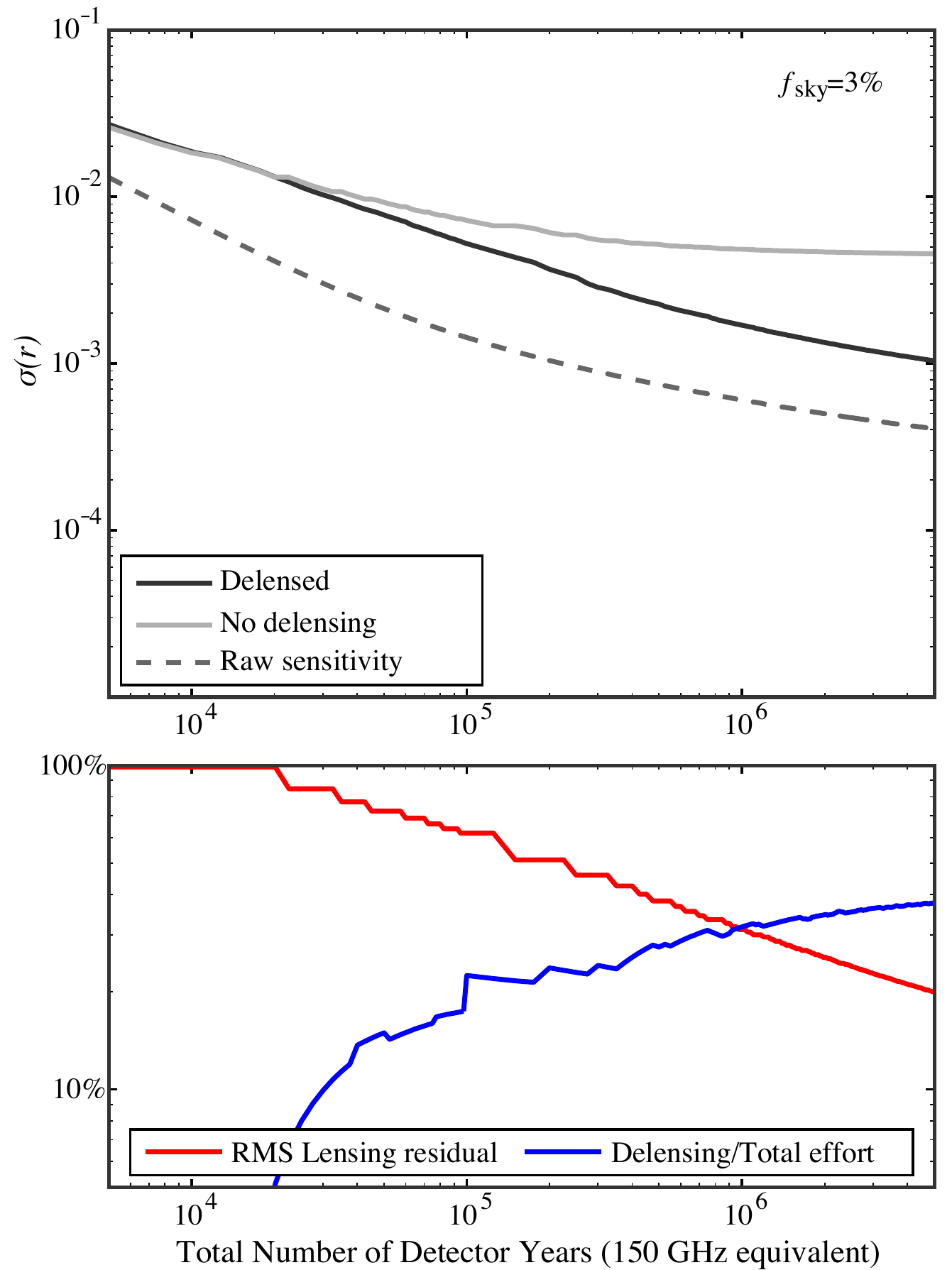}
\end{center}
\caption{(Top Panels)
Uncertainty forecasts on $r$, as a function of the deep survey effort, for a fixed
$f_\mathrm{sky}=0.03$, assuming $r=0$ (left panel) and $r=0.01$ (right panel).
The forecasting procedure is specifically targeted
towards optimizing tensor-to-scalar parameter constraints in the presence of 
Galactic foregrounds and gravitational lensing of the CMB. The optimization 
assumes an amount of achieved delensing that varies with $f_\mathrm{sky}$ and a
level of dust decorrelation. In addition to the ``delensed'' case, two more cases are
 included to quantify the importance of delensing (``no delensing'' case), and
foregrounds + residual lensing (``raw sensitivity'' case), towards achieving the
desired r constrains.
(Bottom panels)
For the ``delensed'' case, an rms lensing residual and the fraction of the total
deep survey effort devoted towards delensing are included as a function of
effort. For a detailed description of the forecasting framework and the
assumptions made, see Section~\ref{sec_specforecast}.}  
\label{fig_rforecast2}
\end{figure}

The preference for small sky area is in tension with other CMB-S4 science goals that prefer large sky areas but have much lower requirements for foreground cleaning.
To balance these goals, we assume that roughly one half of the total CMB-S4 experiment is devoted to a deep survey targeting degree-scale B modes, while the other half is spent on a broad survey.
However, if an important science goal was slightly out of the reach of the default survey, one could consider increasing the effort spent on constraining $r$.
Fig.~\ref{fig_rforecast2} shows the forecasted sensitivity to $r$ as a function of the total effort spent on the deep survey.
With 250,000 detectors operating for four years, CMB-S4 will exceed the $\sigma(r)=0.001$ benchmark, again assuming a true value of $r=0$.

If the true value of $r$ is not zero, the optimum survey strategy and detector effort will change.
As shown in the right panel of Fig.~\ref{fig_rforecast1}, if $r$ is as large as 0.01, then a larger sky area will be needed to improve precision. Thus, the CMB-S4 deep survey must be designed with the flexibility to increase sky area in the event of a detection.

In these forecasts, it is assumed that the degree-scale CMB and foregrounds are measured using small-aperture telescopes, while the delensing is achieved with a separate, large-aperture telescope. 
In all combinations of sky fraction and total detector effort, these forecasts indicate that at least 10--35\% of the total effort must be spent on high-resolution maps that can be used for delensing.
If the large-aperture, high-resolution data can also be used for degree-scale science, the allocation of resources between bands and telescopes would change slightly, and the overall constraints on $r$ would improve. 
If, furthermore, cost per detector were independent of telescope aperture size, an argument could be made to carry out all the science with large-aperture telescopes. 
Neither of these assumptions is clearly supported by current data, however.

Finally, we note that we have validated the results of the primary Fisher forecasting code used in this section with two other codes, one Fisher-based and one map-based. 
For several individual points in sky-fraction and detector-effort parameter space (and using a common set of assumptions about observing bands and noise per detector), the three independent codes return consistent values of $\sigma(r)$.
Worth noting in particular is that the map-based forecasting code does not assume purely Gaussian-distributed foreground emission or translationally invariant foreground properties. Rather, foregrounds are simulated in map space based on currently available data.
The agreement between the map-based approach and the power-spectrum-based approaches provides reassurance that the foreground-mitigation approach in the design of CMB-S4 (in particular the number of observing bands and their placement in frequency) is adequate.

\section{Implications of a detection of primordial gravitational waves}
\label{sec:detection}

The inflationary amplification of vacuum fluctuations of the metric leads to a nearly scale-invariant, very nearly Gaussian tensor power spectrum. This signal is very well characterized by a single parameter defining the (relative) amplitude of tensor fluctuations, $r$. In this section, we consider the consequences of a detection of primordial gravitational waves consistent with this simplest inflationary expectation. Together with implications of an improved upper limit on $r$ presented in the next section, these expectations motivate the threshold level of sensitivity for CMB-S4 and guide the baseline proposals for the instrument in Section~\ref{sec:needs}. 

Of course, in the event of a detection it will be essential to characterize the accuracy with which we can test the standard inflationary prediction. In Section~\ref{sec:beyond_r} we will use the baseline instrument design from Section~\ref{sec:needs} with hypothetical detection levels to forecast constraints on $n_{\rm t}$ and tensor-mode non-Gaussianity. We also use Section~\ref{sec:beyond_r} to discuss alternatives to the inflationary vacuum prediction, including non-vacuum sources during inflation. We will see that non-vacuum scenarios would be distinguishable, as long as $r$ is detected at high significance.

The remainder of this section derives the remarkable implications of a detection of primordial gravitational waves with amplitude accessible by CMB-S4, and with a nearly scale-invariant, nearly Gaussian spectrum.  This would reveal the energy scale of inflation, provide compelling evidence for linearized quantum gravity, and yield strong support for structure in non-linear quantum gravity that accommodates a large field range for the inflaton. 

\subsection{The energy scale of inflation}
\label{sec:scale-of-inflation}
According to the inflationary prediction for the amplitude of primordial gravitational waves, Eq.~(\ref{eq:inf_Pt}), a detection provides a direct measurement of the Hubble scale during inflation. In single-field slow-roll models the Friedmann equation relates the Hubble scale to the potential energy $V$ of the inflaton, $3H^2M_{\rm P}^2\simeq V$. The inflationary prediction for the amplitude of scalar fluctuations, Eq.~(\ref{eq:inf_Pzeta}) can be used to write $H$ in terms of the measured amplitude and the so-far undetermined slow-roll parameter, $\epsilon$. Since $\epsilon$ is directly proportional to the tensor-to-scalar ratio this allows us to express the energy scale of inflation in terms of measured numbers, known constants, and $r$ (all at the pivot scale $k_\ast=0.05$ Mpc$^{-1}$) as
\begin{equation}\label{eq:Vofr}
V^{1/4}=1.04\times 10^{16}{\rm GeV}\left(\frac{r_\ast}{0.01}\right)^{1/4}\,.
\end{equation}
A detection of primordial gravitational waves therefore determines the energy scale of inflation to within a few per cent. 

{\it A detection of primordial gravitational waves by CMB-S4 would reveal a new scale of particle physics, near the GUT scale. If the signal is reasonably scale-invariant and at most weakly non-Gaussian, this scale corresponds to the energy scale of inflation.} 

\subsection{Planckian field ranges and symmetries}
The spectrum of tensor fluctuations depends only on the Hubble parameter $H$ during inflation, while the scalar power depends on both $H$ and the evolution of the homogeneous field sourcing inflation. As a consequence, the tensor-to-scalar ratio $r$ determines the inflaton field range in Planck units (called the ``Lyth bound'' \cite{Lyth:1996im})
\begin{equation}
\label{eq:Lyth}
\frac{\Delta\phi}{M_{\rm P}}=\int_0^{\mathcal{N}_\ast}d\mathcal{N}\,\left(\frac{r}{8}\right)^{1/2}\,,
\end{equation}
where (applying the general equation to the observationally accessible regime) $\mathcal{N}_\ast$ is the number of e-folds between the end of inflation and the moment when the mode with $k_\ast=0.05\,{\rm Mpc^{-1}}$ (corresponding to the CMB pivot scale) exits the horizon. In many common inflationary models $r$ is a monotonic function of $\mathcal{N}$ so that
\begin{equation}
\label{eq:lbound}
\frac{\Delta\phi}{M_{\rm P}}\gtrsim \left(\frac{r_\ast}{8}\right)^{1/2}\mathcal{N}_\ast\gtrsim \left(\frac{r}{0.01}\right)^{1/2}\,.
\end{equation}  
The value of $\mathcal{N}_\ast$ is not well constrained and depends on unknown details of reheating, but $\mathcal{N}_\ast\gtrsim 30$ provides a conservative lower limit, justifying the second inequality in Eq.~(\ref{eq:lbound}). Thus, a tensor-to-scalar ratio $r>10^{-2}$ typically corresponds to a trans-Planckian excursion in field space between the end of inflation and the epoch when the modes we observe in the CMB exit the horizon.

The relationship in Eq.~(\ref{eq:lbound}) is significant because it relates the observed amplitude of linearized metric fluctuations to a property of the full quantum field theory for gravity coupled to the inflaton. The action describing inflation, like the action for any other particle physics phenomena, in general will include terms that encode the effects from degrees of freedom that couple to the inflaton, but are too energetic to be probed directly by physics near the inflationary scale. The field range is a measure of the distance in field space over which the corrections from the unknown physics do not significantly affect the low energy dynamics, since otherwise slow-roll inflation would not persist. In theories of quantum gravity we expect degrees of freedom to enter at the Planck scale or below. A field range exceeding the Planck scale would imply that quantum gravity contributions do not have a significant effect over the naively expected scale. 
A detection of $r$ would therefore provide very strong motivation to better understand how ``large-field inflation" can be naturally incorporated in quantum gravity.

To understand this more quantitatively, recall that unless we work in a UV-complete theory such as string theory, we rely on an effective field theory description of the inflationary epoch. General relativity, viewed as an effective field theory, breaks down as energies approach the Planck scale because interactions between gravitons become strongly coupled. The same is true for matter coupled to general relativity, so that the effective field theory governing the inflationary period will generically have a sub-Planckian cut-off $\Lambda_{\rm UV}<M_{\rm P}$. In fact, in any weakly-coupled UV completion of general relativity the new degrees of freedom must enter well below the Planck scale to ensure weak coupling so that $\Lambda_{\rm UV}\ll M_{\rm P}$. 
According to the bound in Eq.~(\ref{eq:lbound}), a tensor-to-scalar ratio $r>10^{-2}$ (and even somewhat smaller) requires a displacement in field space that is larger than the cut-off of the effective field theory. While this does not invalidate an effective field theory description, it has important consequences. Assuming that the UV-complete theory is known, the effective field theory is obtained by integrating out all modes parametrically heavier than the cut-off $\Lambda_{\rm UV}$ of the single-field model. In the absence of symmetries, we expect the inflaton $\phi$ to couple to heavy degrees of freedom $\chi$ that, once integrated out, will introduce significant structure in the potential for the inflation on scales $\Delta\phi\ll \Lambda_{\rm UV}$. For example, consider the action
\begin{equation}\label{eq:action}
S=\int d^4x\sqrt{-g}\left[-\frac12g^{\mu\nu}\partial_\mu\phi\partial_\nu\phi-\frac12g^{\mu\nu}\partial_\mu\chi\partial_\nu\chi-\frac12m^2\phi^2-\frac12M^2\chi^2-\frac12\mu\phi\chi^2+\dots\right]\,.
\end{equation}
By assumption, the mass of the heavy degrees of freedom to be integrated out is $M\gtrsim\Lambda_{\rm UV}$, and the dots represent various other interaction terms. Generically the dimensionful coupling $\mu$ is also expected to be of order the cut-off, $\mu\sim\Lambda_{\rm UV}$. From the last two terms in Eq.~(\ref{eq:action}), we see that displacements of $\phi$ by a distance comparable to the cut-off may lead to cancellations in the effective mass of the heavy degrees of freedom, and heavy states (in this case $\chi$) may become light if $\phi$ is displaced by a distance large compared to the cut-off. In particular, since $\Lambda_{\rm UV}<M_{\rm P}$ we should not expect potentials that are smooth over super-Planckian distances in a generic low energy effective field theory with cut-off $\Lambda_{\rm UV}<M_{\rm P}$. 

We can only expect potentials suitable for large-field inflation if some mass scales (in the example $m$ and $\mu$) are well below the cut-off, or if dimensionless couplings are small. This occurs naturally if the UV theory respects a weakly broken shift symmetry $\phi\rightarrow\phi+c$, which ensures that quantum corrections from the inflaton and graviton will not introduce large corrections to the inflationary Lagrangian \cite{Linde:2005ht, Kaloper:2011jz, Csaki:2014bua,Kaplan:2015fuy,Choi:2015fiu}. At the level of an effective field theory we can simply postulate such an approximate shift symmetry, but one should keep in mind that we ultimately require the existence of such a symmetry in quantum gravity. 

As the best developed theory of quantum gravity, string theory is a useful framework for exploring mechanisms that allow large-field inflation to be realized, even in the presence of heavy degrees of freedom. Axions are ubiquitous in string theory and provide natural candidates for the inflaton because they enjoy a shift symmetry that is weakly broken by instanton effects and by the presence of branes or fluxes~\cite{Wen:1985jz}. Early field theory models relied on the familiar periodic contributions to the potential generated by instantons to drive inflation~\cite{Freese:1990rb,Adams:1992bn}. In string theory the periods are expected to be sub-Planckian~\cite{Banks:2003sx,ArkaniHamed:2006dz}, while constraints on the scalar spectral index require super-Planckian axion periods, so that a UV completion of these models does not currently exist. The excitement over the initial { BICEP}2 results \cite{Ade:2014xna} led to renewed interest in models in which the inflaton is an axion with a potential that is entirely due to instanton effects and intensified the discussion about the extent to which some means to achieve large-field inflation via multiple axions may be incompatible with basic principles of quantum gravity \cite{Kim:2004rp,Rudelius:2014wla,delaFuente:2014aca,Rudelius:2015xta,Brown:2015iha,Bachlechner:2015qja,Brown:2015lia,Heidenreich:2015wga,Heidenreich:2015nta,Kooner:2015rza}.

In addition to the familiar non-perturbative contributions that break the continuous shift symmetry to a discrete one, the presence of fluxes and branes lead to contributions to the axion potentials that break the discrete shift symmetry as well. As the axion is displaced by one period, one unit of charge is induced, so that the axion field space becomes non-compact. As a consequence, super-Planckian decay constants are not required for super-Planckian excursions in these so-called ``monodromy'' models~\cite{Silverstein:2008sg, McAllister:2008hb, Kaloper:2008fb, Berg:2009tg, Palti:2014kza,McAllister:2014mpa, Marchesano:2014mla, Blumenhagen:2015xpa,Hebecker:2015tzo}. Generically both contributions to the potential are present and these models predict periodic effects at some level, either directly from the periodic features in the potential or from periodic bursts of string or particle production. Unfortunately, the strength of the signal is very model-dependent. Even if one of these models is a good approximation to nature, the periodic features could be undetectably small even for CMB-S4.

\begin{figure}[ht]
\begin{center}
\includegraphics[width=6in]{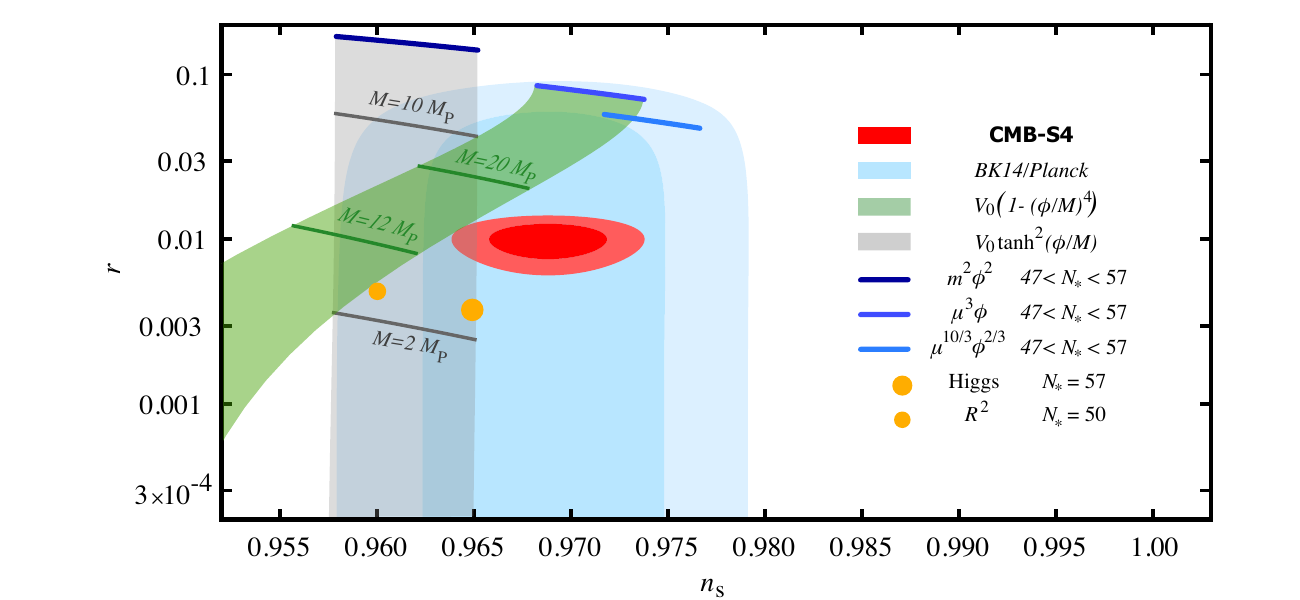}
\end{center}
\caption{Forecast of CMB-S4 constraints in the $n_{\rm s}$--$r$ plane for a fiducial model with $r=0.01$. Constraints 
on $r$ are derived from the expected CMB-S4 sensitivity to the B-mode power spectrum as described in 
Section~\ref{sec:needs}. Constraints on $n_{\rm s}$ are derived from expected CMB-S4 sensitivity to temperature and 
E-mode power spectra as described in Section~\ref{sec:ttee}. Also shown are the current best constraints from a combination of the { BICEP}2/{\em Keck Array} experiments and \planck\ \cite{Array:2015xqh}. Chaotic inflation with $V(\phi)=\mu^{4-p}\phi^p$ for \mbox{$p=2/3,1,2$} are shown as blue lines for $47<N_\star<57$ (with smaller $N_\star$ predicting lower values of $n_{\rm s}$). The Starobinsky model and Higgs inflation are shown as small and large filled orange circles, respectively. The lines show the classes of models discussed in Section~\ref{sec:upperLimits}. The green band shows the predictions for quartic hilltop models, and the gray band shows the prediction of a sub-class of $\alpha$-attractor models~\cite{Kallosh:2013hoa}.
}
\label{fig:nsrp01}
\end{figure}

In writing Eq.~(\ref{eq:lbound}), we have assumed that $r$ is monotonic, or at least of the same order of magnitude throughout the inflationary period. However, one can easily construct models in which $r$ is non-monotonic to weaken the bound~\cite{BenDayan:2009kv,Hotchkiss:2011gz, Chatterjee:2014hna}. In the case of a detection with CMB-S4 of a spectrum that is at least approximately scale-invariant, we can write the weaker bound
\begin{equation}
\frac{\Delta\phi}{M_{\rm P}}\gtrsim\left(\frac{r}{0.3}\right)^{1/2}\,,
\end{equation}
which limits the distance in field space traveled during the time the modes we observe in the CMB exited the horizon. This inequality shows that even if the distance in field space traveled during this period is sub-Planckian, it is not parametrically smaller than $M_{\rm P}$. 
This implies that we cannot avoid the question of the embedding of the inflation model into quantum gravity for $r=0.01$ or even for $r=0.005$, unless we assume that the UV completion of general relativity is strongly coupled. These levels of primordial gravitational waves can be detected at high significance with CMB-S4 as shown in Fig.~\ref{fig_clBBr01} and Fig.~\ref{fig:nsrp01} for a fiducial model of $r=0.01$.

In deriving the primordial power spectra and Eq.~(\ref{eq:Lyth}), we have assumed the Bunch-Davies vacuum state \cite{Bunch:1978yq}. The relation between $r$ and the scale of inflation is modified if we assume that the tensor  modes (and the scalar modes) either do not start in the Bunch-Davies state~\cite{Ashoorioon:2014nta,Collins:2014yua}, or that the evolution during inflation will lead to departures from it. The first option generically introduces a stronger scale-dependence into the tensor spectrum \cite{Aravind:2014axa,Flauger:2013hra} (and additional non-Gaussianity). In addition, this way of achieving observable primordial B modes from a low-scale model has a similar feature to large-field models: one should show that the initial state is not only acceptable from the point of view of low-energy considerations, but can be generated by pre-inflationary physics. The second option (discussed in section~\ref{sec:scale-of-inflation}) leads to non-trivial higher $n$-point functions that are in principle measurable.

{\it A detection of primordial B modes with CMB-S4 would provide evidence that the theory of quantum gravity must accommodate a Planckian field range for the inflaton. Conversely, the absence of a detection of B modes with CMB-S4 will mean that a large field range is not required. The relation between inflaton field range and the amplitude of primordial gravitational waves means that a detection of $r$ would provide an observational window into quantum gravity.}

\section{Implications of an improved upper limit on $r$} 
\label{sec:upperLimits}
As detailed in previous sections, a detection of primordial gravitational waves would have profound implications. However, even excluding the presence of gravitational waves at a level observable by CMB-S4 would have important consequences for the theory of inflation. Current constraints already strongly disfavor models that were plausible candidates, such as chaotic inflation with a quadratic potential~\cite{Ade:2015lrj,Ade:2015tva,Array:2015xqh}. Upper limits from CMB-S4 would rule out large classes of inflationary models. In particular, all models that explain the observed value of $n_{\rm s}$ naturally (in the sense detailed below), with a scale of the characteristic variation of the potential exceeding the Planck scale would be excluded.

We present a version of an argument for the implications of an upper limit on $r$, developed in Refs.~\cite{Mukhanov:2013tua,Roest:2013fha,Creminelli:2014nqa}, which does not rely on the microscopic details of inflationary models. In the limit where the slow-roll parameter $\epsilon\ll1$, Eqs.~(\ref{eq:inf_Pzeta}) and~(\ref{eq:specind}) lead to a differential equation
\begin{equation}\label{eq:epsdiffeq}
\frac{d\ln\epsilon}{d\mathcal{N}}-(n_{\rm s}(\mathcal{N})-1)-2\epsilon=0\,,
\end{equation} 
where $\mathcal{N}$ is the number of e-folds until the end of inflation, and $n_{\rm s}(\mathcal{N})-1$ denotes the spectral index evaluated at the wavenumber of the mode that exits the horizon $\mathcal{N}$ e-folds before the end of inflation. Note that $\epsilon$ is small (but positive) during inflation and $\epsilon\sim 1$ when inflation ends. If $\epsilon$ is a monotonic function of $\mathcal{N}$ this implies $n_{\rm s}(\mathcal{N})-1\leq 0$, in agreement with observations. 

Denoting by $\mathcal{N}_\ast$ the number of e-folds before the end of inflation at which the CMB pivot scale exits the horizon, the departure from a scale-invariant spectrum observed by the \planck\ satellite is $\mathcal{O}(1/\mathcal{N}_\ast)$. While this could be a coincidence, it would find a natural explanation if 
\begin{equation}\label{eq:nsassump}
n_{\rm s}(\mathcal{N})-1=-\frac{p+1}{\mathcal{N}}\,,
\end{equation}
up to subleading corrections in an expansion in large $\mathcal{N}$ for some real $p$. Under this assumption, the general solution to Eq.~(\ref{eq:epsdiffeq}) is
\begin{equation}\label{eq:epssol}
\epsilon(\mathcal{N})=\frac{p}{2\mathcal{N}}\frac{1}{1\pm\left(\mathcal{N}/\mathcal{N}_{\rm eq}\right)^{p}}\,,
\end{equation}
where we have chosen to parameterize the integration constant by $\mathcal{N}_{\rm eq}$ so that the magnitudes of the first and second terms in the denominator become equal when $\mathcal{N}=\mathcal{N}_{\rm eq}$. We take $\mathcal{N}_{\rm eq}>0$ and indicate the choice of sign for the integration constant by ``$\pm$.''

Assuming that the epoch when the modes we observe in the CMB exit the horizon is not special, so that $\mathcal{N}_\ast\gg\mathcal{N}_{\rm eq}$ or $\mathcal{N}_\ast\ll\mathcal{N}_{\rm eq}$, Eq.~(\ref{eq:epsdiffeq}) leads to four classes of solutions:
\begin{eqnarray}
{\rm I.}&&\epsilon(\mathcal{N})=\frac{p}{2\mathcal{N}}\,;\label{eq:classI}\\
{\rm II.}&&\epsilon(\mathcal{N})=\frac{p}{2\mathcal{N}}\left(\frac{\mathcal{N}_{\rm eq}}{\mathcal{N}}\right)^p\hspace{1.92cm}\qquad{\rm with}\qquad \hspace{7.3mm}p>0 \hspace{1.0cm}\qquad{\rm and}\qquad\mathcal{N}_{\rm eq}\ll\mathcal{N}_\ast\,;\label{eq:classII}\\
{\rm III.}&&\epsilon(\mathcal{N})=\frac{|p|}{2\mathcal{N}}\left(\frac{\mathcal{N}}{\mathcal{N}_{\rm eq}}\right)^{|p|}\hspace{1.77cm}\qquad{\rm with}\qquad \hspace{7.3mm}p<0 \hspace{1.0cm}\qquad{\rm and}\qquad\mathcal{N}_{\rm eq}\gg\mathcal{N}_\ast\,;\label{eq:classIII}\\
{\rm IV.}&&\epsilon(\mathcal{N})=\frac{1}{2\mathcal{N}\ln\mathcal{N}_{\rm eq}/\mathcal{N}}+\frac{p}{4\mathcal{N}}+\dots\qquad{\rm with}\qquad |p|\ll\frac{1}{\ln\mathcal{N}_{\rm eq}/\mathcal{N}_\ast}\qquad{\rm and}\qquad \mathcal{N}_{\rm eq}\gg\mathcal{N}_\ast\,.\label{eq:classIV}
\end{eqnarray}

As we explain in what follows, if CMB-S4 does {\it not\/} detect primordial B modes, only class II with $\mathcal{N}_{\rm eq}\lesssim 1$ will remain viable, while the other cases will be disfavored or excluded. We will see that $\mathcal{N}_{\rm eq}$ sets the characteristic scale (in Planck units) over which the potential varies, so that this would amount to excluding all models that naturally explain the spectral index with a characteristic scale that exceeds the Planck scale. 

The value of $\mathcal{N}_\ast$ depends on the post-inflationary history of the Universe. Equation~(\ref{eq:nsassump}) implies that a measurement of the spectral index and its running would determine $p$ and hence $\mathcal{N}_\ast$, but unfortunately a measurement of the running at the level of $(n_{\rm s}-1)^2$ is out of reach for CMB-S4. A given reheating scenario predicts $\mathcal{N}_\ast$, but the space of reheating scenarios is large. Instantaneous reheating leads to $\mathcal{N}_\ast\simeq 57$ for $k_\ast=0.05\,  {\rm Mpc}^{-1}$, while smaller values correspond to less efficient reheating. We will assume $47<\mathcal{N}_\ast<57$ for the following discussion. 

Current constraints on $n_{\rm s}$ and $r$ from Ref.~\cite{Ade:2015tva} disfavor class III at just over $2\,\sigma$ relative to class II. Furthermore, the best-fit point of class III occurs for $p\simeq 0$, where classes I, II, and III are degenerate, so that class III need not be discussed separately. Additionally class IV is disfavored at 2--$3\,\sigma$ relative to class II. As a consequence we focus on classes I and II in what follows.

For class I, constraints from the \planck\ satellite and the { BICEP}2/{\em Keck Array} experiments~\cite{Ade:2015tva} translate into $p=0.32\pm0.16$ at $1\,\sigma$, and favor models with inefficient reheating. The best-fit point in this class is $r=0.044$ and $n_{\rm s}=0.973$, which is currently disfavored relative to class II at 1--$2\,\sigma$. Upper limits on $r$ directly translate into constraints on $p$. A $1\,\sigma$ upper limit on the amount of primordial gravitational waves from CMB-S4 at a level of $r<0.001$ would imply $p<0.013$ and effectively rule out this class as it degenerates into class II in this limit. 

For class II the tensor-to-scalar ratio is naturally smaller than in class I, as long as $p$ is of order unity because $\mathcal{N}_\ast\gg\mathcal{N}_{\rm eq}$. Under the additional assumption that the scaling of Eq.~(\ref{eq:classII}) should be valid until the end of inflation we have $\mathcal{N}_{\rm eq}\simeq 1$. In this case, current data from Ref.~\cite{Ade:2015tva} imply $p=0.67\pm0.24$ after marginalization over $\mathcal{N}_\ast$. The best-fit point occurs at $p=0.83$ and instantaneous reheating, so that in this class the data favors models with efficient reheating. At the best-fit point, $r=0.004$ and $n_{\rm s}=0.968$. An upper limit of $r<0.001$ would disfavor this scenario relative to scenarios with $\mathcal{N}_{\rm eq}\ll 1$ at approximately $2\,\sigma$. The precise significance depends slightly on the true value of the spectral index. Similarly, for an upper limit of $r<0.001$, the regime with $p\ll1$ (and equivalently class I) would be disfavored relative to class II with $\mathcal{N}_{\rm eq}\ll 1$ at $3\,\sigma$. To disfavor the scenario with $\mathcal{N}_{\rm eq}\simeq 1$ at approximately $3\,\sigma$ relative to $\mathcal{N}_{\rm eq}\ll 1$ would require an upper limit of $r\lesssim 5\times 10^{-4}$.  

In summary, in the absence of a detection of primordial gravitational waves, CMB-S4 would place constraints on $n_{\rm s}$ and $r$ that are strong enough to rule out or disfavor all models that naturally explain the observed value of the scalar spectral index in the sense that $n_{\rm s}(\mathcal{N})-1\propto 1/\mathcal{N}$ and in which the behavior of Eqs.~(\ref{eq:classI})--(\ref{eq:classIV}) provides a good approximation until the end of inflation. 

To understand the implications better, let us discuss the models that underlie the classes favored by current data, classes I and II. The  potentials can be obtained from 
\begin{equation}
\frac{d\phi}{d\mathcal{N}}=M_{\rm P}^2\frac{V'}{V}\qquad{\rm and}\qquad \left(\frac{d\phi}{d\mathcal{N}}\right)^2=2\epsilon M_{\rm P}^2\,,
\end{equation}
where $M_{\rm P}$ is the reduced Planck mass.

Class I corresponds to models of chaotic inflation with monomial potentials, $V(\phi)=\mu^{4-2p}\phi^{2p}$,
as already considered in~Ref.~\cite{Linde:1983gd}. The most commonly studied examples were $p=1,2$, both of which are now ruled out or strongly constrained~\cite{Ade:2015tva}. Models with fractional powers $1/3<p<1$ that are still viable candidates have naturally appeared in the study of large-field models of inflation in string theory~\cite{Silverstein:2008sg,McAllister:2008hb,Flauger:2009ab}. If gravitational waves are not observed with CMB-S4, these would be ruled out.

Provided $p\neq 1$, class II corresponds to potentials of the form 
\begin{equation}\label{eq:classIIpot}
V(\phi)=V_0\exp\left[-\left(\frac{\phi}{\Lambda}\right)^{\frac{2p}{p-1}}\right]\,,
\end{equation}
with $\Lambda=\sqrt{\alpha(p)\mathcal{N}_{\rm eq}}M_{\rm P}$, where $\alpha(p)=4p/(1-p)^2$. The parameter $\Lambda$ is closely related to the characteristic scale $M$ over which the potential varies appreciably. For the range of $p$ that corresponds to the observed value of $n_{\rm s}$, it is well approximated by
\begin{equation}\label{eq:potscale}
M=\Lambda\frac{|1-p|}{p}\,.
\end{equation}
For $p>1$ inflation typically occurs when $\phi\ll \Lambda$. In this regime, the potential behaves like a hilltop model, $V(\phi)\simeq V_0(1-\left(\phi/\Lambda\right)^n)$, with $n=2p/(p-1)$. For $0<p<1$ inflation typically occurs for $\phi\gg \Lambda$ and $V(\phi)\simeq V_0(1-\left(\Lambda/\phi\right)^n)$ with $n=2p/(1-p)$. In the limit $p\to0$, in which classes I, II, and III become degenerate, the $\phi$-dependence is logarithmic. 

For the special case $p=1$ the dependence on the inflaton in Eq.~(\ref{eq:classIIpot}) becomes exponential, and in the inflationary regime the potential is well approximated by $V(\phi)\simeq V_0\left(1-\exp\left(-\phi/M\right)\right)$ with $M=\sqrt{\mathcal{N}_{\rm eq}}M_{\rm P}$. There are many examples of models with a potential with this asymptotic behavior for $\phi\gg M$. These include the Starobinsky model~\cite{Starobinsky:1980te}, Higgs inflation~\cite{Salopek:1988qh,Bezrukov:2007ep}, an early example of chaotic inflation~\cite{Goncharov:1983mw}, and the T-model~\cite{Kallosh:2013hoa}.

If only the asymptotic forms of the potentials agree with Eq.~(\ref{eq:classIIpot}), then Eq.~(\ref{eq:nsassump}) will not be exact and the departures from Eq.~(\ref{eq:classIIpot}) will be encoded in the subleading terms that vanish more rapidly than $1/\mathcal{N}$ in the limit $\mathcal{N}\to\infty$. Unfortunately, just like the running of the scalar spectral index, the subleading contributions are typically too small to be detected.

Note that $\mathcal{N}_{\rm eq}$ sets the characteristic scale in field space. For $\mathcal{N}_{\rm eq}$ of order unity, the variation of the inflaton is naturally given in units of the reduced Planck mass, while for $\mathcal{N}_{\rm eq}\ll 1$ the characteristic scale in field space is sub-Planckian. 

This allows us to rephrase the lesson we can draw from an upper limit on $r$ from CMB-S4 as follows:

{\em In the absence of a detection, CMB-S4 would rule out or disfavor all models that naturally explain the observed value of the scalar spectral index and in which the characteristic scale in field space exceeds the Planck scale.}

\begin{figure}[t]
\begin{center}
\includegraphics[width=6in]{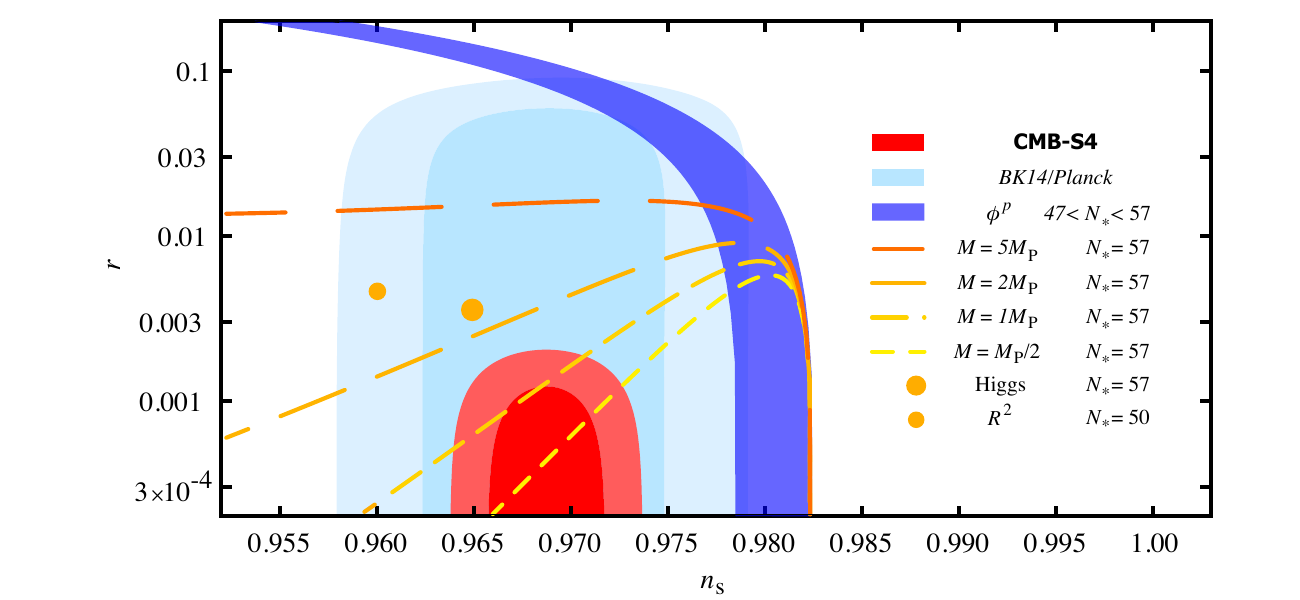}
\end{center}
\caption{Forecast of CMB-S4 constraints in the $n_{\rm s}$--$r$ plane for a fiducial model with $r=0$. Constraints 
on $r$ are derived from the expected CMB-S4 sensitivity to the B-mode power spectrum as described in 
Section~\ref{sec:needs}. Constraints on $n_{\rm s}$ are derived from expected CMB-S4 sensitivity to temperature and 
E-mode power spectra as described in Section~\ref{sec:ttee}. Also shown are the current best constraints from a combination of the { BICEP}2/{\em Keck Array} experiments and \planck\ \cite{Array:2015xqh}.  
The Starobinsky model and Higgs inflation are shown as small and large filled orange circles. The lines show the classes of models discussed in Section~\ref{sec:upperLimits} that naturally explain the observed value of the scalar spectral index for different characteristic scales in the potential (see eq.~\eqref{eq:potscale}), $M=M_{\rm P}/2$, $M=M_{\rm P}$, $M=2\,M_{\rm P}$, and $M=5\,M_{\rm P}$. Longer dashes correspond to larger values of the scale $M$. 
}
\label{fig:nsr0}
\end{figure}

This is shown in Figure~\ref{fig:nsr0}. We see that for $n_s=0.968$, $\sigma(r)=5\times 10^{-4}$ would allow to disfavor characteristic scales that exceed the Planck scale at $95\%$ CL. Unfortunately, because of the scaling $M\propto \sqrt{\mathcal{N}_{\rm eq}}$ it will only be possible to place constraints $M\lesssim M_{\rm P}$, but not $M\ll M_{\rm P}$. It should also be kept in mind that a natural explanation of the value of the scalar spectral index is not guaranteed and its value could be an accident. That a natural explanation is possible is, however, encouraging.

\section{Tensor-mode science beyond $r$}
\label{sec:beyond_r}

If a detection of primordial gravitational waves is made with CMB-S4, the next step would be to understand the possible sources of the signal. The spectrum of B modes from vacuum fluctuations of the metric, amplified by inflation driven by a scalar field, is nearly scale-invariant and very nearly Gaussian. In this section we discuss how well CMB-S4, given a detection, could characterize the shape of the B-mode spectrum, test for significant higher-order correlations involving tensor modes, and test for parity violation. These additional features would demonstrate the degree to which the data supports the expectation from the simplest inflationary models, or whether there is evidence for richer (or non-inflationary) physics. 

\subsection{The shape of the tensor power spectrum}
The vast majority of inflation scenarios predict a red spectrum for gravitational waves, and in the simplest cases the canonical single-field consistency relation fixes $n_{\rm t}=-r/8$. For a single field with a sound speed less than one, or multiple fields, $n_{\rm t}/r<-1/8$ instead \cite{Price:2014ufa}. Allowing the inflaton to couple to higher curvature terms can produce a blue tilt \cite{Baumann:2015xxa}, but in general a very blue tensor spectral index is only possible in alternatives to inflation (see Section~\ref{subsec:alternatives}). 

\begin{figure}[th]
\begin{center}
\includegraphics[width=2.5in]{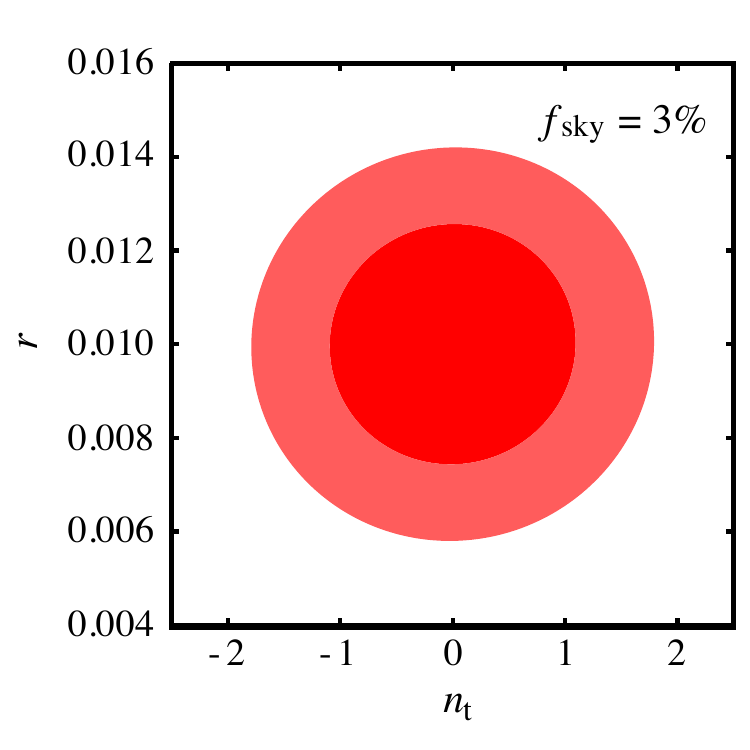}
\includegraphics[width=2.5in]{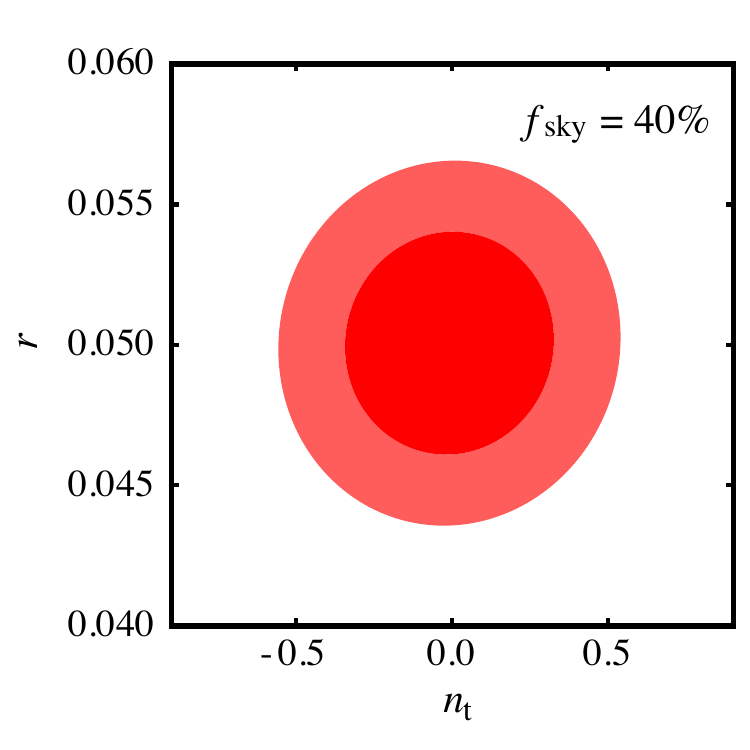}
\end{center}
\caption{ Forecasts for joint constraints on the tensor to scalar ratio $r$, and the tensor spectral index, $n_{\rm t}$ assuming fiducial values of $r=0.01$ (left) or $r=0.05$ (right). The pivot scale is set to $k_{\rm t}=0.0099\, \rm{Mpc}^{-1}$ to break the degeneracy. The forecasts assume $f_{\rm sky}=3\%$ and $f_{\rm sky}=40\%$ for the fiducial values of $r=0.01$ and $r=0.05$, respectively.}
\label{fig:rnT}
\end{figure}

Figure~\ref{fig:rnT} shows the projected contours in the $r$--$n_{\rm t}$ plane for CMB-S4, for fiducial values of $r=0.01$ or $r=0.05$. A test of the canonical single-field consistency relation $n_{\rm t}=-r/8$ is unfortunately out of reach. However, a significant bump in the spectrum, as would be produced if a non-vacuum source of gravitational waves dominates the signal \cite{Namba:2015gja} (see Section \ref{subsec:nonvac} for details) would be detectable. 

An upper limit or detection of a stochastic gravitational wave background from CMB-S4 would, on its own, provide only limited information on the overall shape of its spectrum. However, CMB-S4 combined with direct detection techniques such as pulsar timing and laser interferometry can also place limits on $n_{\rm t}$. Recent analysis shows the complementarity between observations over a wide range of frequencies in constraining the shape of the spectrum \cite{Lasky:2015lej,Meerburg:2015zua}. If indirect CMB limits to the high-frequency part of the spectrum are included (coming from its behavior as an additional source of radiation energy density \cite{Smith:2006nka}) the constraints become even tighter. For example, current upper limits to $n_{\rm t}$ (at 95\% CL) using the Parkes Pulsar Timing Array, LIGO, and indirect constraints which place an upper limit on the number of effective radiative degrees of freedom, $N_{\rm eff}-3.046 < 0.31$ \cite{Pagano:2015hma}, are given as a function of $r$ by \cite{Lasky:2015lej}
\begin{equation}
n_{\rm t} < -0.04 \log_{10} \left(\frac{r}{0.11}\right) + 0.36.
\end{equation}
A measurement of $r$ from CMB-S4, along with these other probes of the stochastic gravitational wave background, would allow us to place a firm upper limit to $n_{\rm t}$.  Constraints to the amount of dark radiation from CMB-S4 will further improve the upper limit to $n_{\rm t}$, with the improvement scaling approximately linearly with the CMB-S4 constraint to $N_{\rm eff}-3.046$ \cite{Meerburg:2015zua} (see Section~\ref{sec:constraintsntNeff}).

\subsection{Probing matter and gravitational interactions at the inflationary scale}\label{subsubsec:Interactions}
\label{subsec:sst}

Information about the spectrum of interacting particles relevant during inflation is contained in correlations beyond the power spectrum. Correlators including at least one B-mode will benefit significantly from the improved sensitivity of CMB-S4 and will probe the particles that contribute to sourcing primordial gravitational waves. In particular, the three point correlation $\langle \zeta(\mathbf{k}_1)\zeta(\mathbf{k}_2)\gamma_\sigma(\mathbf{k}_3) \rangle$ can be constrained using $\langle BTT\rangle$, $\langle BTE\rangle $ and $\langle BEE\rangle$.

The details of the tensor-scalar-scalar correlator are contained in the bispectrum $B_{\zeta\zeta\gamma_\sigma}(\mathbf{k}_1,\mathbf{k}_2,\mathbf{k}_3)$, defined by pulling out the appropriate polarization structure associated with the tensor mode:
\begin{equation}
\label{eq:Bsst}
\qquad \langle \zeta(\mathbf{k}_1)\zeta(\mathbf{k}_2)\gamma_\sigma(\mathbf{k}_3) \rangle = (2\pi)^3 \delta(\mathbf{k}_1+\mathbf{k}_2+\mathbf{k}_3) B_{\zeta\zeta\gamma_\sigma}(k_1,k_2,k_3) e_{ij}(\mathbf{k}_3,\sigma)\hat{k}_1^i \hat{k}_2^j,
\end{equation}
where $e_{ij}(\mathbf{k},\sigma)$ is the transverse-traceless polarization tensor. The amplitude and momentum dependence of the bispectrum $B_{\zeta\zeta\gamma_\sigma}$ can be parametrized by \cite{Meerburg:2016ecv}
\begin{equation}
\label{eq:Bzzg}
B_{\zeta\zeta\gamma_\sigma}(k_1,k_2,k_3)= 16 \pi^4 A_{\rm s}^2 \sqrt{r}f_\mathrm{NL}^{\zeta\zeta\gamma} F(k_1,k_2,k_3).
\end{equation} 
The function $F(k_1,k_2,k_3)$ is often referred to as the ``shape,'' indicating which triangle of momentum modes is most strongly coupled.  For example, the equilateral template most strongly couples equal wavelength modes. The ``local" template has significant coupling in a configuration where one mode is substantially longer wavelength than the other two ($k_1\ll k_2\sim k_3$). The simplest models of inflation produce non-Gaussianity of approximately equilateral shape with an amplitude $f^{\zeta \zeta\gamma }_{\rm NL} = \sqrt{r}/16$ \cite{Maldacena:2002vr,Maldacena:2011nz}. 

In Table~\ref{tab:fnl_forecast2} we show the results of forecasts for $\sqrt{r}f_{\rm NL}$ using local and equilateral templates for $F$ and the $\langle BTT\rangle$ correlation. We anticipate similar constraints for $\langle BTE\rangle $ and $\langle BEE\rangle$. The level of non-Gaussianity predicted by the simplest models of inflation is out of reach. A detection of this correlation would be an immediate indication of some deviation from the simple inflationary paradigm \cite{Bordin:2016ruc,Dimastrogiovanni:2015pla}. There are a few known possibilities that would generate a scalar-scalar-tensor bispectrum with larger amplitude and/or different shape: different symmetry patterns (e.g. solid inflation \cite{Endlich:2012pz} or gauge-flation \cite{Maleknejad:2011jw, Adshead:2016iix}), gravitational waves not produced as vacuum fluctuations, or multiple tensors (e.g.\ bigravity) \cite{Bordin:2016ruc}. Any non-zero tensor amplitude could also be sourced by a higher-order massive spin field that couples to two scalars and one graviton. See for example \cite{Hayden:2016xxa} for a very recent discussion of such signatures. A detection would therefore constitute a clear signature of new physics. 

\begin{table*}[t]
  \begin{center}
    \begin{tabular}{ | c || c | c | c | c |}
      \hline
      Type & {\it Planck} & CMB-S4 & Rel. improvement  \\ \hline \hline
      Local & $\sigma(\sqrt{r}f_{\rm NL}) = 15.2$ & $\sigma(\sqrt{r}f_{\rm NL}) = 0.3$ & 50.7\\ \hline 
      Equilateral &  $\sigma(\sqrt{r}f_{\rm NL}) = 200.5$ & $\sigma(\sqrt{r}f_{\rm NL}) = 7.4$ & 27.1\\ \hline 
      Local ($r = 0.01$) & $\sigma(\sqrt{r}f_{\rm NL}) = 15.2$ & $\sigma(\sqrt{r}f_{\rm NL}) = 0.7$ & 25.3\\ \hline 
      Equilateral ($r = 0.01$) &  $\sigma(\sqrt{r}f_{\rm NL}) = 200.8$ & $\sigma(\sqrt{r}f_{\rm NL}) = 14.7$ & 13.7\\ \hline 
    \end{tabular}
  \end{center}
  \caption{Forecasted constraints on local and equilateral shapes sourced by primordial and equilateral correlations of the form $\langle \gamma \zeta\zeta \rangle$ constrained through $\langle BTT \rangle$. \planck\ forecast is based on Blue Book values \cite{Planck:2006aa}, with $f_{\rm sky} = 0.75$. Constraints were derived using the flat-sky approximation as in Ref.~\cite{Meerburg:2016ecv} with $\ell_{\rm min} = 30$, and with no cosmic variance in B.  We expect similar constraints from $\langle BEE \rangle$ and $\langle BTE \rangle$. For $r = 0.01$ \planck\ is still noise dominated, while CMB-S4 is cosmic variance dominated. }
\label{tab:fnl_forecast2}
\end{table*}

\subsection{Distinguishing vacuum fluctuations from other particle physics sources of B modes}
\label{subsec:nonvac}
Although CMB-S4 constraints on the shape of the tensor spectrum and its Gaussianity cannot test the predictions of the simplest inflation models, they will be able to perform the very important function of distinguishing a primordial but non-vacuum dominant source of B modes. In non-minimal models with additional sectors coupled to the inflaton, excitations and particle production associated with other fields during inflation can source additional primordial gravitational waves~\cite{Cook:2011hg,Senatore:2011sp,Barnaby:2012xt}. The new fields and interactions that generate additional tensor fluctuations also generically contribute to the scalar fluctuations, so the non-vacuum B-mode signal is significantly constrained by the observed scalar power spectrum and its high degree of Gaussianity \cite{Barnaby:2012xt,Meerburg:2012id,Ferreira:2014zia,Mirbabayi:2014jqa,Ozsoy:2014sba}. 

In cases where additional sectors are directly coupled to the inflaton (with stronger than gravitational-strength couplings) \planck\ satellite constraints do not allow for the secondary source signal to have an amplitude competitive with the vacuum signal \cite{Ozsoy:2014sba,Mirbabayi:2014jqa}. However, a model with a significant non-vacuum signal can be constructed if the inflationary sector is only gravitationally coupled to a hidden sector containing a light pseudo-scalar and a gauge field during inflation \cite{Barnaby:2012xt,Peloso:2016gqs}. Fluctuations of the light scalar excite fluctuations of the gauge field, which in turn leads to gravitational wave production. To evade constraints from scalar non-Gaussianity, the source field's potential must be adjusted so that the production of gauge field quanta occurs only around the time the modes contributing to the multipoles relevant for the B-mode search leave the horizon \cite{Namba:2015gja}. Then, at the expense of fine-tuning the scales on which production occurs, there exists a range of values for other parameters that can lead to a gravitational-wave signal competitive with the vacuum fluctuations while remaining consistent with existing \planck\ data \cite{Namba:2015gja,Peloso:2016gqs}. For example, the gravitational waves from gauge-field production could be measured at a level of $r=10^{-1}$ with a vacuum contribution of only $r=10^{-4}$. While in that case the determination of the scale of inflation is affected by less than an order of magnitude, adjusting the parameters of the scenario may allow for more dramatic modifications of the relationship between $r$ and the inflationary energy scale given by Eq.~(\ref{eq:Vofr}). 

When secondary production of this sort is large enough to dominate the signal, the predicted gravitational-wave spectrum differs significantly from that of the vacuum fluctuations in several ways. First, the production mechanism is not continuous (to avoid non-Gaussianity constraints) and so sources a B-mode spectrum that is far from scale-invariant. Second, the tensor spectrum is strongly non-Gaussian. Finally, the gravitational waves resulting from the gauge field come with a definite handedness \cite{Anber:2006xt,Sorbo:2011rz} so parity violating TB and EB correlations would be sourced~\cite{Contaldi:2008yz}, and the angular bispectrum of B modes would be dominated by $\ell_1+\ell_2+\ell_3=$ even modes, which would vanish in any theory that respects parity. Of these, the deviation from a flat spectrum is likely to be detected at highest significance by CMB-S4 \cite{Namba:2015gja,Peloso:2016gqs}. 

{\it Through constraints on the shape of the tensor power spectrum, CMB-S4 will be able to distinguish a signal dominated by an inflationary, but non-vacuum, source of primordial gravitational waves.}

Tests for parity violation and the shape of the spectrum are also of broader interest for inflation models whose particle content departs from the minimal scalar field for reasons other than specifically generating non-vacuum gravitational waves. In particular, scenarios in which non-Abelian gauge fields play a significant role in sourcing inflation are closely related to the models discussed above. In chromo-natural inflation and gauge-flation scenarios \cite{Maleknejad:2011jw,Adshead:2012kp,Adshead:2012qe,Adshead:2013qp,Adshead:2013nka,Dimastrogiovanni:2012st,Dimastrogiovanni:2012ew}, the central ingredient is a homogeneous and isotropic, flavor-space-locked gauge field that helps slow the roll of the inflaton or else is the inflaton itself. For a non-Abelian field with SU(2) symmetry, this means that the three flavor gauge vector potentials are mutually orthogonal in space. The stress-energy of this configuration could leave a unique imprint on a spectrum of primordial gravitational waves, which would be transferred to the B-mode spectrum in the CMB. The non-Abelian nature of the field introduces a preferred handedness onto this medium, leading to an enhancement of left (or right) circularly polarized gravitational waves. Again this would lead to parity-violating EB and TB correlations~\cite{Lue:1998mq,Gluscevic:2010vv} or parity violating higher $n$-point functions. If this process takes place in the post-inflationary environment, the gauge field could further impress a periodic modulation on the gravitational wave spectrum \cite{Bielefeld:2014nza,Bielefeld:2015daa}. Although the basic chromo-natural and gauge-flation models have been ruled out \cite{Namba:2013kia}, these unique features are expected to be generic to any viable variation on these scenarios and would be constrained by CMB-S4. 
As parity violation is difficult to detect at high significance for these scenarios, but is also interesting for post-inflationary physics, forecasts for CMB-S4 can be found in the discussion of cosmic birefringence in Chapter 6.

Post-inflationary phase transitions themselves have also been proposed as a source of nearly scale-invariant gravitational waves detectable through CMB polarization (and direct detection) \cite{Krauss:1991qu,JonesSmith:2007ne,Giblin:2011yh,Figueroa:2012kw,Fenu:2013tea}. Even for a spectrum that matches the inflationary result on small scales, any such signal can in principle be distinguished from the inflationary expectation by the absence of super-horizon correlations at the time of recombination. A framework to extract specifically this part of the signal was proposed in Ref.~\cite{Baumann:2009mq} and could be applied to robustly extract the component of any signal that must come from physics outside of the hot big bang paradigm. Existing forecasts in the literature \cite{Lee:2014cya} indicate that a ground-based survey alone will not be able to detect super-horizon correlations at high significance if $r$ is much below $0.1$. But, if CMB-S4 does make a detection, this physics could be in reach of an eventual satellite mission.

\subsection{Constraining alternatives to inflation}
\label{subsec:alternatives}

Vacuum fluctuations during inflation provide a simple, elegant, and compelling mechanism to create the initial seeds required for structure formation. 
One of inflation's most robust predictions is an adiabatic, nearly
scale-invariant spectrum of scalar density perturbations.
This prediction is in excellent agreement with observations, especially
considering the need to account for a small deviation from exact scale
invariance.  However, it is disputable whether these observations can be
considered a proof that inflation actually occurred (as has been discussed
since inflation was first proposed).
Clearly, a fair evaluation of the status of inflation requires the
consideration of competing theories and the hope to find experimental distinctions between inflation and these alternatives.

Leading alternatives to inflation can be classified into two primary categories based on the 
way in which they account for the observed causality of the scalar density fluctuations.
``Bouncing cosmologies'' rely on an initially cold, large universe and a subsequent phase of
slow contraction. This is then followed by a bounce that leads
to an expanding and decelerating FRW cosmology.  The most well studied examples are provided by 
``ekpyrotic'' or cyclic models \cite{Khoury:2001bz,Khoury:2001wf}
and more recently ``matter bounce'' models \cite{Brandenberger:2012zb,Cai:2014jla,deHaro:2015wda}.
The second class of alternatives to inflation arises from models that invoke a loitering phase of the cosmic expansion prior to the hot big bang---with 
string gas cosmology \cite{Brandenberger:1988aj,Tseytlin:1991xk,Battefeld:2005av} providing an example. 

A detailed critique of these alternatives and their relevance to the science case for a
near-term CMB-based mission was presented in Appendix B of the ``CMBPol
Mission Concept Study'' \cite{Baumann:2008aq}.  
Since that publication, these alternative approaches to inflation have received considerable attention;
however, as science drivers for the CMB-S4 mission there are two important points to re-emphasize.
\begin{itemize}
\item These alternatives invoke novel and incompletely understood physics to solve the problems associated with standard big bang cosmology.  This implies important theoretical challenges that have to be addressed carefully before the models mature into compelling alternatives to inflation.

\item Most or all of the alternatives to inflationary cosmology predict negligible tensors on CMB scales.
This strengthens the case for considering B modes as a ``smoking gun'' of inflation.  It should be considered an important
opportunity to use CMB observations to constrain all known alternatives to inflation.
\end{itemize}

One property that is shared by many (if not all) alternatives to inflation is that they require a violation of the Null Energy Condition (NEC). 
Such a violation typically implies the existence of catastrophic instabilities and/or fine-tuning of initial conditions.
This presents an important challenge for alternatives to inflation, but it does not imply that alternatives are impossible to realize. 
An example of a stable bounce violating the NEC was put forward in Ref.~\cite{Creminelli:2006xe} and then used in the ``new ekpyrotic'' scenario in Refs.~\cite{Buchbinder:2007ad,Creminelli:2007aq}. Although this model is consistent at the level of effective field theory, it is not clear whether it is possible to find a UV completion for it. 
This is a very important issue because the quantization of the new ekpyrotic theory, prior to the introduction of a UV cutoff and a UV completion, leads to a catastrophic vacuum instability \cite{Kallosh:2007ad}.  Similar challenges arise in models like string gas cosmology where NEC violation is required to exit the loitering phase to a radiation-dominated universe
\cite{Brustein:1994kw,Kaloper:1995ey,Kaloper:1995tu,Kaloper:2007pw}.  Whether such obstacles can be overcome is an area of ongoing research.

However, despite the theoretical challenges in understanding the background evolution, it has been argued that many 
of the observational predictions of such alternative models are independent of these issues.
Most notable is that all known alternative constructions seem to predict negligible tensor modes on large scales.
This was an early prediction of ekpyrotic models, and appears true as well for the more recently 
studied matter bounce models when constraints on scalar non-Gaussianity are also taken into consideration \cite{Quintin:2015rta}. It is not yet clear if an observable amplitude of non-vacuum primordial gravitational waves could be sourced during the contracting phase, but see Ref.~\cite{Ben-Dayan:2016iks} for some recent work in that direction.

{\it A detection of primordial gravitational waves with a spectrum consistent with vacuum fluctuations would rule out all currently proposed alternatives to inflation.}

\subsection{Constraints on the graviton mass}

Theories of massive gravity come in many flavors (see e.g.\ Refs.~\cite{Dubovsky:2004sg,Hinterbichler:2011tt}), and their predictions in the scalar sector differ significantly. However, by definition, the dispersion relation for the graviton in all of them is
\begin{equation}
\omega^2=p^2+m_g^2\,,
\end{equation}
where $p$ is the physical momentum and $m_g$ the possibly time-dependent graviton mass. As a consequence, gravitational waves necessarily have frequencies $\omega>m_g$. A detection of primordial B-mode polarization on angular degree scales may be considered as a detection of gravitational waves with frequencies $\omega\sim H_{\rm rec}$ through the quadrupole they produce in the primordial plasma, where $H_{\rm rec}\simeq 3\times 10^{-29}$~eV is the Hubble parameter at recombination. A detection then implies a model-independent bound $m_g<H_{\rm rec}$ or 
\begin{equation}
m_g< 3\times 10^{-29}{\mbox{ eV}}\,.
\end{equation}
If the graviton mass is time-dependent, this should be interpreted as a constraint on its mass around the time of recombination.

Because the perturbations in the primordial plasma before and around recombination are linear, the effect of the graviton mass is straightforward to incorporate by a simple modification of the field equation for tensor metric perturbations so that the above argument can be made more quantitative. The equation of motion for the transverse-traceless metric perturbation $\gamma$ takes the same form as for a minimally coupled massive scalar field
\begin{equation}
\label{massive}
\ddot{\gamma}_k(t)+3{\dot a\over a} \dot{\gamma}_k(t)+\left(m_g^2+\left(\frac{k}{a}\right)^2\right) \gamma_k(t)=0\,.
\end{equation}
Here $k$ is the comoving momentum of the metric perturbation, and the background cosmological metric is
\begin{equation}
ds^2= -dt^2+a^2(t)d{\bf x}^2\,.
\end{equation}
The consequences of this modification are discussed in detail in Ref.~\cite{Dubovsky:2009xk}. The most important consequence is that superhorizon modes start to oscillate around the time $t_m$ when $H(t_m)=m_g$, and their amplitude subsequently redshifts as $a^{-3/2}$. In contrast, in the massless case all modes remain frozen until they enter the horizon. This results in a suppression of the amplitude of primordial B modes for $m_g\gg H_{\rm rec}$, and a detection of B modes would rule out this possibility. For masses around $H_{\rm rec}$, there is no suppression, but the angular power spectra are modified by the presence of a graviton mass, and a detection of primordial B-mode polarization would allow a measurement of the graviton mass. 
 
Weak lensing currently constrains the mass of the graviton to be $m_g<6\times 10^{-32}\,{\rm eV}$. This bound assumes that the dispersion relation of the scalar modes is modified. The limits discussed here are weaker but have the advantage that they are model-independent and directly constrain tensor modes. For comparison, the current model-independent bounds on the graviton mass arise from the indirect detection of $\sim 3\times 10^{-5}$~Hz gravitational waves through the timing of the Hulse-Taylor binary pulsar~\cite{Finn:2001qi}, and the bound on the difference in arrival times for gravitational waves with different frequencies in the recent direct detection of astrophysical gravitational waves with LIGO~\cite{Abbott:2016blz}. The resulting bounds are $m_g\lesssim 10^{-19}{\mbox{ eV}}$ and \mbox{$m_g\lesssim 10^{-22}{\mbox{ eV}}$}, respectively.  

{\it A detection of B-mode polarization consistent with expectations in the context of general relativity would improve current model-independent bounds on the mass of the graviton by more than six orders of magnitude.} 

We note that this improvement is calculated assuming measurements of the degree-angular-scale B modes only. Measurements of B-mode polarization on the largest angular scales would further strengthen the bound.

\section{Improved constraints on primordial density perturbations}
\label{sec:scalar}
All current data are consistent with primordial density perturbations that are adiabatic, Gaussian, and nearly scale-invariant. With its sensitivity and angular resolution, CMB-S4 will significantly improve current constraints on the scale dependence of the primordial power spectrum of scalar perturbations, on departures from Gaussianity, and on departures from adiabaticity. In fact, it will measure anisotropies in both the temperature and E-mode polarization of the CMB to cosmic variance over the entire range of multipoles that is not contaminated by foregrounds. As a consequence, it will place the strongest constraints achievable by any ground-based CMB experiment on observables that benefit from the number of modes measured, such as the primordial power spectrum and higher-order correlations.

\subsection{The scalar power spectrum}
The density perturbations are close to scale-invariant but not exactly so. In the context of $\Lambda$CDM, \planck\ has measured the scalar spectral index to be $n_{\rm s}=0.9677\pm0.0060$ and has established $n_{\rm s}-1<0$ at more than $5\,\sigma$. Realistic configurations of CMB-S4 will roughly decrease the uncertainty on the spectral index by a factor 2. To be specific, assuming a configuration without a site in the northern hemisphere, so that $40\%$ of the sky can be used after masking, a white noise level of $1\,\mu$K-arcmin in temperature and a modest angular resolution of $3$ arcmin, CMB-S4 will improve current constraints to $\sigma(n_{\rm s})=0.0019$. A configuration that includes a site in the northern hemisphere so that $60\%$ of the sky could be retained after masking the same noise levels and angular would further improve the constraints to $\sigma(n_{\rm s})=0.0017$. These improvements will provide valuable constraints on the space of inflationary models. 

As mentioned in section~\ref{sec:upperLimits}, a measurement of the running of the scalar spectral index with a precision of a few parts in ten thousand would allow a measurement of $p$ in Eq.~(\ref{eq:nsassump}), or equivalently $\mathcal{N}_\ast$. Such precision cannot be achieved with CMB-S4. For typical configurations of CMB-S4 the constraints on the running would improve to $\sigma(n_{\rm run})=0.002-0.003$. 

Models of inflation that achieve super-Planckian inflaton displacements from repeated circuits of a sub-Planckian fundamental period may give rise to oscillatory features in the spectrum of primordial perturbations. The features may arise either from instanton effects or from periodic bursts of particle or string production. A search for such features is well motivated, even though the amplitude is model dependent and may be undetectably small. A detection would provide clues about the microscopic origin of the inflaton, while the absence of a detection can constrain the parameter space of these models in interesting ways. Again assuming a configuration with sites only in the southern hemisphere with a noise level of $1\,\mu$K-arcmin and an angular resolutions between $1$ and $3$ arcmin, CMB-S4 would tighten the constraints on the amplitude of features in the primordial power spectrum by a factor of about 2.

Other physical effects during inflation could lead to weak structure in the observed power spectrum, e.g. by changing the equation of state during inflation \cite{Achucarro:2014msa}. Because of the stringent constraints on the minimum number of e-folds, such modifications cannot last very long and associated features only affect a small range of scales. It is therefore unlikely that CMB-S4 will significantly improve constraints on these type of features unless they are on very small scales.

\subsection{Higher-order correlations of scalar modes}
\label{subsec:scalarNG}

Unlike the scalar and tensor power spectra, higher-order correlations of the scalar modes are directly sensitive to the dynamics and field content responsible inflation (and alternatives).  While non-Gaussian correlations are small in conventional single-field slow-roll inflation, there exist many other possibilities for the nature of inflation that give strikingly different predictions when we move beyond the power spectrum.  The constraints on non-Gaussianity from the \wmap\ and \planck\ satellites currently place the most stringent limits on a wide range of mechanisms for inflation; however, these measurements are not sufficiently sensitive to suggest a particular mechanism is favored by the data.  As our understanding of inflation is continually refined, there is an associated need to improve our understanding of the underlying dynamics directly through constraints on higher order correlations.  

The space of non-Gaussian signals from inflation can broadly be grouped into two conceptual categories that generate distinguishable features in the correlation functions. These are signals that: (1) indicate non-trivial self-interactions of the effective inflaton fluctuation; or (2) indicate interactions with degrees of freedom other than the inflaton. These two categories can be further divided up. For the first category self-interactions that respect the time-translation invariance during inflation lead to qualitatively different predictions from self-interactions that violate it. For the second category the signatures qualitatively depend on the mass of the additional degrees of freedom. Additional light degrees of freedom can fluctuate significantly and may have large self-interactions. Heavy degrees of freedom do not fluctuate appreciably but may come into existence by quantum fluctuations and decay into inflaton quanta, generating non-Gaussian correlations. Alternatively, they may become excited by the dynamics of the inflaton and their backreaction on the inflationary dynamics may lead to non-Gaussian correlations. 

Constraints on non-Gaussianity are often expressed in terms of the correlator of three scalar modes, described by the bispectrum $B_{\zeta}(\mathbf{k}_1,\mathbf{k}_2,\mathbf{k}_3)$, defined by
\begin{equation}
\label{eq:Bsss}
\langle\zeta(\mathbf{k}_1)\zeta(\mathbf{k}_2)\zeta(\mathbf{k}_3)\rangle=(2\pi)^3\delta(\mathbf{k}_1+\mathbf{k}_2+\mathbf{k}_3)B_{\zeta}(\mathbf{k}_1,\mathbf{k}_2,\mathbf{k}_3)\,.
\end{equation}
The structure of particle interactions relevant for inflation provides both a general organizing principle for this functional space and several specific, well motivated forms of the bispectrum that can be explicitly compared to data. Here we briefly review the classification of scalar non-Gaussianity from inflation and then present forecasts for improvements from CMB-S4 on a few of the standard bispectral templates. We also comment on non-Gaussian signatures that are especially relevant for large-field inflation.

All bispectra that come from fluctuations of the field that drives inflation (``single-clock'' scenarios) most strongly couple Fourier modes of similar wavelengths. The ``squeezed limit'' of these bispectra (the coupling of modes $k_1\ll k_2\sim k_3$) is very restricted. 
A large fraction of the parameter space for scenarios involving interactions during inflation that respect the underlying shift symmetry (i.e.\ are approximately scale-invariant) is captured by two templates for the bispectrum, the so-called equilateral~\cite{Babich:2004gb} and orthogonal shapes~\cite{Senatore:2009gt}. 
This may include scenarios in which inflaton fluctuations have non-trivial self-interactions~\cite{Silverstein:2003hf,ArkaniHamed:2003uz,Alishahiha:2004eh,Chen:2006nt,Cheung:2007st,Senatore:2009gt} or couplings between the inflaton and other (potentially massive) degrees of freedom~\cite{Chen:2009zp,Tolley:2009fg, Cremonini:2010ua, Achucarro:2010da,Baumann:2011nk,Barnaby:2011pe,Arkani-Hamed:2015bza}.  While many of these models lead to different shapes in detail, the signal-to-noise ratio is often dominated by equilateral configurations.  One of the important features of these shapes is that single-field slow-roll inflation necessarily produces $f_{\rm NL}^{\rm equil} < 1$~\cite{Creminelli:2003iq} and therefore any detection of $f_{\rm NL}^{\rm equil} \geq 1$ would rule out this very wide class of popular models.  Furthermore, a detection would imply that inflation is a strongly coupled phenomenon and/or involved more than one field~\cite{Baumann:2014cja,Alvarez:2014vva,Baumann:2015nta}.  These possibilities could be distinguished, in principle, with further observations.  Current constraints on the equilateral and orthogonal shapes are $f_{\rm NL}^{\rm equil} = -4 \pm 43$ and $f_{\rm NL}^{\rm ortho} = -26 \pm 21$, both (68\% CL)~\cite{Ade:2015ava}. In single-field inflation, the amplitude of the non-Gaussianity typically suggests a new energy scale, $M_{\rm s}$, such that $f_{\rm NL}^{\rm equil} \propto A_{\rm s}^{-1/2} \, (H/M_{c_{\rm s}})^2$~\cite{Cheung:2007st,Baumann:2011su}. At this energy scale self-interactions become strongly coupled and current limits translate into $M_{c_{\rm s}}>\mathcal{O}(10)H$.  In the presence of additional hidden sectors, the amplitude of non-Gaussanity scales with the strength of the coupling between the inflaton and these additional fields, usually suppressed by an energy scale $\Lambda$.  Current limits give $\Lambda  > (10{-}10^{5}) H$~\cite{Green:2013rd,Assassi:2013gxa}, with the variation depending mostly on the dimension of the operator coupling the two sectors.  For $r > 0.01$, these constraints require some of the interactions to be weaker than gravitational.  The improvements from CMB-S4 would further tighten existing constraints on a wide variety of interactions of the inflaton with itself and any other fields that are excited during inflation.    

When light degrees of freedom other than the inflaton contribute to the observed scalar fluctuations, a much wider degree of coupling between modes of very different wavelengths is allowed. Historically, the most well-studied template of this type comes from the ``local'' model, which couples short wavelength modes $k_2\sim k_3$ with equal strength to all long wavelength modes $k_1$. A detection of this shape (and any non-trivial squeezed limit coupling more generally) would rule out all models of single-clock inflation \cite{Creminelli:2004yq}. In addition, such a signal would open the door to significant cosmic variance on all scales from coupling of fluctuations within our observed volume to any super-Hubble modes \cite{Nelson:2012sb,LoVerde:2013xka,Nurmi:2013xv}. Indeed, there would be room for a significant shift between the observed amplitude of scalar fluctuations (and so the observed $r$) and the mean value of fluctuations on much larger scales \cite{Bonga:2015urq}. Any scenario that predicts local non-Gaussianity together with fluctuations on scales much larger than our observed volume predicts a probability distribution for our observed $f_{\rm NL}^{\rm local}$, but many well-motivated scenarios also predict a small mean value; these include the simplest modulated reheating scenario \cite{Zaldarriaga:2003my} and ekpyrotic cosmology \cite{Lehners:2009ja}, both of which predict mean values of $f_{\rm NL}^{\rm local}\sim5$. 
Currently the strongest constraints on the local shape come from the \planck\ 2015 temperature and polarization analysis that finds $f_{\rm NL}^{\rm local} = 0.8 \pm 5.0$~\cite{Ade:2015ava}. Conservatively assuming a configuration without a site in the northern hemisphere, leaving around $40\%$ of the sky unmasked, and assuming a white noise level of $1\,\mu$K-arcmin in temperature and an angular resolution of $1$ arcmin, the improvement expected of CMB-S4 over current limits is slightly more than a factor of 2. This is not sufficient to reach the interesting theoretical threshold around $|f_{\rm NL}^{\rm local}|\lesssim 1$~\cite{Alvarez:2014vva}, but will still reduce the space of viable models or hint at a detection. CMB-S4 could, for example, provide hints for the mean level of non-Gaussianity expected from modulated reheating scenario or ekpyrotic cosmology at roughly the $2\,\sigma$ level. The simplest curvaton scenario, which predicts $f_{\rm NL} = -5/4$ \cite{Lyth:2001nq}, will unfortunately be out of reach. Large-scale structure surveys (e.g.\ \cite{Dore:2014cca}) may eventually achieve constraints $\sigma_{f_{\rm NL}}\sim\mathcal{O}(1)$; those observations of the inhomogeneities in the late Universe would be very complementary to the results of CMB-S4.

Table~\ref{tab:fnl_forecast} shows the forecasted constraints on the local, equilateral, and orthogonal shapes from CMB-S4. Our convention for extracting a normalization of the amplitude, $f_{\rm NL}$, follows \cite{Babich:2004gb} and \cite{Ade:2013ydc}, i.e., 
\begin{equation}
B_{\zeta}(k_1,k_2,k_3) = \frac{3}{5} (4 \pi^4) 2 A_{\rm s}^2 f_{\rm NL} F(k_1,k_2,k_3) ,
\end{equation}
with e.g. 
\begin{equation}
F^{\rm local}(k_1,k_2,k_3) = \frac{1}{k_1^{4-n_{\rm s}}k_2^{4-n_{\rm s}}} + {\rm 2\;perms.}
\end{equation}
Local type non-Gaussianities benefit from large scales, and as much as $40\%$ of the signal is lost if the modes $\ell<30$ are not available. 
Ideally, large-scale information from \planck\ \cite{Ade:2015ava} should be included to place the best constraints on local non-Gaussianities. Including \planck\ low-$\ell$ modes (using $f_{\rm sky} = 0.75$ \cite{Ade:2015ava} to determine the noise level, and $f_{\rm sky}=0.4$ for the maximal overlap) we can improve the forecasted bounds on local type non-Gaussianities by a factor of 2.5. Equilateral and orthogonal non-Gaussianities are not affected by excluding the lowest multipoles. Note that our forecast, using \planck\ Blue Book\cite{Planck:2006aa}
values, deviates slightly from the actual bounds on non-Gaussianities obtained in Ref.~\cite{Ade:2015ava}. The expected factor of improvement over {\it Planck}-only is somewhere between 2.1 and 2.5 for all shapes considered. Information saturates beyond $\ell_{\rm max} = 4000$ for all shapes for an experiment with a $1^\prime$ beam. 

While the improvement relative to \planck\ is somewhat modest, these are likely to be the strongest constraints on the equilateral and orthogonal templates that will be available for the foreseeable future.  As limits on non-Gaussianity provide a unique and fundamental insight into the nature of inflation, this increased sensitivity would provide a non-trivial improvement in our understanding of the Universe.  Ultimately, more dramatic improvements will require going beyond the CMB, but CMB-S4 will remain the strongest constraint on the equilateral and othorgonal templates until reliable and competitive constraints from large scale structure become a reality.  

\begin{table*}[t]
  \begin{center}
    \begin{tabular}{ | c || c | c | c | c |}
      \hline
      Type & {\it Planck} actual (forecast) & CMB-S4 & CMB-S4 + low-$\ell$ {\it Planck} & Rel. improvement \\ \hline \hline
      Local & $\sigma(f_{\rm NL}) = 5\, (4.5)$ & $\sigma(f_{\rm NL}) = 2.6$ &  $\sigma(f_{\rm NL}) = 1.8$ & 2.5\\ \hline 
      Equilateral &  $\sigma(f_{\rm NL}) = 43\,(45.2)$ & $\sigma(f_{\rm NL}) = 21.2$ &  $\sigma(f_{\rm NL}) = 21.2$ & 2.1\\ \hline 
      Orthogonal &  $\sigma(f_{\rm NL}) = 21\, (21.9)$ & $\sigma(f_{\rm NL}) = 9.2$ &  $\sigma(f_{\rm NL}) = 9.1$ & 2.4\\ \hline 
    \end{tabular}
  \end{center}
  \caption{Constraint forecasts for several well-motivated non-Gaussian shapes using T and E modes. We show both the actual \planck\ results and what our forecast predicts given \planck\ Blue Book values, with $f_{\rm sky} = 0.75$. The table shows that we need to include low-$\ell$ information from \planck\ for local type non-Gaussianities. CMB-S4 is assumed to have $f_{\rm sky} = 0.4$, T-noise = 1$\,\mu$K-arcmin and E-noise = $\sqrt{2}\,\mu$K-arcmin and a beam of $1^\prime$, and $\ell_{\rm min} = 30$. The relative improvement factor compares forecasted CMB-S4 to forecasted \planck\ uncertainties.}
  \label{tab:fnl_forecast}
\end{table*}

Perhaps of special interest for CMB-S4 are non-Gaussian signatures that would be expected in models of large-field inflation. For example, in models in which the inflaton is an axion, there is only an approximate discrete shift symmetry. In that case instanton contributions to the potential and periodic bursts of particle or string production naturally lead to periodic features in the bispectrum. If moduli in the underlying string constructions do not evolve appreciably, instanton contributions lead to oscillations with a constant amplitude in the logarithm of $k$. In general, moduli evolve during inflation and cause a drift in the frequency and a scale-dependent amplitude~\cite{Flauger:2014ana}. At present, these shapes have not yet been constrained systematically. Often these contributions will lead to counterparts in the power spectrum and are expected to be detected there first~\cite{Behbahani:2011it}, but this need not be the case~\cite{Behbahani:2012be}. A first attempt has been made~\cite{Ade:2015ava} to look for resonant and local features in the bispectrum, and a more dedicated analysis is underway. Since features in the power spectrum and the bispectrum generally contain correlated parameters \cite{Meerburg:2009ys,Achucarro:2010da, Flauger:2010ja,Meerburg:2015yka,Achucarro:2012fd,Palma:2014hra}, statistical methods have been developed to use constraints from both the power spectrum and the bispectrum to further limit the model space \cite{Fergusson:2014hya,Fergusson:2014tza,Meerburg:2015owa}. Signatures of higher-order massive spin fields \cite{Arkani-Hamed:2015bza,Chen:2015lza} would also lead to a bispectrum with decaying features, which will not be present in the power spectrum.

\section{Spatial curvature}
\label{sec:curvature}

Despite the fact that inflation drives the spatial curvature to zero at the level of the background evolution, it predicts small, but non-zero curvature for a typical observer. The curvature measured in a Hubble patch receives contributions from long wavelength perturbations and is expected to be $|\Omega_k|\lesssim10^{-4}$. A measurement exceeding this expectation would contain important information about the process responsible for inflation. In particular, if $|\Omega_k|$ is found to be considerably larger than this value, it would tell us that the inflaton was not slowly rolling when scales slightly larger than our observable horizon exited the horizon. Furthermore, observations of large negative $\Omega_k$ would falsify eternal inflation, while observation of positive and large $\Omega_k$ would be consistent with false vacuum eternal inflation~\cite{Guth:2012ww,Kleban:2012ph}.

Current constraints on this parameter from the CMB alone are $\Omega_k= 0.005^{+0.016}_{-0.017}$ \cite{Ade:2015xua}. Including baryon acoustic oscillation (BAO) data tightens the bound to $\Omega_k=0.000\pm0.005$. 
In the context of a 1-parameter extension of $\Lambda$CDM that includes curvature, the constraints on $\Omega_k$ only weakly depend on the resolution and sensitivity for the range considered for CMB-S4. For sensitivities between $1$ and $3\,\mu K$-arcmin and a resolution between $1$ and $3$ arcmin, CMB-S4 together with low-$\ell$ \planck\ data will place $1\,\sigma$ limits of $3\times10^{-3}$.
Adding the DESI measurements of the BAO standard ruler (at redshifts of 0 to 1.9) significantly reduces the uncertainty to a combined $1\,\sigma$ limit of around $7.1\times10^{-4}$. CMB-S4 is therefore unable to measure curvature levels typical of slow-roll eternal inflation and any detection of curvature by CMB-S4 would have profound implications for the inflationary paradigm.

\section{Isocurvature}
Measurements of CMB temperature and polarization power spectra indicate that the primordial initial conditions are adiabatic, that is, spatial entropy fluctuations vanish:
\begin{align}
S_{i \gamma}\equiv \frac{\delta n_{i}}{n_{i}}-\frac{\delta n_{\gamma}}{n_{\gamma}} =0.
\end{align}
Here the species label $i$ can denote baryons, cold dark matter (CDM), or neutrinos. Number densities are denoted by $n_{i}$ and perturbations in them by $\delta n_{i}$.

This can also be expressed in a gauge-invariant way for any two species $i,j$:
\begin{equation}
S_{ij}=3(\zeta_i-\zeta_j),
\end{equation}
where $\zeta_i = -\Psi - H \delta\rho_i/\dot{\rho}_i$.

Adiabatic perturbations are produced if the initial perturbations in all species are seeded by the inflaton. If fluctuations are also sourced by a second field, the initial conditions are a mixture of adiabatic and entropy (or isocurvature) perturbations, for which $S_{i\gamma}\neq 0$. These initial conditions determine the acoustic peak structure and large-scale amplitude of CMB anisotropies, as well as large-scale structure statistics \cite{Bond:1984fp,Kodama:1986fg,Kodama:1986ud,Hu:1994jd,Moodley:2004nz,Bean:2006qz}. Observations can thus probe additional light fields present during inflation. 
\label{sec:isosec}
Each species can carry isocurvature perturbations in its density (e.g.\ Refs~\cite{Bucher:1999re,Bucher:2004an,Moodley:2004nz}).\footnote{Neutrinos can also carry velocity isocurvature, but this mode is not well motivated in inflationary models.} Indeed, the modes of the perturbation evolution equations correspond to adiabatic, CDM density isocurvature (CDI), baryon density isocurvature (BDI), neutrino density isocurvature (NDI), and neutrino velocity isocurvature (NVI) initial conditions. 

Data from \wmap\ \cite{Dunkley:2008ie}, \planck\ \cite{Planck:2013jfk,Ade:2015lrj}, and other experiments \cite{Enqvist:2000hp,MacTavish:2005yk} indicate that perturbations are predominantly adiabatic. The limits can be stated in terms of the fractional primordial power in each isocurvature mode:\begin{equation}
\beta\equiv \frac{P_{S_{i\gamma}}(k)}{P_{S_{i\gamma}}(k)+P_{\zeta\zeta}(k)}.
\end{equation}

In this section we focus on two specific scenarios for isocurvature: the curvaton; and compensated isocurvature perturbations (CIPs). Disucssion of axion-type isocurvature is deferred to Section~\ref{axion_iso}.

The curvaton scenario is an alternative to single-field inflationary models in which a sub-dominant second field $\sigma$ acquires vacuum fluctuations during inflation, becomes more important later, sources $\zeta$, and then decays \cite{Mollerach:1989hu,Mukhanov:1990me,Moroi:2001ct,Lyth:2001nq,Lyth:2002my}. Curvaton candidates include sneutrinos, string moduli, and others \cite{Postma:2002et,Kasuya:2003va,Ikegami:2004ve,Mazumdar:2004qv,Allahverdi:2006dr,Papantonopoulos:2006xi,Mazumdar:2010sa,Mazumdar:2011xe}. Depending on whether a species $i$ (or its set of quantum numbers) is produced by, before, or after curvaton decay, perturbations in $i$ are offset from $\zeta$, leading to isocurvature perturbations \cite{Lyth:2001nq,Lyth:2002my,Gordon:2002gv}:

\begin{eqnarray}
S_{i \gamma}=\left\{\begin{array}{ll}-3\zeta-3(\zeta_{\gamma}-\zeta),&\mbox{if $i$ is produced before $\sigma$ decay,}\\3\left(\frac{1}{r_{\rm D}}-1\right)\zeta-3(\zeta_{\gamma}-\zeta),&\mbox{if $i$ is produced by $\sigma$ decay},\\ -3(\zeta_\gamma-\zeta),&\mbox{if $i$ is produced after $\sigma$ decay},\end{array}\right.\label{eq:strew}.
\end{eqnarray} Here $\zeta_{i}$ is the density perturbation in $i$ on surfaces of constant curvature. The parameter $r_{\rm D}$ is the fractional energy density in the curvaton when it decays. 

The mixture of isocurvature modes is determined by whether or not baryon number, lepton number, and CDM are produced before, by, or after curvaton decay. Curvaton-type isocurvature is distinct from axion isocurvature, because it is correlated (or anti-correlated) with $\zeta$. If lepton number is produced by curvaton decay, the lepton chemical potential $\xi_{\rm lep}$ is important in setting the amplitude of NDI modes \cite{Lyth:2002my,Gordon:2003hw,DiValentino:2011sv}:
\begin{equation}
S_{\nu \gamma}=
-\frac{135}{7}\left(\frac{\xi_{\rm lep}}{\pi}\right)^2\zeta_{\gamma}.\end{equation}
There are $27$ distinct curvaton decay scenarios, since baryon number, lepton number, and CDM could each be produced before, by, or after curvaton decay. Viable models are those in which one of baryon number or CDM is produced by curvaton decay, and those in which \textit{both} baryon number and CDM are produced after curvaton decay. For curvaton-decay scenarios, we use the notation ($b_{y_{\rm b}}$, $c_{y_{\rm c}}$, $L_{y_{\rm L}}$), where the subscripts run over $y_i\in \{{\rm before, by, after}\}$. Here $b$ denotes baryon number, $c$ denotes CDM, and $L$ denotes lepton number. For example, $(b_{\rm before}, c_{\rm by}, L_{\rm by})$ is a model in which baryon number is produced before curvaton decay, CDM by curvaton decay, and lepton number by curvaton decay.

Current isocurvature limits favor values of $r_{\rm D}\simeq 1$, except for models in which baryon number is produced by curvaton decay and CDM before (or vice versa), which favor central values of $r_{\rm D} \simeq 0.16$ ($r_{\rm D} \simeq 0.84$). 

The current limits \cite{Smith:2015bln} on $r_{\rm D}$ are shown in Table~\ref{limits_rd}, along with a forecast of CMB-S4's sensitivity to $r_{\rm D}$ via isocurvature modes. There is dramatic improvement in the $(b_{\rm by},c_{\rm before},L_{\rm by})$ and $(b_{\rm before},c_{\rm by},L_{\rm by})$ scenarios because of the accompanying NDI perturbations. One unusual case is the $(b_{\rm after},c_{\rm after}, L_{y_{\rm L}})$ scenario. Here the isocurvature component just constrains the degenerate combination \cite{Smith:2015bln} $\chi_{\rm D} \equiv [1+\xi_{\rm lep}^2/(\pi^2) \left(1/r_D -1\right)]^{-1}$, while the independent constraint to $\xi_{\rm lep}^{2}$ is driven by the CMB limit on the effective number of relativistic degrees of freedom ($N_{\rm eff}$).

\begin{table}[htbp!]
\begin{center}
\begin{tabular}{| c || c | c |}
\hline
{\rm Isocurvature scenario} &  \planck & CMB-S4 \\ \hline \hline

  & $ \Delta r_D/r_{D}^{\rm adi}$ &$ \Delta r_D/r_{D}^{\rm adi}$\\\hline
$(b_{\rm by},c_{\rm before},L_{y_{\rm L}})$ & $0.03$&$0.005$\\
$(b_{\rm before},c_{\rm by},L_{y_{\rm L}})$ &  $0.01$ &$0.004$\\
$(b_{\rm by},c_{\rm after},L_{y_{\rm L}})$ &  $0.04$&$0.01$\\
$(b_{\rm after},c_{\rm by},L_{y_{\rm L}})$ & $0.008$&$0.002$\\
$(b_{\rm by},c_{\rm by},L_{y_{\rm L}})$ &  $0.007$&$0.002$\\ \hline\hline
& $\Delta \chi_{\rm D}/\chi_{\rm }^{\rm adi}$&$\Delta \chi_{\rm D}/\chi_{\rm }^{\rm adi}$ \\\hline
$(b_{\rm after},c_{\rm after},L_{y_{\rm L}})$ & $0.003$&$0.0004$ \\ \hline \hline
 &  $\Delta \xi^{2}_{\rm lep}$ &$\Delta \xi^{2}_{\rm lep}$\\\hline
$(b_{\rm by},c_{\rm before},L_{\rm by})$ &$0.02$ &$0.002$\\
$(b_{\rm before},c_{\rm by},L_{\rm by})$ &$0.4$  & $0.04$\\
$(b_{\rm by},c_{\rm after},L_{\rm by})$ &$0.3$  &$0.04$\\
$(b_{\rm after},c_{\rm by},L_{\rm by})$ & $0.3$&$0.04$\\
$(b_{\rm by},c_{\rm by},L_{\rm by})$ & $0.3$ & $0.04$\\
$(b_{\rm after},c_{\rm after},L_{\rm by})$ & $0.3$ & $0.04$\\
\hline
\end{tabular}
\end{center}
\caption{Isocurvature constraints on $r_{\rm D}$ and $\xi_{\rm lep}^{2}$, both at ($95\%$ CL) using \planck\ TT+BAO+LowP data \cite{Smith:2015bln} in viable curvaton decay-scenarios, and Fisher forecasts for CMB-S4 sensitivity. 
\label{limits_rd}}
\end{table}

Depending on the scenario, forecasting shows that the CMB-S4 sensitivity to curvaton-sourced isocurvature should improve on current limits by a factor of 2--4. In models with nearly-canceling CDM and baryon isocurvature perturbations, CMB-S4 limits to neutrino isocurvature drive an improvement in the sensitivity to the lepton asymmetry from $\Delta \xi_{\rm lep}^{2}\simeq 0.015$ to $\Delta \xi_{\rm lep}^{2}\simeq 0.003$. This dramatic improvement would make CMB limits comparably sensitive to BBN probes of $\xi_{\rm lep}^{2}$ (for this decay scenario).

If baryon number is produced by curvaton decay, but CDM is produced before (or vice versa), a relatively large compensated isocurvature perturbation (CIP) is produced between the baryons and CDM, that is
\begin{equation}
S_{bc}=\frac{\delta n_{\rm b}}{n_{\rm b}}-\frac{\delta n_{\rm c}}{n_{\rm c}}\neq 0.
\end{equation} 
Curvaton-generated CIPs are proportional to $\zeta$, $S_{\rm bc}=A\zeta$, where $A\simeq 17$ in the $(b_{\rm by}, c_{\rm before}, L_{ y_{\rm L}})$ scenario and $A\simeq -3$ for $(b_{\rm by}, c_{\rm before}, L_{ y_{\rm L}})$. For CIPs, the initial relative densities of baryons and CDM vary, but with no additional overall matter or radiation density fluctuation.

CIPs are relatively unconstrained at the linear level of the CMB power spectrum (see Ref.~\cite{Munoz:2015fdv} for an exception), but would induce non-Gaussianities in the CMB \cite{Grin:2011nk,Grin:2011tf,Grin:2013uya,He:2015msa}. As with weak gravitational lensing \cite{Hu:2001kj}, the CIP field $\Delta(\hat{n})$ can be reconstructed using CMB data. We find that at CMB-S4 sensitivity \cite{He:2015msa}, the threshold for a $95\%$ CL detection is $A\simeq 10$, and so a CIP test of the $(b_{\rm by}, c_{\rm before}, L_{ y_{\rm L}})$ scenario is within reach of CMB-S4. This is a significant improvement over \planck\ sensitivity, which at $95\%$ CL is $A\simeq 43$. Uncorrelated CIPs are less  motivated theoretically. Updating the analysis of Ref.~\cite{He:2015msa} with current parameters \cite{Ade:2015lrj} and CMB-S4 specifications, we find that the sensitivity of CMB-S4 to a scale-invariant (SI) angular power spectrum of uncorrelated CIPs is $\Delta_{\rm cl}=0.003$ at the $95\%$ CL. Here $\Delta_{\rm cl}$ is the rms CIP amplitude on cluster scales. This is a significant improvement over the upper limit of $\Delta_{\rm cl}\leq 0.077$ from \wmap\ \cite{Grin:2013uya}, or the forecasted \planck\ \cite{Ade:2015lrj} (including polarization) sensitivity of $\Delta_{\rm cl}\leq 0.015$ \cite{He:2015msa}. 

A complementary constraint on uncorrelated CIPs can be derived from an independent search for their second-order effect on the CMB power spectrum \cite{Munoz:2015fdv}. With a CMB-S4 experiment, the power spectra would be sensitive to a CIP amplitude of $\Delta_{\rm cl} = 0.026$ at the 95\% CL, a factor of 3 better than the corresponding limit from the \planck\ analysis (see Ref.~\cite{Munoz:2015fdv}).

\section{Microwave Background Anomalies}

Several features have been observed in maps of CMB temperature at
relatively low $\ell$ or large angular scales.  Some of these
so-called ``anomalies'' have been seen in both \wmap\ and \planck\
data, and there is little doubt that they are real features on the CMB sky;
However, there is much less agreement about their statistical significance,
largely because of the difficulty of assessing the effects of a posteriori
choice when determining the probability of such features
(see, e.g.\ Refs.~\cite{Bennett:2010jb,Ade:2015hxq} and \cite{Schwarz:2015cma} for
different views).  Nevertheless, there is
the potential for any one of these departures from statistical isotropy to
tell us about fundamental physics, for example about the beginning of inflation.
Hence it is extremely interesting to perform follow-up studies using modes
beyond those of CMB temperature.  Deep mapping of CMB E modes provides the most
promising way of further probing these anomalies.

Some of the identified features (such as the lack of power in $C_\ell^{TT}$
for $\ell\lesssim30$ or the apparent correlation in the first few multipoles)
are unlikely to be probed by CMB-S4 because they only affect the very largest scales.
However, a measurement of E-mode polarization on scales $\ell\gtrsim 30$ and over the 
part of the sky accessible from Chile would allow CMB-S4 to explore two specific anomalies

\begin{itemize}
  \item a hemispherical asymmetry or dipolar modulation of the large-scale
  power and variance (with an apparently normal southern sky  and an anomalous northern sky) 
  \item the presence of a particularly large region of low temperature,
  the ``cold spot.''
\end{itemize}

The first \cite{Eriksen:2003db, Ade:2015hxq} appears to extend to $\ell_{\rm max}\simeq65$, while the second \cite{Vielva:2003et, Ade:2015hxq} has
structure up to about $5^\circ$ in scale. 

For the asymmetry, the question will be whether there could be a modulation
of the polarized CMB sky, and whether it is over the same range of
multipoles as in the temperature data. The ability to search for a signal in 
polarization from a Chilean site was characterized by \cite{O'Dwyer:2016lna} 
(see also \cite{Bunn:2016kwh}).
If the origin is a modification of the inflationary power
spectrum in $k$ (e.g.\ Ref.~\cite{Gordon:2005ai}),
then there should in fact be a slightly different projection
into $\ell$ space, which could be explored. 

For the cold spot, if it is simply a large fluctuation
on the last-scattering surface, then the correlated part of the E modes can
be predicted from the temperature profile, and one can test whether the prediction
is consistent with the measured E-mode map; i.e., one can see if the E-mode map, after subtraction
of the temperature-correlated part, is consistent with the expected level of temperature-uncorrelated E-mode fluctuations. Inconsistency would be evidence for a more interesting origin for the cold spot. 

\section{Cosmic Strings}

Multi-field inflationary scenarios that end with phase transitions \cite{Hindmarsh:1994re,Vilenkin:1981iu,Kofman:1995fi,Tkachev:1998dc,Jeannerot:1995yn,Jeannerot:2003qv,Rocher:2004my} and models of brane-inflation in string theory \cite{Sarangi:2002yt,Jones:2003da,Copeland:2003bj} generically predict some level of vector and tensor modes actively sourced by topological defects. In particular, either a breaking of a $U(1)$ symmetry or the production of fundamental strings at the end of inflation can lead to ``cosmic strings'' whose B-mode spectrum is primarily generated by vector modes, peaks on small scales ($\ell\sim 600$--1000), and is more similar in shape to the lensing B-mode signal than to the vacuum spectrum. CMB-S4 should be able to distinguish even a small contribution from such sources \cite{Urrestilla:2008jv}, but the precise bounds from non-detection are related to the precision with which the lensing signal can be removed. Estimates in~\cite{Seljak:2006hi,Avgoustidis:2011ax} indicate that an experiment like CMB-S4 should significantly improve the limit on cosmic string tension beyond the current bounds from the CMB ($G\mu\lesssim10^{-7}$)~\cite{Ade:2013xla,Ade:2015ava,Ade:2015xua} and may be competitive with direct detection limits from the stochastic gravitational wave background~\cite{Arzoumanian:2015liz}. 
In addition, the spectra of different types of defects have different shapes, and should be distinguishable \cite{Urrestilla:2007sf,Avgoustidis:2011ax}. Measuring the location of the main peak would provide valuable insights into fundamental physics. For example, in the case of cosmic superstrings the position of the peak of the B-mode spectrum constrains the value of the fundamental string coupling $g_s$ in string theory \cite{Avgoustidis:2011ax}.

Cosmic strings can at most contribute O(1\%) to the total CMB temperature anisotropy~\cite{Ade:2013xla,Lizarraga:2014xza,Lazanu:2014eya}, however, they can still generate observable B modes. As shown in \cite{Moss:2014cra}, the bounds on cosmic strings obtained solely from the POLARBEAR \cite{Ade:2014afa} and { BICEP}2~\cite{Ade:2014xna} B-mode spectra are comparable to those from temperature spectra. 
Forecasts of the predicted constraints on cosmic strings using the StringFast code \cite{Foreman:2011uj}, based on the CMBACT simulations \cite{Pogosian:1999np} of a general string network, allows for the correlation length of the strings, the ``wiggliness'' (which controls the small-scale structure of the string network) and the string rms velocity. StringFast allows for fast computation of the relevant string spectra, and includes the contribution to the string spectrum from scalar, vector and tensor modes, which are most relevant for the string B modes \cite{Foreman:2011uj}.

In keeping with the methodology of recent results, we compute the string spectrum with a value of the string tension ($G\mu=1.97\times10^{-6}$) that allows strings to make up all the TT power at $\ell=10$, and then use the fraction of the spectrum at that multipole $f_{10}$ as the forecast parameter. 
The Fisher projections for \planck\ around a fiducial model of $f_{10}=0.01$ are $f_{10}<0.032$. This corresponds to $G\mu < 3.5\times 10^{-7}$ at $95\%$~CL and is consistent with the \planck\ constraints.
We assume a fiducial model for the string fraction $f_{10}$, ``wiggliness'' $\alpha_\mathrm{str}$, string velocity $v_\mathrm{str}$ and correlation length $\xi_\mathrm{str}$ of $0.01, 1.05, 0.4$ and $0.35$ respectively, in keeping with the model assumed in \cite{Foreman:2011uj}. We consider models where only the string fraction is varied, and the additional model where the small-scale structure of the string network is varied. The constraints are summarised in Table~\ref{tab:string_forecast} for a baseline resolution of 1 arcmin and baseline noise level of 1 $\mu$K/arcmin. The error on the string fraction for a \mbox{2 arcmin} beam is weaker only by a few percent relative to the nominal case. In addition, the constraints are not strongly improved with the addition of BAO data, or with a more improved measurement of $\tau$.

\begin{table}[htbp!]\label{tab:string_forecast}
  \begin{center}
    \begin{tabular}{ | c || c | c  |}
      \hline
       Model & {\it Planck} & CMB-S4 ($1^\prime$ resolution)  \\ \hline \hline
       Fixed $\alpha_\mathrm{str}:$ & $\sigma(f_{10})= 0.015$ & $\sigma(f_{10})=1.06\times 10^{-3}$  \\ \hline
       Varying $\alpha_\mathrm{str}:$ & $\sigma(f_{10})= 0.017$ & $\sigma(f_{10})=1.85\times 10^{-3}$ \\
        & $\sigma(\alpha_\mathrm{str})= 5.55$ & $\sigma(\alpha_\mathrm{str})=0.64$ \\\hline
    \end{tabular}
  \end{center}
  \caption{Forecasts constraints on the fraction of power in cosmic strings at $\ell=10$ (around a fiducial model of $f_{10}=0.01$) and on the ``wiggliness'' of the string network, $\alpha$, for a fiducial value of $\alpha=1.05$. The \planck\ forecast is based on Blue Book values, with $f_{\rm sky} = 0.75$. CMB-S4 will yield an order of magnitude improvement in constraints on cosmic string parameters.}
\end{table}

\section{Primordial Magnetic Fields}
\label{sec:PMF}
The origin of the microgauss ($\mu$G) strength magnetic fields in galaxies and galaxy clusters is one of the long-standing puzzles in astrophysics \cite{Durrer:2013pga}. It is challenging to explain such fields based solely on the dynamo mechanism, at least in the absence of an initial seed field. However, if magnetic fields were present in the early Universe, they would remain frozen in the cosmic plasma and collapse with the rest of the matter to form the Galactic fields \cite{Grasso:2000wj}, or at least provide the seeds for the dynamo. A primordial magnetic field (PMF) could be produced in the aftermath of cosmic phase transitions \cite{Vachaspati:1991nm} or in specially designed inflationary scenarios \cite{Turner:1987bw,Ratra:1991bn}. Detecting their signatures in the CMB temperature and polarization would decisively prove their primordial origin. Aside from explaining the Galactic fields, bounds on PMF have profound implications for our understanding of the early Universe.  They help constrain theories of inflation \cite{Bonvin:2011dt}, models of the QCD and electroweak phase transitions \cite{Caprini:2007xq}, and baryogenesis \cite{Vachaspati:2001nb}.

The PMF affects CMB in several ways. Magnetic stress-energy induces scalar, vector, and tensor mode perturbations in the metric, and the Lorentz force generates vorticity in the photon-baryon fluid \cite{Subramanian:1998fn,Mack:2001gc,Lewis:2004ef,Shaw:2009nf,Paoletti:2010rx}. Dissipation of PMF on small scales dumps energy into the plasma, which produces spectral distortions and affects the recombination history \cite{Kunze:2014eka}.  Finally, Faraday rotation (FR) of CMB polarization converts some of the E modes into B modes \cite{Kosowsky:2004zh,Pogosian:2011qv}.

A stochastic PMF has two potentially observable frequency-independent contributions to the B-mode spectrum \cite{Shaw:2009nf}. One comes from the passive, or uncompensated tensor mode, which is generated by the PMF before neutrino decoupling. For nearly scale-invariant PMF, the spectrum of this component is indistinguishable from the inflationary gravity wave signal, while the amplitude is proportional to $B^4_{1\rm{Mpc}} [\ln(a_\nu / a_{\rm{PMF}})]^2$ \cite{Lewis:2004ef}, where $B_{1\rm{Mpc}}$ is the PMF strength smoothed over one megaparsec, $a_\nu$ is the scale factor at neutrino decoupling, and $a_{\rm{PMF}}$ is the scale factor at which PMF was generated. The other is the PMF vector mode which peaks at $l \simeq 2000$, with the precise peak position dependent on the PMF spectrum. The vector-mode contribution is independent of $a_{\rm{PMF}}$.

\planck\ data limit the magnetic field strength to $B_{1 {\rm Mpc}}<4.4$ nanogauss (nG) at the $95\%$ confidence level \cite{Ade:2015cva}. Similar bounds were recently obtained by POLARBEAR \cite{Ade:2015cao} based on their B-mode spectrum alone. CMB-S4 with a $1$ arcmin resolution and 1.4\,$\mu$K-arcmin polarization sensitivity can improve the $95\%$ CL bound on $B_{1 {\rm Mpc}}$ to 0.6\,nG based on the PMF vector mode contribution to the B-mode spectrum.

Comparable bounds can be obtained from the mode-coupling correlations induced by FR. The mode-coupling is the same as in the case of birefringence discussed in Section~\ref{sec-biref}, with the FR angle depending on frequency as $\nu^{-2}$ \cite{Yadav:2012uz,De:2013dra,Pogosian:2013dya}. Unlike the CMB anisotropies sourced by the stress-energy of the PMF, which scale as a square of the PMF strength (so that the CMB spectra scale as $B^4_{1 {\rm Mpc}}$), the FR contribution is linear in the PMF strength. Thus, despite the fact that the FR angles are typically very small at CMB frequencies, FR offers a significantly larger gain of constraining power with the improved sensitivity and resolution \cite{Pogosian:2013dya} than the PMF sourced vector and tensor mode signatures. A 150-GHz Stage IV experiment can detect mode-coupling correlations sourced by a scale-invariant PMF of 0.6\,nG strength at 95\%CL without any subtraction of the weak lensing B modes or the Galactic FR contribution. The impact of subtracting the Galactic FR is negligible at PMF strengths above 0.3\,nG \cite{De:2013dra}, however, removing the weak lensing contribution can improve the 95\% CL bound to 0.4\,nG. This will be a significant improvement on the FR based 95\% CL bound of 93\,nG obtained by POLARBEAR  \cite{Ade:2015cao}, based on measurements at 150\,GHz.

\section{Summary}
\label{sec:summary}
CMB-S4 is an ideal tool to test the inflationary paradigm and competing theories for the origin of structure in the early Universe. Its exquisite sensitivity will allow a detection of degree-scale B modes in the CMB or achieve upper limits on the amount of B-mode polarization that improve current constraints on the tensor-to-scalar ratio by over an order of magnitude. In particular, it is sensitive enough to detect the level of B-mode polarization predicted in a wide range of well-motivated inflationary models. In doing so, it would provide invaluable information about physics at energy scales far outside the capabilities of any terrestrial particle physics experiment. In the absence of a detection it would exclude large classes of inflationary models. Furthermore, with sufficient angular resolution, CMB-S4 will measure anisotropies in both the temperature and E-mode polarization of the CMB to cosmic variance limits over the entire range of multipoles that is not contaminated by unresolved foregrounds, and it will extend our window to the early Universe by almost one $e$-fold beyond the reach of current experiments. As a consequence, it will provide the best constraints achievable by any ground-based CMB experiment on any observable that benefits from the number of modes measured, such as the shape of the primordial power spectrum---including the spectral index, its running and the presence of any features---and higher-order correlations.

 
\chapter{Neutrinos}

\def\beq{\begin{equation}}
\def\eeq{\end{equation}}

\def\bea{\begin{eqnarray}}
\def\eea{\end{eqnarray}}

\def\Neff{N_{\rm eff}}
\def\Nf{N_{\rm eff}}
\def\gs{g_{\star}}
\def\Mpl{M_{\rm P}}
\newcommand{\nucl}[3]{ \ensuremath{ \phantom{\ensuremath{^{#1}_{#2}}} \llap{\ensuremath{^{#1}}} \llap{\ensuremath{_{\rule{0pt}{.75em}#2}}} \mbox{#3} } }

\def\lsim{\raise-.75ex\hbox{$\buildrel<\over\sim$}}

\bigskip

\begin{quotation}

Neutrinos are the least explored corner of the Standard Model of particle physics.  The 2015 Nobel Prize recognized the discovery of neutrino oscillations, which shows that they have mass. However, the overall scale of the masses and the full suite of mixing parameters are still not measured.  Cosmology offers a unique view of neutrinos; they were produced in large numbers in the high temperatures of the early universe and left a distinctive imprint in the cosmic microwave background and on the large-scale structure of the universe. Therefore, CMB-S4 and large-scale structure surveys together will have the power to detect properties of neutrinos that supplement those probed by large terrestrial experiments such as short- and long-baseline as well as neutrino-less double beta decay experiments.

Specifically, while long baseline experiments are sensitive to the differences in the masses of the different types of neutrinos, CMB-S4  will probe the sum of all the neutrino masses. The current lower limit on the sum of neutrino masses imposed by oscillation experiments is \mbox{$\sum m_\nu=58\,\rm{meV}$}. CMB-S4, in conjunction with upcoming baryon acoustic oscillation surveys, will measure this sum with high significance. Once determined, the sum of neutrino masses will inform the prospects for future neutrino-less double beta decay experiments that aim to determine whether neutrinos are their own anti-particle. Furthermore, an upper limit below \mbox{$\sum m_\nu=105\,\rm{meV}$} would disfavor the inverted mass hierarchy. Finally, CMB-S4 is particularly sensitive to the possible existence of additional neutrinos that interact even more weakly than the neutrinos in the Standard Model. These so-called {\it sterile} neutrinos are also being vigorously pursued with short baseline experiments around the world. So the combination of CMB-S4, large scale structure surveys, and terrestrial probes adds up to a comprehensive assault on the three-neutrino paradigm.

\end{quotation}

\section{Introduction}

Direct interactions between neutrinos and observable matter effectively ceased about one second after the hot big bang.  Nevertheless, the total energy density carried by neutrinos was comparable to other components through recent cosmological times.  As a result, the gravitational effect of the neutrinos is detectable both at the time of recombination and in the growth of structure at later times~\cite{Abazajian:2013oma}, leaving imprints in the temperature and polarization spectra as well as in CMB lensing.

CMB-S4 can improve our understanding of neutrino physics in regimes of interest for both particle physics and neutrino cosmology.  For neutrinos, arguably the most important parameters measurable with CMB-S4 will be the sum of the neutrino masses ($\sum m_\nu$), the effective number of neutrino species ($\Neff$) and the helium fraction ($Y_p$).  These three parameters are highly constrained within the Standard Model of particle physics and will be precisely measured with a CMB-S4 experiment:
\begin{itemize}
\item $ \sum m_\nu \gtrsim \, 58$ meV is the lower bound guaranteed by observations of solar and atmospheric neutrino oscillations.  A CMB experiment with $\sigma(\sum m_\nu) < 20$ meV would guarantee a detection of $\sim3\,\sigma$.  At this level, one can detect the overall scale of the neutrino masses even for the normal mass hierarchy.
\item $\Neff \approx  3.046$ and $Y_p \approx 0.2311 + 0.9502 \, \Omega_b h^2$ are predicted by standard neutrino decoupling and big bang nucleosynethesis (BBN).  $\Neff$ is a measure of the total radiation energy density at recombination while $Y_p$ is sensitive to the radiation density and the neutrino distribution at BBN. 
\end{itemize}
Current CMB data already provides a robust detection of the cosmic neutrino background at $\sim10 \sigma$.  A CMB-S4 experiment will provide an order of magnitude improvement in sensitivity in $\Neff$ that opens a new window back to the time of neutrino decoupling and beyond.  $\Neff$ and $Y_p$ are also sensitive to any additional light particles beyond the Standard Model, such as sterile neutrinos or axions.  The implications of these constraints will be explored in more detail in Chapter 4, where we will discuss the observational signatures and forecasts for future measurements of $\Neff$ and $Y_p$.  

In section~\ref{sec:neureview} we review neutrino cosmology and the motivation for studying neutrino masses with cosmological probes.  Section~\ref{sec:mnuobs} discusses the various cosmological signatures of neutrino mass, emphasizing the CMB specifically, including forecasts of CMB-S4 sensitivity to $\sum m_\nu$.  Section~\ref{sec:lab} discusses the relationship between cosmological and lab-based neutrino measurements, including a discussion of sterile neutrinos.  Section~\ref{sec:neuscenarios} explores some possible scenarios for neutrino physics with CMB-S4 and neutrino experiments.

Unless otherwise stated, we will work in units where $c=\hbar = 1$.  We will also set $k_B=1$ such that temperature, $T$, carries units of energy ($T = 1 \, {\rm K} \to T= 8.62 \times 10^{-5} \, {\rm eV}$).  The scale factor, $a$, is normalized such that $a(z=0) =1$.  

\section{Review of Neutrino Cosmology}\label{sec:neureview}

Cosmological measurements of neutrinos depend on our detailed understanding of the cosmic history, starting with their decoupling at high temperatures ($T \lesssim 10$ MeV) through to their contribution to the growth of structure at late times.  In this Section we will give an overview of these epochs with an eye towards to the measurements of $\sum m_\nu$, $\Neff$ and $Y_p$ to be discussed in this chapter and the next.  

\subsection{Neutrino Physics Basics}

Measurements of $ \sum m_\nu $ by CMB-S4 will be interesting within the context of the broader neutrino experimental program. Specifically, the sum of the neutrino mass fixes one of the many parameters relevant to neutrino physics.  Neutrino flavor oscillations are described by a model where the neutrino flavor eigenstates are a mixture of massive neutrino eigenstates. The mixing is parameterized by the Pontecorvo-Maki-Nakagawa-Sakata (PMNS) matrix,
\[ \left( \begin{array}{c} \nu_e \\ \nu_{\mu} \\ \nu_{\tau} \\ \end{array} \right) = 
\left( \begin{array}{ccc} U_{e1} & U_{e2} & U_{e3} \\ U_{\mu1} & U_{\mu2} & U_{\mu3} \\ U_{\tau1} & U_{\tau2} & U_{\tau3} \\
\end{array} \right) \cdot
\left( \begin{array}{c} \nu_1 \\ \nu_2 \\ \nu_3 \\ \end{array} \right),
\]
where $\nu_{i}$, $i=1,2,3$ are the neutrino mass eigenstates. $U_{\rm PMNS}$ depends upon six real parameters: three mixing angles, $\theta_{12}$,  $\theta_{23}$, $\theta_{13}$ that
correspond to the three Euler rotations in a 3--dimensional space, and three phases, $\delta$, $\alpha_1$, $\alpha_2$. A suitable parametrization is
\[ U_{\rm PMNS}=
\left( \begin{array}{ccc} c_{12}c_{13} & s_{12}c_{13} & s_{13} e^{-i\delta} \\ 
-s_{12}c_{23} - c_{12}s_{13}s_{23} e^{i\delta} & c_{12}c_{23} - s_{12}s_{13}s_{23} e^{i\delta} & c_{13}s_{23} \\
s_{12}s_{23} - c_{12}s_{13}c_{23} e^{i\delta} & -c_{12}s_{23} - s_{12}s_{13}c_{23} e^{i\delta} & c_{13}c_{23} \\
\end{array} \right) \cdot
\left( \begin{array}{ccc} 1 & 0 & 0 \\ 0 & e^{i\alpha_1/2} & 0 \\ 0 & 0 & e^{i(\alpha_2/2)} \\ 
\end{array} \right)
\]
where $c_{ij} \equiv \cos\theta_{ij}$ and $s_{ij} \equiv \sin\theta_{ij}$. 
The phases $\delta$ ($\equiv \delta_{CP}$) and $\alpha_1$, $\alpha_2$ are Dirac--type and Majorana--type $CP$ violating phases, respectively.

\begin{figure}[h!]
\centering 
\includegraphics[width=0.7\textwidth]{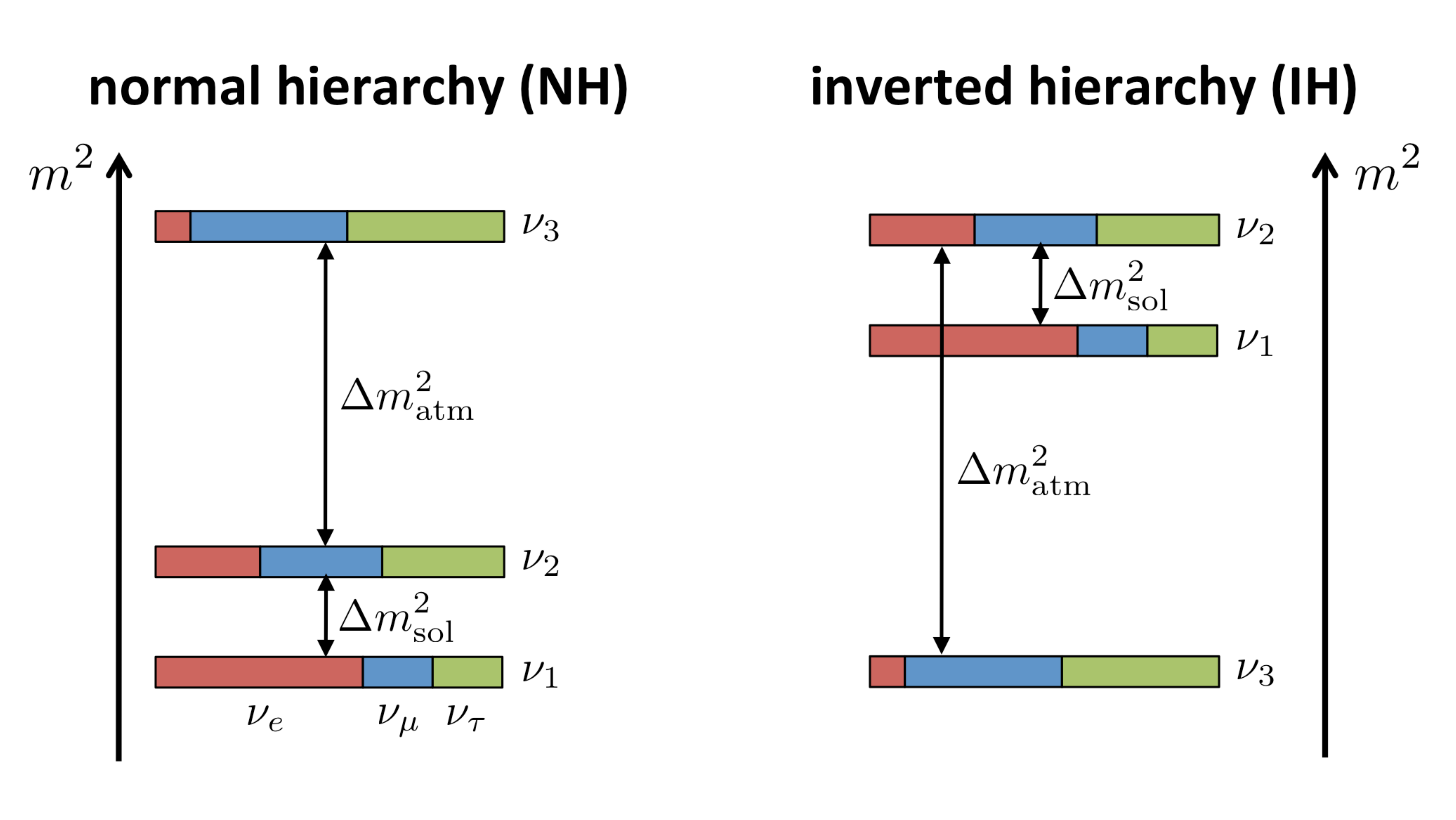}
\caption{Cartoon illustrating the two distinct neutrino mass hierarchies. The colors indicate the fraction of each distinct flavor contained in each mass eigenstate.}
\label{fig:NeutrinoMixHierarch}
\end{figure}

Experiments have measured the three mixing angles of $U_{\rm PMNS}$ and the two mass splittings, $\Delta m^2_{21}$ (the ``solar'' mass splitting) and $\Delta m^2_{32}$ (the ``atmospheric'' mass splitting), but fundamental aspects of neutrino mass and mixing are yet to be settled. These include:
\begin{itemize}
\item measuring the absolute mass scale,
\item determining the mass ordering (see Fig.~\ref{fig:NeutrinoMixHierarch}),
\item searching for Lepton number violation (i.e., determining whether neutrinos are Majorana particles) and
\item observing CP violation (measuring $\delta_{CP}$).
\end{itemize}
Exploring these goals is the focus of current and upcoming neutrino experiments. CMB-S4 will measure $ \sum m_\nu $ with sufficient sensitivity to be relevant to these open issues.  Most unambiguously, $\sum m_\nu$ determines  the absolute mass scale of neutrinos.  In some circumstances, this may also determine the mass ordering.  These goals are complementary to the program for lab-based neutrino measurements, as we will discuss in Section~\ref{sec:lab}.

\subsection{Thermal History of the Early Universe} \label{ThermalHistory}
Cosmological measurements of $\sum m_\nu$ rely on our detailed understanding of thermal history of the Universe, particularly the origin of the cosmic neutrino background.  In this section, we will give a sketch of the thermal history of the standard hot big bang universe when the temperature of the plasma was falling from about $10^{11}$~K to about $10^8$~K following Section 3.1 of \cite{Weinberg:2008zzc}.  For other reviews see \cite{Dolgov:2002wy,Agashe:2014kda}.  During this era, there are two events of particular interest: neutrinos decoupled from the rest of the plasma, and a short time later electrons and positrons annihilated, heating the photons relative to the neutrinos.  Our task is to follow how these events impact the evolution of the energy densities of the photons and neutrinos.

For massless particles described by the Fermi-Dirac or Bose-Einstein distributions, the energy density is given by
\beq
	\rho(T) =
	\Bigg\{\begin{array}{l c l}
        g\frac{\pi^2}{30}T^4 &  &{\rm Boson}\\
       \frac{7}{8}g\frac{\pi^2}{30} T^4 &  &{\rm Fermion}
        \end{array}
\eeq
where $g$ counts the number of distinct spin states.  The entropy density for massless particles is given by
\begin{equation}
	s(T) = \frac{4\rho(T)}{3T} \, .
\end{equation}
It is convenient to define a quantity $\gs$ which counts the spin states for all particles and antiparticles, with an additional factor $\frac{7}{8}$ for fermions.  With this definition, the total energy density and entropy density of the Universe during radiation domination are given by
\bea
	\rho(T) &=& \gs\frac{\pi^2}{30}T^4 \, , \nonumber \\
	s(T) &=& \frac{4}{3}\gs\frac{\pi^2}{30}T^3 \, .
\eea
In an expanding universe, the first law of thermodynamics implies that for particles in equilibrium, the comoving entropy density is conserved
\begin{equation}
	a^3s(T) = \mathrm{const} \, .
\end{equation}
One straightforward consequence of this conservation is that for radiation in free expansion, the temperature evolves as the inverse of the scale factor
\begin{equation}
	T\propto \frac{1}{a} \, .
\end{equation}
Let us now apply this to the physics of the early universe.

At a temperature of $10^{11}$~K ($T\sim10$~MeV), the Universe was filled with photons, electrons and positrons, and neutrinos and antineutrinos of three species, all in thermal equilibrium with negligible chemical potential, along with a much smaller density of baryons and dark matter both of which are unimportant for the present discussion.  As the temperature of the plasma dropped below about $10^{10}$~K (about 1 second after the end of inflation), the rate of collisions between neutrinos and electrons and positrons could no longer keep up with the expansion rate of the Universe, neutrinos fell out of equilibrium and began a free expansion.  Electrons and positrons remained in equilibrium with photons and their number densities fell with decreasing temperature, so that they were effectively gone by the time the temperature reached $ T \sim 10\,{\rm keV}$.  We will simplify the discussion by assuming that neutrinos decoupled instantaneously before electron-positron annihilation and comment below how a more detailed calculation modifies the results.  Non-zero neutrino masses can safely be neglected here as long as $m_\nu \lesssim1$~keV which is guaranteed by current observational bounds.

From this point on, we will distinguish the temperature of neutrinos $T_\nu$ from that of the photons $T_\gamma$.  Before neutrino decoupling, frequent interactions kept neutrinos and photons in equilibrium, ensuring they had a common temperature.  After the Universe became transparent to neutrinos, the neutrinos kept their relativistic Fermi-Dirac distribution with a temperature which decreased as the inverse of the scale factor.  The photons, on the other hand, were heated by the annihilation of the electrons and positrons.  Comoving entropy conservation allows us to compute the relative temperatures at later times.

After neutrino decoupling, but before electron positron annihilation, the thermal plasma contained two spin states of photons, plus two spin states each of electrons and positrons, which means that during this period,
\begin{equation}
	g_{\star}^{\mathrm{before}} = 2 + \frac{7}{8}(2+2) = \frac{11}{2} \, .
\end{equation}
After electron positron annihilation, only the two spin states of photons remained, and so
\begin{equation}
	g_\star^{\mathrm{after}} = 2 \, .
\end{equation}
Since $T_\nu\propto a^{-1}$ during this period, we can express the condition of comoving entropy conservation as follows
\begin{equation}
	\frac{g_{\star}^{\mathrm{before}} \, T_{\gamma,\mathrm{before}}^3}{T_{\nu,\mathrm{before}}^3} = \frac{g_\star^{\mathrm{after}} \, T_{\gamma,\mathrm{after}}^3}{T_{\nu,\mathrm{after}}^3} \, .
\end{equation}
Using the fact that $T_{\gamma,\mathrm{before}} = T_{\nu,\mathrm{before}}$, we find as a result
\begin{equation}
	\frac{T_{\gamma,\mathrm{after}}}{T_{\nu,\mathrm{after}}} = \left(\frac{11}{4}\right)^{1/3} \, .
\end{equation}
We find that in the instantaneous neutrino decoupling limit, the annihilation of electrons and positrons raised the temperature of photons relative to that of neutrinos by a factor of $(11/4)^{1/3}\simeq1.401$.

After electron positron annihilation, assuming three species of light neutrinos and antineutrinos, each with one spin state, the radiation density of the Universe is
\begin{equation}
	\rho_r = \frac{\pi^2}{30}\left[2T_\gamma^4 + 6\frac{7}{8}T_\nu^4\right] = \frac{\pi^2}{15}\left[1+3\,\frac{7}{8}\left(\frac{4}{11}\right)^{4/3}\right] T_\gamma^4 \, .
\end{equation}
It is conventional to define a quantity $\Nf$ which gives the radiation energy density in terms of the effective number of neutrino species as
\begin{equation}\label{eq:rho_r_Neff}
	\rho_r = \frac{\pi^2k_B^4}{15\hbar^3 c^3}\left[1+\frac{7}{8}\left(\frac{4}{11}\right)^{4/3}\Nf\right] T_\gamma^4 \, ,
\end{equation}
where we have restored $k_B$, $c$ and $\hbar$ for completeness.  In the instantaneous neutrino decoupling approximation described above, we found $\Nf = 3$.  In the real Universe, however, decoupling of neutrinos is not instantaneous, and the residual coupling of neutrinos at the time of electron positron annihilation increases $\Nf$ by a small amount in the Standard Model.

Unlike photon decoupling at temperature $k_B T \sim 0.2$ eV, active neutrino decoupling at $T \sim 10 \, {\rm MeV} - 0.1 \, {\rm MeV}$ takes place over many tens of Hubble times, with the result that we expect distortions in the relic neutrino energy spectra relative to the thermal relativistic Fermi-Dirac distribution. Standard Model Boltzmann neutrino transport calculations show that these distortions change $\Neff$ at the percent level, with the current best estimate predicting $\Nf = 3.046$ \cite{Mangano:2005cc}.   This result is largely due to (1) the incomplete decoupling of neutrinos during electron-positron annihilation and (2) QED plasma effects.  While both effects have been calculated independently quite accurately, there is some theoretical uncertainty in this quantity at the level of about $10^{-3}$ due to the various numerical approximations that are made in the calculations when both effects are included simultaneously (see e.g.~\cite{Grohs:2015tfy} for discussion).

\subsection{Neutrino Mass and Structure Formation}
\label{ssec:numasstheoryreview}

The relic neutrinos were relativistic at the time of their decoupling through to recombination.  As the Universe expanded and cooled the neutrino momenta redshifted as $p_\nu\propto 1/a$ and eventually the energy of most relic neutrinos came to be dominated by their rest mass, rather than their momentum. The energy density in nonrelativistic neutrinos therefore contributes to the matter budget of the Universe today. The neutrinos, however, were relativistic for much of the history of the Universe so their gravitational clustering is qualitatively different from that of cold dark matter (CDM) particles. This difference can be used to distinguish the neutrino and cold dark matter contributions to the matter density \cite{Hu:1997mj, Lesgourgues:2006nd, Abazajian:2011dt}.  In this section, we review how neutrino mass affects the evolution of the neutrino energy density and the gravitational clustering of matter in the Universe. 
 
As discussed more detail in the Chapter 4, cosmic background neutrinos have been detected indirectly through their contribution to the energy density in radiation in the early Universe. The current CMB constraints from $\Neff$ are in excellent agreement with the Standard Model expectation of three species of neutrinos and antineutrinos each described very nearly by a relativistic thermal Fermi-Dirac distribution \cite{Ade:2015xua}. The distribution function for each species of neutrinos and antineutrinos (neglecting here the small non-thermal distortions discussed above) is given by
\beq
f_\nu(p) = \frac{1}{e^{a p/T_{\nu 0}} +1} \ ,
\eeq
where $T_{\nu 0} \approx 1.68 \times 10^{-4}\,{\rm eV}$ ($1.95K$) is the temperature today. Note that with Standard Model physics the spectral shape of the neutrino phase space distribution is preserved with the expansion so relic neutrinos have retained the relativistic Fermi-Dirac momentum distribution inherited from decoupling even as the individual neutrinos became non-relativistic. 

The neutrino energy density is given by
\beq
\rho_\nu = \sum_i \int\frac{d^3 {\bf p}}{(2\pi )^3} \frac{\sqrt{p^2 + m_{\nu i}^2}}{e^{ap/T_{\nu 0}} +1}
\eeq
where $m_{\nu i}$ are the three neutrino mass eigenstates.  For $T_{\nu}/a \gg m_{\nu i}$ the neutrino energies are dominated by their momenta and the total energy density behaves like radiation
\bea
\rho_{\nu}\Bigg|_{{\rm {\tiny early}}}  &\approx& \frac{7 \pi^2}{40}( T_{\nu 0})^4 \frac{1}{a^4} \\
& \propto& a^{-4} \nonumber
\eea
While for $T_{\nu 0}/a \ll m_{\nu i}$ the energy density behaves like matter
\bea
\label{eq:rhonumassive}
\rho_{\nu} \Bigg|_{{\rm {\tiny late}}}  &\approx& \sum_i m_{\nu i} \bar{n}_\nu \\
&\propto& a^{-3} \nonumber
\eea
where $\bar{n}_\nu$ is the number of neutrinos and antineutrinos in each mass eigenstate
\beq
\bar{n}_\nu =\int\frac{d^3 {\bf p}}{(2\pi )^3} \frac{2}{e^{ap/T_{\nu 0}} +1} \approx \frac{113}{a^3}\,{\rm cm}^{-3}\,.
\eeq

For a neutrino of mass $m_{\nu i}$ the transition between these two regimes ($k_B T_\nu(a) \sim m_{\nu i} c^2$) occurs at redshift $z_{\rm nr} \sim 300 (m_{\nu i}/0.05 {\rm eV})$. Using Eq.~(\ref{eq:rhonumassive}) the fractional energy density in neutrinos today can be written as
\beq
\Omega_\nu h^2 \approx \frac{\sum_i m_{\nu i}}{93\,{\rm eV}}\,.
\eeq
The individual masses of the neutrino states are unknown but neutrino oscillation data specifies the square of two mass splittings $\Delta m_{12}^2 = 7.54 \times 10^{-5}$~eV, $|\Delta m_{13}^2|\approx 2.4 \times 10^{-3}$~eV \cite{Agashe:2014kda}. These mass splittings, in combination with the neutrino number density, give a lower limit on the contribution of neutrinos to the cosmic energy budget
\beq
\Omega_\nu h^2 \,  \gtrsim \, 0.0006\,.
\eeq
At $z \ll z_{\rm nr}$ the matter density of the Universe, which enters into the Hubble equation, is the sum of the CDM, baryon, and massive neutrino energy densities $\Omega_m = \Omega_c + \Omega_b + \Omega_\nu$. Whereas, at $z\gg z_{\rm nr}$ the matter density is solely made up of the baryon and CDM parts while neutrinos contribute to the radiation density. 

Neutrinos do not participate in gravitational collapse until late times when they have become nonrelativistic. Prior to this transition, the neutrinos {\em free-stream} out of gravitational wells, leaving the CDM and baryons behind  \cite{Bond:1983hb, Ma:1996za, Hu:1997vi, Hu:1997mj}. Primordial fluctuations in the neutrino density are therefore damped away on scales smaller than the horizon at $z_{\rm nr}$. In comoving units, this scale corresponds to a wave number
\beq
k_{\rm nr} \equiv a_{\rm nr} H(a_{\rm nr}) \approx 0.003 \left(\frac{\Omega_m}{0.3}\frac{m_{\nu}}{0.05\, {\rm eV}}\right)^{1/2} h/{\rm Mpc}\,.
\eeq
Once the neutrinos are non-relativistic, their finite velocity dispersion still prevents them from clustering on scales smaller than the typical distance a neutrino travels in a Hubble time, $v_\nu /H(a)$ where $v_\nu \approx 3.15 T_{\nu 0}/(a m_\nu)$ the mean neutrino velocity. In analogy with the Jeans criterion for gravitational collapse, the neutrino free-streaming scale is defined by \cite{Bond:1983hb, Lesgourgues:2006nd}
\beq
k_{\rm fs}(a) \equiv \sqrt{\frac{3}{2}}\frac{aH(a)}{v_\nu(a)} \approx 0.04\, a^2 \sqrt{\Omega_m a^{-3} + \Omega_\Lambda}\left(\frac{m_\nu}{0.05 {\rm eV}}\right) h/{\rm Mpc}
\eeq
in comoving coordinates. 

\begin{figure}[t]
\begin{center}
\includegraphics[width = 0.8\textwidth]{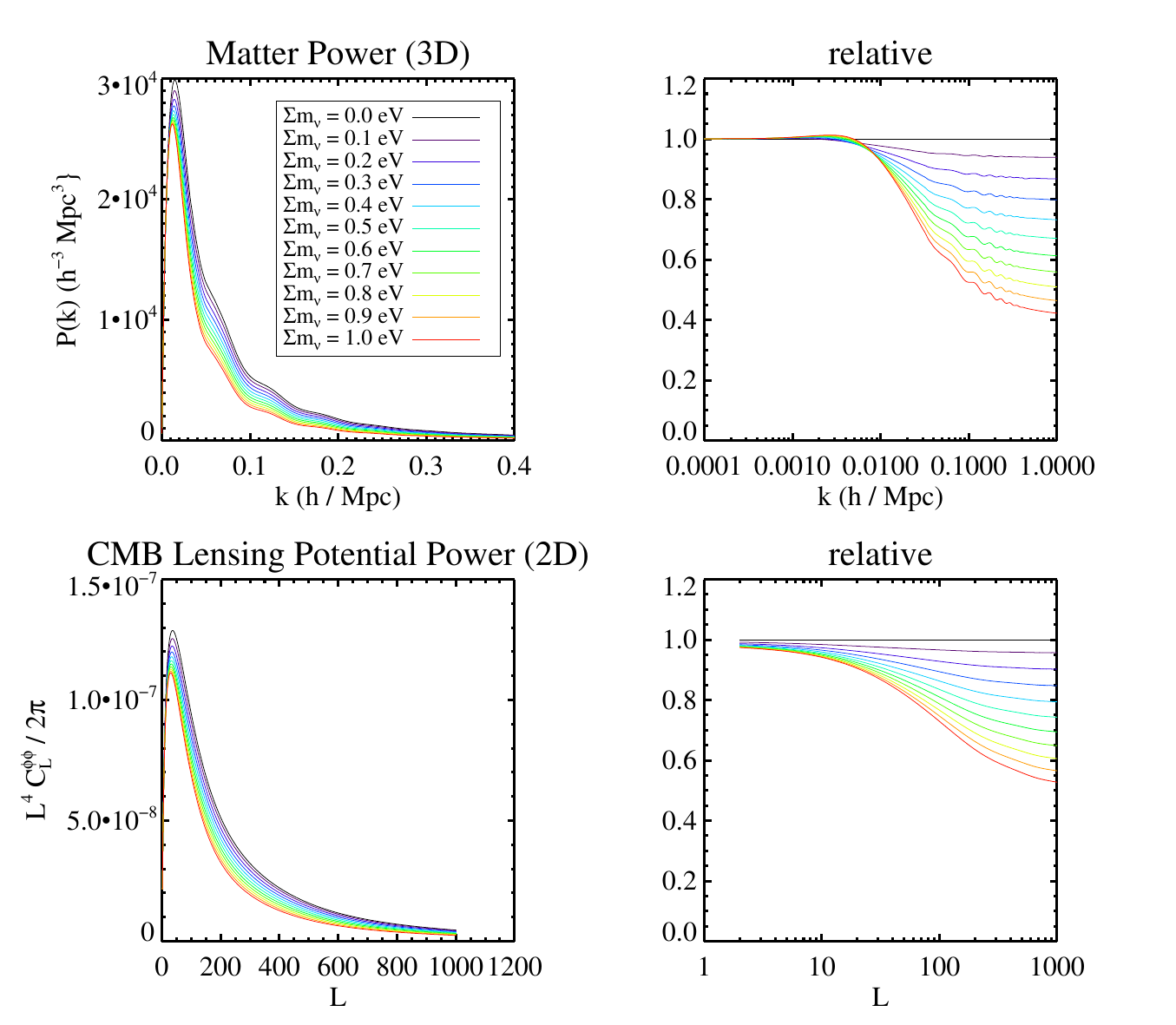} 
\caption{ The effect of massive neutrinos on the matter power spectrum and CMB lensing power spectrum.  {\it Top Left:} The effect of neutrino mass on the matter power spectrum. {\it Top Right:} The change to the matter power spectrum relative to the case with massless neutrinos. {\it Bottom Left:} The projected matter power spectrum observed through CMB lensing shows the same suppression with neutrino mass. {\it Bottom Right:} The relative change to the lensing potential power spectrum.} 
\label{fig:lensingsuppression}
\end{center}
\end{figure}

On scales larger than $k_{\rm nr}$ (adiabatic) perturbations in the density of neutrinos, baryons, and CDM are coherent and can be described by a single perturbation to the total matter density $\delta_m= \delta \rho_m/\rho_m$. On smaller scales where the neutrino perturbations have decayed, only the perturbations to the CDM and baryons remain so that $\delta_{m} = \delta_{cb}\, (\Omega_{c} + \Omega_b)/\Omega_m $. The remaining CDM and baryon perturbations also grow more slowly because the neutrino energy density contributes to the expansion rate, but not to the source potentials. These two effects cause a suppression in the amplitude and the growth rate of matter perturbations with wavenumbers $k > k_{\rm fs}$ relative to a universe with massless neutrinos (and also relative to density perturbations with $k<k_{\rm nr}$). The net change in the amplitude of perturbations with $k > k_{\rm nr}$ primarily depends on the fractional energy density in massive neutrinos (keeping $\Omega_c+\Omega_b$ fixed) but retains a small sensitivity to the individual neutrino masses through a dependence on $a_{\rm nr}$. A plot of the suppression in the matter power spectrum at small scales due to neutrino mass is shown in the top left panel of Figure~\ref{fig:lensingsuppression}.

An estimate of the effect of massive neutrinos on the growth of structure can be made by studying the evolution of matter perturbations in the two regimes $k\ll k_{\rm fs}$ and $k\gg k_{\rm fs}$.  In the synchronous gauge, linear perturbations to the matter density with wavenumbers $k \ll k_{\rm fs}$ evolve as
\beq
\label{eq:ddotdeltalarge}
\ddot{\delta}_m + 2 H(a) \dot\delta_m - \frac{3}{2}\Omega_mH_0^2 a^{-3}\delta_m = 0\quad {\rm for }\quad k \ll k_{\rm nr}\,,
\eeq
which has solutions $\delta_m \propto a, \, a^{-\frac{3}{2}}$ during the matter dominated era. 

On scales where the neutrino perturbations have decayed, perturbations to matter density are just in the CDM and baryon components 
\beq
\delta_m(k\gg k_{\rm fs})\approx  (\delta\rho_c + \delta\rho_b)/\rho_m = (1-f_\nu)\delta_{cb}\,,
\eeq 
where $f_\nu = \Omega_\nu/\Omega_m$ and $\delta_{cb} = (\delta\rho_c + \delta\rho_b)/(\rho_c + \rho_b)$, but the neutrino energy density still contributes to the Hubble friction. In this limit, linear perturbations to the  CDM and baryon density evolve as
\beq
\label{eq:ddotdeltasmall}
\ddot{\delta}_{m} + 2 H(a)\dot{\delta}_{m} - \frac{3}{2}\Omega_{cb}H_0^2 a^{-3} \delta_{m} =0\quad {\rm for}\quad  k \gg k_{\rm fs} \,,
\eeq
where $\Omega_{cb} = \Omega_c + \Omega_b$ and $\Omega_{cb} < \Omega_m$ for a cosmology with massive neutrinos. Equation (\ref{eq:ddotdeltasmall}) has the approximate solutions during the matter dominated era of $\delta_{cb} \propto a^{1-\frac{3}{5}f_\nu}, a^{-\frac{3}{2} + \frac{3}{5}f_\nu}$ for $f_\nu  \ll 1$. \

The matter dominated solutions give a simple estimate of the net effects of massive neutrinos on the amplitude of matter perturbations. For fixed $\Omega_c h^2$, the evolution of perturbations in a cosmology with $f_\nu \neq 0$ is the same as a cosmology with $f_\nu =0$ up until $a_{\rm nr}$. After $a_{\rm nr}$, the perturbations with $k\gg k_{\rm fs}$ grow more slowly (according to Eq.~(\ref{eq:ddotdeltasmall}), the growing mode solution grows as  $\propto a^{1-\frac{3}{5}f_\nu}$) than those with $k\ll k_{\rm fs}$ (according to Eq.~(\ref{eq:ddotdeltalarge}), $\propto a$).  At scale-factor $a$ during the matter dominated era, the total difference in growth or perturbations with $k\gg k_{\rm fs}$ is roughly
\beq
\frac{\delta_{cb}(k \gg k_{\rm fs}, a | f_\nu)}{\delta_{cb}(k \gg k_{\rm fs}, a | f_\nu=0)} \sim \left(\frac{a}{a_{\rm nr}}\right)^{-\frac{3}{5}f_\nu}\,.
\eeq
The resulting difference in the amplitude of the matter power spectra is then 
\beq
\frac{P_{mm}(k \gg k_{\rm fs}, a  |f_\nu)}{P_{mm}(k \gg k_{\rm fs}, a |f_\nu =0)}\sim (1-2f_\nu)\frac{P_{cc}(k \gg k_{\rm fs} , a |f_\nu)}{P_{cc}(k \gg k_{\rm fs}, a |f_\nu =0)}\sim \left(1-2f_\nu -\frac{6}{5}f_\nu\ln\left(a/a_{\rm nr}\right)\right)\,.
\eeq
On the other hand, the evolution of the large scale modes is identical,
\beq
\frac{P_{mm}(k \ll k_{\rm fs}, a  |f_\nu)}{P_{mm}(k \ll k_{\rm fs}, a |f_\nu =0)}=1\,,
\eeq
where $P_{mm}$ is the power spectrum of the total matter fluctuations (CDM, neutrino, and baryon) and $P_{cc}$ is the power spectrum of just the CDM and baryons.  The above expression overestimates the effect of neutrino mass by assuming the transition from relativistic to non-relativistic is instantaneous. It also ignores the effects of the cosmological constant at late times. Using the true evolution of $\delta_{cb}$ through $a_{\rm nr}$ and allowing for the cosmological constant gives 
\beq
\frac{P_{mm}(k \gg k_{\rm fs} |f_\nu)}{P_{mm}(k \gg k_{\rm fs} |f_\nu =0)} \approx 1- 6 f_\nu
\eeq
at $a=1$. Note that this expression assumes fixed $\Omega_ch^2$, $\Omega_bh^2$ so that matter-radiation equality is not changed by neutrino mass and that and $\Omega_\Lambda = 1-\Omega_m$ is fixed by adjusting $h$ so that the onset of cosmological constant domination is also unchanged. Alternatively, assuming fixed $\Omega_m$ and decreasing $\Omega_{cb}$ to account for $\Omega_\nu$ makes matter-radiation equality, which occurs while the neutrinos are relativistic, slightly later so that the suppression is increased to
\beq
\frac{P_{mm}(k \gg k_{\rm fs} |f_\nu)}{P_{mm}(k \gg k_{\rm fs} |f_\nu =0)} \approx (1-8f_\nu)\,.
\eeq

\section{Cosmological Measurements of Neutrino Mass} \label{sec:mnuobs}

As explained in the previous section,  the signature of massive neutrinos manifests through
the energy density $\Omega_\nu$, which is related to the mass through
\beq
\Omega_\nu h^2 \simeq \, \frac{\sum m_\nu}{93 \, {\rm eV}} \, \gtrsim  \, 0.0006  \ .
\eeq
The lower limit on $\Omega_\nu h^2$ is a reflection of the lower limit on the sum of the masses, $\sum m_\nu \, \gtrsim  \, 58 \, {\rm meV}$, that is determined from neutrino oscillation experiments \cite{Agashe:2014kda}.  This sets a clear observational target for future observations.

Any probe of $P_{mm}$ at late times is, in principle, sensitive to the sum of the neutrino masses.  The question we will be most interested in is whether a given probe is sensitive to the lower limit, $\sum m_\nu = 58 \, {\rm meV}$ (or $\Omega_\nu h^2 = 0.0006$) under realistic circumstances.  In this subsection, we will discuss the two methods through which CMB-S4 can directly constrain the neutrino mass, CMB lensing and SZ cluster abundances.  We will also compare these observables to other cosmological probes of the neutrino mass from upcoming large scale structure surveys such as DESI and LSST.

\subsection{CMB Lensing}\label{sec:neulens}

Likely the cleanest probe of the neutrino mass in the CMB is through gravitational lensing~\cite{Kaplinghat:2003bh}, which directly measures the matter distribution along the line of sight.  A measurement of the CMB lensing power spectrum provides a two-dimensional projection of the three-dimensional matter power spectrum, see Figure~\ref{fig:lensingsuppression}.  To be concrete, in the Limber approximation (see e.g.~\cite{Lewis:2006fu, LoVerde:2008re}), the lensing power spectrum is given by
\bea
\label{eq:CellPhiPhi}
C_\ell^{\phi \phi} &=& \frac{8\pi^2}{\ell^2} \int^{\chi_\star}_0 \chi d\chi P_\Psi(\ell /\chi ; \eta_0 - \chi)  \frac{(\chi_\star -\chi)^2}{\chi_\star \chi} \\
P_\Psi (k;\eta)&=& \frac{9 \Omega_m^2 (\eta) H^4(\eta)}{8 \pi^2} \frac{P_{mm}(k;\eta)}{k}
\eea
where $\chi$ ($\chi_\star$) is the co-moving distance (to the last scattering surface) and $\eta$ ($\eta_0$) is conformal time (today).   More details regarding CMB lensing, including current and future measurements, will be discussed in Chapter 7.

For the purposes of the neutrino mass measurement, the advantage of lensing over other probes is that it is largely free of astrophysical uncertainties.  As we see from the lensing power spectrum, we are directly sensitive to the matter power spectrum (rather than a biased tracer) and the relevant scales are in the linear regime where modeling should be reliable.

The primary challenges for the lensing measurement are degeneracies with other cosmological parameters.  The two primary degeneracies in $\Lambda$CDM are
\begin{itemize}
\item Optical depth, $\tau$: The suppression of small scale power at low redshift requires a reliable measurement of the amplitude of the power spectrum at high redshift.  In principle, this is measured by the primary CMB anisotropies, but the overall normalization is degenerate with $\tau$ for $\ell \gtrsim 20$.  A precise measurement of $\tau$ is therefore crucial to calibrate the suppression at low redshifts.  Such a measurement will likely come from $\ell \lesssim 20$ polarization data from other CMB experiments and/or CMB-S4.  It should be emphasized that CMB-S4 sensitivity is not needed for a sufficiently precise measurement of $\tau$ and such a measurement could be performed by a Stage III experiment.

\item $\Omega_m h^2$ : The amount of lensing is controlled by the total amount of matter.  Therefore, we can compensate for a suppression from neutrinos by increasing the matter power spectrum.  This degeneracy will be broken by DESI BAO measurements of the expansion history.
\end{itemize}
In addition to degeneracies in $\Lambda$CDM there can be degeneracies with possible extensions.  Most notably:
\begin{itemize}
\item $\Neff$: The energy density of neutrinos after they become non-relativistic is given by $\rho_\nu \simeq m_\nu n_\nu$ where $n_\nu$ is the number density.  Therefore, we only measure the mass if we know the number density to sufficient accuracy.  Fortunately, as we will discuss in the next chapter, measurements of the neutrino energy density from the primary CMB will be sufficiently accurate as to make this degeneracy insignificant under plausible assumptions.
\end{itemize}
In principle, a measurement of the free streaming scale directly in the matter power spectrum would separate the neutrino mass from most other physical quantities.  Unfortunately, given current limits on the neutrino mass, the change to the shape of the lensing potential power spectrum is not expected to drive future constraints.  This is due to the fact that the presence of neutrino mass mostly changes the overall amplitude of the lensing power spectrum at moderate and high $L$, and cosmic variance limits the constraining power of the lensing power spectrum at low $L$; see Figure~\ref{fig:lensingsuppression} and also Figure~\ref{mnuS4errors}.

{\it Status of current observations} -- \planck\ has provided a strong constraint of $\sum m_\nu < 0.194$ eV when combining both temperature and polarization data with the CMB lensing power spectrum and external data.  A weaker constraint of $\sum m_\nu < 0.492$ eV can be derived using only the temperature and polarization data.  This constraint arises through the effect of massive neutrinos on the primordial TT and EE power spectra.  For sufficiently large masses, the neutrinos do not behave as radiation around the time of recombination which impacts the damping tail and locations of the acoustic peaks.  Improvements in the limits on the sum of the neutrino masses will be driven primarily by lensing given that current limits imply that the neutrinos are effectively massless from the point of view of the primary CMB anisotropies.  External data (BAO) will continue to be important in breaking the degeneracy with $\Omega_m$.  

\subsection{Other Cosmological Probes}

While CMB lensing is likely the cleanest probe of neutrino mass, there are a number of other promising cosmological tools. In the following section, we discuss the CMB-S4 lensing data in context with other cosmological datasets. In particular, we discuss opportunities to cross-correlate CMB-S4 maps with external cosmological datasets to constrain neutrino mass with new probes and to ameliorate systematics in non-CMB measurements of large-scale structure. The abundance of galaxy clusters is a particularly well-developed probe and CMB-S4 cluster catalogs have the potential to make a huge impact. The section after, \ref{sssec:mnucluster} is devoted to a discussion of galaxy clusters as a probe of neutrino mass and the unique role of CMB-S4 cluster data.

\subsubsection{CMB Measurements in Context with Other Datasets}
\label{sssec:mnuexternal}
Current and future large-scale structure surveys, such as BOSS, DES, DESI, LSST, Euclid, and WFIRST, provide maps of the distribution of mass and galaxies in the late Universe.\footnote{While these surveys measure structure by a variety of means (the distribution of galaxies or quasars, weak gravitational lensing, and the Lyman-$\alpha$ forest, for example) we refer to all of them as galaxy surveys.} Large-scale structure datasets are primarily sensitive to the neutrino mass scale via two means: (i) the suppression of the matter power spectrum, which can be inferred from weak gravitational lensing \cite{Tereno:2008mm}, fluctuations in the number of galaxies \cite{Xia:2012na, Cuesta:2015iho}, or fluctuations in the opacity of intervening gas \cite{Palanque-Delabrouille:2014jca}, for example, and (ii) the change in the growth rate of matter perturbations which is inferred from redshift-space distortions (RSD)\cite{Beutler:2014yhv}. The first effect, the suppression in the matter power spectrum, is the same effect tested by CMB lensing.  The primary qualitative difference between information from galaxy surveys and the CMB is that galaxy surveys measure structure at multiple epochs in cosmic history whereas the CMB provides a map of the integrated mass distribution out to the surface of last scattering. The LSS information content in galaxy surveys is therefore greater than the CMB, but interpreting the data can be considerably more complex because the structure measured in galaxy surveys is typically more nonlinear and the relationship between the galaxy and mass distribution is less well-understood (in fact, massive neutrinos can make this even more complicated \cite{LoVerde:2014pxa}). Importantly, many of the observational and astrophysical systematics in the CMB and galaxy surveys are different, so the two approaches to measuring the neutrino mass scale are complementary. A summary of the forecasted constraints on $\sum m_\nu$ from galaxy surveys is given in Table \ref{table:numassLSS} \cite{Font-Ribera:2013rwa}. 
\begin{table}[t!]
\begin{center}
\begin{tabular}{|c|c|} 
\hline
    				  Datasets 			& $\sigma_{\sum m_\nu}$(eV) \\
				  \hline
\planck\ + DES lensing and galaxy clustering 		& 			0.041		\\
\hline
\planck\ + DESI Lyman-$\alpha$ Forest + BAO           &			0.098		\\
\hline
\planck\ + DESI Galaxy Power spectrum + BAO         &			0.024		\\
\hline
\planck\ + LSST Lensing and Galaxy Clustering         &   0.02					\\
\hline
\end{tabular}
\caption{Forecasted constraints on neutrino mass from future galaxy surveys in combination with \planck\ CMB from \cite{Font-Ribera:2013rwa}.  These forecasts assume \planck\ Blue Book priors for $\tau$, which is a stronger assumption than we apply to our forecasting, which assumes only the current \planck\ precision on $\tau$. A stronger prior on $\tau$ makes forecasted constraints on $\sum m_\nu$ much stronger. See related discussion in Section~\ref{sec:nuforecasts}. }
\label{table:numassLSS}
\end{center}
\end{table}

There are a number of synergistic opportunities between CMB-S4 and external cosmological datasets. For instance, the large-scale structure measured by galaxy surveys gravitationally lenses the CMB so there is a physical correlation between the CMB lensing convergence and e.g. the galaxy distribution or weak lensing shear maps inferred from galaxy surveys. Constraints on neutrino mass can therefore be tightened by cross-correlating maps of structure from galaxy surveys with the lensing information from the CMB (e.g.\ \cite{Takeuchi:2013gpa, Pearson:2013iha}). Or, the CMB can be cross-correlated with galaxy survey data to constrain neutrino mass via the mean pair-wise momentum of galaxy clusters \cite{Mueller:2014dba}. Additionally, CMB-S4 data indirectly aid measurements of neutrino mass from galaxy surveys because the CMB data can be used to calibrate systematics in weak lensing shear data \cite{Das:2013aia}.

\subsubsection{Galaxy Cluster Abundance}
\label{sssec:mnucluster}
Galaxy clusters form from rare high peaks in the matter density field. A galaxy cluster of mass $M$ forms from a region of size $R\sim \left(M/(4/3\pi \bar\rho_m)\right)^{1/3}$, which is smaller than the neutrino free streaming scale for even the most massive galaxy clusters so long as $m_{\nu i} \, \lsim \, 0.1$ eV (e.g. \cite{LoVerde:2013lta}). The neutrino free-streaming therefore slows the growth of structure on cluster scales, suppressing the abundance of galaxy clusters.

The number density of clusters with mass $M$ can be expressed by (e.g. \cite{Tinker:2008ff,Bhattacharya:2010wy}), 
\beq
\label{eq:clmfcn}
\frac{dn}{dM}(M,z) = \frac{\rho}{M}\frac{d\ln \sigma^{-1}}{dM} f(\sigma, z)
\eeq
where $\sigma = \sigma(M,z)$ is the variance of linear perturbations in CDM and baryons on mass scale $M$ given by
\beq
\label{eq:sigmaM}
\sigma^2(M, z) = \int \frac{dk}{k} \frac{4\pi}{(2\pi)^3} P_{cc}(k, z) |W(kR)|^2
\eeq
where $R = (3M/(4\pi \rho_{cb}))^{1/3}$, $P_{cc}(k)$ is the power spectrum of CDM and baryons, and $W(kR) = 3(\sin(kR)/(kR)^3 - \cos(kR)/(kR)^2)$ is a top-hat window function \cite{Costanzi:2013bha,LoVerde:2014rxa}. The cluster abundance is extremely sensitive to $\sigma(M, z)$, and therefore $\sum m_\nu$ via the suppression in $P_{cc}$ discussed in section~\ref{ssec:numasstheoryreview}.

Current constraints on neutrino mass from cluster abundance, in combination with the primary CMB and BAO, are $\sum m_\nu \lsim 0.2$--$0.3$\,eV at $95\%$ confidence \cite{Hasselfield:2013wf,Mantz:2014paa,Ade:2015fva,deHaan:2016qvy}. To date, the constraints have been driven by the difference between (or consistency of) the matter power spectrum amplitude measured at late times, from clusters, and at early times, from the CMB. An additional signal is present internally to the cluster data, namely the time-dependent influence of massive neutrinos on cluster growth through Eqs.~\ref{eq:clmfcn} and \ref{eq:sigmaM}, which can potentially provide tighter constraints \cite{Wang:2005vr}.

Making these measurements of cluster abundance and growth require cluster surveys with well understood selection functions extending to high redshift ($z\sim2$). Galaxy clusters can be identified from CMB data via the thermal Sunyaev-Zel'dovich (tSZ) effect, the frequency shift of CMB photons that have scattered off of electrons in the hot, intra-cluster gas.  CMB-S4 is projected to detect a nearly mass-limited sample of 40,000, 70,000 and 140,000 clusters 3, 2 and 1 arc-minute configurations, motivating the need for higher angular resolution.

Given such a survey, the primary systematic limitation on cluster measurements of the neutrino mass, through either of the approaches above, is cluster mass estimation. This can be usefully broken into two problems: that of absolute mass calibration, i.e.\ our ability to measure cluster masses without bias on average, and relative mass calibration. The former is required for accurate inference of the power spectrum amplitude, while the latter significantly boosts cosmological constraints by providing more precise measurements of the shape and evolution of the mass function. Precise relative mass information can be provided by X-ray observations of the intracluster medium using existing ({\it Chandra}, XMM-{\it Newton}) and future (eROSITA, ATHENA) facilities; for the most massive SZ-selected clusters found in existing CMB surveys, this work is already well advanced \cite{deHaan:2016qvy, Andersson:2010vy}. 

For absolute mass calibration, the most robust technique currently is galaxy-cluster weak lensing, which (with sufficient attention to detail) can provide unbiased results \cite{Corless:2009hi,Becker:2010xj}. At redshifts $z\lsim1$, residual systematic uncertainties in the lensing mass calibration are at the $\sim7\%$ level currently \cite{Applegate:2012kr}, and reducing these systematics further is the focus of significant effort in preparation for LSST and other Stage 4 data sets, with 1--2\% mass calibration at low redshifts seen as an achievable goal \cite{Abate:2012za}. Lensing of the CMB by clusters has also been measured \cite{Madhavacheril:2014slf,Baxter:2014frs,Melin:2014uaa}, and provides an additional route to absolute mass calibration. CMB-cluster lensing is particularly important for calibrating clusters at high redshifts ($z\gtrsim1$) where ground-based galaxy-cluster lensing becomes inefficient. CMB data with sufficient resolution and depth (especially in polarization) can potentially provide a percent-level mass calibration at these redshifts~\cite{Hu:2007bt}, comparable to galaxy-cluster lensing at lower redshifts.

As with dark energy studies, there are strong synergies between the SZ cluster catalog and CMB-cluster lensing information that CMB-S4 can provide when combined with external cluster surveys. The complementarity of CMB-S4 and LSST cluster data is particularly strong because the two datasets will primarily be sensitive to different redshift ranges. LSST will provide highly mass-complete cluster catalogs out to redshifts $\sim1.2$ and competitive lensing measurements for clusters at $z\lsim 1$. CMB-S4 will cleanly select the most massive clusters at all redshifts of interest for cluster cosmology through the SZ effect. Further, CMB lensing data can provide the key mass calibration for clusters at the highest redshifts detected by either survey. These high redshift clusters are key because they will provide the longest possible lever arm for detecting the time-dependent impact of neutrino mass on the growth of clusters.  Because their systematic uncertainties are not identical, the combination of galaxy-cluster (from e.g. LSST) and CMB-cluster lensing information can straightforwardly produce tighter constraints on the power spectrum amplitude than either alone. The prospects for Stage 4 cluster data sets to definitively detect the neutrino mass are thus strong \cite{Mantz:2014paa,Wang:2005vr}.

\subsection{Forecasts}\label{sec:nuforecasts}

For a fiducial value of $\sum m_\nu \approx$~58 meV, the target thresholds for 2$\sigma$ and 3$\sigma$ detections are $\sigma(\sum m_\nu) = 30$~meV and $20$~meV respectively.  Our goal is to explore the degree to which CMB-S4 can reach these targets with realistic forecasts.  We will assume this fiducial value throughout this discussion.  Of course, if it turns out that $\sum m_\nu > 58$~meV, it will make the prospects for detection more favorable.  

The most concrete constrains on $\sum m_\nu$ from the CMB are derived from the CMB lensing power spectrum, as seen in Equation~(\ref{eq:CellPhiPhi}).  Our sensitivity to $\sum m_\nu$ is limited by the error in the reconstruction of $\phi$ from the $T$ or $E/B$ maps.  The noise for the lensing reconstruction is discussed in chapter 7 and the noise curves for both the $TT$- and $EB$-estimators are show in Figure~\ref{n0s_s4}.  There are many reasons to prefer the $EB$-estimator which motivates a sensitivity $< 5 \, \mu K$-arcmin.

\begin{figure}[ht]
\begin{center}
\includegraphics[scale=0.4]{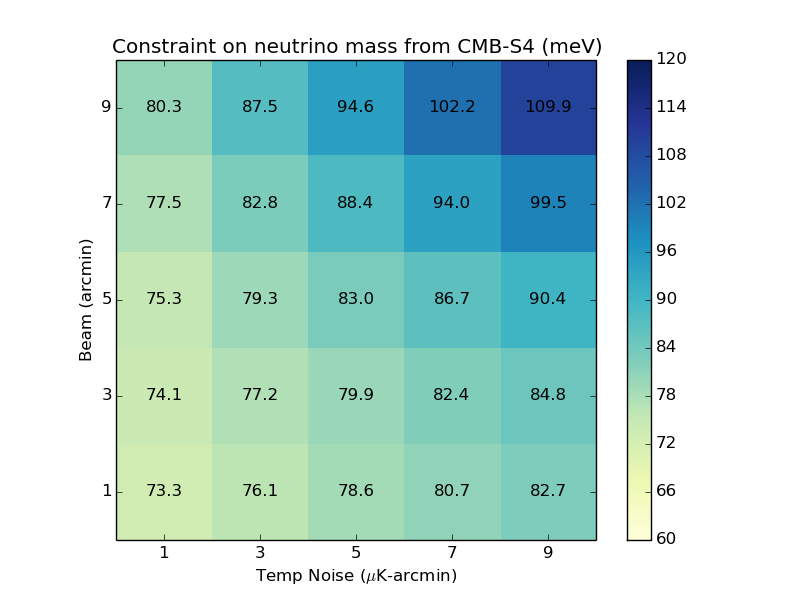}
\includegraphics[scale=0.4]{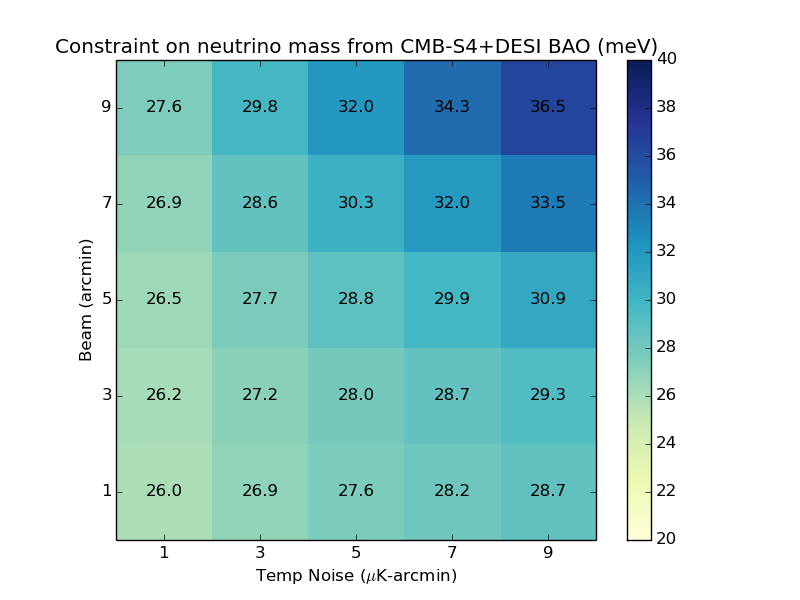}  
\includegraphics[scale=0.4]{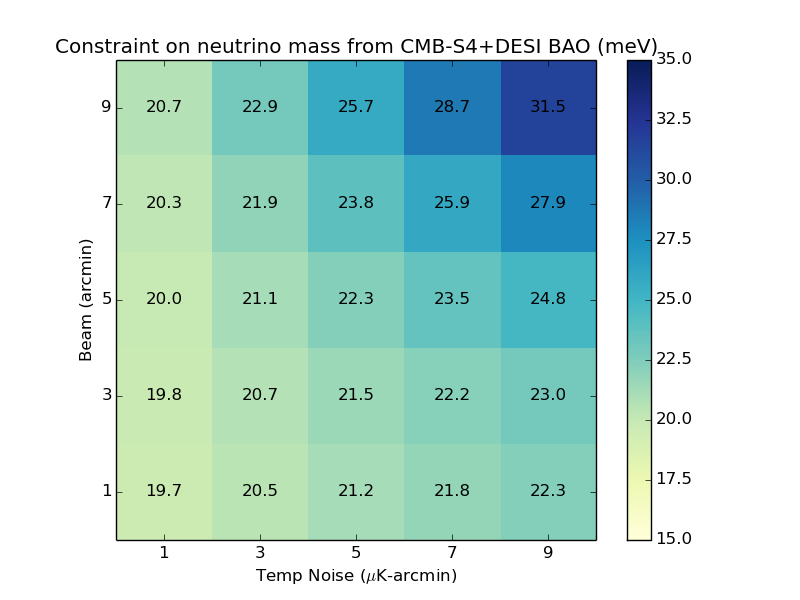}

\caption{ Forecasts for $\sigma(\sum m_\nu)$ assuming $\Lambda$CDM + $\sum m_\nu$.  All three figures vary beam size in arcmin and effective detector noise in $\mu$K-arcmin. {\it Top Left:} CMB-S4 alone with an external prior on $\tau = 0.06 \pm 0.01$. {\it Top Right: } CMB-S4 plus DESI BAO  with an external prior on $\tau = 0.06 \pm 0.01$.  {\it Bottom:} Forecasts assuming a prior $\tau = 0.06 \pm 0.006$, corresponding to the \planck\ Blue Book expected sensitivity.}
\label{fig:mnuforecast}
\end{center}
\end{figure}

Forecasts for CMB-S4 with and without DESI BAO are shown in Figure~\ref{fig:mnuforecast}, following the methodology outlined in Section~\ref{sec:Forecasting}.  As we explained in Section~\ref{sec:neulens}, the main signature of neutrino mass in CMB lensing is degenerate with both $\tau$ and $\Omega_m h^2$.  The sensitivity is therefore strongly dependent on the constraints on these parameters both internally and with external data.  We can see the effect of the measurement of $\Omega_m h^2$ by comparing the results with and without DESI BAO.  In particular, we note that the constraints improve by a factor of three by including DESI.

The most significant limitation to measuring $\sum m_\nu$ with CMB-S4+DESI BAO is the uncertainty of $\tau$.  We will conservatively assume that $\ell \geq 30$ for CMB-S4 and therefore we do not constrain $\tau$ directly.  Under such circumstances, the constraint on $\tau$ must come from external data, either currently from \planck\ or from future observations.  We will consider the possibility of reaching $\ell < 30$ with CMB-S4 in the same category as other future experiments that hope to improve the measurement of $\tau$.

The current best constraint on $\tau$ comes from \planck~\cite{Adam:2016hgk}, roughly
corresponding to an external prior of
 $\tau = 0.06 \pm 0.01$.  We see in Figure~\ref{fig:mnuforecast} that
 this prior is sufficient to reach $\sigma(\sum m_\nu) < 30$ meV for a wide range of experimental configurations.  On the other hand, we also note that there is little improvement with decreased noise or beamsize as we saturate at $\sigma(\sum m_\nu) \sim 26$ meV,  even if we increase $f_{\rm sky} >0.4$.  Of course, the reason is that we are limited by the $\tau$-degeneracy.

\begin{figure}[h!]
\begin{center}
\includegraphics[scale=0.6]{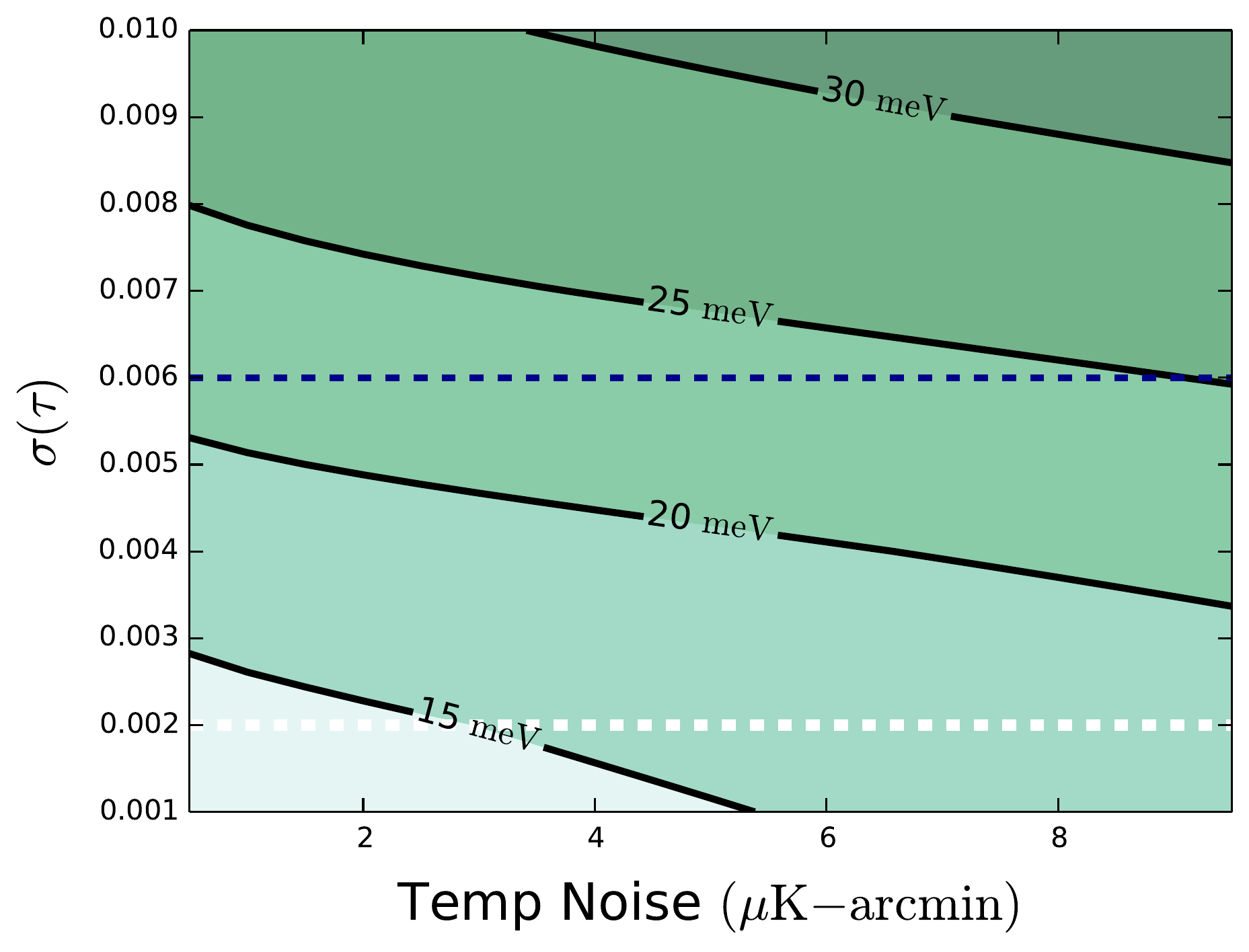}

\caption{ Forecasts for $\sigma(\sum m_\nu)$ assuming $\Lambda$CDM + $\sum m_\nu$ using CMB-S4 and DESI BAO.  We vary sensitivity in $\mu$K-arcmin and $\tau$-priors, $\tau = 0.06 \pm \sigma(\tau)$ with the contours showing 1$\sigma$ errors for $\sum m_\nu$.  We fixed the resolution using 1' beams and set $f_{\rm sky} =0.4$.  The white and blue dashed lines correspond to the low-$\ell$ cosmic variance limit and \planck\ Blue Book values respectively. }
\label{fig:mnu_tau}
\end{center}
\end{figure} 

In order to reach the 3$\sigma$ target, $\sigma(\sum m_\nu) < 20$ meV,
one needs a better measurement of $\tau$.
As shown in bottom panel of Figure~\ref{fig:mnuforecast},
we can reach $\sigma(\sum m_\nu) \approx 20$\,meV for a variety of plausible
configurations of CMB-S4 with \planck's designed reach in sensitivity, a
measurement at the level of $\sigma(\tau) = 0.006$.
 However, as before, we see that there are only moderate improvements coming from lower noise or smaller beams.  A similar limitation applies to other cosmological probes, as seen in Table~\ref{table:numassLSS}, which also saturate at a similar sensitivity.

More generally, improved measurements of $\tau$ and $H_0$ may become
available before, during or after CMB-S4.  We therefore also examine impacts of measurements even further in the future in evaluating the value of the legacy data from CMB-S4.
There are ground-based CMB
instruments~\cite{Essinger-Hileman:2014pja,Oguri:2015uhi}
designed to observe
very large angular scales, with possible reach to constrain $\tau$;
the CLASS experiment is forecasted to reach
$\sigma(\tau) \simeq 0.004$~\cite{CLASSSPIE2016}, for example.
Space missions~\cite{Matsumura:2016sri,Kogut:2011xw} are
proposed to constrain the primordial gravitational
waves through the so-called reionization bump during the 2020s; they are designed to
reach sensitivity well beyond that required to achieve a cosmic-variance
limited $\tau$ measurement, $\sigma(\tau) \sim 0.002$.
Measurement of reionization through 21-cm hyperfine transition can,
in principle, determine $\tau$ beyond the accuracy of the
cosmic-variance limited CMB measurement~\cite{Liu:2015txa}.

Small-scale Stage Three CMB probes may themselves provide a window into improving the current $\tau$ bounds. As discussed in \cite{Calabrese:2014gwa, Galli:2014kla} the $EE$ polarisation spectrum will soon allow one to remove the CMB as `foreground' from temperature measurements and yield a multi-$\sigma$ detection of the kSZ effect, thereby improving our constraint on the optical depth by leveraging the relationship between $\tau$ and the amplitude of the $kSZ$ spectrum. Projects errors on the optical depth are $\sigma(\tau) =  0.006,$ from Stage Three data alone, provided the CMB-S3 experiments can probe down to multipoles of $\ell \simeq 10$. 
Any of these measurements should provide significant improvements in CMB-S4's sensitivity to the neutrino mass. Allowing for such possibilities, we consider the impact of stronger priors on $\tau$.  Figure~\ref{fig:mnu_tau} shows the 1$\sigma$ errors on $\sum m_\nu$ for varying $\tau$-priors.  
We see that $\sigma(\sum m_\nu) < 15$ meV, corresponding to 4$\sigma$ detection of the minimum neutrino mass, can be reached with a cosmic-variance-limited optical depth constraint and a CMB-S4 survey noise level below 2.5 $\mu$K-arcmin.

\section{Relation to Lab Experiments}\label{sec:lab}

\subsection{Determining the neutrino mass scale}
\begin{figure}[h!]
\centering
\includegraphics[width=1.0\textwidth]{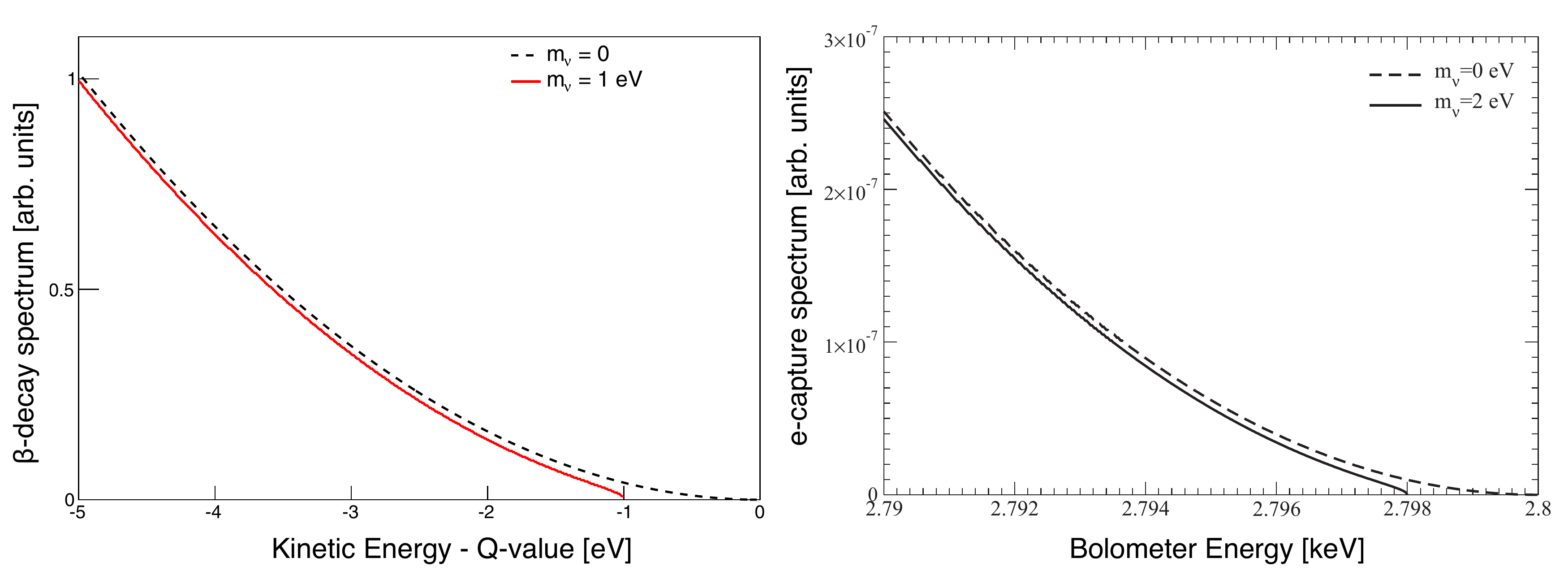}
\caption{Shown on the left is the kinematic suppression of the
  $\beta$-decay spectrum in the presence of a zero or 1\,eV $\beta$-decay
  neutrino mass. On the right are the bolometer spectra for endpoint
  energies in electron capture for effective zero and 2\,eV neutrino
  masses.}
\label{fig:kinematic_mass}
\end{figure}

As discussed above, CMB-S4 will make a cosmological measurement of $\sum m_\nu$ and thus the neutrino mass scale. This approach complements terrestrial measurements of the neutrino mass using radioactive decay. These kinematic measurements of neutrino mass focus on one of two processes, beta-decay or electron-capture, where the decay spectra near the decay endpoint is particularly sensitive to the mass of the neutrino (see Fig.~\ref{fig:kinematic_mass}).

Current kinematic measurements from Mainz~\cite{Kraus:2004zw} and Troitsk~\cite{Aseev:2011dq} limit the electron antineutrino mass to $< 2.0$~eV. The KATRIN experiment~\cite{Angrik:2005ep} will begin taking data in 2016 and is expected to improve this limit by a factor of ten. 

Within the standard neutrino mass and cosmological paradigm, the kinematic and cosmological measurements of the neutrino mass are connected through the PMNS matrix. Thus, the combination of cosmological and terrestrial neutrino mass measurements tests our cosmological neutrino model. A discrepancy could point to new physics (e.g. modified thermal history through neutrino decay).

Improving kinematic measurements beyond KATRIN's $0.2$~eV limit will require new technology since \mbox{KATRIN} will be limited by the final state spectrum of the source itself, specifically rotational-vibrational states of molecular Tritium. One of the new approaches is a calorimetric measurement of the electron-capture spectrum of $^{163}$Ho. The calorimetric measurement of the $^{163}$Ho endpoint is insensitive to the details of the source configuration and may provide an avenue for eventually surpassing the KATRIN sensitivity. Interestingly, upcoming experiments such as ECHO~\cite{Eliseev:2015pda}, HOLMES \cite{Ceriale:2015mtn}, and NuMECS~\cite{Croce:2015kwa} utilize multiplexed superconducting detectors, the same technology baselined for the CMB-S4 experiment. Another promising direction for direct neutrino mass measurement is the frequency-based technique employed by the Project-8 experiment~\cite{Asner:2014cwa}. Project-8 aims to measure the beta-decay spectrum of Tritium by measuring the frequency of cyclotron radiation emitted by the decay electrons when trapped in a magnetic field. An exciting aspect to this frequency-based technique is the potential to trap atomic Tritium which is not subject to the rotational-vibrational excitations of molecular Tritium. A spectroscopic measurement using atomic Tritium could eventually achieve sensitivities of $<0.04$~eV. 

\subsection{Lepton number violation: Majorana vs. Dirac neutrinos}
One of the more interesting connections between cosmological measurements of neutrino mass and terrestrial experiments is the complementarity between cosmological neutrino mass measurements and the search for neutrinoless double beta decay (NLDBD). NLDBD is a hypothetical decay mode of certain nuclei where two neutrons convert to two protons and two electrons with no emission of neutrinos. The observation of NLDBD would be transformational demonstrating that neutrinos are Majorana particles and revealing a new lepton-number-violating mechanism for mass generation. This new physics could potentially explain both the smallness of neutrino masses and matter-antimatter asymmetry in the Universe.

Initial results from the current generation of NLDBD searches limit the NLDBD half life, $T^{0\nu}_{1/2}$, to be larger than  $\sim10^{26}$~years~\cite{Agostini:2013mzu,Auger:2012ar,Artusa:2014lgv,KamLAND-Zen:2016pfg}.  Planning and technology development is already underway for next generation ``ton-scale'' NLDBD searches which would achieve sensitivities of $10^{27}-10^{28}$~years~\cite{Geesaman:2015fha}.

\begin{figure}[h!]
\centering \includegraphics[width=0.8\textwidth]{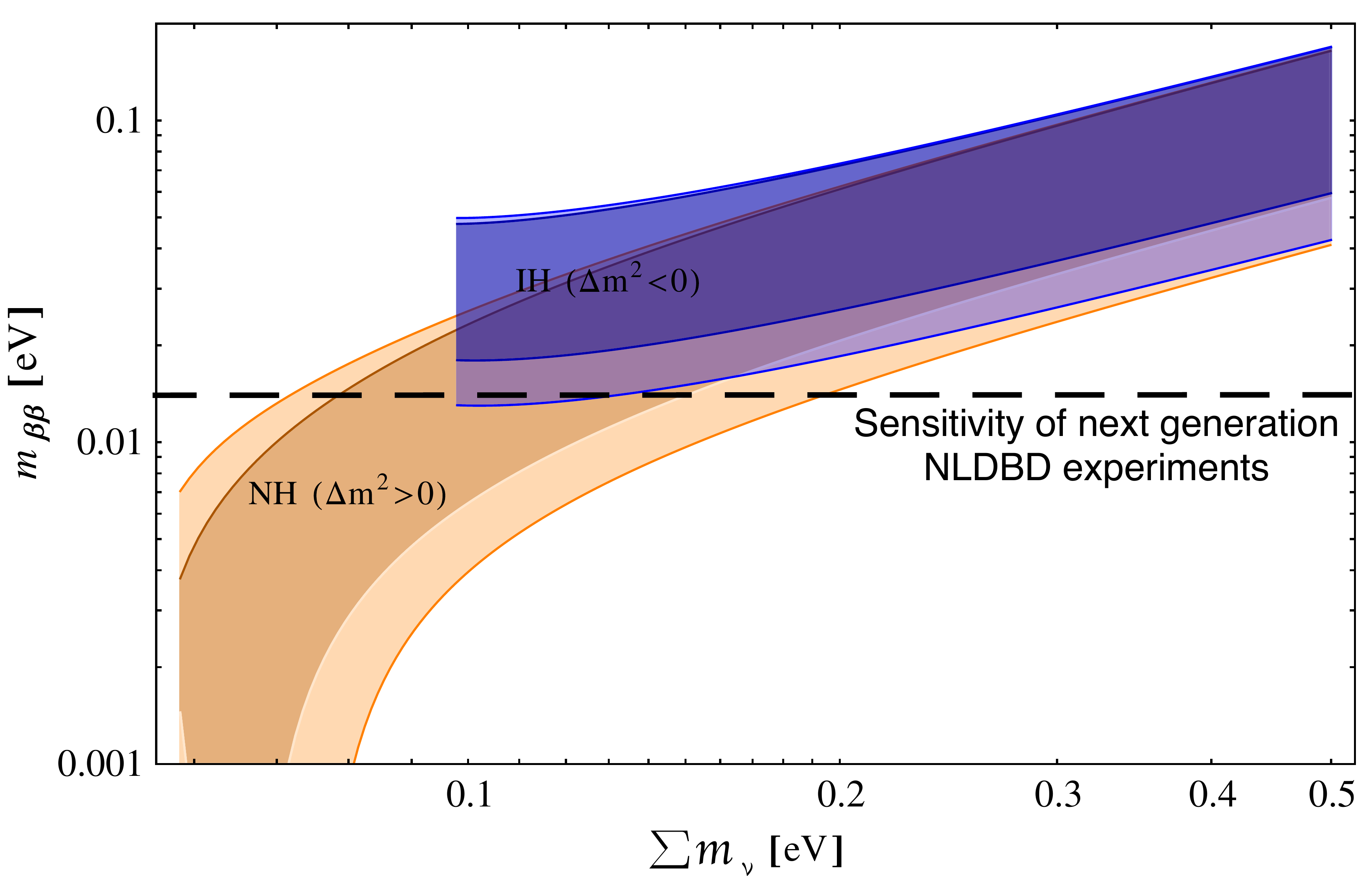}
\caption{Plot of effective neutrino mass versus $\sum m_\nu$ in the scenario where NLDBD is mediated by light neutrino exchange. The blue band corresponds to inverted ordering and the orange band corresponds to normal ordering. Next generation ``ton-scale'' NLDBD searches will have sensitivities down to $m_{\beta\beta}>15$~meV (dashed line). Figure from~\cite{Dell'Oro:2014yca}}
\label{fig:NLDBD}
\end{figure}

We can illustrate the connection between NLDBD searches to cosmological determinations of neutrino mass by examining the simplest case where NLDBD is mediated by exchange of light Majorana neutrinos. Within the context of this mechanism, we can define an ``effective neutrino mass,'' $m_{\beta\beta}$, given by 

\beq
m_{\beta\beta}^{2} = ( \sum_i U_{ei}^{2}m_{\nu i} )^{2}
\label{eq:mbb}
\eeq
where $m_{\nu i}$ are the light neutrino masses and $U_{ei}$ is the usual PMNS mixing matrix including two unknown Majorana phases. The NLDBD half-life is then given by 

\beq
(T^{0\nu}_{1/2})^{-1} = G^{0\nu}\cdot (M^{0\nu} )^{2}\cdot m_{\beta\beta}^2,
\eeq

where $G^{0\nu}$ is a phase space integral and $M^{0\nu}$ is the nuclear matrix element. In this simple scenario, the signal from NLDBD experiments can be directly related to other measures of neutrino mass. Figure~\ref{fig:NLDBD} illustrates this relationship between the effective neutrino mass and the lightest neutrino mass including limits and sensitivities of current and next generation NLDBD searches.

\begin{figure}[h!]
\centering \includegraphics[width=0.70\textwidth]{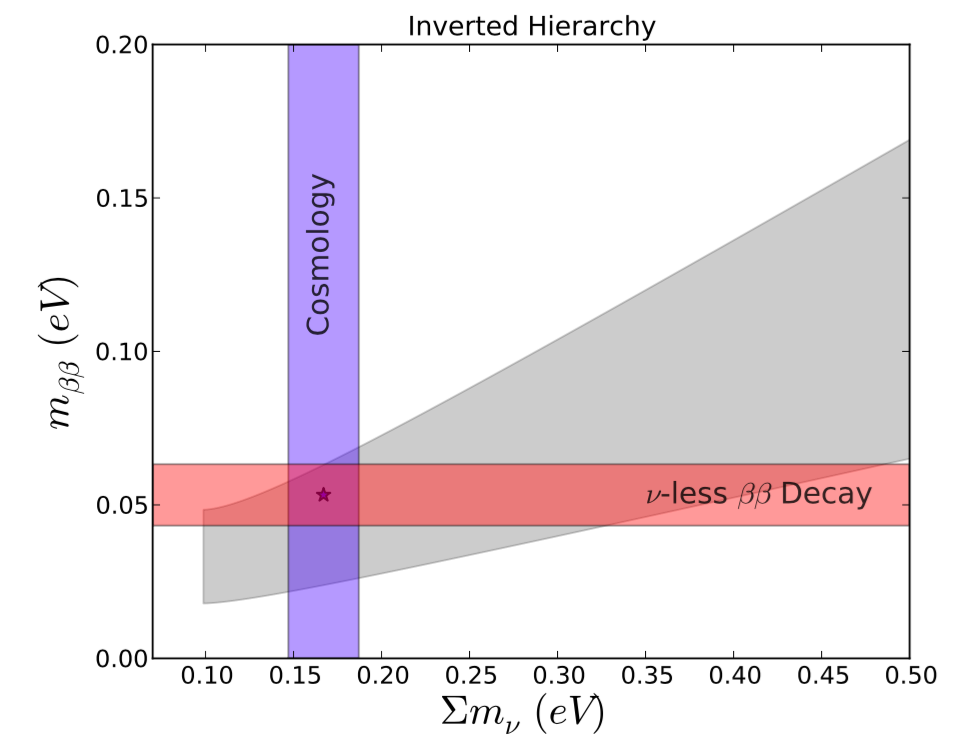}
\caption{Relationship between effective neutrino mass as measured by NLDBD experiments versus $\sum_i m_{\nu i}$ as measured by cosmology for the inverted hierarchy. The gray band corresponds to a region allowed by existing measurements where the width of the band is determined by the unknown Majorana phase. }
\label{fig:MajoranaPhase}
\end{figure}

The complementarity between cosmological neutrino mass measurement and NLDBD can be understood by considering scenarios where NLDBD experiments either observe or fail to observe NLDBD. In the absence of a signal in next generation NLDBD searches, a cosmological measurement constraining $\sum m_\nu > 100$~meV (corresponding to either the inverted hierarchy or a minimum neutrino mass of 50~meV) would strongly point to neutrinos being Dirac particles (see Fig.~\ref{fig:NLDBD}). On the other hand, if NLDBD is observed, equation~\ref{eq:mbb} shows that cosmological measurements of $\sum m_\nu$ are sensitive to the Majorana phases. For example, Fig.~\ref{fig:MajoranaPhase} shows that in the inverted mass hierarchy cosmological measurements together with NLDBD measurements can constrain one of the Majorana phases. Perhaps even more interesting would be the situation where cosmological and NLDBD measurements violate equation~\ref{eq:mbb} indicating new physics beyond the simple model of light Majorana neutrino mediated decay.

\subsection{Neutrino mass ordering and CP violation}
In the case of normal ordering with non-degenerate neutrino mass, the CMB-S4 measurement of $\sum m_\nu$ will provide a 2--4$\sigma$ determination of the neutrino mass ordering. Fully characterizing neutrino mass ordering and CP violation is one of the goals of the terrestrial neutrino physics program~\cite{Patterson:2015xja}. The upcoming reactor neutrino experiment JUNO~\cite{An:2015jdp} is scheduled to start data taking around $\sim$ 2020 and will have $\sim$ 2--3$\sigma$ sensitivity to neutrino ordering after six-years of operation. Future experiments measuring atmospheric neutrino oscillations (e.g Hyper-K~\cite{Abe:2015zbg}, DUNE~\cite{Goodman:2015gmv}, KM3NeT/ORCA~\cite{Adrian-Martinez:2016fdl}) can also resolve the neutrino mass ordering. For example, KM3NeT/ORCA forecasts a 3$\sigma$ measurement of the mass ordering by around 2023. Accelerator neutrino experiments are the only known method for exploring neutrino CP violation and in some cases, are also sensitive to neutrino mass ordering. In a $\sim 5$-year timescale, the currently operating NO$\nu$A experiment~\cite{Adamson:2016tbq} may determine the neutrino ordering at the 2--3$\sigma$ level, provided that $\delta_{CP}$ falls into a favorable range. Hyper-K will measureme $\delta_{CP}$, though it requires external input regarding the neutrino ordering (e.g. from Hyper-K atmospheric neutrinos or from cosmology). The next-generation US-based long-baseline neutrino-oscillation experiment, DUNE, is planned to start operation around 2024, and will measure both the neutrino mass ordering (at the 2--4$\sigma$ level) and $\delta_{CP}$. External input on neutrino ordering from other sources such as CMB-S4 would provide a strong consistency check of DUNE results and test the three-neutrino paradigm.

In the scenario where the neutrino mass specrtum is normally ordered and non-degenerate, CMB-S4 would be a strong complement to terrestrial experiments by providing a measurement of neutrino ordering that is independent of oscillation parameters and $\delta_{CP}$. Under all circumstances, the combination of CMB-S4 with terrestrial determinations of neutrino ordering will provide a definitive measurement of the neutrino mass spectrum.

\begin{figure}[h!]
\centering \includegraphics[width=0.55\textwidth]{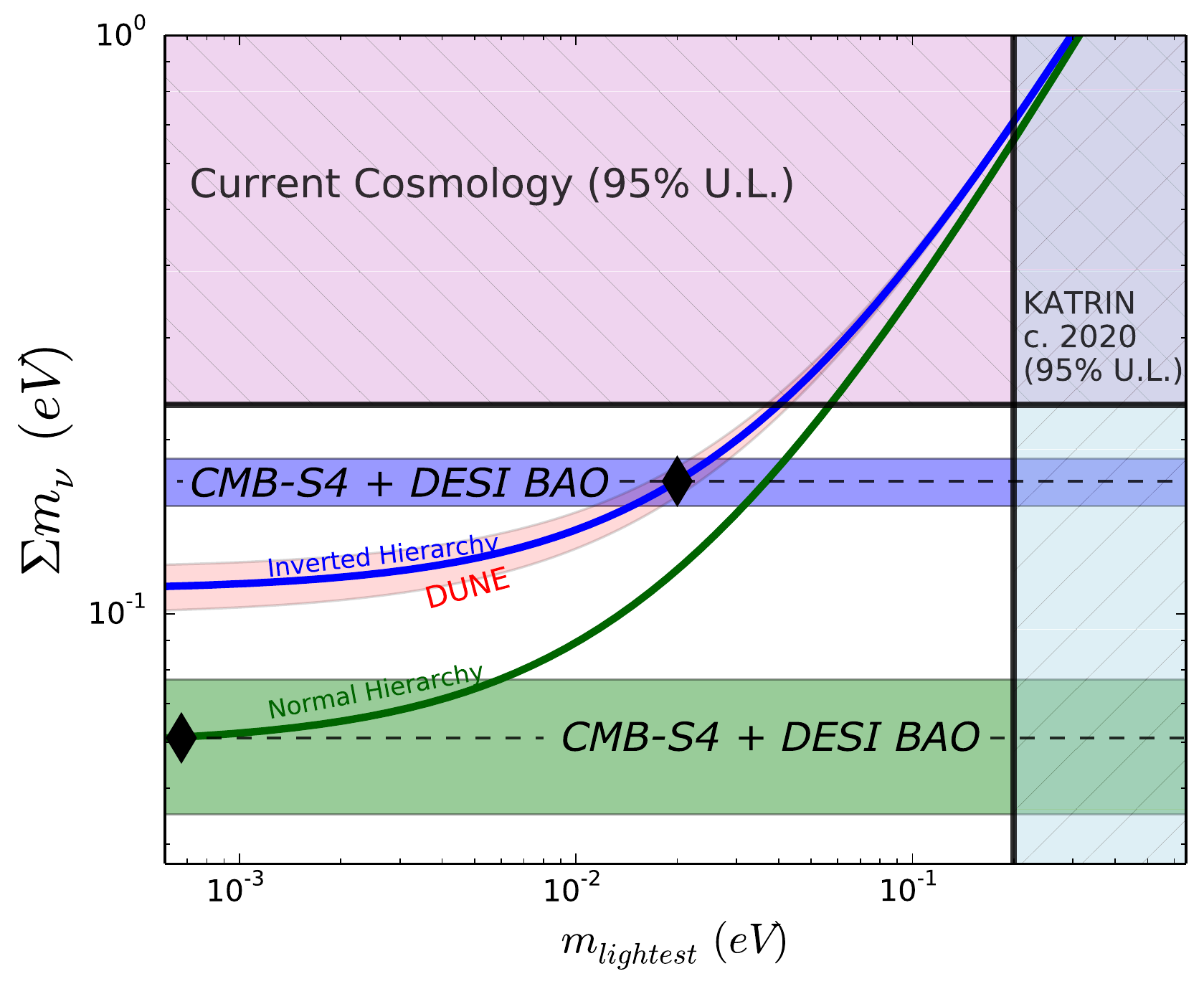}
\caption{Shown are the current constraints and forecast sensitivity of
  cosmology to the neutrino mass in relation to the neutrino mass
  hierarchy.  In the case of an ``inverted ordering,'' with an
  example case marked as a diamond in the upper curve, the CMB-S4 (with DESI BAO prior)
  cosmological constraints would have a very high-significance
  detection, with $1\sigma$ error shown as a blue band.  In the case
  of a normal neutrino mass ordering with an example case marked as
  diamond on the lower curve, CMB-S4 would detect the lowest
  $\sum m_\nu$ at $\gtrsim 3 \sigma$. Also shown is the
  sensitivity from the long baseline neutrino experiment (DUNE) as the
  pink shaded band, which should be sensitive to the neutrino
  hierarchy. Figure adapted from the Snowmass CF5 Neutrino planning document.
 }
\label{fig:neutrino-noose}
\end{figure}

\subsection{Sterile Neutrinos}
\label{sec:sterile_neutrinos}

Mechanisms of introducing neutrino mass often include sterile
neutrinos, with both Majorana and Dirac terms potentially
contributing (e.g., Ref.~\cite{Langacker:2011bi}):
\bea
\mathcal{L}_D &=& -m_D\left(\bar\nu_L\nu_R + \bar\nu_R\nu_L\right) \\
\mathcal{L}_M &=& -\frac{1}{2}m_T\left(\bar\nu_L\nu_L^c + \bar\nu_L^c\nu_L\right) 
-\frac{1}{2}m_S\left(\bar\nu_R\nu_R^c +\bar\nu_R^c\nu_R\right) =
-\frac{1}{2}m_T\left(\bar\nu_a\nu_a\right) -\frac{1}{2}m_S\left(\bar\nu_s\nu_s\right),
\eea
where $\nu_a \equiv \nu_L + (\nu_L)^c$ and $\nu_S \equiv \nu_R +
(\nu_R)^c$ are active and sterile Majorana two component spinors,
respectively. The mass $m_T$ can be generated by a Higgs triplet,
i.e., $m_T = y_T\langle \phi^0_T\rangle$, or from a higher-dimensional
operator involving two Higgs doublets with coefficients
$C/\mathcal{M}$. For dimension 5 operators, this becomes the Type-I
seesaw mechanism, where both Majorana and Dirac terms are present and
$m_S \gg m_D$.

A number of recent neutrino oscillation experiments have reported anomalies
that are possible indications of four or more neutrino mass eigenstates. The
first set of anomalies arose in short baseline oscillation experiments.
First, the Liquid Scintillator Neutrino Detector (LSND) experiment observed
electron antineutrinos in a pure muon antineutrino
beam \cite{Athanassopoulos:1997pv}. The MiniBooNE Experiment also observed
an excess of electron neutrinos and antineutrinos in their muon
neutrino beam \cite{Aguilar-Arevalo:2013pmq}. Two-neutrino oscillation
interpretations of these results indicate mass splittings of $\Delta m^2
\approx 1\rm\ eV^2$ and mixing angles of $\sin^2 2\theta \approx
3\times 10^{-3}$ \cite{Aguilar-Arevalo:2013pmq}. Another anomaly
arose from re-evaluations of reactor antineutrino fluxes that
indicate an increased flux of antineutrinos and a lower
neutron lifetime. This commensurately increased the predicted antineutrino events
from nuclear reactors by 6\%, causing previous agreement of
reactor antineutrino experiments to have a $\approx$6\% deficit
\cite{Mention:2011rk,Huber:2011wv}. Another indication consistent with
sterile neutrinos was observed in radio-chemical gallium experiments for solar
neutrinos. In their calibrations, a 5-20\% deficit of the measured
count rate was found when intense sources of electron neutrinos from
electron capture nuclei were placed in proximity to the
detectors. Such a deficit could be produced by a $m_S >1\rm\ eV$ sterile
neutrino with appreciable mixing with electron neutrinos
\cite{Bahcall:1994bq,Giunti:2010zu}. Some simultaneous fits to the
short baseline anomalies and reactor neutrino deficits, commensurate
with short baseline constraints, appear to prefer at least two extra
sterile neutrino states \cite{Conrad:2012qt,Kopp:2013vaa}, but see~\cite{Giunti:2015mwa}. Because such neutrinos have relatively
large mixing angles, they would be thermalized in the early Universe
with a standard thermal history, and affect primordial nucleosynthesis
\cite{Abazajian:2002bj} and CMB measurements of $\Neff$.  These implications will be discussed in the next chapter.

To accomodate $m_S = \mathcal{O}({\rm eV})$ with some mixing between active
and sterile states in the neutrino mass generation mechanism discussed
above requires mixing between active and sterile states with the same
chirality, which does not occur for pure Majorana or Dirac mass cases
or for the conventional seesaw mechanism. One proposed mechanism is
the minimal mini-seesaw ($m_T =0$ and $m_D\ll
m_S\sim\mathcal{O}(\mathrm{eV})$,
e.g. Ref.~\cite{deGouvea:2011zz,Donini:2012tt}). In such models, the
sterile neutrinos can have the appropriate masses and mixings to
accommodate the short baseline anomalies. For standard thermal
histories, these sterile neutrinos are typically fully thermalized
\cite{Abazajian:2002bj}. However, it is possible they are partially
thermalized in two extra neutrino models \cite{Jacques:2013xr}.

For the simple case of a fully thermalized short baseline sterile neutrino at the minimal mass scale from recent reactor limits~\cite{Adamson:2016jku}, CMB-S4 would have the sensitivity of approximately $\sqrt{|\delta m^2_{41}|}/\sigma(\Sigma m_\nu) \approx 40 \sigma$ to the presence of a massive thermalized sterile neutrino and $1/\sigma(N_{\rm eff}) \approx 37 \sigma$ sensitivity to the thermalized extra radiation-like energy density.  While one does not need the power of CMB-S4 to place exclusions on this scenario, the increased sensitivity would further push the tension into the domain where additional physics is needed, above and beyond sterile neutrinos, to simultaneously explain the observations.  

Interestingly, there are combinations of CMB plus LSS
datasets that are in tension, particularly favoring a smaller amplitude of
fluctuations at small scales than that predicted in zero neutrino mass
models. This would be alleviated with the
presence of massive neutrinos, extra neutrinos, or both. In particular,
cluster abundance analyses \cite{Wyman:2013lza,Ade:2015fva} and weak lensing analyses
\cite{Battye:2013xqa} indicate a lower amplitude of
fluctuations than zero neutrino mass models \cite{Giusarma:2014zza}. Baryon Acoustic
Oscillation measures of expansion history are affected by the presence
of massive neutrinos, and nonzero neutrino mass may be indicated 
\cite{Beutler:2014yhv}, though 2015 \planck\ results do not show a preference for models with massive or extra neutrinos
\cite{Ade:2015xua}. 

There is a potential emergence of both laboratory and cosmological
indications of massive and, potentially, extra neutrinos. However, the
combined requirements of the specific masses to produce the short
baseline results, along with mixing angles that require thermalized
sterile neutrino states, are inconsistent at this point with
cosmological data sets
\cite{Joudaki:2012uk,Archidiacono:2013xxa}. The tensions present between data sets are
not highly significant at this point ($\lesssim 3\sigma$), and there
are a significant set of proposals for short baseline oscillation
experiment follow up \cite{Abazajian:2012ys}. 

In fact, we know little about the sterile neutrino sector, though its possible existence is in part motivated by the experimental establishment that neutrinos have nonzero rest masses as discussed above. CMB-S4 could shed light on the sterile neutrino mass and vacuum flavor mixing parameters invoked to explain the experimental neutrino anomalies. It must be kept in mind that sterile neutrinos might have different masses and much smaller vacuum mixing with active neutrinos species. Telltale signatures in $\Neff$, $\sum m_{\nu}$, and $Y_p$ can allow CMB-S4 to probe this larger parameter space.  These measurements will be discussed in the next chapter.

\section{Detection Scenarios for Neutrino Physics} \label{sec:neuscenarios}

As discussed in Section~\ref{sec:lab}, the measurements of the absolute scale of the neutrinos masses from the lab and from cosmology are complementary in that they are sensitive to different parameters.  In principle, there are a variety of possible scenarios where detections are made both in cosmology and in the lab.  However, given current constraints, most scenarios that involve mostly conventional neutrino physics will result in upper limits from the lab based measurements and a detection of $\sum m_\nu$ and/or $\Delta\Neff \equiv \Neff - \Neff^{\rm SM}$ with $\Neff^{\rm SM} \approx 3.046$.  We will discuss $\Neff$ is greater detail in the next chapter but here it serves as a measurement of the total number of thermal neutrinos which calibrates the cosmological mass measurement. A plausible list of detection scenarios are shown in Table~\ref{table:neutrinoscenarios}:

\begin{itemize}
\item Conventional neutrino mass scenarios imply Majorana masses with a normal or inverted hierarchy.  The normal hierarchy with $\sum m_\nu \simeq 58$~meV is perhaps the most conventional as it reflects the same hierarchical / non-degenerate masses that appear in the charged fermions of the Standard Model.  This scenario is only detectable in the near term via cosmology due to the small size of the neutrino masses.  Somewhat more exotic is the case of a Dirac mass, as it predicts the existence of new light states.

\item The more exotic possibility is that there could be sterile neutrinos that are consistent with a variety of anomalies, as discussed in Section~\ref{sec:sterile_neutrinos}.  In this case, we would observe a correlated signature in both a excess in $\sum m_\nu$ and $\Delta \Neff$ due to the presence of thermalized sterile neutrinos in addition to the active neutrinos.  The sterile neutrino parameters that are most consistent with the anomalies in short-baseline experiments are already in tension with cosmology but would be detected at high significance if these models describe our Universe.  

\item Given the current cosmological constraints on $\sum m_\nu$, detections of $m_\beta$ and $m_{\beta \beta}$ in near term experiments would require a significant change to the thermal history.  In particular, a detection of a Majorona mass at the $0.25 \, {\rm eV}$ level would predict a $\sum m_\nu$ that is already excluded by cosmology.  Making the current (or future) limit consistent then requires a mechanism that satisfies both the present bound on $\sum m_\nu$ and the current constraints on $\Neff$.

\item There are a variety of scenarios that produce $\Delta \Neff  \neq 0$ without changing neutrino physics.  In this case, the neutrino physics may follow a conventional pattern like the normal hierarchy.  In principle, one would distinguish scenarios where there is no change to the number density of neutrinos (dark radiation) from scenarios where the neutrinos are diluted or enhanced by a change to the thermal history (late decay) as the interpretation of $\sum m_\nu$ depends on the neutrino number density.  However, given that current measurements allow for $< 10$ percent change to the neutrino number density, we would need to detect $\sum m_\nu$ at 10$\sigma$ to be sensitive to such a change.  Nevertheless, dark radiation and changes to the thermal history can make correlated predictions for other experiments as we will discuss in the next Chapter.

\item The neutrino sector may be rich in new physics and CMB-S4 provides us with significant new discovery potential. Finding evidence for sterile neutrinos or a primordial lepton number larger than the baryon number, but well below the present BBN bound, $\sim 0.1$, would be a signal event for particle physics. Sterile neutrinos with masses of order $\sim 1\,{\rm eV}$ and relatively large vacuum mixing with active species ($\sin^2 \theta \sim 10^{-3}$), like those invoked to explain the neutrino anomalies, would have relic densities and relic energy spectra comparable to those of the active neutrinos and therefore are easy targets for CMB constraint or detection as discussed above. However, a primordial lepton number $> {10}^{-4}$ would suppress the production of these sterile neutrinos in the early Universe. Moreover, sterile neutrinos with tiny vacuum mixing could escape laboratory detection, yet still might acquire small but significant population in the early Universe through lepton number-induced resonant conversion of active neutrinos~\cite{Abazajian:2005gj}. Both of these cases could leave telltale signatures in CMB observables, specifically in $\Neff$, $\sum m_\nu$, and $Y_p$. The pattern of changes in these observables, compared to results from high precision neutrino transport and flavor oscillation physics coupled to weak and nuclear reactions, can give distinctive markers for sterile neutrino mixing physics (see e.g. \cite{Smith:2006uw, Grohs:2015tfy}).

\end{itemize}
\begin{table}[t!]
\begin{center}
\begin{tabular}
{| l | c c c c | p{5cm} | }\hline Scenario & $m_{\beta \beta}$ & $m_{\beta}$&  $\sum m_\nu$ & $\Delta \Neff$ & Conclusion \\
\hline 
Normal hierarchy & $< 2\sigma$ & $< 2\sigma$  & $60 \, {\rm meV}$ & 0 & Normal neutrino physics; no evidence for BSM
\\[.2cm]
Dirac Neutrinos & $< 2\sigma$ & $< 2\sigma$  & $350  \, {\rm meV}$ & 0 & Neutrino is a Dirac particle \\[.2cm]
Sterile Neutrino & $< 2\sigma$ & $< 2\sigma$   & $350  \, {\rm meV}$ & $>0$ & Detection of sterile neutrino consistent with short-baseline \\
\hline
Diluted Neutrinos & $ 0.25 \, {\rm eV}$ & $ 0.25 \, {\rm eV}$  & $<150  \, {\rm meV}$ & $< 0$ & Modified thermal history (e.g. late decay) \\[.2cm]
Exotic Neutrinos & $ 0.25 \, {\rm eV}$ & $ 0.25 \, {\rm eV}$  & $<150  \, {\rm meV}$ & $0$ & e.g. Modified thermal history; (e.g. neutrino decay to new particle) \\[.2cm]
Excluded & $ 0.25 \, {\rm eV}$ & $ 0.25 \, {\rm eV}$  & $500  \, {\rm meV}$ & $0$ & Already excluded by cosmology \\
\hline
Dark Radiation & $< 2\sigma$ & $< 2\sigma$  & $60  \, {\rm meV}$ & $>0$ & Evidence for new light particles; normal hierarchy for neutrinos
\\[.2cm]
Late Decay & $< 2\sigma$ & $< 2\sigma$  & $60  \, {\rm meV}$ & $<0$ & Energy-injection into photons at temperature $T \lesssim 1$ MeV \\
\hline 
\end{tabular}
\caption{Relation between neutrino experiments and cosmology.  We include the measurement of the Majorona mass via NLDBD ($m_{\beta \beta}$) or a kinematic endpoint ($m_\beta$) compared to the cosmological measurement of the sum of the masses $\sum m_\nu$ and the CMB measurement of $\Neff$.  Here $< 2 \sigma$ indicates an upper limit from future observations.  For Section~\ref{sec:lab}, one can use $\sigma(m_{\beta\beta}) \approx 0.075 \, {\rm eV} $ and $\sigma(m_\beta) \approx 0.1 \, {\rm eV} $ for observations on the timescale of CMB-Stage IV.  For $\Delta\Neff$ the use of $\gtrless 0$ indicates a significant deviation from the Standard Model value.}
\label{table:neutrinoscenarios}
\end{center}
\end{table}

 
\chapter{Light Relics}

\def\beq{\begin{equation}}
\def\eeq{\end{equation}}

\def\bea{\begin{eqnarray}}
\def\eea{\end{eqnarray}}

\def\Neff{N_{\rm eff}}
\def\Nf{N_{\rm eff}}
\def\gs{g_{\star}}
\def\Mpl{M_{\rm P}}

\def\lsim{\raise-.75ex\hbox{$\buildrel<\over\sim$}}

\bigskip

\begin{quotation}

\end{quotation}
    
\section{Introduction}

The cosmic neutrino background is detected at high significance in the CMB through the measurement of the total energy density in radiation prior to recombination, often parameterized by $\Neff$.  In the CMB, $\Neff$ is a measure of the gravitational influence of free streaming radiation that is decoupled from the photon-baryon fluid.  In addition to probing neutrino physics, $\Neff$ is therefore equally a probe of any light dark-sector relics.  In this chapter, we will explore the broad implications of measurements of $\Neff$ and CMB implications for light particles more generally.

Plausible configurations of CMB-S4 are capable of reaching some extremely interesting targets relevant to our cosmological history with significant implications for particle physics. These are most easily characterized in terms of the change to $\Neff$, $\Delta\Neff$, due to a single additional species that was in thermal equilibrium with the Standard Model that decouples at some temperature $T_{F}$.  This picture leads to two important theoretical targets that are within reach of CMB-S4:
\begin{itemize}
\item $\Delta \Neff \geq 0.047$ is predicted for models containing additional light particles of spin $1/2$, $1$ and/or $3/2$ that were in thermal equilibrium with the particles of the Standard Model at any point back to the time of reheating.  A CMB experiment reaching $\sigma(\Neff) \lesssim 0.02-0.03$ would be sensitive to all models in this very broad class of extensions of the Standard Model at 2$\sigma$, which includes any thermal population of gravitinos and dark photons.  
\item $\Delta \Neff \geq  0.027$ is predicted for models containing additional light particles of spin 0 that were in thermal equilibrium with the Standard Model.  A CMB experiment reaching $\sigma(\Neff) \lesssim 0.02-0.03$ would be sensitive to all such models at 1$\sigma$, which includes a wide range of models predicting axions and axion-like particles. 
\end{itemize}
At these levels of sensitivity, CMB-S4 can reach a number of compelling targets for beyond the Standard Model (BSM) physics.  Even in the absence of a detection, CMB-S4 would place constraints that can be orders of magnitude stronger than current probes of the same physics.

In addition to precise constraints on $\Neff$, another advance of CMB-S4 will be an {\it independent} high precision measurement of the primordial helium abundance, $Y_p$. First of all, $Y_p$ and $\Neff$ probe the density of radiation at well separated times in our cosmic history, a few minutes and 380,000 years after reheating respectively, which provides a window onto non-trivial evolution in the energy density of radiation in the early Universe.  Furthermore, these two quantities probe neutrino and BSM physics in related, but different ways, allowing even finer probes of BSM physics, especially in the neutrino sector.  In particular, helium gives an integrated measure of the expansion rate in the early Universe convolved with the weak interaction rates that interconvert neutrons and protons. Helium therefore gives us a handle on neutrino energy distribution functions in a way that $\Neff$ alone does not.  In combination, these two parameters yield even further insights into our cosmological history at very disparate times.  

In Section~\ref{sec:Neff} we review the motivation for studying $\Neff$ as a probe of BSM physics, emphasizing the important theoretical targets, we explain how the CMB is sensitive to free streaming radiation, and we present the forecasts for CMB-S4.  We emphasize the unique impact $\Neff$ has on the CMB that makes it distinguishable from other extensions of $\Lambda$CDM.   Section~\ref{sec:BSMneff} discusses the implications for a variety of well-motivated models, including axions and gravitinos.  In Section~\ref{sec:bbn}, we discuss the relation between CMB and BBN based constraints, and we forecast our ability to measure $Y_p$ and $\Neff$ simultaneously with the CMB.  In Section~\ref{sec:neffscenarios}, will discuss possible detection scenarios involving light fields, $\Neff$ and $\sum m_\nu$.

\section{New Light Species at Recombination}\label{sec:Neff}

The angular power spectrum of the cosmic microwave background at small angular scales is very sensitive to the radiation content of the early Universe, usually parametrized by a quantity $\Nf$ which was defined in Eq.~(\ref{eq:rho_r_Neff}).  In the Standard Models of cosmology and particle physics, $\Nf$ is a measure of the energy density of the cosmic neutrino background.  More generally, however, $\Nf$ receives contributions from all forms of radiation apart from photons which are present in the early Universe.  Due to its sensitivity to $\Nf$, the CMB can be used as a tool to probe aspects of the physics of the Standard Model and beyond which are difficult to measure through other means.  Here we will give an overview of the key observational targets and signatures of $\Neff$ in the CMB.  In particular, we will focus on changes to the radiation density that occur from beyond the Standard model physics, parameterized in terms of $\Delta \Neff \equiv \Neff - \Neff^{\rm SM}$ with $\Neff^{\rm SM} \approx 3.046$.

\subsection{Natural Target}

A measurement of the value of $\Nf$ can provide deep insights into the early Universe.  Most significantly, it is an observational window onto the conditions at very early times, well before recombination.  Even within the Standard Model, $\Nf$ provides an observational handle on the thermal history back to about one second after reheating through the decoupling of neutrinos.  The true power of measuring $\Nf$, however, comes from the realization that it is sensitive not just to the neutrinos of the Standard Model, but it in fact receives contributions from all forms of radiation apart from photons present in the early Universe and is thus a probe of new physics.

Collider experiments are known to provide a measurement of the number of neutrino species (or more precisely the number of species of fermions coupling to the $Z$ boson with mass below $m_Z/2$) and find very close agreement with three families of light active neutrinos \cite{ALEPH:2005ab}.  Cosmological measurements of $\Nf$ provide complementary constraints and are sensitive to the total energy density of radiation whether it consists of active neutrinos or other light species.

If the measured value of $\Nf$ exceeds the Standard Model prediction, it would be an indication that there is additional radiation in the early Universe or that the thermal history is modified.  Additional radiation which contributes to $\Nf$ is often referred to as dark radiation.  There is a large number of possible sources for dark radiation, including axions~\cite{Brust:2013xpv,Salvio:2013iaa,Kawasaki:2015ofa,Baumann:2016wac}, sterile neutrinos~\cite{Abazajian:2001nj,Strumia:2006db,Boyarsky:2009ix}, gravitational waves \cite{Boyle:2007zx,Stewart:2007fu,Meerburg:2015zua}, dark photons~\cite{Ackerman:2008gi,Kaplan:2011yj,CyrRacine:2012fz}, and many more \cite{Cadamuro:2010cz,Weinberg:2013kea,Arkani-Hamed:2016rle}.  It is also possible that the measured value of $\Nf$ could be found below the Standard Model prediction.  This can happen if for example photons are heated after neutrinos decouple \cite{Steigman:2013yua,Boehm:2013jpa}.

\begin{figure}[t!]
\begin{center}
\includegraphics[width=0.65\textwidth]{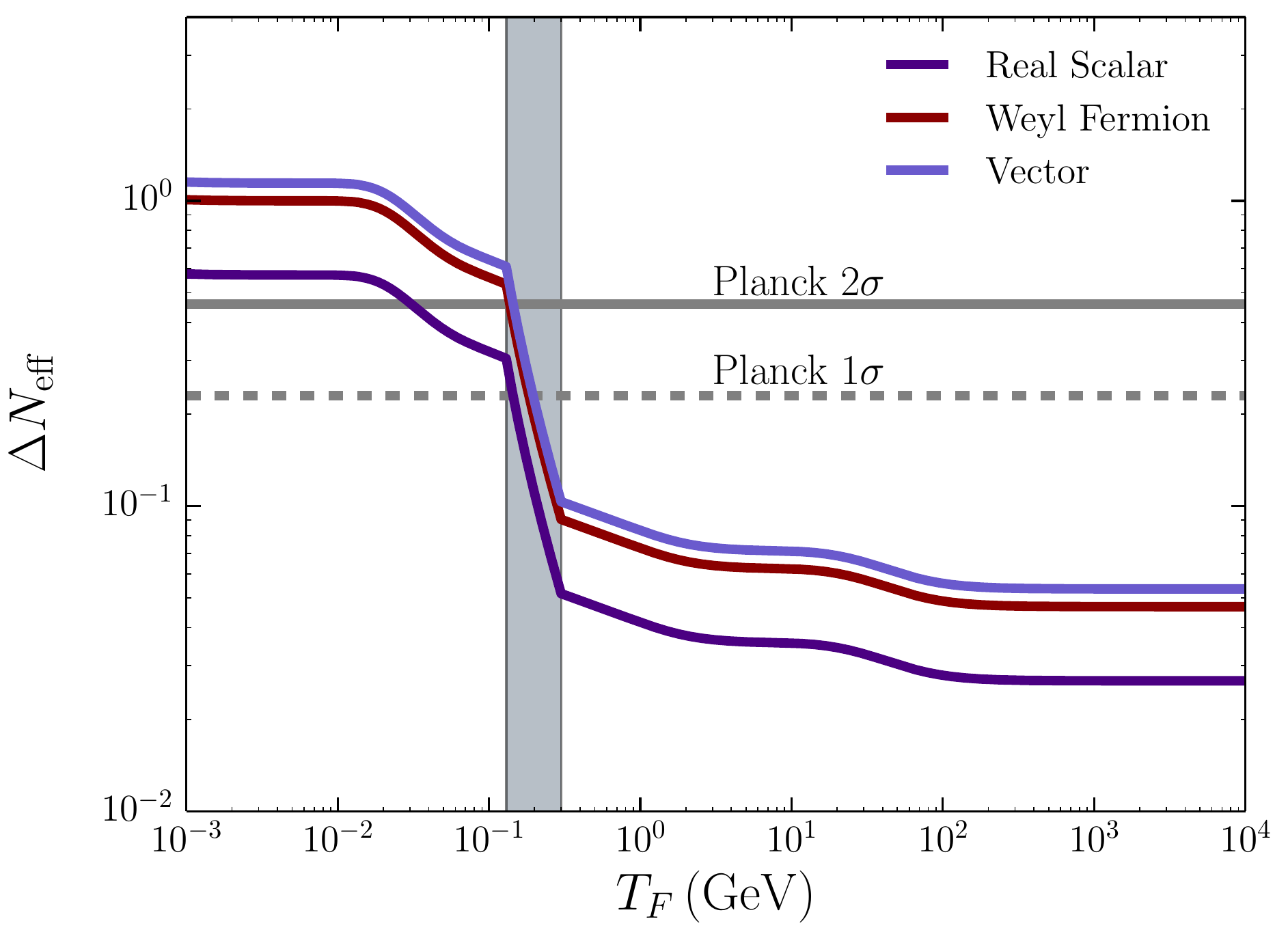}
\caption{Contribution to $\Neff$ from a massless field that was in thermal equilibrium with the Standard Model at temperatures $T> T_{F}$.  For $T_{F} \gg m_{\rm top}$, these curves saturate with $\Delta \Neff > 0.027$.   The dashed and solid grey lines correspond to the 1$\sigma$ and 2$\sigma$ sensitivity of \planck, using $\sigma(\Neff) = 0.23$. Temperatures in the grey region correspond to the QCD phase transition.}
\label{fig:Neff_thermal}
\end{center}
\end{figure}

One of the features that makes $\Neff$ a compelling theoretical target is the degree to which broad classes of models fall into two basic levels of $\Delta\Neff$.  As illustrated in Figure~\ref{fig:Neff_thermal}, any species that was in thermal equilibrium with the Standard Model degrees of freedom produces a characteristic correction to $\Neff$ that depends only on its spin and its freeze-out temperature.  For freeze-out after the QCD phase transition, one finds $\Delta \Neff \gtrsim 0.3$.  Freeze-out before the QCD phase transition instead produces $\Delta \Neff > 0.027$.  The first category has been tested by the data from the \planck\ satellite.  The second category, which is sensitive to freeze-out temperatures as high as the reheating temperature, falls into the level of sensitivity attainable by CMB-S4.

The contributions to $\Neff$ from hot thermal relics are relatively easy to understand from the discussion of neutrino decoupling in Section~\ref{ThermalHistory}.  After freeze-out, the temperature of a relativistic species redshifts like $a^{-1}$ and therefore is only diluted relative to photons when energy is injected.  The annihilation of heavy Standard Model particles into light Standard Model Particles conserves the comoving entropy of the plasma and therefore, the diluted temperature of a relic before neutrino decoupling is given by
\beq
\left( \frac{T_{\rm relic}}{T_{\nu}} \right)^3 = \frac{\gs(T_{\nu-{\rm decoupling}})}{\gs( T_F) }= \frac{43/4}{\gs(T_F)} \, ,
\eeq
where $T_F$ is the temperature where the relic decouples (freezes-out) and $\gs(T)$ is defined as in Section~\ref{ThermalHistory} to be the number of independent spin states including an additional factor of $\frac{7}{8}$ for fermions that are relativistic at temperature $T$.  Combining this with Eq.~(\ref{eq:rho_r_Neff}), this implies that the change to $\Neff$ due to a thermal relic with $g$ independent spin states is
\begin{equation}\label{eq:DeltaNeff_thermal}
	\Delta \Neff =
	\Bigg\{\begin{array}{l c l}
         \frac{4g}{7}\left(\frac{43/4}{\gs(T_F)}\right)^{4/3}&  &{\rm Boson}\\
        \frac{g}{2}\left(\frac{43/4}{\gs(T_F)}\right)^{4/3} &  &{\rm Fermion} \, .
        \end{array}
\eeq
  The order of magnitude difference in $\Delta \Neff$ before and after the QCD phase transition comes from an order of magnitude drop in $\gs$ below the QCD scale.  At temperatures well above to top mass, the Standard Model gives $\gs = 106.75$.  We can then see from Eq.~(\ref{eq:DeltaNeff_thermal}) that the minimum value of $\Delta \Neff$ for a single real scalar is 0.027, for a Weyl fermion is 0.047, and for a light vector boson is 0.054.

Even a measurement of $\Nf$ which agrees with the Standard Model prediction to high precision would be very interesting due to the constraints it would place on physics beyond the Standard Model.  Some specific implications for sterile neutrinos, axions, and other popular models will be discussed below.  Broadly speaking, constraining $\Delta \Neff$ at the $10^{-2}$ level would constrain or rule out a wide variety of models that are consistent with current cosmological, astrophysical, and lab-based constraints.  Furthermore, because of the sharp change in $\Delta \Neff$ at the QCD phase transition, the improvement from current constraints to projections for CMB-S4 can be quite dramatic.

For the minimal scenario of a single real scalar, reaching $\sigma(\Neff) \sim 1\times 10^{-2}$ would push the constraint on freeze-out temperatures from electroweak scale to the reheat temperature.  This broad reach to extremely high energies and very early times demonstrates the discovery potential for a precision measurement of $\Nf$ with the CMB.  Furthermore, the CMB power spectrum has the ability to distinguish among certain types of dark radiation based on the behavior of its density perturbations  \cite{Chacko:2015noa,Baumann:2015rya}.  This point will be discussed further below.

A measurement with a slightly larger error on $\Nf$ would still be extraordinarily valuable for higher spin fields, multiple light scalars, and modifications to the thermal history up to the electroweak scale.  In particular, massless fermions and vectors have two helicity states which imply contributions of $\Delta \Neff  \geq 0.047$ and $\Delta \Neff  \geq 0.054$ respectively.  As a result, a less sensitive instrument is still capable of probing physics back to reheating since it could detect or rule out the existence of light thermal relics with non-zero spin.  In addition, there is good reason to think dark sectors could contain multiple light fields which could appear at any level of $\Delta \Neff$ above the minimum contribution from a single scalar field.  Still, the most dramatic jumps in discovery potential occur at the critical values $\Delta \Neff = 0.027, 0.047$, and $0.054$.  

Many extensions of the Standard Model include additional massive particles that increase $g_\star$ at high temperatures ($T \gg 100$ GeV).  The annihilation of these additional particles in the early Universe will further diluted any new light particles (by increasing $\gs(T_F)$ in Equation~\ref{eq:DeltaNeff_thermal}) and can allow $\Delta \Neff < 0.027$.  This possibility is redundant with the uncertainty of the reheating temperature, given that this additional dilution only occurs at temperatures above the masses of these new particles.  Furthermore, we require many new particles to significantly alter the predictions for $\Delta \Neff$.  For example, to reduce the minimum contribution to $\Delta \Neff$ by a factor of two, we must double the degrees of freedom in $g_\star$.  While such large numbers of new particles are common in extensions of the Standard model (e.g. the MSSM), it is also common that these extensions come with many new light particles as well~\cite{Arvanitaki:2009fg,Arkani-Hamed:2016rle}.   Assuming that we have only a single additional degree of freedom was a conservative assumption in this sense.

\subsection{Observational Signatures}\label{sec:Neffsignatures}

Cosmic neutrinos and other light relics play two important roles in the CMB that are measured by $\Neff$.  They contribute to the total energy in radiation which controls the expansion history and, indirectly, the damping tail of the power spectrum.  The fluctuations of neutrinos and any other free streaming radiation also produces a constant shift in the phase of the acoustic peaks.  These two effects drive both current and future constraints on $\Neff$.  

The effect of neutrinos on the damping tail drives the constraint on $\Neff$ in the CMB in $\Lambda$CDM + $\Neff$.  The largest effect is from the mean free path of photons, which introduces a suppression $e^{-(k/k_d)^2}$ of short wavelength modes, with~\cite{Zaldarriaga:1995gi}
\beq
k_d^{-2} =\int \frac{da}{a^3 \sigma_T n_e H} \frac{R^2+ \frac{16}{15}(1+R)}{6(1+R)^2} \ ,
\eeq
where $R$ is the ratio of the energy in baryons to photons, $n_e$ is the density of free electrons, and $\sigma_T$ is the Thomson cross-section.  The damping scale is sensitive to the energy density in all radiation through $H \propto \sqrt{\rho_{\rm radiation}}$ during radiation domination (which is applicable at high $\ell$), and is therefore sensitive to $\Neff$ or any form of dark radiation.  From this discussion, we can also see the origin of the degeneracy between $\Neff$ and $n_e$, the latter of which may be altered by the primordial helium fraction, $Y_p$.

In reality, the effect on the damping tail is subdominant to the change to the scale of matter-radiation equality and the location of the first acoustic peak~\cite{Hou:2011ec}.  As a result, the effect of neutrinos on the damping tail is more accurately represented by holding the first acoustic peak fixed.  This changes the sign of the effect on the damping tail, but the intuition for the origin of the effect (and degeneracy) remains applicable.

In addition to the effect on the Hubble expansion, perturbations in neutrinos affect the photon-baryon fluid through their gravitational influence.  The contributions from neutrinos are well described by a correction to the amplitude and the phase of the acoustic peaks in both temperature and polarization~\cite{Bashinsky:2003tk}.  The phase shift is a particularly compelling signature as it is not degenerate with other cosmological parameters~\cite{Bashinsky:2003tk,Baumann:2015rya}.  This effect is the result of the free-streaming nature of neutrinos that allows propagation speeds of effectively the speed of light (while the neutrinos are relativistic).  Any gravitationally coupled free-streaming light relics will also contribute to the amplitude and phase shift of the acoustic peaks.

E-mode polarization will play an increasingly important role for several reasons.  First of all, the acoustic peaks are sharper in polarization which makes measurements of the peak locations more precise, and therefore aid the measurement of the phase shift.  The second reason is that polarization breaks a number of degeneracies that would also affect the damping tail~\cite{Baumann:2015rya}.

{\it Status of current observations} -- \planck\ has provided a strong constraint on $\Neff = 3.15 \pm 0.23$ when combining temperature, low-$\ell$ polarization and BAO data.  The addition of high-$\ell$ polarization data is expected to improve the constraint on $\Neff$ and reduce the impact of the degeneracy with $Y_p$ (preliminary results from \planck\ give $\Neff = 3.04 \pm 0.18$ when combining temperature, polarization and BAO).  Recently, the phase shift from neutrinos has also been established directly in the \planck\ temperature data~\cite{Follin:2015hya}.  This provides the most direct evidence for presence of free-streaming radiation in the early Universe, consistent with the cosmic neutrino background.

\subsection{Forecasts}\label{sec:neff_forcast}

As we described in Section~\ref{sec:Neffsignatures}, the most concrete signatures of $\Neff$ from CMB-S4 are derived from the damping tail at high-$\ell$ in both $T$ and $E$.  In addition, there is the effect of the shift in the phase of the acoustic peaks that is present at $\ell > 500$.  Within $\Lambda$CDM these parameters are robust to degeneracies, having already accounted for the degeneracy of the location of the first acoustic peak with $H_0$.  The damping tail is degenerate with other extensions, like $Y_p$, as we will explore in Section~\ref{sec:NeffBBNfore}.  In this chapter, forecasts will vary $Y_p$ with $\Neff$ to be consistent with the predictions of BBN for the given value of $\Neff$.  

The forecasts for $\Neff$ will follow the methodology outlined in Section~\ref{sec:ttee}.  Unless otherwise stated, we will work with delensed spectra and the associated lens-induced covariances as explained in that section and in~\cite{Green:2016}.  This is particularly relevant to the phase shift, as delensing is expected to sharpen the acoustic peaks.  Realistic modeling of delensing is important as our forecasts are quite close to our theoretical targets and unlensed spectra will underestimate $\sigma(\Neff)$. 

Forecasts for $\sigma(\Neff)$ under various experimental configurations are given in Figure~\ref{fig:Neffbeam}.  Since the damping tail extends to very high-$\ell$, constraints on $\Neff$ are sensitive to $\ell_{\rm max}$ and the beamsize of the experiment.  We assume $\ell_{\rm max} = 5000$ except for $C^{TT}_{\ell}$ where we assume $\ell_{\rm max} = 3000$ due to foregrounds.  One can see the impact of the highest-$\ell$ modes by looking at the effect of the beamsize.  One noticeable feature is that to be in the vicinity of our threshold targets, one needs both high angular resolution (better than 2') and low noise (less than 2 $\mu$K-arcmin).

The other important parameter for $\Neff$ is the sky fraction of the survey, $f_{\rm sky}$.  Since the signature of neutrinos depends on the detailed measurement of the shape of the power spectra, constraints depend on the number of observed modes.  At fixed noise levels this implies that \mbox{$\sigma(\Neff) \propto f^{-1/2}_{\rm sky}$}.  Of course, with a fixed number of detectors the noise-equivalent temperature, $s$, is proportional to \mbox{$s \propto f_{\rm sky}^{1/2}$}.  Therefore, at fixed effort, increasing $f_{\rm sky}$ also increases the temperature noise.  Yet, for the range of $s$ of interest, it turns out that $\sigma(\Neff)$ grows more slowly than linear in $s$, and we see that there is a net improvement in $\sigma(\Neff)$ by increasing $f_{\rm sky}$ at a fixed number of detectors.  In other words, when we increase $f_{\rm sky}$, the increase in $s$ is less important than the gain in the number of modes.  This trend is shown in the left panel of Figure~\ref{fig:Neff_fsky}.

\begin{figure}[t!]
\begin{center}
\includegraphics[height=6.5cm]{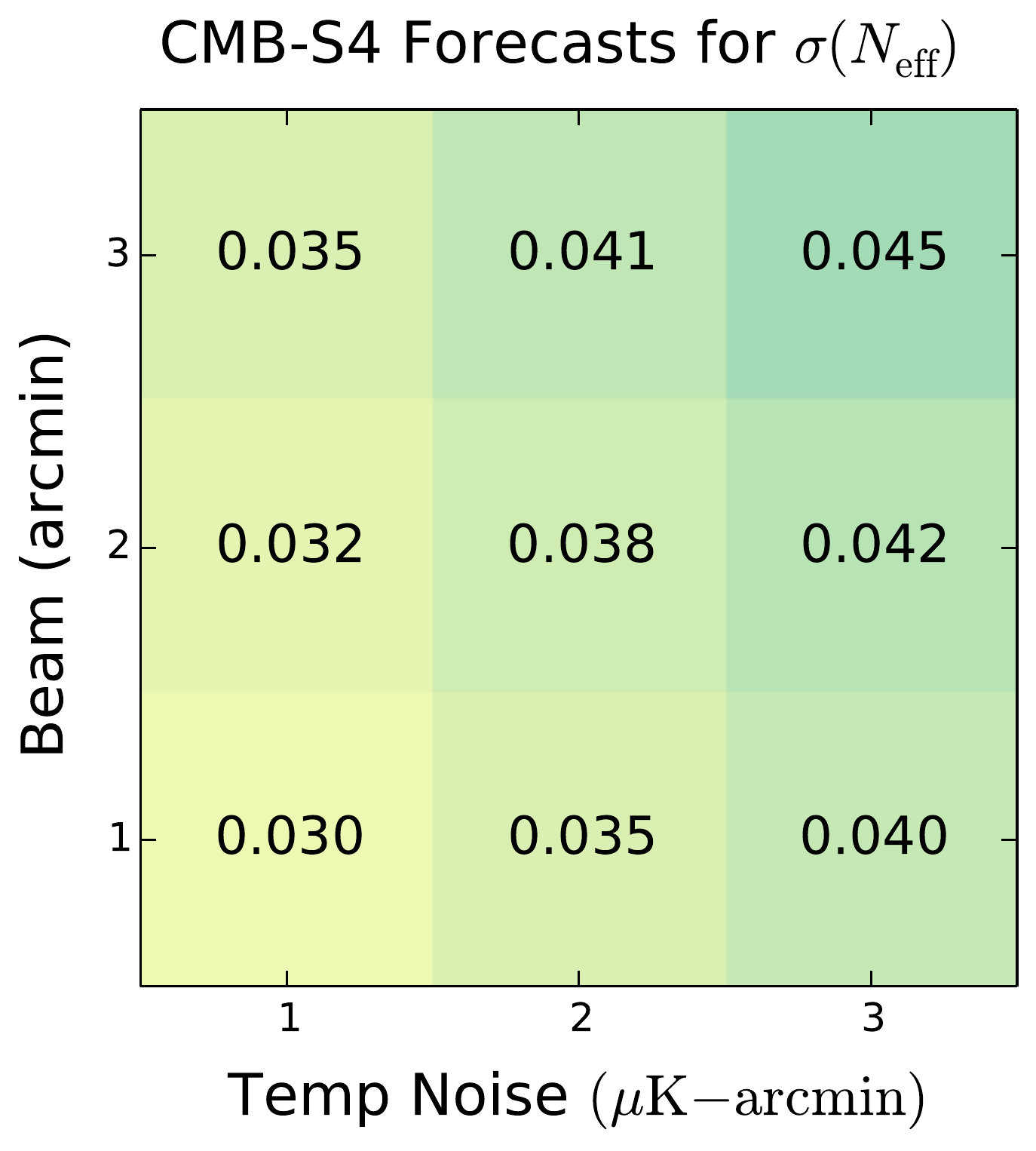}\hskip 1cm
\includegraphics[height=6.5cm]{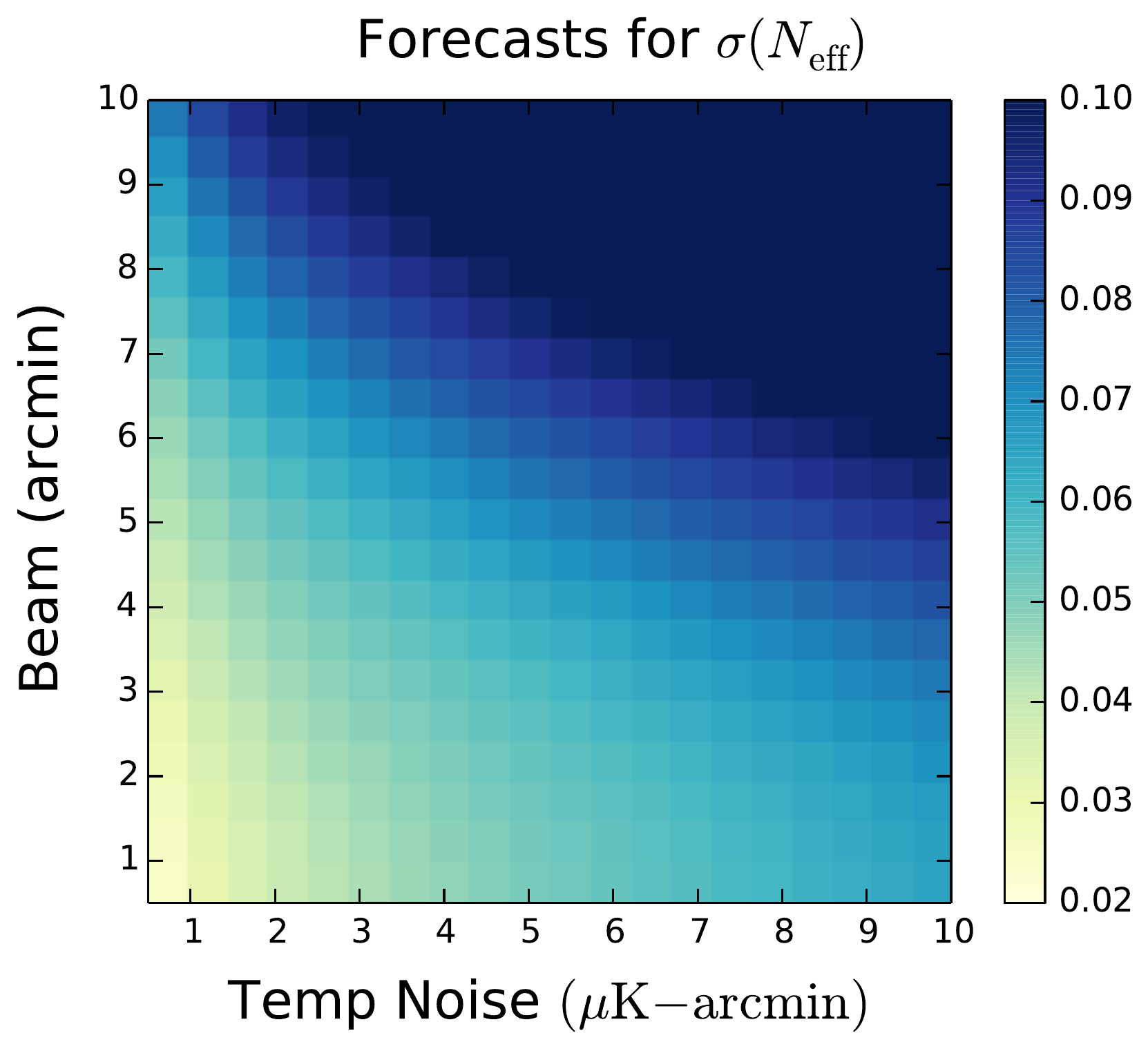}  

\caption{Forecasts for $\sigma(\Neff)$ with varying beamize in arcmin and temperature noise.  These forecasts assume $f_{\rm sky} = 0.4$ and vary $Y_p$ with $\Neff$ to be consistent with BBN.    The color scale is the same for both panels.  {\bf Left:} Specific forecasts, including delensing, for various CMB-S4 configurations. {\bf Right:} A wide range of beam sizes and sensitivities are used to show the need for the high resolution and sensitivity of CMB-S4 to be close to our thresholds.  }
\label{fig:Neffbeam}
\end{center}
\end{figure}

\begin{figure}[t!]
\begin{center}
\includegraphics[width=0.45\textwidth]{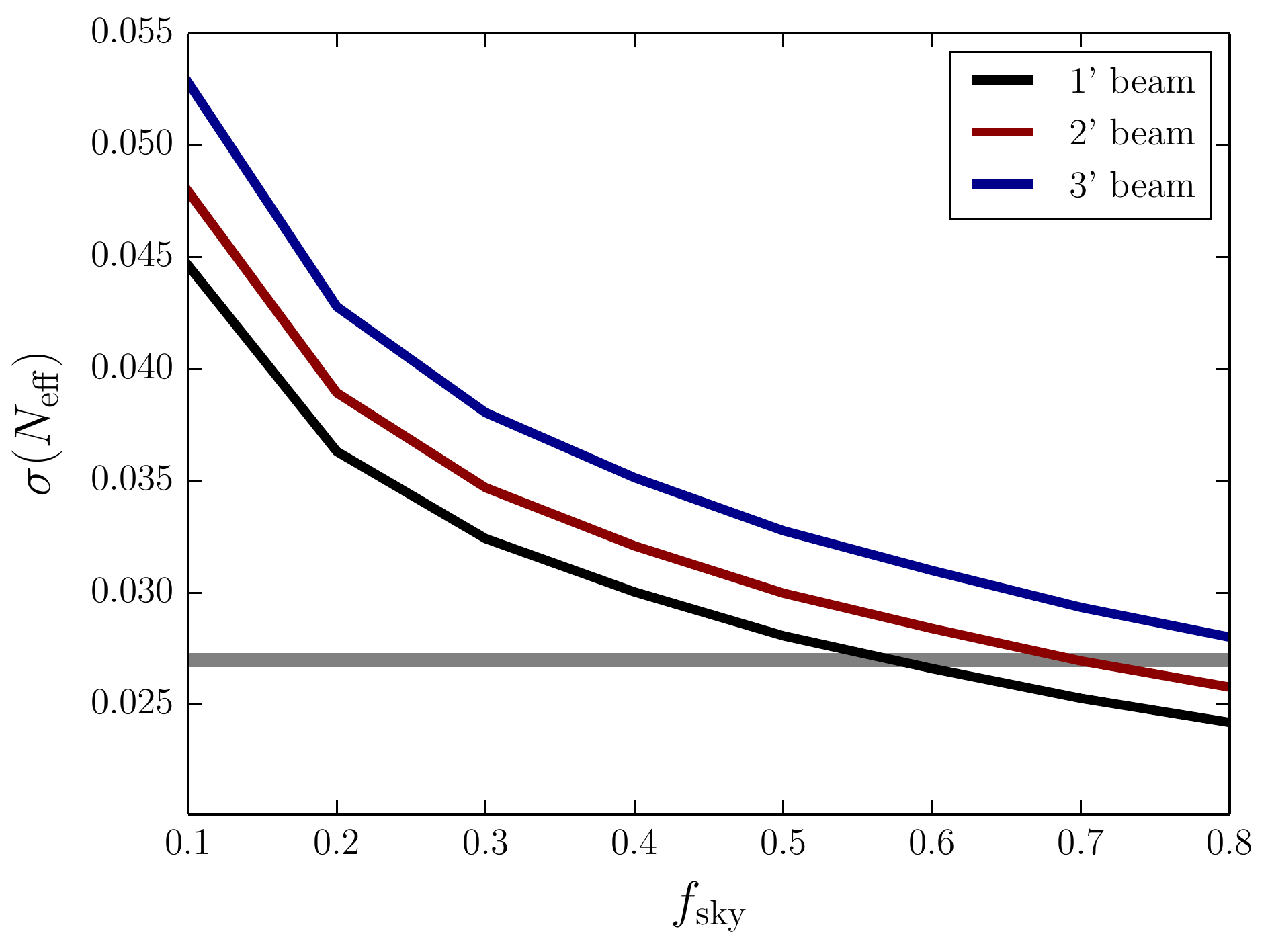}
\includegraphics[width=0.45\textwidth]{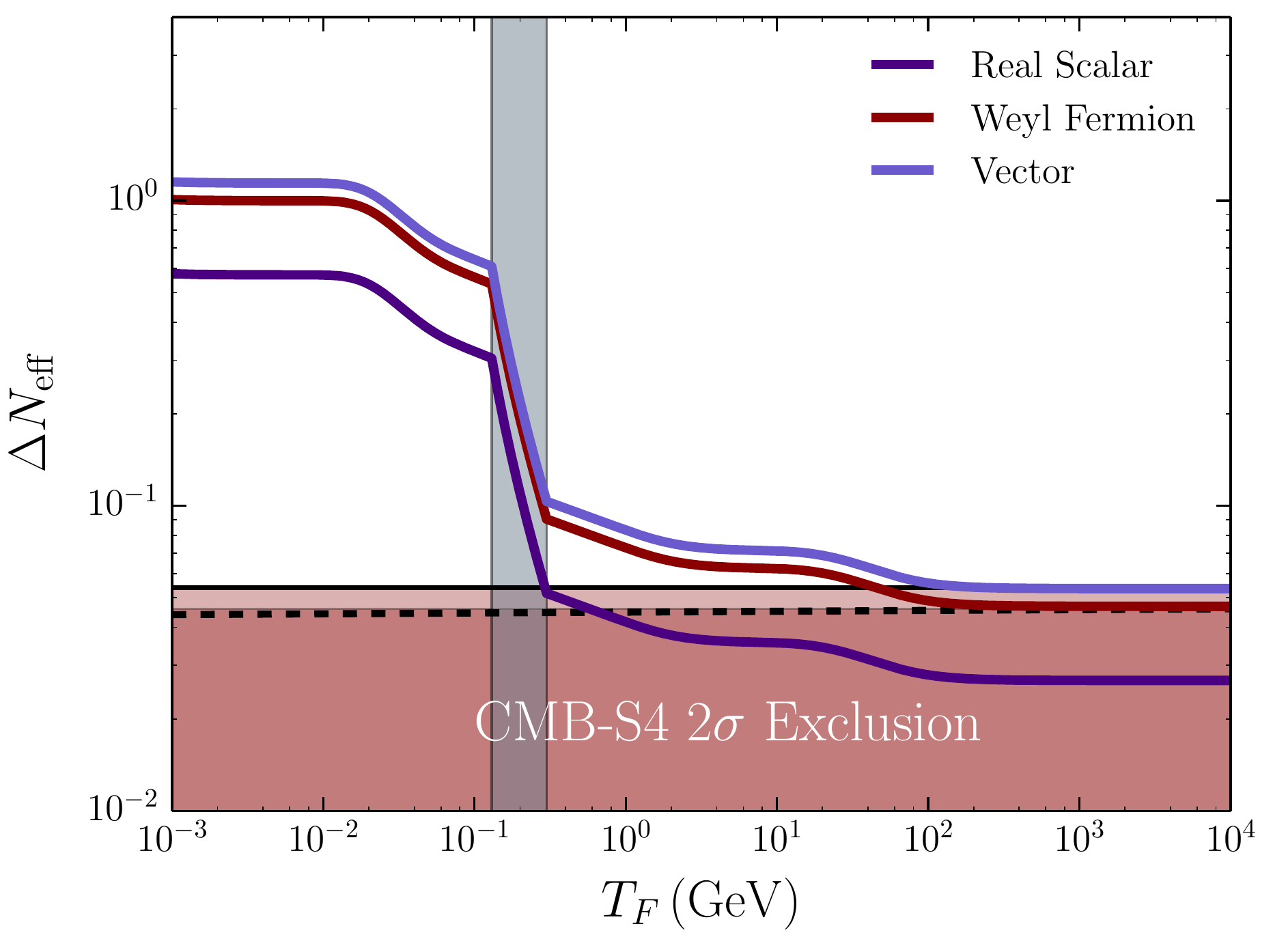}
\caption{{\bf Left:} Forecasts for $\sigma(\Neff)$ as a function of sky fraction.  The sensitivity has been normalized to 1 $\mu$K-arcmin for $f_{\rm sky} =0.4$ and is scaled according to $S \propto f_{\rm sky}^{1/2}$ for different sky fractions.  The grey line shows the value of $\sigma(\Neff) =0.027$ which is the 1$\sigma$ sensitivity any scalar thermal relic, or equivalently, 2$\sigma$ sensitivity to any vector thermal relic.  For a fixed number of detectors, we see that $\sigma(\Neff)$ is minimized by increasing sky fraction.  {\bf Right:} Same as Figure~\ref{fig:Neff_thermal} showing plausible 2$\sigma$ limits from CMB-S4 in red, assuming 1' beams and 1 $\mu$K-arcmin temperature noise.  The light red region with solid boundary and darker red with dashed boundary are for $f_{\rm sky}=0.5$ and $f_{\rm sky} =0.7$ respectively.  These modest increases in sky fraction can have a significant impact with regards to the theoretical thresholds for vectors or Weyl fermions.}
\label{fig:Neff_fsky}
\end{center}
\end{figure}

For sufficiently large sky fraction, the thresholds for the light fermions and vectors are accessible at 2$\sigma$ for plausible experimental configurations, or equivalently, 1$\sigma$ for the minimum threshold for a light scalar.  Specifically, to reach $\sigma(\Neff) = 0.027$ with $s = 1$ $\mu$K-arcmin, we need $f_{\rm sky} \geq 0.5$ and $f_{\rm sky} \geq 0.6$ for 1' and 2' beams respectively.  In the right panel of Figure~\ref{fig:Neff_fsky}, we show how 2$\sigma$ limits available with CMB-S4 translate into limits on the freeze-out temperature of a single additional species.  While \planck\ is only sensitive to physics after the QCD phase transition (Figure~\ref{fig:Neff_thermal}), CMB-S4 can reach times before the QCD phase transition for all spins, and back to reheating for spins 1/2 and 1.  

\begin{figure}[t!]
\begin{center}
\includegraphics[width=0.5\textwidth]{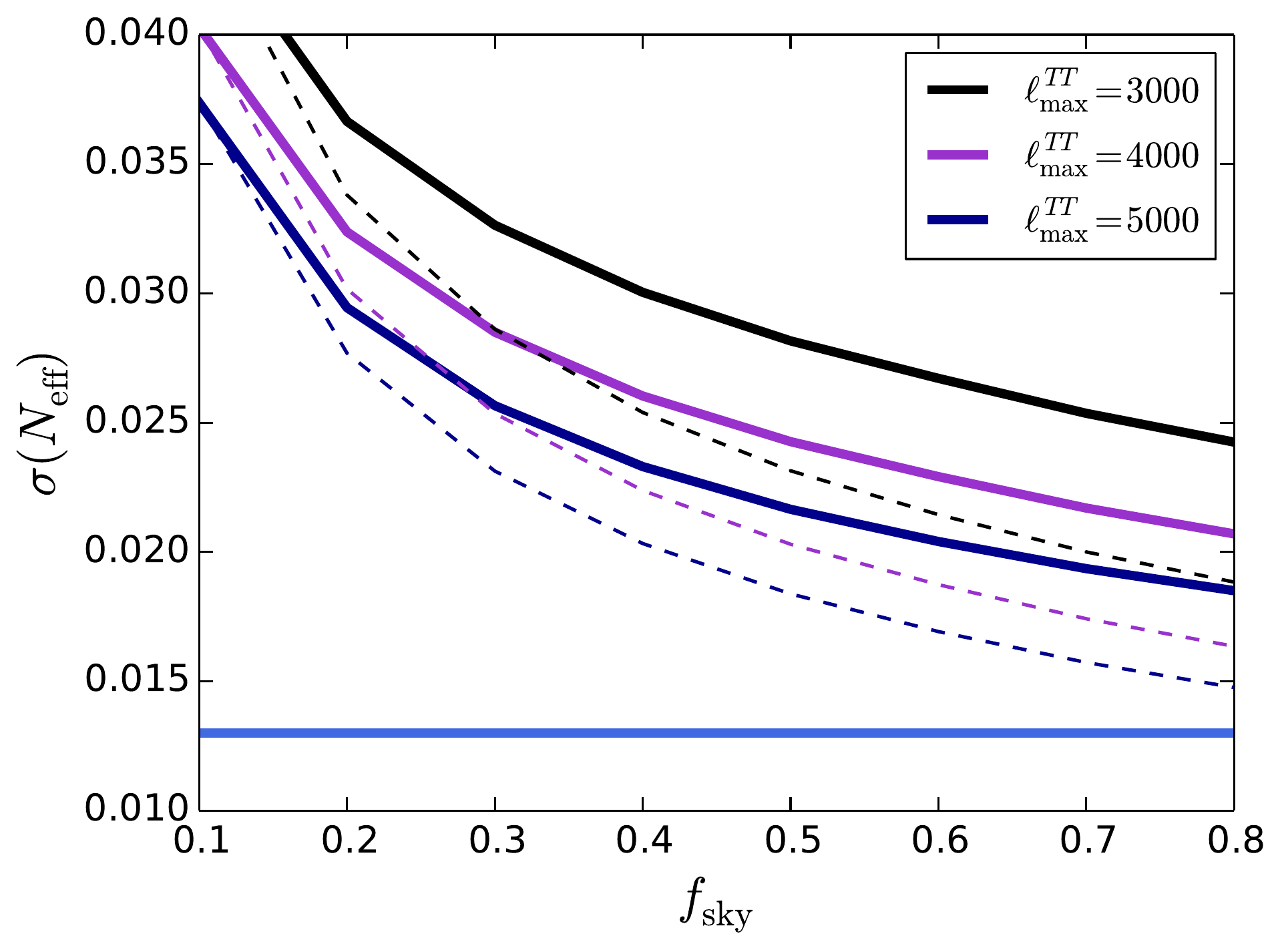}
\caption{Forecasts for $\sigma(\Neff)$ as a function of sky fraction for various choices of $\ell_{\rm max}^{TT}$, with $\ell_{\rm max} = 5000$ for all other spectra.  For solid curves, the sensitivity has been normalized to 1 $\mu$K-arcmin for $f_{\rm sky} =0.4$ and is scaled according to $S \propto f_{\rm sky}^{1/2}$ for different sky fractions.  The dashed curves indicate a fixed temperature noise of $0.5$ $\mu$K-arcmin.  The horizontal blue line shows the value of $\sigma(\Neff) =0.013$ which is the 2$\sigma$ threshold for sensitivity to any light thermal relic.  While these forecasts are optimistic with regards to foreground removal, they do show the underlying modes would significantly improve our sensitivity to $\Neff$ .
}
\label{fig:Neff_LTT}
\end{center}
\end{figure} 

This raises the question whether we can reach these targets at higher significance.  In particular, we might hope to reach 2$\sigma$ sensitivity for the minimum contribution of a light scalar, $\Delta\Neff = 0.027$.  To assess whether the CMB contains enough information to achieve this, we consider an idealized situation ignoring foregrounds. Figure~\ref{fig:Neff_LTT} shows the constraints for different $\ell_{\max}^{TT}$. We see an improvement of 10-15 percent and 20-25 percent for $\ell_{\rm max}^{TT} =4000$ and $\ell_{\rm max}^{TT} =5000$ respectively. In practice, unresolved point sources will make it challenging to extract information from the temperature data for $\ell>3000$. We also show the results for a fixed temperature noise of $0.5$ $\mu$K-arcmin to show how close one can get to the threshold by increasing the sensitivity.

\begin{figure}[ht]
\begin{center}
\includegraphics[width=0.45\textwidth]{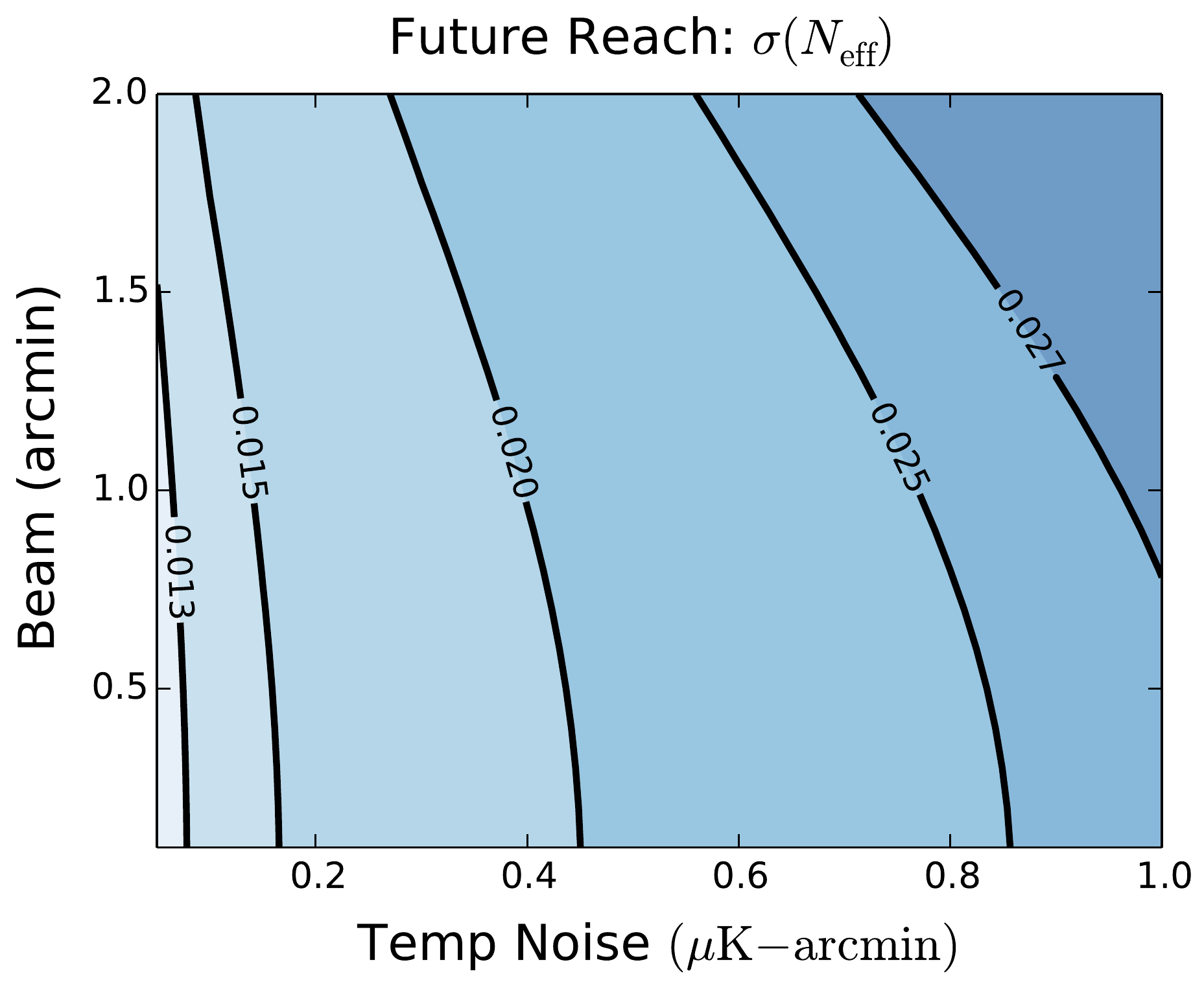}
\includegraphics[width=0.45\textwidth]{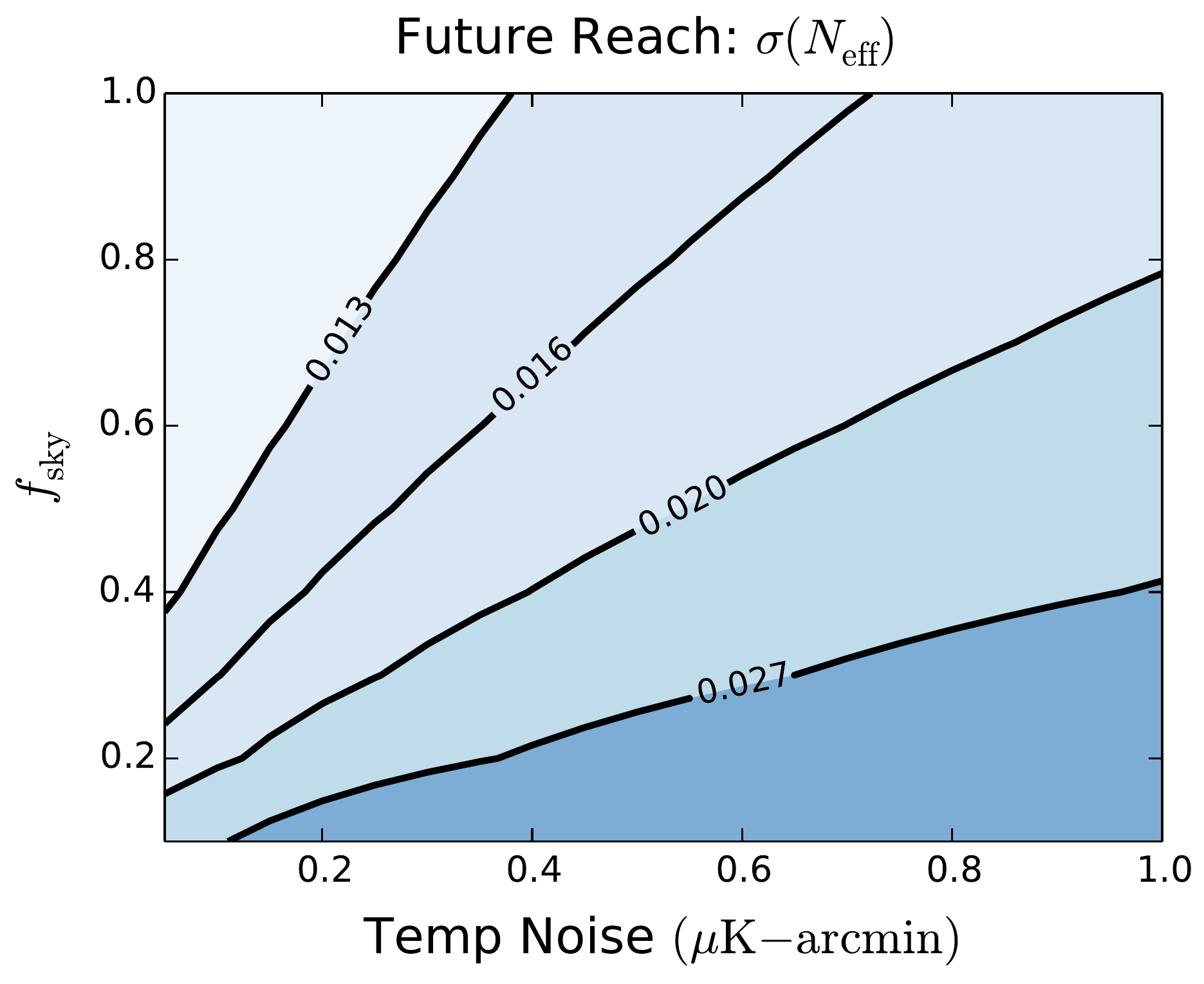}
\caption{Forecasts for $\sigma(\Neff)$ assuming a futuristic survey, using unlensed spectra with $\ell^{TT}_{\rm max} =3000$ and $\ell_{\rm max} = 8000$ for all other spectra {\bf Left:}  Forecast varying the beamize in arcmin (') and temperature noise in $\mu$K-arcmin, assuming $f_{\rm sky} = 0.4$.   {\bf Right:} Forecast varying $f_{\rm sky}$ and temperature noise in $\mu$K-arcmin, assuming a 1' beam.  We see that archiving a $\sigma(\Neff) < 0.013$ pushes to low noise and large $f_{\rm sky}$ but there is little gain from smaller beamsize.}
\label{fig:Neff_lownoise}
\end{center}
\end{figure}

To convincingly reach $\sigma(\Neff) <0.013$, the most straightforward strategy is to design a more sensitive experiment.  Figure~\ref{fig:Neff_lownoise} shows the improvements that are possible in $\sigma(\Neff)$ for futuristic values of the temperature noise as a function of beamsize and $f_{\rm sky}$.  The combination of large $f_{\rm sky}$ and noise below $0.4$ $\mu$K-arcmin can reach the target, while smaller beamsize appears to have little impact.  We can compare these values to the cosmic variance limits for various choices of $\ell_{\rm max}$, as shown in Table~\ref{tab:CVL}.  Here we have used that foregrounds in polarization will be more favorable at high-$\ell$ than they are in temperature, so we assume a fixed $\ell_{\rm max}^{TT} =3000$.  In principle, there is enough information in the CMB to reach 5$\sigma$ for $\Delta \Neff =0.027$, although it is clearly very challenging to reach to 2$\sigma$ (and beyond).

\begin{table}[t!]
\begin{center}
\begin{tabular}{l || c c} 
 \toprule
    		$\ell_{\rm max}$    			&   $\sigma_{\rm cv} (\Neff)$-unlensed 		& $\sigma_{\rm cv}(\Neff)$-lensed		 \\ [0.5ex]
 \hline
5000	 &.010	&	.012 \\
6000	&.0080	&	.011 \\
7000	&.0068	&	.010 \\
8000	&.0059	&	.096 \\
9000	&.0050	&	.0087	 		  \\
    \bottomrule
\end{tabular}
\caption{Cosmic variance limit for $\sigma(\Neff)$ for various values of $\ell_{\rm max}$ with $\ell_{\rm max}^{TT} =3000$.  We see that with $\ell_{\rm max} =5000$ it is difficult to reach beyond 3$\sigma$ in terms of the threshold $\Delta \Neff = 0.027$.    }
\label{tab:CVL}
\end{center}
\end{table}

Finally, we have made little use of additional data available from LSS surveys like DESI and LSST.  In principle, the information about $\Neff$ that appears in the CMB is also contained in the matter power spectrum.  Furthermore, late time measurements can also break degeneracies that could be indirectly limiting the constraints on $\Neff$ or other parameters.  Preliminary forecasts show only modest improvements by including this data, but LSS may still hold promise for enhancing the impact of CMB-S4 for light relics.

\section{Implications for Light Particles}\label{sec:BSMneff}

Contributions to $\Delta \Neff$ from light particles depends sensitively on the spin.  For thermal relics, each degree of freedom contributes equally, so the signatures scale like the effective degrees of freedom of a given particle.  In addition, the couplings that would thermalize these particles depend both on the spin of the particle and the spin of the particle(s) in the Standard Model to which it couples.  Furthermore, these couplings can lead to non-thermal production mechanisms that also contribute to $\Delta \Neff$. This section explores the implications of a CMB-S4 measurement of $\Neff$ for a number of well motivated models, organized by the spin of the relevant light particle from axions (spin 0) to gravitons (spin 2).  

\subsection{Axions}
\label{sec:neffaxions}

Light particles of spin-0 (scalar fields) are highly constrained by naturalness.  In the absence of a symmetry, one would expect them to be heavy and thus a poor candidate for dark radiation.  However, (pseudo) Goldstone bosons are naturally light and they appear generically from spontaneous breaking of some high energy global symmetry.  A ubiquitous  example in beyond the Standard Model physics are axions and/or axion-like particles (ALPs).  Axions have been introduced to solve the strong CP problem~\cite{Peccei:1977hh}, the hierarchy problem~\cite{Graham:2015cka}, and the naturalness of inflation~\cite{Freese:1990rb}.  Furthermore, they appear generically in string theory, in large numbers, leading to the qualitative phenomena described as the string axiverse~\cite{Arvanitaki:2009fg}.  They may even be tied to the origin of the breaking of the flavor and baryon/lepton number symmetries of the Standard Model.

At low energy, the mass of the ALP is protected by an approximate shift symmetry of the general form $a \to a + c$ where $a$ is the axion and $c$ is a constant (for non-abelian Goldstone bosons, this transformation will include higher order terms in $a$).  We will define an ALP to be any such particle for which all of the couplings of the axion to the Standard Model respect such a symmetry.  This symmetry may be softly broken with an explicit mass term, although this is highly restricted in the case of the QCD axion (i.e.~axions meant to solve the strong CP problem).

Two very common features of models containing ALPs are that the ALPs are typically light (in many cases, $m \ll 1$~eV) and their interactions are suppressed by powers of the (typically large) decay constant $f_a$.  These two features make ALPs a particularly compelling target for cosmology and $\Neff$ specifically~\cite{Brust:2013xpv,Salvio:2013iaa,Kawasaki:2015ofa,Baumann:2016wac}.  Because of their small masses, ALPs will often behave as relativistic species in the early Universe.  Furthermore, because their production rate will scale as $T^{2n +1} / f_a^{2n}$ for some $n \geq 1$, they are likely to be thermalized at high temperatures.  Given that $\Delta \Neff > 0.027$ under such circumstances, a CMB experiment with sensitivity at this level will be sensitive to a very wide range of ALP models.   In the absence of a detection of $\Delta \Neff$, we can place constraints on the axions couplings to the Standard Model.  

Two couplings of particular interest for axion phenomenology are the coupling to photons and gluons, 
\beq
\frac{1}{4} \frac{1}{\Lambda_\gamma} a \tilde F_{\mu \nu}F^{\mu\nu} \ , \qquad \qquad \frac{1}{4} \frac{1}{\Lambda_g} a \tilde G_{\mu \nu}G^{\mu\nu}  \ ,
\eeq
These couplings typically appear as the consequence of chiral anomalies and such that $\Lambda_{\gamma, g} \propto f_a$.  The coupling of the axion to gluons makes the solution to the strong-CP problem possible.  The coupling to photons is somewhat model dependent but typically arises in conjunction with the gluon coupling.  In addition to or instead of these couplings, a variety of other possible couplings to matter may also be included.

\begin{figure}[t]
\centering \includegraphics[width=0.7\textwidth]{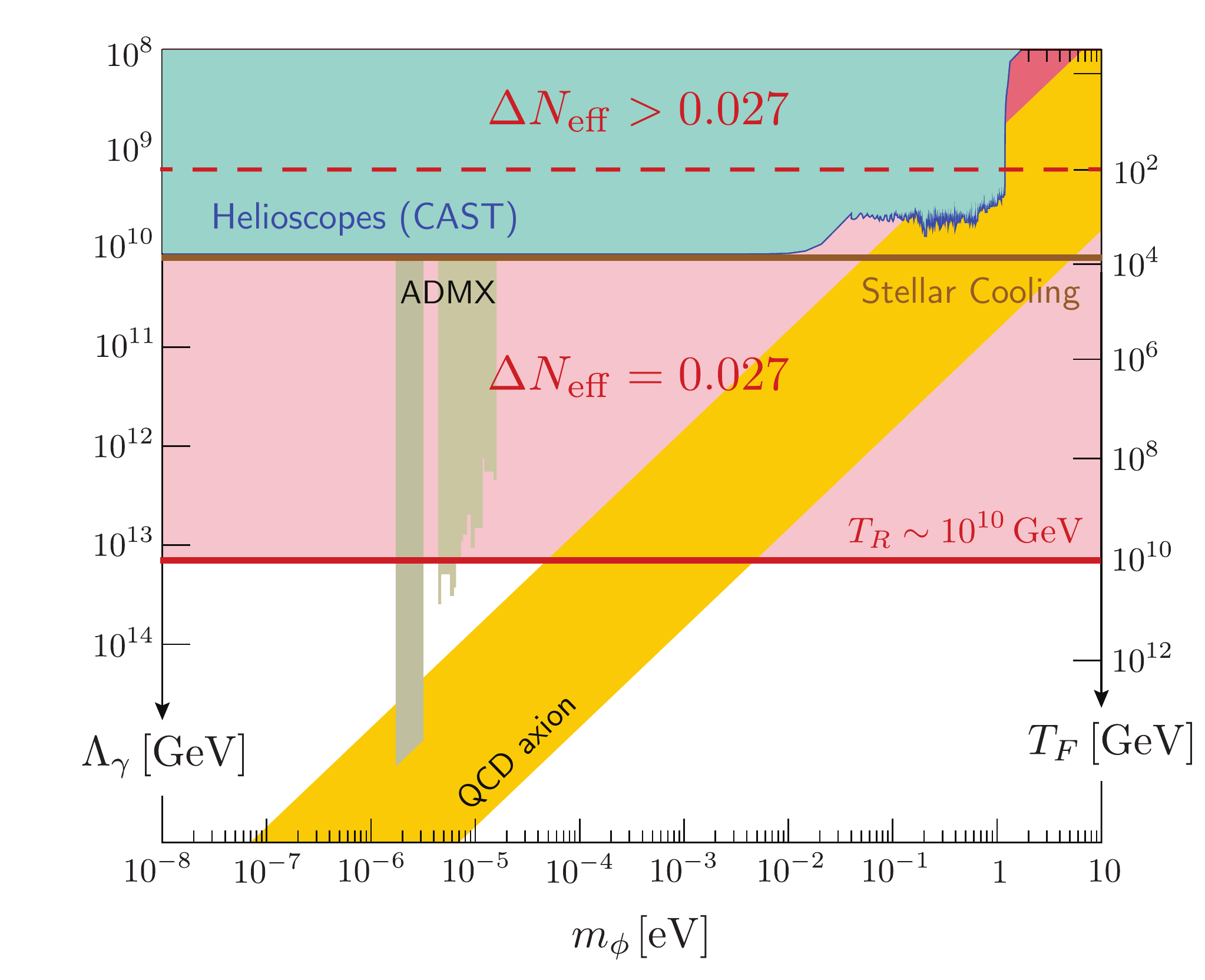}
\caption{Sensitivity to the axion-photon coupling ($\Lambda_{\gamma}$) as a function of the axion mass, $m_a$ in eV.  Also shown is the freeze-out temperature ($T_F$) in GeV for a given value of the coupling.  The region in light red illustrates the axion parameter space that predicts $\Delta \Neff = 0.027$ for a plausible reheat temperature of $T_{\rm R} \sim 10^{10}$ GeV, from the requirement $T_F \leq T_{\rm R}$.  Also shown are the existing experimental and astrophysical constraints.  While the freeze-out of thermal axions is independent of the mass, experimental probes of the coupling are strongly mass dependent.  For $T_F < 10^2$~GeV the axion coupling would predict $\Delta \Neff > 0.027$, as qualified by Figure~\ref{fig:Neff_thermal}.  We see that sensitivity to $\Delta \Neff= 0.027$ can be orders or magnitude more sensitive to axions than existing probes. The yellow band shows a plausible range of photon couplings for the QCD axion (i.e.~axions that can solve the strong CP problem).  Figure adapted from~\cite{Baumann:2016wac}.}
\label{fig:axionphoton}
\end{figure}

The coupling of axions to matter has an additional feature that it can bring axions into thermal equilibrium at both high and low temperatures.  Specifically, the lowest-dimension coupling of an axion to charged matter takes the form
\beq
{\cal L} = -\frac{\partial_\mu a}{\Lambda_\psi}  \bar \psi_i ( \gamma^\mu g^{ij}_V + g_A^{ij} \gamma^5 ) \psi_j
\eeq
where $\psi_{i}$ is any of the charged fermions of the Standard Model and $i,j$ label the three generations of fermions of with the same charges.  Above the scale of electroweak symmetry breaking (EWSB), this coupling leads to an abundance of axions with $\Delta \Neff = 0.027$.  Through freeze-out, we are again very sensitive to $\Lambda_\psi$ at levels that vastly exceed current limits.  In addition, below the scale of EWSB, this coupling can bring the axions into thermal equilibrium at low temperatures (freeze-in) down to the mass of fermion in the coupling, $T_F \lesssim m_{\psi}$.  Freeze-in will produce $\Delta \Neff \approx 0.05$ and is therefore easier to detect.  For reheating temperatures well above the electroweak scale, the sensitivity of freeze-out exceeds that of freeze-in, although both are far more sensitive than current limits on axion couplings to  second and third generation fermions.

The coupling to matter is motivated also by the approximate $U(3)^5$ flavor symmetry of the Standard Model.  It is natural for such couplings to arise if the axion is a Goldstone boson that results from spontaneous breaking of this symmetry (or a sub-group).  Given the non-abelian nature of the flavor symmetry, these scenarios can often lead to many axions (also known as familons).  Under such circumstances, the contribution to freeze-out is given by 
\beq
\Delta \Neff = N_a \times 0.027
\eeq
where $N_a$ is the number of axions or number of broken generators of the symmetry group.  It is easy to find scenarios where $N_a\sim {\cal O}(10)$ which is at the current level of sensitivity.

{\it Status of current observations} -- Current constraints on ALPs arise from a combination of experimental~\cite{Graham:2015ouw}, astrophysical~\cite{Raffelt:2012kt}, and cosmological~\cite{Marsh:2015xka} probes.  Current cosmological constraints are driven by several effects that depend on the mass of the axion.  For axion masses greater than 100 eV, stable thermal ALPs are easily excluded because they produce dark matter abundances inconsistent with observations.  By including the free-streaming effects of thermal QCD-axions,  Planck data~\cite{DiValentino:2015wba} combined with local measurements provide the constraint $m_a < 0.525$ eV (95 \% CL).  At larger masses, ALPs become unstable and can be constrained by the change to $\Neff$ from energy injection as well as from spectral distortions and changes to BBN~\cite{Cadamuro:2011fd,Millea:2015qra}.

{\it Implications for CMB-S4} -- Sensitivity to $\Delta \Neff =0.027$ is sufficient to probe the entire mass range of ALPs down to $m_a =0$ under the assumption that they thermalized in the early Universe.  Interpreting such bounds in terms of the couplings of axions is more complicated~\cite{Brust:2013xpv} and can depend on assumptions about the reheating temperature.  For high (but plausible) reheat temperatures of $10^{10}$ GeV, CMB-S4 would be sensitive to $\Lambda_{\gamma}, \, \Lambda_{g} \,  > \, 10^{13} \, {\rm GeV}$~\cite{Baumann:2016wac} as illustrated in figures~\ref{fig:axionphoton} and~\ref{fig:axiondipole}.  These projected limits exceed current constraints and future probes for a range of possible axion masses (including the QCD axion).

The implications for the couplings to matter for the contribution from freeze-out are similar.  The freeze-in contribution of $\Delta \Neff \gtrsim 0.05$ that is easier to exclude experimentally still produces the limits~\cite{Baumann:2016wac}
\beq
\Lambda_{\psi_i}  \ >\ \left\{ \begin{array}{ll} \displaystyle 1.3\times 10^8 \, {\rm GeV} \left(\frac{g_{*,i}}{g_{*,\tau}}\right)^{\!-1/4} \left(\frac{m_i}{m_\tau}\right)^{\!1/2} & \quad i=\text{leptons}, \\[10pt]
\displaystyle 2.1\times 10^9 \, {\rm GeV} \left(\frac{g_{*,i}}{g_{*,t}}\right)^{\!-1/4} \left(\frac{m_i}{m_t}\right)^{\!1/2} & \quad i=\text{quarks}.
\end{array} \right.
\eeq
where $g_{*,i}$ is the number of degrees of freedom at temperature $T = m_i$.  For second and third generation fermions, these limits would exceed current bounds by several orders of magnitude.

The freeze-in contribution from a single axion can be excluded at 2$\sigma$ with CMB-S4, as seen in Figure~\ref{fig:Neff_fsky} (since freeze-in for most fermions is equivalent to a freeze-out temperature at or below 1 GeV). The threshold of $\Delta \Neff =0.027$ is achievable at 1$\sigma$ but more challenging at 2$\sigma$, as discussed in Section~\ref{sec:neff_forcast}. Models with two or more axions could be excluded at 2$\sigma$ even with the more conservative configurations.

\begin{figure}[h!]
\centering \includegraphics[width=0.7\textwidth]{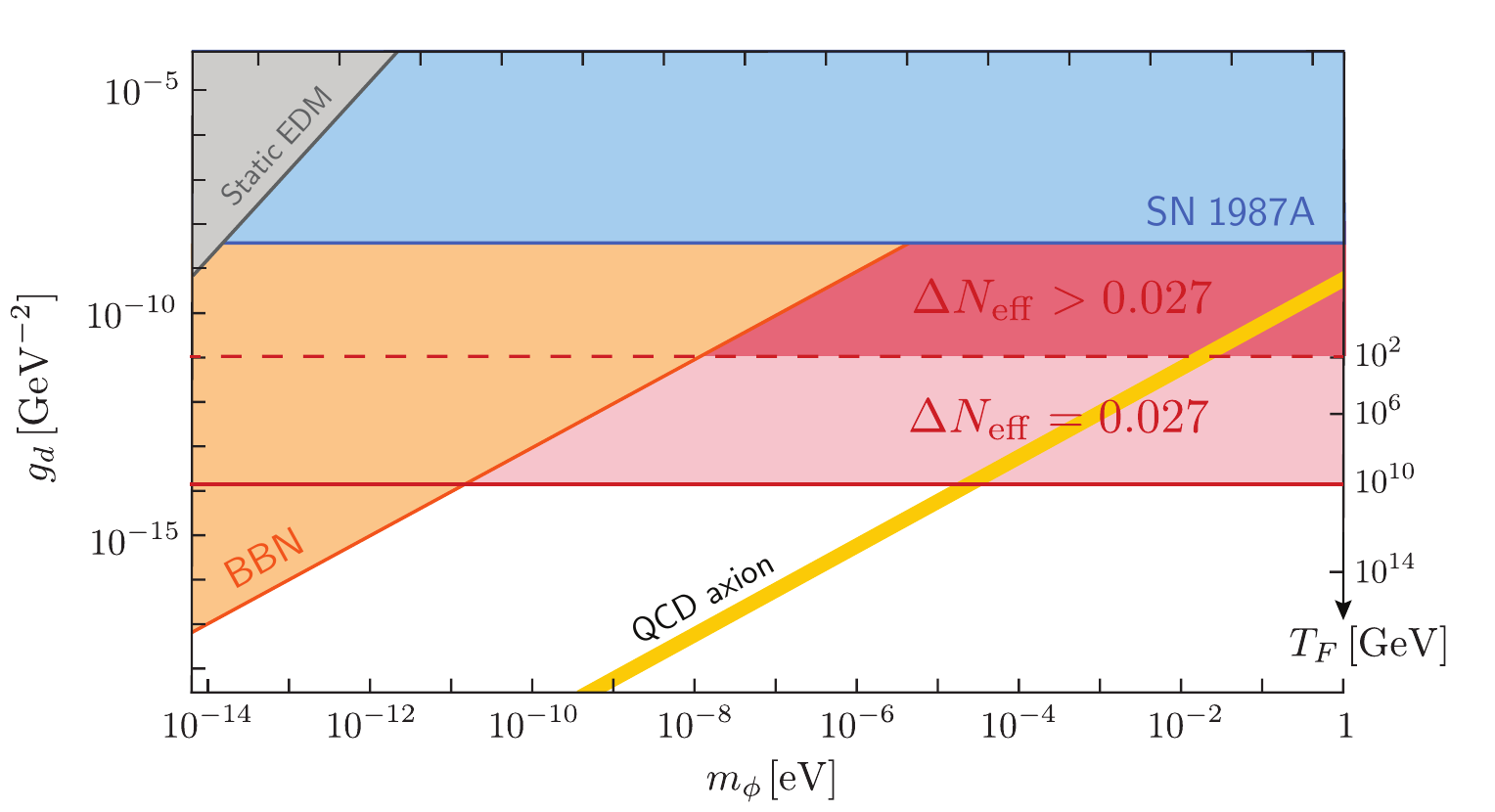}
\caption{Sensitivity to the axion-gluon coupling via the neutron dipole ($g_d$) as a function of the axion mass, $m_a$ in eV.  Also shown is the freeze-out temperature ($T_F$) in GeV for a given value of the coupling.   The region in light red illustrates the axion parameter space that predicts $\Delta \Neff = 0.027$ for a plausible reheat temperature of $T_{\rm R} = 10^{10}$ GeV, from the requirement $T_F \leq T_{\rm R}$.  The region in darker red  shows $T_F < 10^2$~GeV  and predicts $\Delta \Neff > 0.027$, as qualified by Figure~\ref{fig:Neff_thermal}.  Also shown are the existing experimental, astrophysical and cosmological constraints.  We see that cosmology is more sensitive than current limits even for freeze-out temperatures below the eletro-weak phase transition and therefore could be probed by $\Delta \Neff > 0.027$.  The yellow band shows the predictions for the QCD axion, including the theoretical uncertainty in the relationship between the gluon coupling $\Lambda_g$ and $g_d$.  Figure adapted from~\cite{Baumann:2016wac}.}
\label{fig:axiondipole}
\end{figure}

\subsection{Light Fermions and Vectors}

A general approach to interpreting $\Neff$ constraints in terms of massless fields of arbitrary spin was undertaken in~\cite{Brust:2013xpv}.  One identifies the symmetry that is required to explain  the small mass of the particle, and then writes the most general interactions with the Standard Model consistent with the symmetry.  The couplings will take the form
\beq
{\cal L} \supset \sum_{\Delta_h, \Delta_s} \frac{1}{\Lambda^{\Delta_h+ \Delta_s-4}} \,  {\cal O}_{h, \Delta_h} \cdot {\cal O}_{s, \Delta_s}
\eeq
where ${\cal O}_{h,\Delta_h}$ is an operator of dimension $\Delta_h$ constructed only from hidden sector fields (and similarly for ${\cal O}_{s, \Delta_s}$ and Standard Model fields).  The total operator must be a scalar under both the Lorentz transformations and the symmetry that protects the mass of the hidden sector field(s).  The bounds on axions discussed in Section~\ref{sec:neffaxions} are one such example, where the axion is protected by a shift symmetry.  For the purpose of this discussion, we have classified all scalars of this type as axions.

For a single Weyl fermion, $\chi$, the leading couplings to the Standard Model are through the anapole moment and four-fermion interactions
\beq
{\cal L} \supset \frac{\chi^{\dagger} \bar \sigma^{\mu} \chi}{\Lambda_{\chi}^2} \Big( d_a  \, \partial^\nu F_{\mu \nu} + d_f \,  \bar \psi \gamma_\mu \psi \Big) \ ,
\eeq
where we have chosen one of several four-fermion interactions for illustration and $d_f, d_a$ are order one numbers.  Current experimental constraints from LEP and the LHC limit $\Lambda_\chi \gtrsim \, 1$ TeV.  Similar bounds are set by \planck\ by excluding the contributions to $\Delta \Neff$ from freeze-out after the QCD phase transition.  If we are sensitive to the minimal contribution from a Weyl fermion of $\Delta \Neff = 0.047$, then for a reheat temperature of $T_{\rm reheat} \sim 10^{10}$ GeV, we would be sensitive to $\Lambda_\psi \lesssim 10^{12}$ GeV.  We see that for an order of magnitude improvement in sensitivity to $\Neff$, we get as much as a nine orders of magnitude improvement in sensitivity to $\Lambda_\psi$.  

For a hidden Dirac fermion $X$ with a $U(1)$ global symmetry, the leading\footnote{The dipole operator preserves the same $U(1)$ symmetry that allows a Dirac mass.  It is unclear if one can UV-complete this model with a small Dirac mass.} coupling is through an effective dipole interaction.  A similar interaction also permits a hidden $U(1)$ gauge boson, $A'_\mu$, to couple to Standard Model fermions.
\beq
{\cal L} \supset  \frac{1}{\Lambda_X} \bar X \sigma_{\mu\nu} X  F^{\mu \nu} + \frac{1}{\Lambda_{A'}^2} F'_{\mu \nu} \, H \bar \psi \sigma_{\mu\nu} \psi    \ ,
\eeq
where $F'_{\mu \nu}$ is the field strength of $A'_\mu$.  We have included the Higgs field, $H$, in the coupling to Standard Model fermions as it is required by gauge invariance above the scale of EWSB.  Stellar cooling provides a strong constraint of $\Lambda_X \gtrsim 10^9$ GeV and $\Lambda_{A'} \gtrsim 10^5$ GeV.  

Freeze-out above the scale of EWSB for a Dirac fermion produces $\Delta \Neff = 0.094$ and $\Delta\Neff = 0.054$ for a hidden photon.  For a reheating temperature of $10^{10}$ GeV, we are sensitive to $\Lambda_X \lesssim 10^{13}$ GeV and $\Lambda_{A'}  \lesssim 10^{11}$ GeV respectively.  The Dirac fermion will be accessible with CMB-S4 and will improve on the stellar cooling constraint for a reheat temperature $T_{\rm R} > 10^4$ GeV.

{\it Implications for CMB-S4} -- The projected sensitivity to the freeze-out temperatures of fermions and vectors with CMB-S4 is illustrated in Figure~\ref{fig:Neff_fsky}.  Conservative configurations of CMB-S4 can place 2$\sigma$ exclusions on the freeze-out temperatures in the tens of GeV, which would improve limits on the couplings to light fermions or vectors, $\Lambda_{\chi, A'}$, by two orders of magnitude. For even a modest increase of $f_{\rm sky}$, we can reach the thresholds of $\Delta \Neff = 0.047, 0.054$ at 2$\sigma$, which would be sensitive to freeze-out back to the time of reheating.  Reaching these thresholds could translate into nine orders of magnitude improvement in $\Lambda_{\chi, A'}$, depending on the reheat temperature.  Thermal relic Dirac fermions, non-abelian gauge fields or multiple families of fermions or vectors would be detectable with high significance in most configurations of CMB-S4.

\subsection{Gravitinos}

One of the most popular extensions of the Standard Model is supersymmetry, which is motivated both by naturalness and gauge coupling unification.  Although the most generic possibilities are under significant tension for the LHC, there are still a variety of possibilities consistent with low-scale supersymmetry.

One of the universal predictions of supersymmetry is the existence of a spin-3/2 partner to the graviton, the gravitino.  The gravitino mass is determined by the absolute scale of supersymmetry breaking, 
\beq
m_{\frac{3}{2}} = \frac{|F|}{\sqrt{3} M_{\rm P}}
\eeq
where $|F|$ is the order parameter for the scale of SUSY breaking (the vacuum expectation value of the auxiliary field $F$).  This result does not depend on details of the mechanism of SUSY breaking unlike the super-partners of the rest of the Standard Model particles.

The typical coupling strength of the gravitino is the same as the graviton, $8 \pi G = \Mpl^{-2}$.  However, the strength of the coupling to the helicity-1/2 component of the gravitino is enhanced by $\Mpl^2 /F$.  This is simply the statement that the goldstino of SUSY breaking is coupled with strength $F^{-1}$ (but is `eaten' by the gravitino).  Due to the enhanced coupling, the gravitino can be in thermal equilibrium with the Standard Model at plausible temperatures in the early Universe.  The gravitino therefore behaves just like a Weyl fermion in Figure~\ref{fig:Neff_thermal}.

For $m_{3/2} \lesssim 10$ eV, hot relic gravitinos free stream on the scale of the CMB and therefore lead to observable signatures.  Current data from \planck\ already requires that $m_{3/2} < 4.7$ eV from a combination of the primary CMB and CMB lensing~\cite{Osato:2016ixc}.  To probe lower masses, note that for $m_{3/2} < {\cal O}(1)$ eV a gravitino will behave as free-streaming radiation from the point of view of the CMB.  One finds that for these low values of the mass, gravitinos contribute a shift to $\Neff$,
\beq
\left(\Delta \Neff\right)_{3/2} \, \gtrsim \, 0.057 \ .
\eeq
This number is somewhat larger than the minimum value of $\Delta \Neff =0.047$ for the helicity-1/2 component because the effective coupling becomes large as $m_{3/2} \to 0$.  As a result for $m_{3/2} < 1$ eV, the gravitinos decouple below 100 GeV.  At $\sigma(\Neff) \sim 0.03$, CMB-S4 can rule out all low-scale SUSY breaking models allowed by current cosmology.  

\subsection{Gravitational Waves}\label{sec:constraintsntNeff}

Since gravitational waves are massless and free-streaming, any gravitational waves which were present in the early Universe naturally contribute to the total radiation energy density, and can therefore be constrained with $\Neff$ \cite{Boyle:2003km,Boyle:2007zx,Stewart:2007fu,Meerburg:2015zua,Lasky:2015lej}.

Let us briefly review how to compute the energy density of a stochastic background of gravitational waves, following the treatment of \cite{Isaacson:1968zza,Misner:1974qy,Watanabe:2006qe,Maggiore:1900zz}.  We will take the metric of spacetime to be given by a background component $\bar{g}_{\mu}$ described by the flat Friedmann-Robertson-Walker metric and a perturbation $\delta g_{\mu\nu}$.  We will take the characteristic frequency of $\delta g_{\mu\nu}$ to be much higher than that of $\bar{g}_{\mu\nu}$.  In particular, we will focus on gravitational waves whose wavelengths are much shorter than scales of cosmological interest.

The Ricci tensor can be expanded in powers of $\delta g$ as
\begin{equation}\label{eq:Ricci_GW}
	R_{\mu\nu} = \bar{R}_{\mu\nu} + R_{\mu\nu}^{(1)} + R_{\mu\nu}^{(2)} + \cdots \, .
\end{equation}
We are interested in determining how spacetime is curved by the presence of small scale gravitational waves, or in other words, how $\bar{R}_{\mu\nu}$ is affected by terms containing $\delta g_{\mu\nu}$.  Since $R_{\mu\nu}^{(1)}$ is linear in $\delta g_{\mu\nu}$, it contains only high frequency components, while on the other hand, $R_{\mu\nu}^{(2)}$ has both low and high frequency parts.  

The high frequency part of Einstein's equations governs how gravitational waves propagate in a curved background, and is not necessary here.  For the low frequency part, we can take an average over several cycles of the high frequency modes, which then gives
\begin{equation}\label{eq:Einstein_Eq_GW}
	\bar{R}_{\mu\nu} = -\left\langle R_{\mu\nu}^{(2)}\right\rangle + 8\pi G \left\langle T_{\mu\nu}-\frac{1}{2}g_{\mu\nu}T\right\rangle \, .
\end{equation}
We can then read off the effective energy-momentum tensor of small scale fluctuations
\begin{equation}\label{eq:EM_Tensor_GW}
	T_{\mu\nu}^\mathrm{GW} = -\frac{1}{8\pi G} \left \langle R_{\mu\nu}^{(2)}-\frac{1}{2}\bar{g}_{\mu\nu}R^{(2)} \right\rangle + \mathcal{O}(\delta g^3) \, .
\end{equation}
If we take our perturbation to be of the form $\delta g_{ij} = a^2 h_{ij}$ with $h_{ij,j} = 0$ and $h_{ii} = 0$, we can compute the energy density of gravitational waves explicitly in the transverse traceless gauge
\begin{equation}\label{eq:Energy_GW}
	\rho_\mathrm{GW} = T_{00}^\mathrm{GW} = \frac{1}{32 \pi G}\delta^{ik}\delta^{j\ell}\left\langle \dot{h}_{ij} \dot{h}_{k\ell} \right\rangle + \mathcal{O}(\delta g^3) \, .
\end{equation}

We will define the gravitational wave power spectrum as
\begin{equation}\label{eq:GW_power_spectrum}
	\left\langle h_{ij}(\eta,\mathbf{x})h^{ij}(\eta,\mathbf{x})\right\rangle \equiv \int  d \log k \, \mathcal{P}_{\rm t}(k) \left[\mathcal{T}(\eta,k)\right]^2 \, ,
\end{equation}
where $\mathcal{P}_{\rm t}(k)$ is the primordial power spectrum of gravitational waves and $\mathcal{T}(\eta,k)$ is the tensor transfer function.  The energy density of gravitational waves is then given by
\begin{equation}\label{eq:GW_energy_density}
	\rho_\mathrm{GW} = \frac{1}{32\pi G a^2}\int  d\log k \, \mathcal{P}_{\rm t}(k) \left[\mathcal{T}'(\eta,k)\right]^2 \, ,
\end{equation}
where the prime denotes a derivative with respect to conformal time $\eta$.

Direct searches for the stochastic gravitational wave backgound are often quoted in terms of the normalized energy density per logarithmic scale
\begin{equation}\label{eq:Omega_GW}
	\Omega_\mathrm{GW}(k) \equiv \frac{8\pi G}{3H_0^2}\frac{d\rho_\mathrm{GW}}{d\log k} = \frac{\mathcal{P}_{\rm t}(k)}{12H_0^2a_0^2} \left[\mathcal{T}'(\eta_0,k)\right]^2 \, .
\end{equation}
Constraints on $\Neff$ provide an integral constraint on the spectrum of gravitational waves since waves of all frequencies contribute to the total energy density.

If for example, we take the primordial gravitational wave power spectrum to be a simple power law of the form $\mathcal{P}_{\rm t}(k) = A_{\rm t} \left(\frac{k}{k_\star}\right)^{n_{\rm t}}$ (as in Eq.~(\ref{eq:power_spectra_power_law})), we can use the constraint on $\Neff$ to place bounds on the tensor spectral tilt $n_{\rm t}$.  The contribution to $\Neff$ can then be approximated for $n_{\rm t}>0$ as \cite{Meerburg:2015zua}
\begin{equation}\label{eq:Neff_GW}
	\Delta \Neff \simeq \left(3.046 + \frac{8}{7}\left(\frac{11}{4}\right)^{4/3}\right)\frac{A_t}{24n_t}\left(\frac{k_\mathrm{UV}}{k_\star}\right)^{n_t} \, ,
\end{equation}
where $k_\mathrm{UV}$ represents the ultraviolet cutoff of the primordial tensor power spectrum.  While this constraint does not probe the regime of great interest for inflationary models, it does provide a useful constraint on alternatives to inflation which predict positive tensor tilt.

\section{Big Bang Nucleosynthesis}\label{sec:bbn}

Primordial light element abundances have for many decades been an interesting observational test of hot big bang cosmology.  Predicting the formation of light elements in the early Universe, a process known as big bang nucleosynthesis (BBN), was initiated in the early days of the development of the hot big bang model of cosmology~\cite{Alpher:1948ve}.  BBN is a process that unfolded during the first three minutes of our current phase of expansion, involving all four fundamental forces, and has long provided a useful constraint on physics beyond the Standard Model.  Primordial light element abundances resulting from BBN are sensitive to the baryon-to-photon ratio and the expansion rate in the early universe (which is determined by $\Neff$).  CMB-S4 will provide the best available constraints on both of these quantities as well as an improved measurement of the primordial helium abundance which is not subject to the astrophysical systematics which dominate current errors.  Therefore, CMB-S4 will significantly improve our ability to check the consistency of BBN with the results of a single experiment, and will open the possibility for the discovery of new physics affecting the process of BBN. 

\subsection{Standard Big Bang Nucleosynthesis} \label{StandardBBN}
In this section, we will briefly review the physics of big bang nucleosynthesis in the Standard Model.  For more extensive reviews see for example \cite{Weinberg:2008zzc,Agashe:2014kda,Cyburt:2015mya}.

The salient features of the early Universe at the BBN epoch are that: (1) the entropy per baryon is high (the baryon-to-photon ratio $\eta\approx6.1\times {10}^{-10}$); and (2) the expansion rate is driven by radiation.
At temperatures well above $T \sim 1\,{\rm MeV}$ the strong, electromagnetic, and even the weak interaction are in chemical and thermal equilibrium. As the Universe expands and the temperature drops the rates of neutrino scattering processes fall and, eventually, these no longer effect efficient exchange of energy between neutrinos and the photon-electron/positron plasma; this is weak decoupling. Likewise, the charged current lepton capture processes that interconvert neutrons and protons ($\nu_e+n\rightleftharpoons p +e^-$, $\bar\nu_e+p \rightleftharpoons n +e^+$, $n\rightleftharpoons p +e^-+\bar\nu_e$) at high temperature can maintain chemical equilibrium for the neutron-to-proton ratio, but as the Universe expands this ratio decreases and falls out of chemical equilibrium; this is referred to as weak freeze-out. Both the weak decoupling and weak freeze-out processes are not sharp in time/temperature and, in fact both freeze-outs overlap in time and both occur over many Hubble times. The strong interaction and nuclear abundances are in thermal and chemical equilibrium, referred to as nuclear statistical equilibrium, or NSE,  at high temperature. As the temperature falls, rates for individual nuclear reaction processes slow down and so abundances drop out of NSE. In broad brush, the alpha particle abundance goes up extremely quickly at $T \sim 80\,{\rm keV}$ and this effectively locks up nearly all the free neutrons extant at that epoch. As a consequence, the primordial helium abundance is determined by the neutron-to-proton ratio at this epoch, and therefore encodes both the expansion history and the weak interaction history, i.e., it is dependent on $\Neff$ and the details of neutrino energy distribution functions, neutrino degeneracy parameters, etc. Once the alpha particles form, the deuterons fall out of NSE. Their abundance is then modified by out-of-equilibrium nuclear reactions, principally $D(p, \gamma)^3{\rm He}$. The deuterium abundance is then mostly sensitive to the baryon-to-photon ratio, though weak interactions and neutrino physics do play a role in setting this abundance.

Standard BBN is a one parameter model, depending only on the baryon to photon ratio $\eta$ (since $\Neff$ is fixed in the Standard Model).  The theory predicts several abundances which can be used to fix $\eta$ and check the consistency of the theory, or alternatively, to constrain new physics.  Current observations of the primordial $\nucl{ }{ }{D}$ and $\nucl{4}{ }{He}$ abundances agree quite well with the predictions of standard BBN, however measurements of $\nucl{7}{ }{Li}$ do not.  It is unclear whether this disagreement points to a problem with the astrophysical determination of the primordial abundance or a problem with the standard theory.  The cosmological lithium problem remains unsolved \cite{Fields:2011zzb}.  From here on, however, we will ignore the lithium problem and focus on how measurements of the other abundances (primarily $\nucl{ }{ }{D}$ and $\nucl{4}{ }{He}$) can be used to constrain the physics of the early Universe.


\subsection{Beyond the Standard Model}\label{BSM}
Moving beyond standard BBN, measurements of primordial abundances have the ability to constrain many deviations from the standard thermal history and the Standard Model of particle physics.  Because BBN is sensitive to all fundamental forces, changes to any force can in principle impact light element abundances.  Of primary interest for our purpose is that BBN is sensitive to the expansion rate between about one second and a few minutes after reheating.  The expansion rate is in turn determined by the radiation content of the Universe during this period, and thus BBN is sensitive to $\Neff$.  

As discussed above, the expansion rate during this epoch plays a role in setting the ratio of neutrons to protons and the amount of time free neutrons have to decay.  Additional radiation compared to the Standard Model gives a higher expansion rate, which causes weak freeze-out to occur at a higher temperature and gives less time for free neutron decay, leading to a larger primordial $\nucl{4}{ }{He}$ abundance.  The neutrino interaction rates also depend weakly on the distribution function of electron neutrinos, though this effect is subdominant to the dependence on $\Neff$ for small non-thermal distortions \cite{Serpico:2004gx}.

Historically, $Y_p$ had provided the best constraint on $\Neff$.  Recent advancements in the determination of primordial deuterium abundance have made constraints on $\Neff$ from deuterium competitive with those from $Y_p$ \cite{Cooke:2013cba}.  The precision with which primordial abundances constrain $\Neff$ is now comparable to that of constraints the CMB power spectrum, and there is no evidence for deviation from the Standard Model \cite{Ade:2015xua}.


\subsection{CMB Probes of BBN}\label{Complementarity}
The CMB can be used to quite precisely constrain the baryon-to-photon ratio $\eta$ by measurement of the baryon fraction of the critical density, which is related to $\eta$ by
\begin{equation}
	\Omega_b h^2 \simeq \frac{\eta\times10^{10}}{274} \, .
\end{equation}
Using the value of $\eta$ determined from CMB measurements as an input for BBN makes standard BBN a theory without free parameters which agrees very well with all observations (apart from the aforementioned disagreement with the observed lithium abundance).  The CMB and the primordial light element abundances are sensitive to the baryon density measured at different times.  While BBN is sensitive to the baryon-to-photon ratio up to a few minutes after reheating, the CMB is sensitive to the baryon density at much later times, closer to recombination about 380,000 years later.  Combining constraints from BBN and CMB on the baryon fraction therefore allows constraints on models where the photon or baryon density changes between these times.
Similarly, light abundance yields are sensitive to the expansion rate (and therefore $\Neff$) at an earlier epoch than is probed by the CMB.  This allows the combination of light element abundances and direct CMB constraints on $\Neff$ to be combined in order to gain insight into the thermal history of the early universe.  If it were measured for example that $\Nf^{\mathrm{BBN}}<\Nf^{\mathrm{CMB}}$, this could be explained by the late decay of some unstable particles \cite{Fischler:2010xz,Menestrina:2011mz,Hooper:2011aj}.  Alternatively, if observations revealed that $\Nf^{\mathrm{BBN}}>\Nf^{\mathrm{CMB}}$, this might signal late photon heating \cite{Cadamuro:2010cz,Millea:2015qra}.

The power spectrum of the CMB is also directly sensitive to $Y_p$.  Since helium recombines earlier than hydrogen, the density of helium present at the time of recombination affects the free electron density, and thereby affects the damping tail of the CMB (though in a way which can be distinguished from the effects of $\Neff$) \cite{Bashinsky:2003tk,Hou:2011ec,Follin:2015hya,Baumann:2015rya}.  The degeneracy between $Y_p$ and $\Neff$ will be more strongly broken with the precise CMB polarization observations that CMB-S4 will provide.

CMB-S4 will provide constraints on $\Neff$ which are about an order of magnitude better than the current best constraints, and will also improve on the measurement of $Y_p$ compared to the current best astrophysical measurements.  Importantly, CMB observations of $Y_p$ are not subject to the astrophysical systematics which dominate the current error bars.  By providing high precision measurements of the baryon-to-photon ratio $\eta$, the expansion rate as determined by $\Neff$, and the primoridal helium abundance $Y_p$, CMB-S4 will provide a very thorough check of our understanding of the early Universe and provide the opportunity to discover physics beyond the Standard Model.

The primordial abundance of deuterium is an additional independent prediction of BBN and is measurable in isotope-shifted hydrogen absorption lines from Lyman limit systems along lines of sight to high redshift quasars \cite{1976A&A....50..461A,Pettini:2012ph,Cooke:2013cba}. It may be possible to improve the precision of such measurements with future 30-m-class telescopes like TMT, EELT, and GMT. CMB-S4 measurements of $\Neff$ and primordial helium, coupled with these high precision deuterium abundance measurements, then hold out the promise of a new probe of neutrino sector and other physics beyond the Standard Model.

\subsection{Forecasts}\label{sec:NeffBBNfore}

\begin{figure}[t!]
\begin{center}
\includegraphics[width=0.48\textwidth]{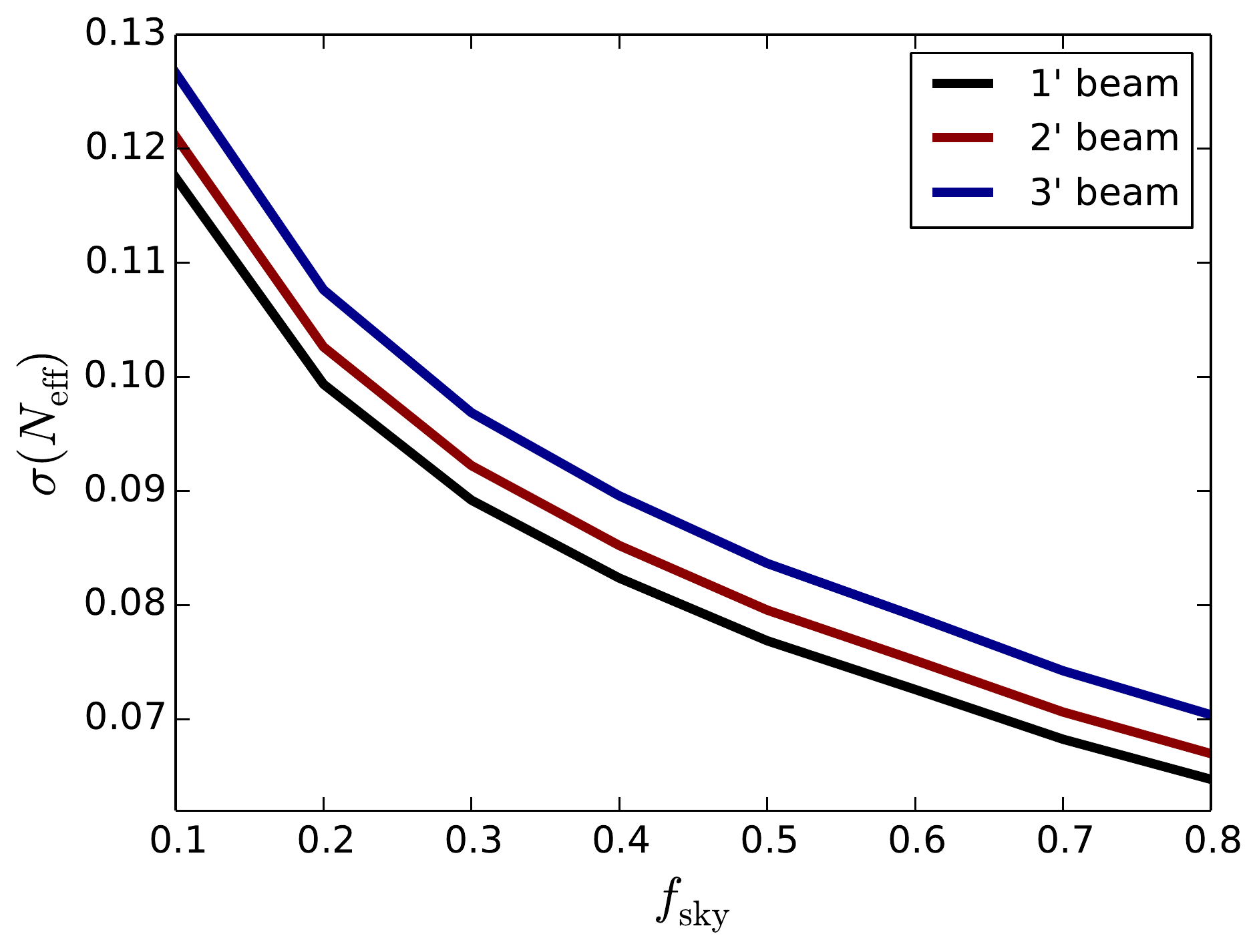}
\includegraphics[width=0.5\textwidth]{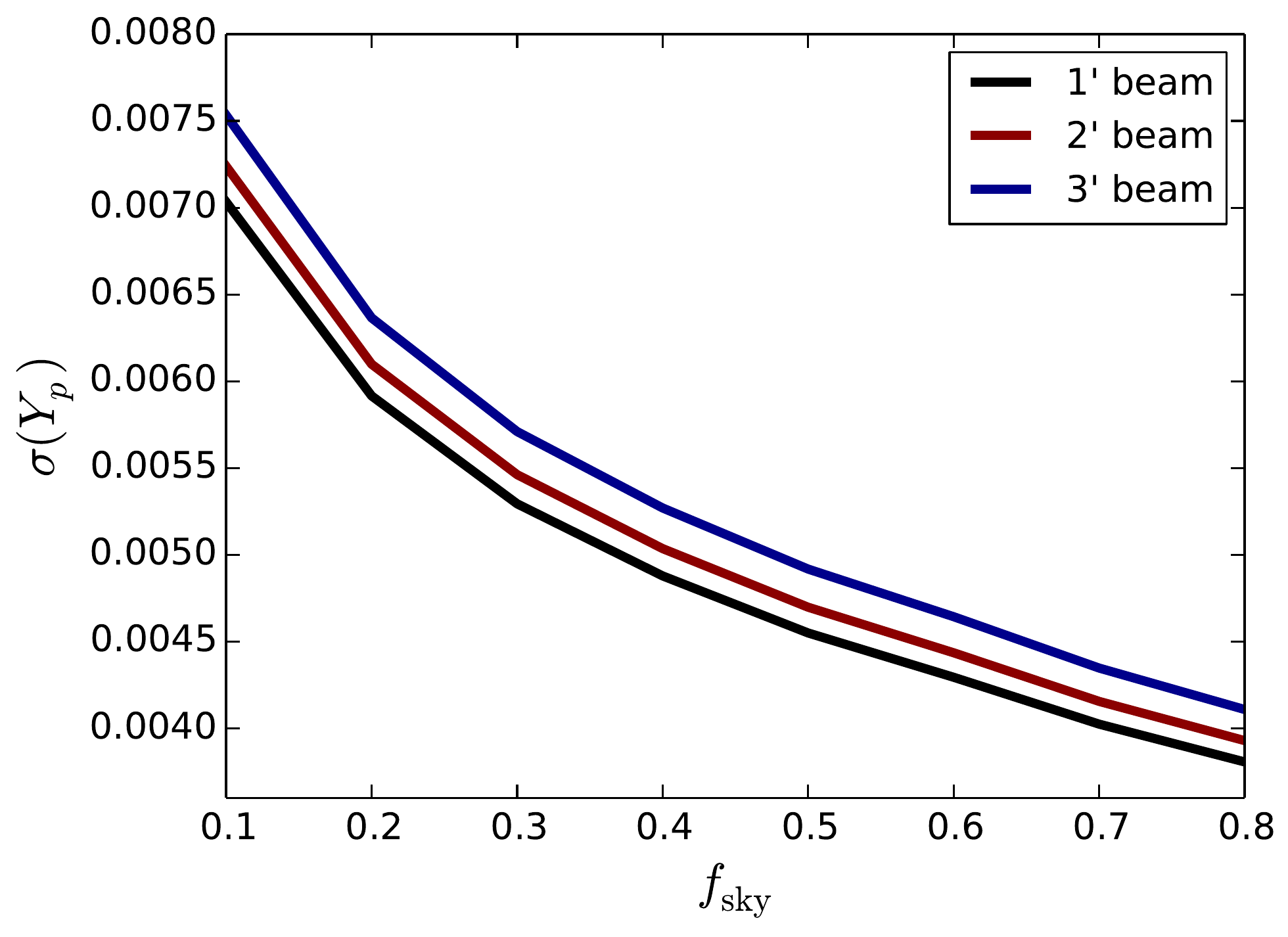}
\caption{{\bf Left:} Forecasts for $\sigma(\Neff)$, marginalized over $Y_p$, as a function of sky fraction for beam sizes of $1',2'$ and $3'$.  {\bf Right :} Forecasts for $\sigma(Y_p)$, marginalized over $\Neff$, as a function of sky fraction for beam sizes of $1',2'$ and $3'$.  For both figures, the sensitivity have been normalized to 1 $\mu$K-arcmin for $f_{\rm sky} =0.4$ and is scaled according to $S \propto f_{\rm sky}^{1/2}$ for different sky fractions.}
\label{fig:YpNeff_fsky}
\end{center}
\end{figure} 

Constraints on the combined $Y_p$-$\Neff$ parameter space are a useful probe of both the physics of BBN and recombination and possible evolution in between.  It is therefore useful to consider joint constraints on these two parameters available from the CMB alone.  Forecasts here will follow the same procedures as in Section~\ref{sec:neff_forcast}.  In particular, all spectra are delensed unless otherwise stated.

\begin{table}[t!]
\begin{center}
\begin{tabular}{l ccc} 
 \toprule
    		Temperature Noise -- Beamsize		    			& $1'$  		& $2'$  		& $3'$  		 \\ [0.5ex]
 \midrule
   1 $\mu$K-arcmin & 0.082 / 0.0049		& 0.085 /  0.0050 		& 0.090 / 0.0053		 		  \\
  2  $\mu$K-arcmin & 0.093 / 0.0054		& 0.096 / 0.0056		& 0.10 / 0.0058	 		  \\
   3  $\mu$K-arcmin & 0.10 / 0.0058		& 0.10 / 0.0059		& 0.11 / 0.0062		 		  \\
    \bottomrule
\end{tabular}
\caption{Forecasts for $\sigma(\Neff)$ / $\sigma(Y_p)$ with varying beamize in arcmin (') and temperature noise. These forecasts allow both $Y_p$ and $\Neff$ to vary (along with the parameters of $\Lambda$CDM).  We assume $f_{\rm sky} = 0.4$. }
\label{tab:YpNeffbeam}
\end{center}
\end{table}

Like the case of $\Neff$-only forecasts, constraints on $Y_p$ and $\Neff$ are primarily sensitive to $f_{\rm sky}$ as shown in Figure~\ref{fig:YpNeff_fsky}.  Although both parameters are sensitive to the damping tail and we also see the significant impact of varying the beamsize from $1'$ to $3'$.  

Breaking the degeneracy between $Y_p$ and $\Neff$ is important for producing independent constraints on both parameters.  Both $Y_p$ and $\Neff$ alter the damping tail in a way that can cancel exactly between the two.  For this reason, we expect the phase shift in the locations of the acoustic peaks induced by free-streaming particles to help break the degeneracy.  CMB lensing somewhat weakens this effect by smearing the acoustic peaks while delensing the spectra sharpens the peaks, leading to a better measurement of the peak locations.  In Figure~\ref{fig:YpNeff_2d} we show the projected 1$\sigma$ contours with and without delensing to illustrate improvement delensing can make in breaking this degeneracy.  
\newpage
\begin{figure}[th!]
\begin{center}
\includegraphics[width=0.5\textwidth]{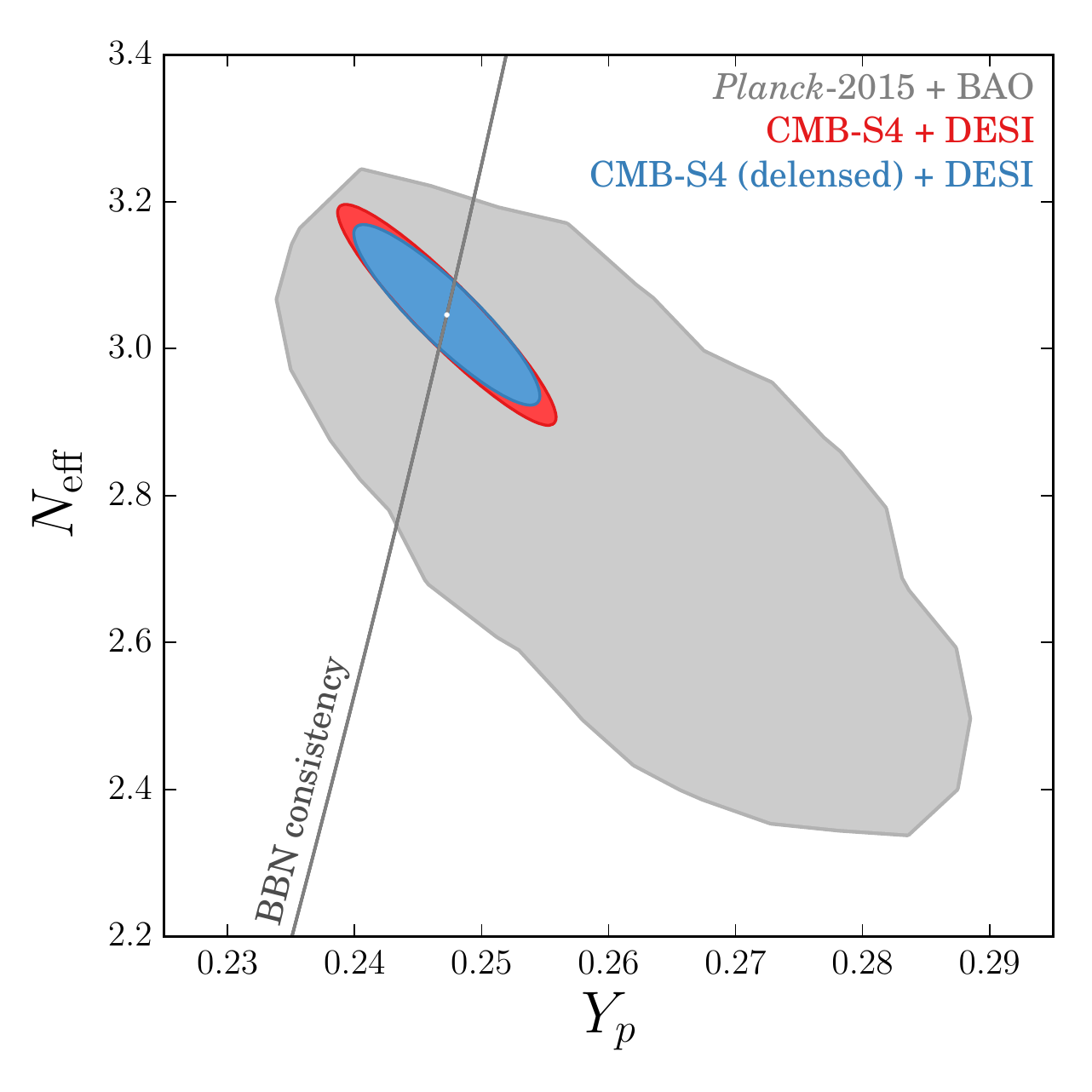}
\caption{ Projected 1$\sigma$ contours in $\Neff$ and $Y_p$ for $f_{\rm sky}=0.4$ with temperature noise of 1 $\mu$K-arcmin and 1' beams.  We show current \planck\ 2015 constraints, with current BAO, along with forecasts for CMB-S4 and DESI BAO, with and without performing delensing~\cite{Green:2016} on CMB-S4 $E$ and $T$ spectra.  We see that delensing primarily shrinks the contours along the degeneracy, which is consistent with expectations from the phase shift in the locations of the acoustic peaks.  Also plotted is the range of values of $Y_p$ predicted by BBN as a function of $\Neff$ (which is assumed to be constant over the relevant periods).}
\label{fig:YpNeff_2d}
\end{center}
\end{figure}

\section{Detection Scenarios for Labs and Cosmology}~\label{sec:neffscenarios}

Experimental efforts searching for light particles are underway in a number of different domains.  There are a number of possible situations where a discovery could be made in cosmology and/or the lab that could inform each other.

In this section, we will discuss plausible theoretical interpretations of a number of such scenarios.  Since there are numerous ways to produce $\Delta\Neff$, these scenarios are not necessarily the only interpretations possible, but are natural interpretations within well studied theoretical frameworks.   

\subsection{Dark Sectors and Particle Physics}

Deviations from $\Neff = 3.046$ can arise from a wide variety of changes to the particle content and thermal history of the Universe.  In most cases, the physics responsible fundamentally requires a coupling of new particles to the Standard Model in regimes where they often can, in principle, be detected by other means.  Cosmology is a very broad tool for searching for physics beyond the Standard Model, but it is also very complementary to more targeted searches.  A list of plausible detection scenarios is shown in Table~\ref{table:darkscenarios}:

\begin{table}[t!]
\begin{center}
\begin{tabular}
{| l | c | p{5cm} | p{5cm} | }\hline Scenario & $\Delta \Neff$ & Experimental Input & Conclusion \\
\hline 
Axions & $\geq 0.027$  & Direct detection of axions & Lower-limit on the reheat temperature
\\[.2cm]
Low-reheating & $0$  & Direct detection of axions & Upper-limit on the reheat temperature
\\[.2cm]    
Stellar Cooling & $\geq 0.027$  & Anomalous Stellar cooling (e.g. white dwarfs) & Evidence for new light particle ; Spin determined by $\Delta \Neff$.  
\\[.2cm]    
Gravitinos & $\geq 0.057$  & LHC evidence for SUSY & Upper-limit on scale of SUSY breaking
\\[.2cm]  
Late Decays & $< 0$  & Spectral Distortions observed ($\mu$ or $y$) & Evidence for new massive particle; energy injection
\\[.2cm]
Evolving $\Neff$ & $> 0$  & Primordial abundances (BBN) consistent with $\Delta \Neff =0$ & Radiation density changed between BBN and Recombination
\\[.2cm]  
\hline 
\end{tabular}
\caption{Relation between particle physics experiments and cosmology.}
\label{table:darkscenarios}
\end{center}
\end{table}

\begin{itemize}
\item Evidence for new massless particles from either experiments or astrophysical observations have immediate implications for cosmology.  Any non-cosmological probe of light particles necessarily requires a coupling of the field to the Standard Model.  A measurement suggesting the strength of the coupling for this particle implies an upper-limit on the freeze-out temperature.  One can then look for the contribution to $\Delta \Neff$ associated with the particle.  If no such contribution is detected, we could place an upper limit on the reheating temperature (or a lower limit for a detection).  Axions present the simplest such examples, as there are a number of experiments that could directly detect axion dark matter (such as ADMX and CASPEr).  A detection in either experiment would predict $\Delta\Neff \geq 0.027$ unless reheating was at sufficiently low temperatures.  The inferred bound in either experiment can be read off of Figures~\ref{fig:axionphoton} or~\ref{fig:axiondipole}.

\item A more complicated example is if a deviation for typical stellar cooling is observed, as has been suggested for white dwarfs.  In this case, there are a number of models that could produce the necessary additional cooling, but would predict $\Delta\Neff \geq 0.027$.  The precise value of $\Delta \Neff$ depends on the spin of the particle and nature of the coupling (which cannot be unambiguously inferred from cooling).  Evidence for additional dark radiation would provide strong support that there is anomalous cooling and would imply a spin for the new light particle through the observed $\Delta \Neff$ using Figure~\ref{fig:Neff_thermal}.

\item Any supersymmetric interpretation of a discovery at the LHC  would require the existence of a gravitino (in order to make gravity consistent with supersymmetry).  There are many such scenarios where there would be no direction indication of a gravitino and/or its mass from colliders.  In this context, a cosmological detection of $\Delta \Neff \sim 0.06$ would have a natural interpretation as a light gravitino but would place a strong upper limit on the absolute scale of supersymmetry breaking of roughly $10^5$ GeV.  

\item There are a variety of possibilities where we may observe $\Delta \Neff \neq 0$ which results from the decay of a massive particle after neutrino decoupling.  A change to $\Neff$ after BBN would imply that $\Neff$ as measured in the CMB could differ significantly from the value inferred from primordial abundances~\cite{Fischler:2010xz}.  From the CMB, we can measure $Y_p$ and $\Neff$ simultaneously which implies such a signature can even be internal to the CMB.  Similarly decays to photons after BBN can also produce $\mu$- or $y$-distortions to the CMB spectrum which could be correlated with $\Delta \Neff < 0$.
\end{itemize}

\subsection{Dark Sectors and Neutrino Mass}

\begin{figure}[t!]
\begin{center}
\includegraphics[width=0.65\textwidth]{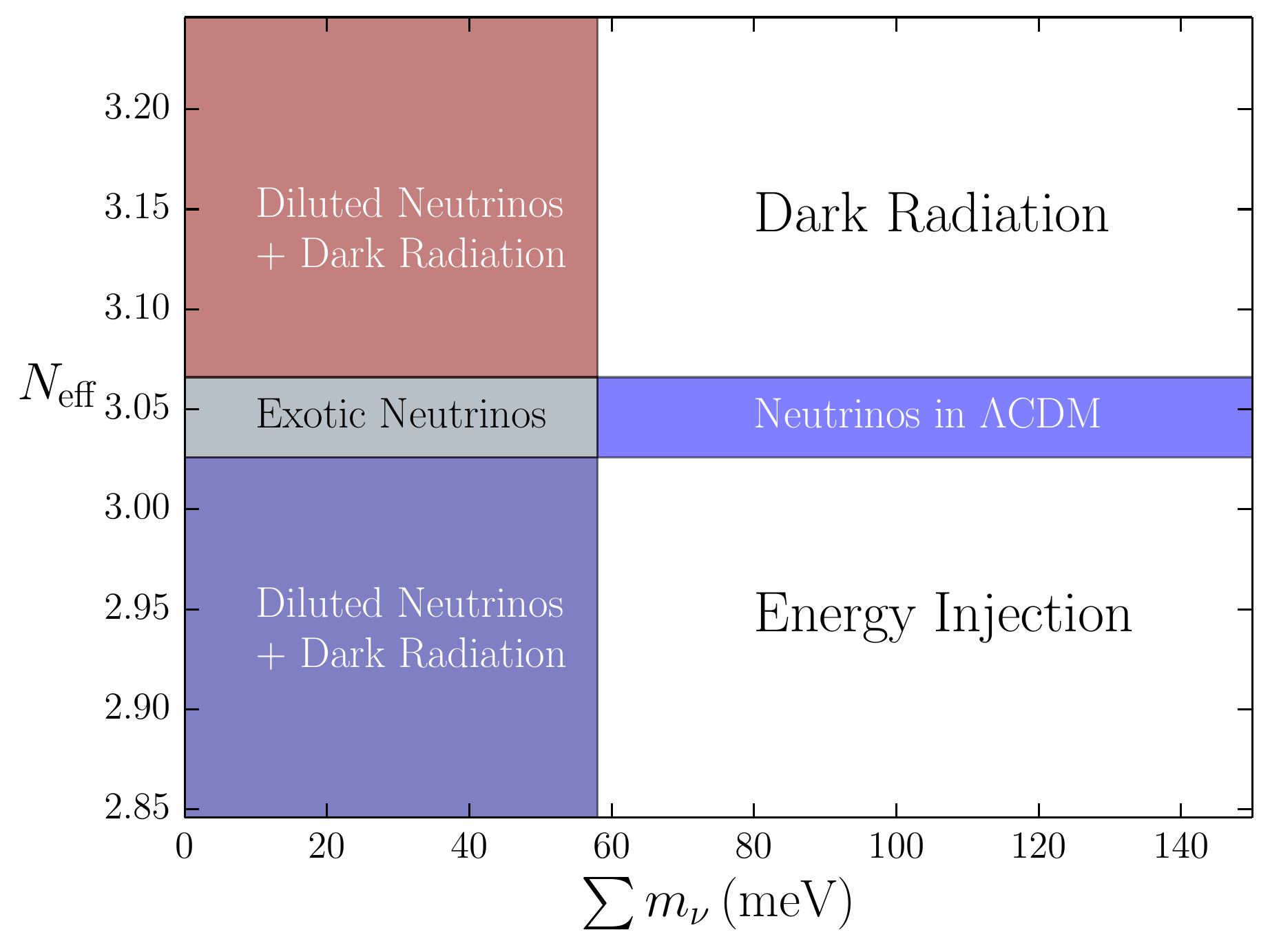}
\caption{ Physical mechanisms behind different regions in the space $\Neff$ - $\sum m_\nu$.   }
\label{fig:MnuNeff}
\end{center}
\end{figure}

CMB-S4 will provide compelling sensitivity in the $\Neff$-$\sum m_\nu$ plane.  From cosmology alone a measurement of $\sum m_\nu \gtrsim 58$ meV is consistent with conventional neutrino physics and therefore does not point to more exotic beyond the Standard Model physics without $\Neff$.  However, an upper limit or detection of $\sum m_\nu < 58$ meV would provide evidence of unconventional cosmology on its own and combined with $\Neff$ may give future insight into possible modifications to the cosmological history and/or neutrino physics necessary to accommodate such observations.   An illustration of possible scenarios is shown in Figure~\ref{fig:MnuNeff} and can be compared directly to the forecasted region in Figure~\ref{fig:Neff_Mnu}:

\begin{itemize}

\item Conventional neutrino physics predicts $\sum m_\nu \gtrsim 58$ meV.  A measurement of $\Delta \Neff < 0.02$ (i.e. consistent with zero) is the expectation from $\Lambda$CDM.  As described in the previous subsection, $\Delta \Neff> 0$ suggests dark radiation or additional energy in neutrinos and $\Delta \Neff < 0$ would imply energy injection into photons after neutrino decoupling.  The amount of energy injection that is consistent with the current constraint from Planck, $\Neff = 3.15 \pm 0.23$, is insufficient to significantly alter $\sum m_\nu$, as it would correspond to at most a five percent change in $\Omega_\nu h^2$.

\begin{figure}[t!]
\begin{center}
\includegraphics[width=0.65\textwidth]{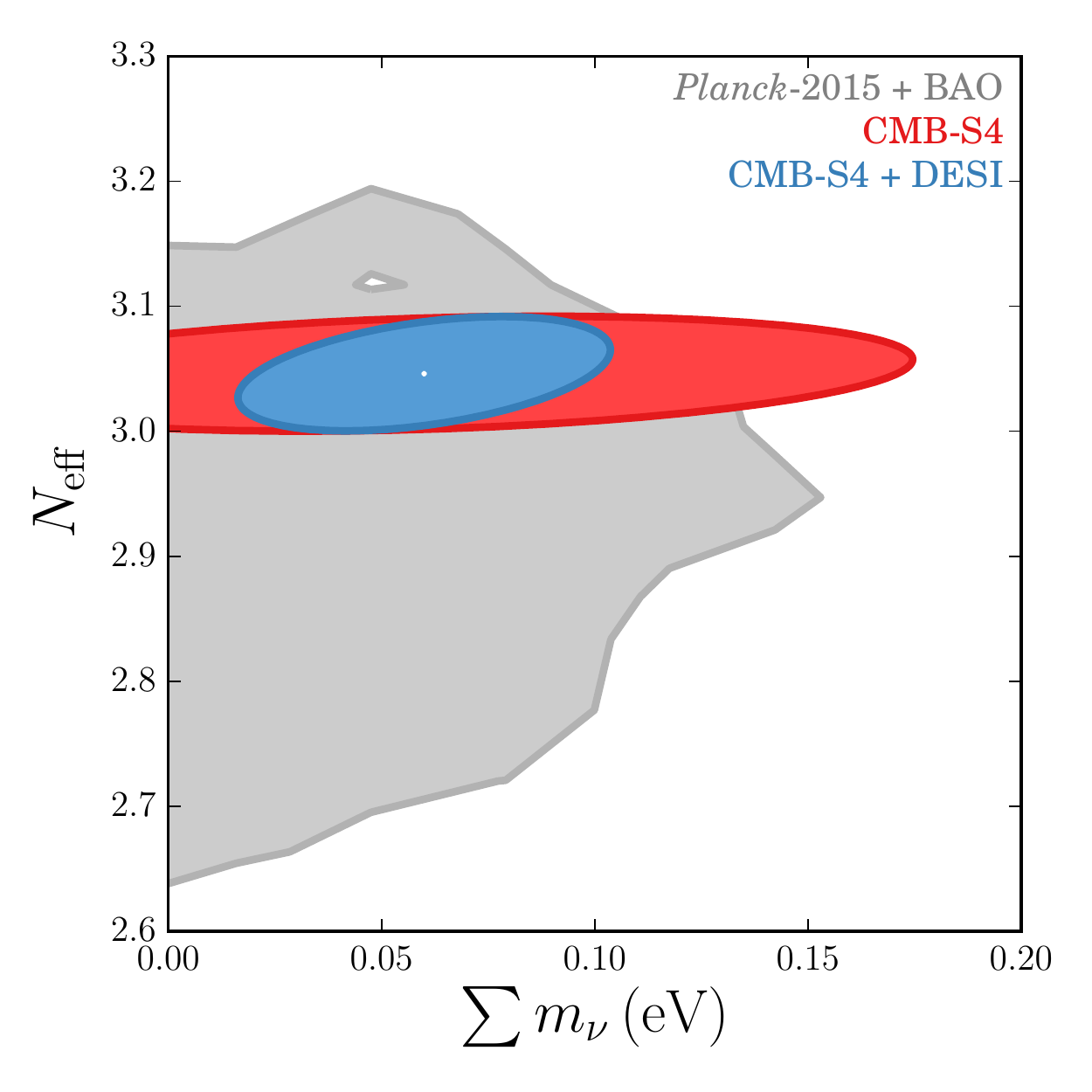}
\caption{Forecasts in the 2d parameter space $\sigma(\Neff)$ and $\sigma(\sum m_\nu)$.  These constraints assume $f_{\rm sky} = 0.4$ and  1 $\mu$K-arcmin noise.  A prior of $\tau = 0.06 \pm 0.01$ was also assumed. }
\label{fig:Neff_Mnu}
\end{center}
\end{figure} 

\item If $\sum m_\nu < 58$ meV, the most natural explanation is that the number density of neutrinos was lowered due to a change to the standard cosmological evolution.  However, in order to significantly lower the number density to explain such an observation, the neutrino number would have to be changed dramatically which, on its own, would be in contradiction with current $\Neff$ constraints.  Therefore, to satisfy existing constraints on $\Neff$ from \planck, some other form of radiation (i.e. other than Standard neutrinos) would also be required to make $|\Delta \Neff| < 0.2$.  Seeing a deviation of the form $\Delta \Neff \gtrless 0$ would give additional evidence for a modification of the thermal history.

\item If $\sum m_\nu < 58$ meV and $\Delta\Neff = 0$, then it suggests that either the mass for the neutrinos is generated after the CMB (late mass) or that the heavy neutrinos decayed to a lighter specifics in some novel way.  This situation would be unusual in that the limits on $\sum m_\nu$ would suggest deviations from the Standard thermal history without any other hints.  Presumably this scenario would be scrutinized heavily to check that the amplitude of the power spectrum is normalized correctly.  Finally, one might also allow for a delicate cancelation between the dilution of the neutrinos and the additional dark energy to be consistent with $\Delta \Neff =0$.

\end{itemize}

\chapter{Dark Matter}

\bigskip

\begin{quotation}

\end{quotation}

\section{Dark Matter Annihilation}

One of the leading candidates for dark matter is a weakly interactive massive particle (WIMP). If dark matter consists of WIMPs, we would expect these particles to self-annihilate. The annihilation of dark matter produces a shower of very energetic particles, that injects energy into the Universe, ionizing the matter in it.

This extra source of ionization has distinctive effects on the CMB: it suppresses the CMB temperature and polarization fluctuations on small angular scales, and it enhances the CMB polarization fluctuations on large angular scales due to the extra scattering of photons off free electrons \cite{Chen:2003gz,Padmanabhan:2005es}.
CMB temperature and polarization spectra can constrain the parameter
$p_{\rm ann}=f_{\rm eff}\langle\sigma v\rangle/m_{\rm DM}$, where $f_{\rm eff}$ is the fraction of energy
deposited into the plasma, $\langle\sigma v\rangle$ is the velocity-weighted
cross section, and $m_{\rm DM}$ is the mass of the dark matter particle.
Current constraints from \planck\ temperature and polarization data exclude
dark matter masses below $16$ GeV at the 2$\sigma$ level \cite{Ade:2015xua}, assuming that
$20\%$ of the energy is deposited in the plasma \cite{Madhavacheril:2013cna}. CMB-S4 is expected to tighten these constraints by a factor of $2$ to $3$ for $f_{\rm sky}=[0.4-0.6]$ \cite{Wu:2014hta}. 
Ref. \cite{Wu:2014hta} found that the main factor that improves the limit in $m_{\rm DM}$ is the sky coverage $f_{\rm sky}$. This is because the constraints are mostly sample variance limited. Fig. \ref{fig:DM_annihilation} shows the dependence on $f_{\rm sky}$, and the small dependence on detector number and beam size.

\begin{figure}[t]
\begin{center}
\includegraphics[width=0.7\textwidth]{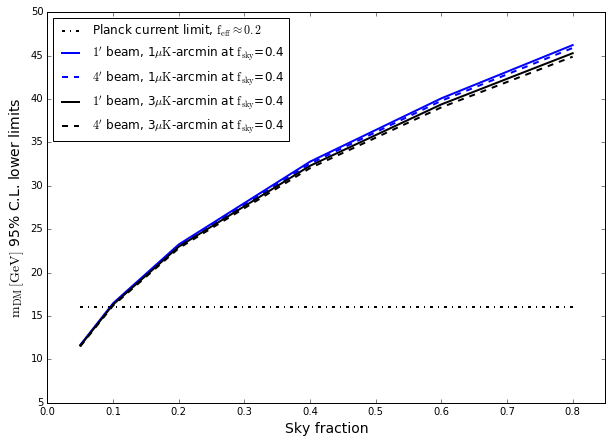}
\caption{95 \% CL lower limit on $m_{\rm DM}$ in GeV as a function of sky coverage, $f_{\rm sky}$.
The blue/black lines correspond to a noise level of $1$/$3$ $\mu$K-arcmin in temperature. The solid/dashed lines correspond to 1/4 arcmin beams.
The dashed/dotted lines show the current limit from \planck\ temperature and polarization data for a thermal cross section. All the cases plotted here correspond to a case where $20\%$ of the energy is absorbed by the plasma.}\label{fig:DM_annihilation}
\end{center}
\end{figure}

Dark-matter annihilation also leads to growing ionization fraction perturbations and amplified small-scale cosmological perturbations, leaving an imprint on the CMB bispectrum \cite{Dvorkin:2013cga}.

\section{Other types of Dark Matter Interactions}
Near the epoch of CMB last scattering, dark matter accounts for about $65\%$ of the energy budget of the Universe, making the CMB a particularly good probe of the dark matter sector. Of particular relevance to CMB-S4 studies, the presence of new dark matter interactions with light degrees of freedom \cite{Goldberg:1986nk,Carlson:1992fn} can leave subtle imprints on the temperature and polarization CMB power spectra. The introduction of such non-minimal dark matter models has been primarily (but not exclusively) motivated in the literature by potential shortcomings of the standard cold dark matter scenario at small sub-galactic scales \cite{deBlok:1997zlw,Klypin:1999uc,Moore:1999nt}. While these issues are far from settled, they motivate the search for other non-minimal dark matter signatures in complementary data sets (such as the CMB) that could indicate whether or not dark matter can be part of the solution. 

\subsection{Dark Matter-Baryon Scattering}

A possible dark matter scenario is that in which dark matter scatters off baryons in the early Universe. 
In this scenario, there is a drag force produced by the baryons on the dark matter fluid, which affects the CMB temperature and polarization power spectra and the matter power spectrum.
Ref. \cite{Dvorkin:2013cea} performed a model-independent analysis on the dark matter-baryon interactions using CMB temperature data from the \planck\ satellite, and Lyman-$\alpha$ forest data from the Sloan Digital Sky Survey as tracer of the matter fluctuations. This analysis suggests that the constraints could improve significantly with better temperature data on small scales, and additional polarization data on large and small scales. Therefore, an experiment such as CMB-S4 would have a large impact on these constraints.

\subsection{Dark Matter-Dark Radiation Interaction}
Dark matter interacting with light (or massless) dark radiation  has been put forward \cite{Buen-Abad:2015ova,Lesgourgues:2015wza} as a potential solution to the small discrepancy between the amplitude of matter fluctuations inferred from CMB measurements and those inferred from cluster number counts and weak lensing measurements. CMB-S4 measurements of the lensing power spectrum have the potential to significantly improve constraints on dark matter interacting with light degrees of freedom in the early Universe.

The key equations governing the evolution of cosmological fluctuations for this broad class of non-minimal dark matter models are presented in Ref.~\cite{Cyr-Racine:2015ihg}. Essentially, the new dark matter physics enters entirely through the introduction of dark matter and dark radiation opacities, which, similarly to the photon-baryon case, prohibit dark radiation free-streaming at early times and provides a pressure term that opposes the gravitational growth of dark matter density fluctuations. The impact of this new physics on CMB fluctuations has been studied in detail in Ref.~\cite{Cyr-Racine:2013fsa} and we briefly review it here. First, the presence of extra dark radiation mimics the presence of extra neutrino species and affects the expansion history of the Universe, possibly modifying the epoch of matter-radiation equality, the CMB Silk damping tail, and the early integrated Sachs-Wolfe effect. However, unlike standard free-streaming neutrinos, the dark radiation forms a tightly-coupled fluid at early times, leading to distinct signatures on CMB fluctuations which include a phase and amplitude shift of the acoustic peaks (see e.g. Ref.~\cite{Bashinsky:2003tk,Cyr-Racine:2013jua,Follin:2015hya}). Second, the dark radiation pressure prohibits the growth of interacting dark matter fluctuations on length scales entering the causal horizon before the epoch of dark matter kinematic decoupling. This weakens the depth of gravitational potential fluctuations on these scales, affecting the source term of CMB temperature fluctuations. Finally, the modified matter clustering in the Universe due to nonstandard dark matter properties will affect the lensing of the CMB as it travels from the last-scattering surface to us. For interacting dark matter models that are still allowed by the current \planck\ data, this latter effect is where CMB-S4 can significantly improve the constraints on these non-minimal theories.

Given the large array of possible dark matter theories to constrain, it is useful to use the effective theory of structure formation (ETHOS) \cite{Cyr-Racine:2015ihg} to systematically parametrize the deviations from standard cold dark matter. Within ETHOS, the impact of having all or a fraction of dark matter interacting with dark radiation can be captured with a handful of ``effective'' parameters which entirely determine the structure of the linear matter power spectrum. For CMB-S4, the most relevant parameters that can be constrain are the amplitude of the interaction cross section between dark matter and dark radiation (parametrized by $a_n$), the fraction of interacting dark matter (in multi-component scenarios), and the amount of dark radiation present in the Universe.\\

\begin{figure}[htbp!]
$\begin{array}{ll}
\includegraphics[width=0.5\columnwidth]{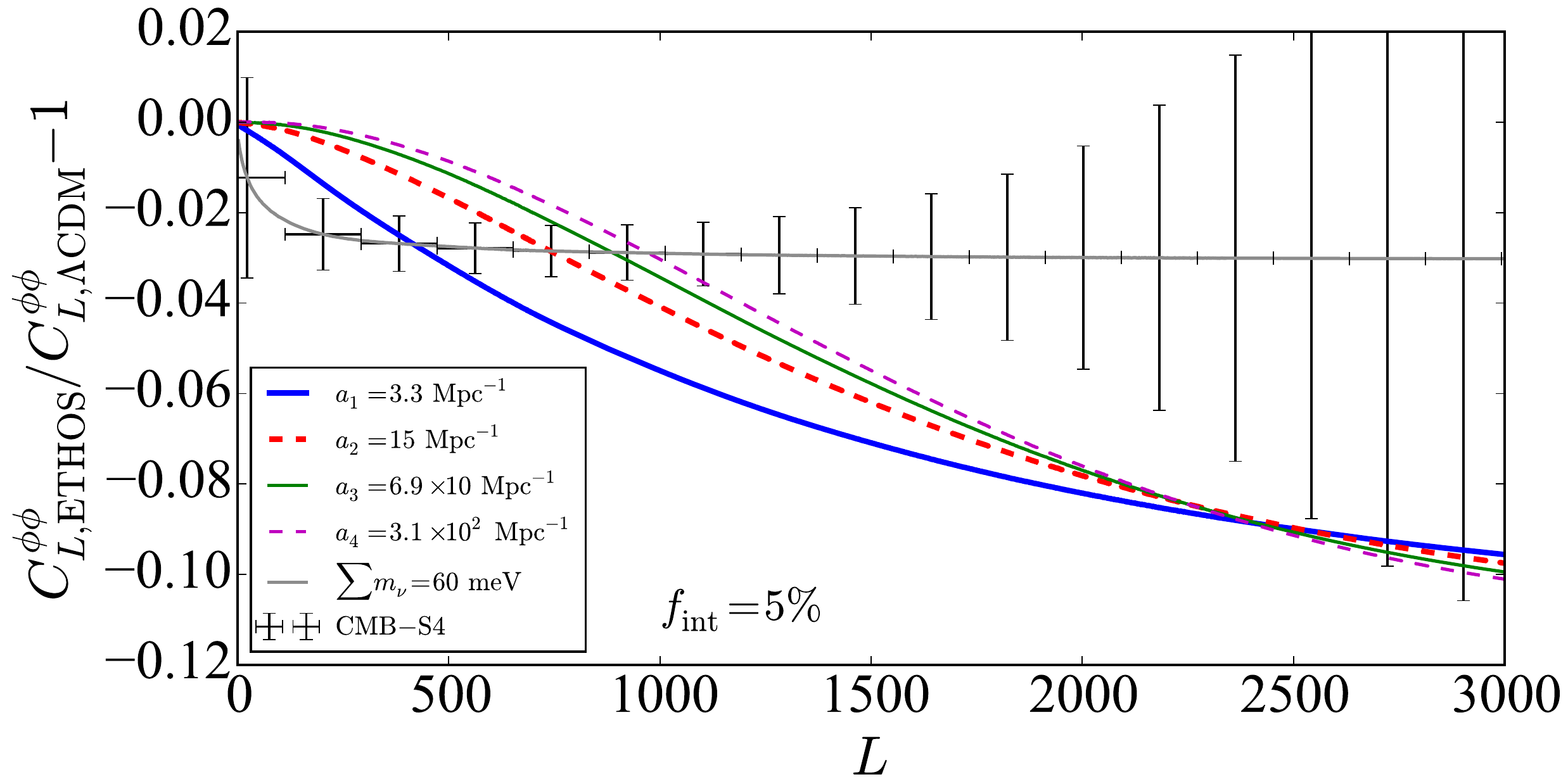}& 
\includegraphics[width=0.5\columnwidth]{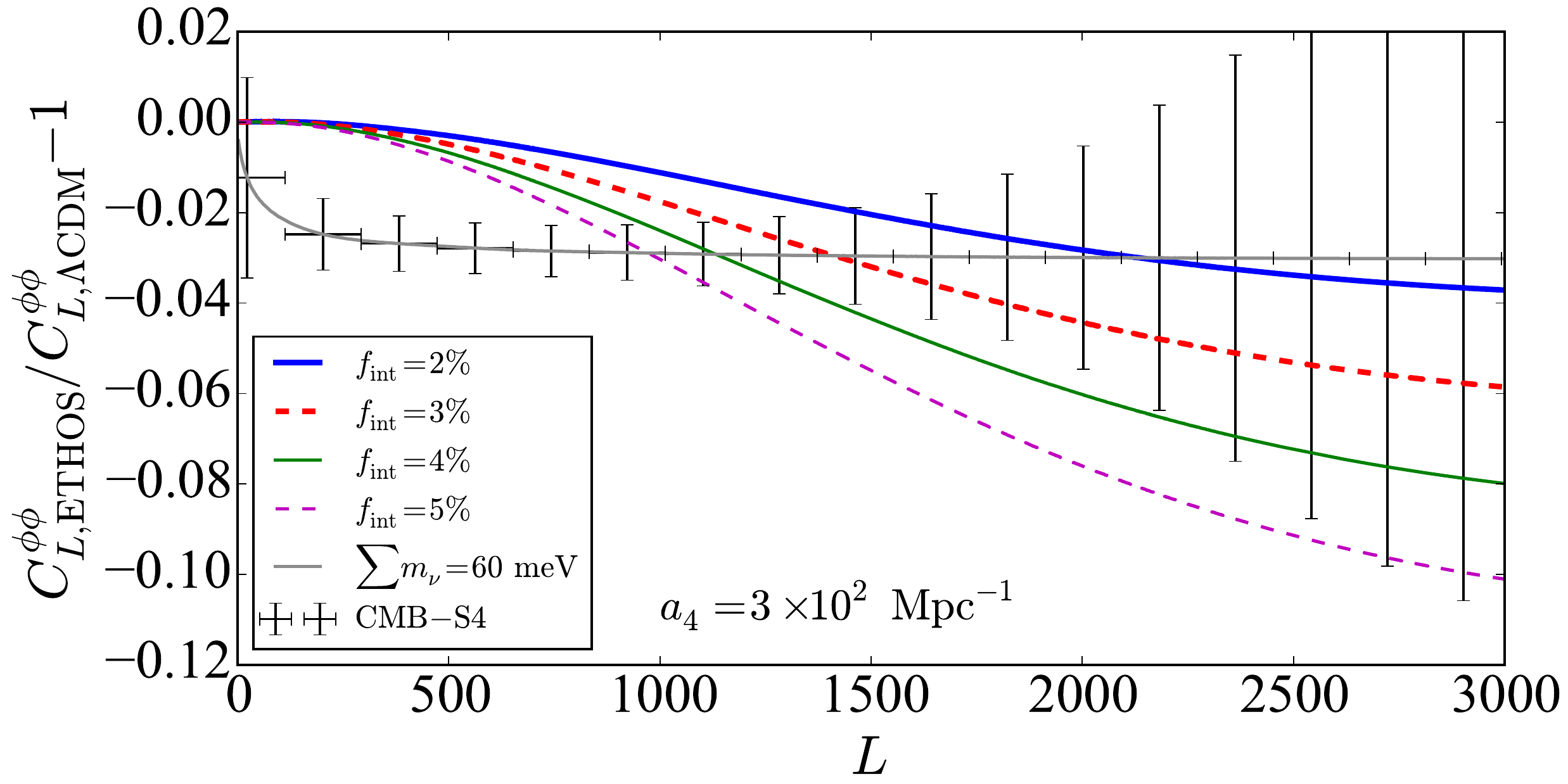}
\end{array}$
\begin{center}
\caption{\textbf{Left panel}: Fractional difference of the CMB lensing spectrum between a standard $\Lambda$CDM model (with massless neutrinos) and four different ETHOS models with opacity coefficients $a_n$ given in the legend. In all models shown, $5\%$ of the dark matter is allowed to interact with dark radiation. For comparison, we also display a standard massive neutrino model with $\sum m_\nu =0.06$ eV. \textbf{Right panel}: Similar to the top panel, but we now vary the fraction of dark matter that can interact with dark radiation, for a fixed opacity coefficient of $a_4 = 3\times 10^2$ Mpc$^{-1}$. }\label{fig:Cls_phi_PIDM}
\end{center}
\end{figure}

We illustrate in Fig.~\ref{fig:Cls_phi_PIDM} the impact of different interacting dark matter models on the CMB lensing power spectrum. In the top panel, we show four partially-interacting dark matter models parametrized by their opacity coefficient $a_n$ and for which only $5\%$ of the total amount of dark matter is interacting. We display the fractional difference between the interacting dark matter models and a standard $\Lambda$CDM model with vanishing neutrino mass. For comparison, we also illustrate the difference for a standard massive neutrino $\Lambda$CDM model with $\sum m_\nu = 0.06$ eV.  Interestingly, the damping of the lensing power spectrum has a different shape than that caused by massive neutrinos. Given the expected performance of CMB-S4 in measuring the lensing power spectrum, all the models illustrated there (which are currently allowed by \planck\ data) could be ruled out, significantly improving our knowledge about interacting dark matter. The right panel of Fig.~\ref{fig:Cls_phi_PIDM} is similar, but illustrates how the fractional difference in the CMB lensing power spectrum is affected as the fraction of interacting dark matter is varied from $2$ to $5$ per cent. Again, this illustrates that CMB-S4 can provide very tight constraints on the fraction of interacting dark matter.

Since non-standard dark matter models primarily affect the large CMB lensing multipoles, the constraining power of CMB-S4 on interacting dark matter is largely independent of the specific choice of $L_{\rm min}$. We foresee that the main difficulty in constraining non-standard dark matter theories with CMB-S4 will be the proper modeling of non-linearities in the matter power spectrum, which are quite important for $L > 500$. We note that recent progress has been made in this direction \cite{Vogelsberger:2015gpr}.

\section{Axion Dark Matter}
The QCD axion and other axion-like particles (ALPs), if stable on cosmological timescales, can contribute to the DM density. Along with thermal WIMPs, they are a well-motivated DM candidate (see Ref.~\cite{Marsh:2015xka} for a recent review). Ultralight axions (ULAs) are non-thermally created via vacuum realignment and have a distinctive phenomenology as a dark matter candidate. 
Vacuum realignment axions are cold, and do not contribute to $N_{\rm eff}$. However, if these axions have couplings to ordinary matter, as described in Section~\ref{sec:neffaxions}, then a second, relativistic population of axions is additionally created.

We consider axions within a range of masses $10^{-33}~\mathrm{eV}\leq m_{a}\leq 10^{-20}~\mathrm{eV}$ but do not assume any particular coupling to the Standard model particles.  We can compare these assumptions to those used in Section~\ref{sec:neffaxions}; the contribution of thermal axions to $\Neff$ applies to any mass $m_{a}\lesssim 1~\mathrm{eV}$, including the well known QCD axion, but depends in detail on the couplings to the Standard Model particles and on the reheat temperature.  In this sense, cosmological constraints on axion dark matter are orthogonal (complimentary) in the space of masses and couplings to the constraints on a thermal population of axions.

The ULAs we consider here are motivated by string theory and are associated with the geometry of the compact spatial dimensions. These axions can contribute either to the dark matter or dark energy budget of the Universe depending on their particular mass, which sets the time at which the axions begin to coherently oscillate (since the Hubble term provides the friction in the axion equations of motion).

The current best constraints on ULAs from the primary CMB TT power, and the WiggleZ galaxy redshift survey were made in \cite{Hlozek:2014lca}. Our fiducial value for the axion energy density is chosen to be consistent with these constraints. We discuss the potential of CMB-S4 to constrain the total allowed energy density of axions in addition to the usual dark matter and dark energy components. In addition, in Section~\ref{axion_iso}, we explain that CMB-S4 could place strong bounds on the energy scale of inflation, $H_I$ through the uncorrelated axion isocurvature generated in models with $H_I/2\pi<f_a.$

\subsection{Constraints on cold axion energy density \label{sec:uladiabat}}

The degeneracies of the axions with other cosmological parameters, such as $N_\mathrm{eff}$ or $m_\nu$, vary depending on the axion mass (see Fig.~\ref{fig:axions}, right panel). Dark energy-like axions with masses around $10^{-33}~{\rm eV}$ change the late-time expansion rate and therefore the sound horizon, changing the location of the acoustic peaks. This has degeneracies with the matter and curvature content. 
Heavier axions ($m_a \gtrsim 10^{-26}~{\rm eV}$) affect the expansion rate in the radiation era and reduce the angular scale of the diffusion distance, leading to a boost in the higher acoustic peaks, which has a degeneracy with $N_{\rm eff}$. 

In both of these cases, improved errors on the temperature and polarization power spectrum, coupled with constraints on the Hubble constant (for the lightest axions) from baryon acoustic oscillations, lead to improvements in the error on allowed axion energy density of a factor of three from these spectra alone. 

In the matter power spectrum, and thus CMB lensing power, light axions suppress clustering power, suggesting a degeneracy with effects of massive neutrinos that must be broken to make an unambiguous measurement of neutrino mass using the CMB. The above-mentioned effects in the expansion rate break this degeneracy for some axion masses. There remains a significant degeneracy between axions and massive neutrinos $m_a=3\times 10^{-29}\text{ eV}$ and $\Sigma m_\nu=60\text{ meV}$. Effort should be made to break this degeneracy and distinguish the effects of non-thermal axions from massive neutrinos for an unambiguous detection of neutrino mass using the CMB. 

\begin{figure}[t] 
\begin{center} 
\includegraphics[width=0.49\textwidth]{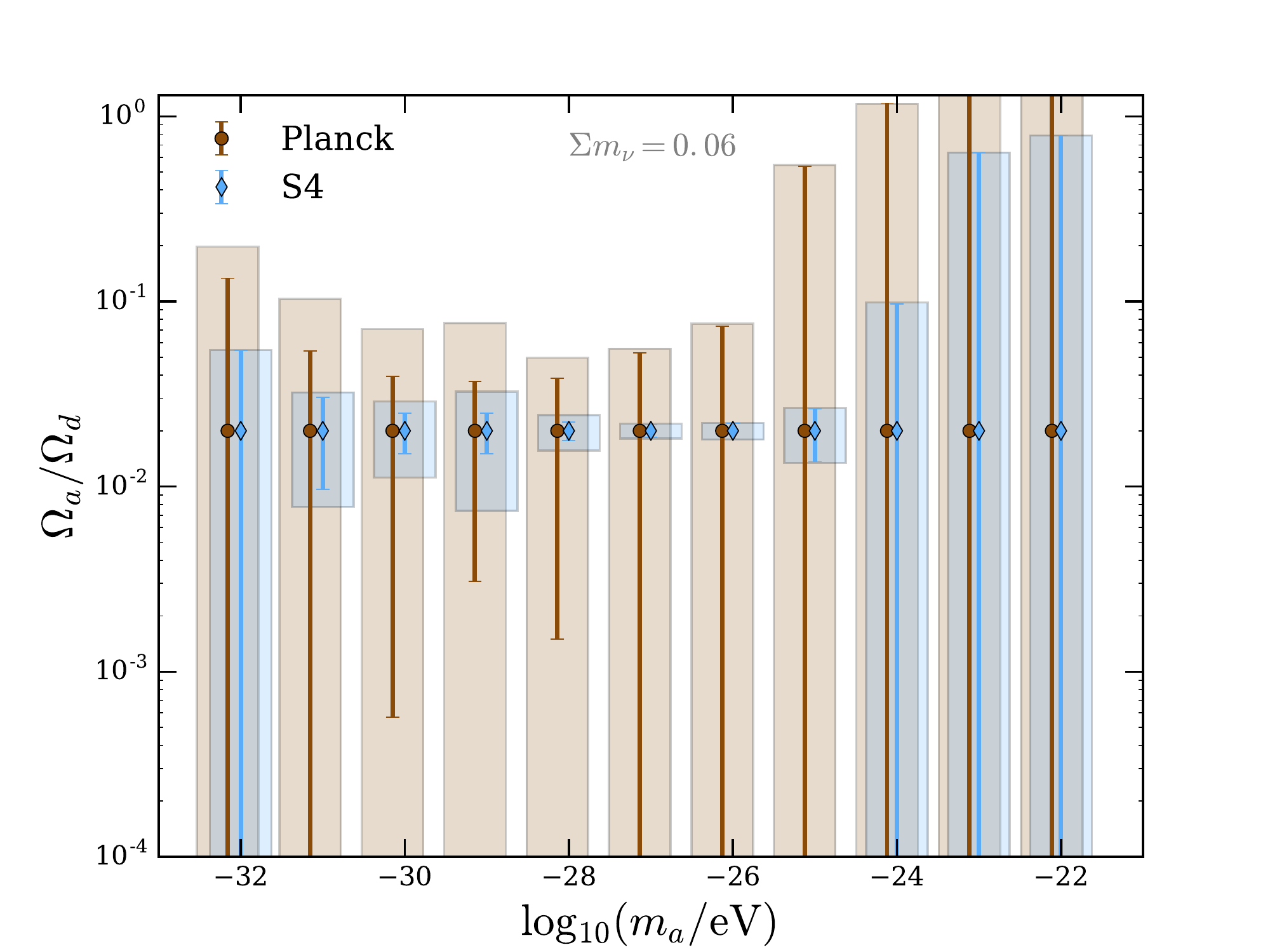}
\includegraphics[width=0.49\textwidth]{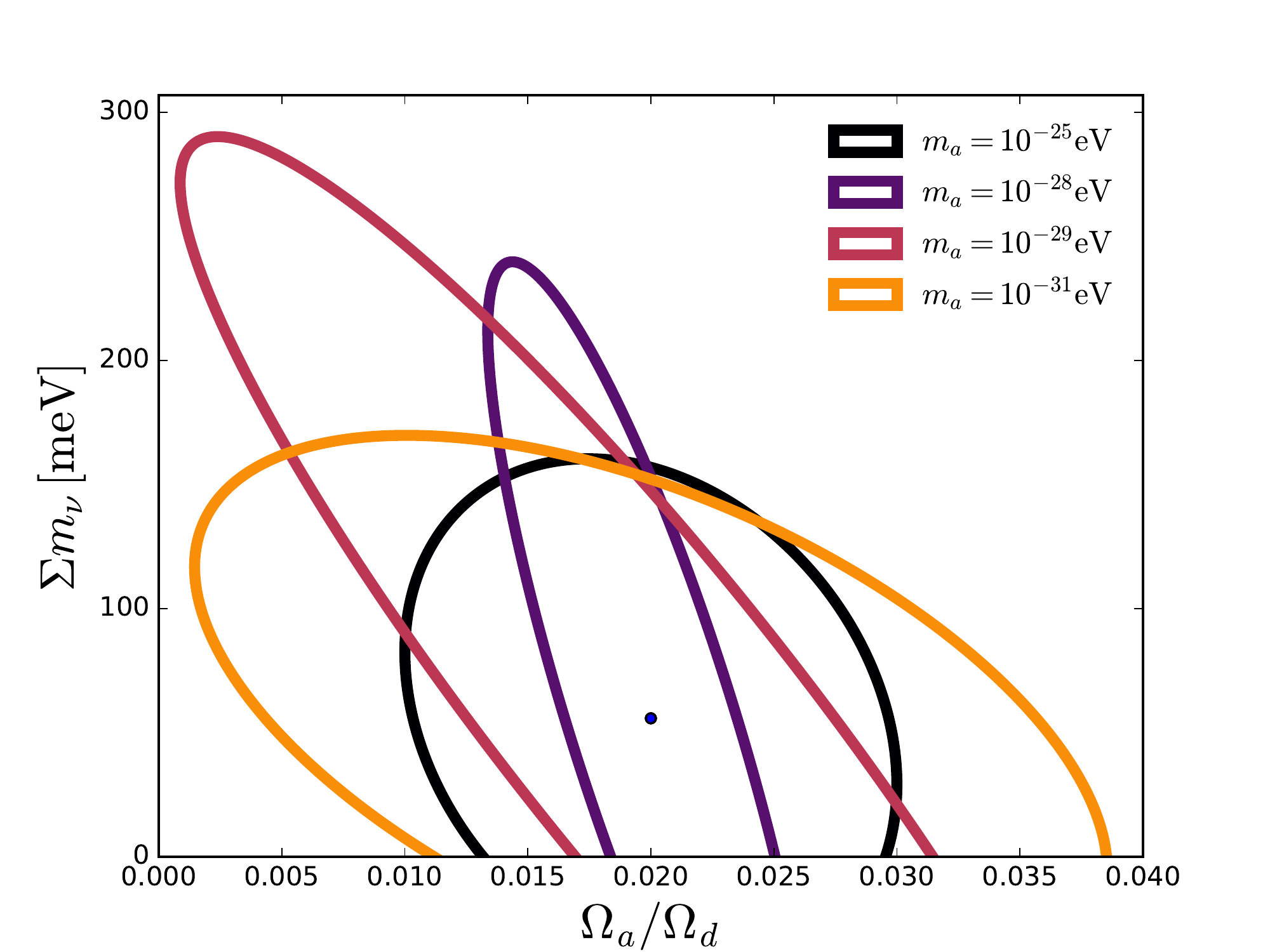}
\caption{\textbf{Left:} Constraints on the axion energy density as a function of axion mass at fixed neutrino mass $\Sigma m_\nu = 0.06~\rm{eV}$ (errorbars) and marginalising over the neutrino mass (shaded error bars), for \planck\ Blue Book constraints (brown) and a CMB-S4-like survey (blue). Over the `fuzzy' dark matter region ($-28 < \log(m_a/\mathrm{eV}) < -25$, CMB-S4 allow for percent-level constraints on an axion component, improving significantly on current constraints. For the same range of masses, the degeneracy is weakest with massive neutrinos - the bands shown when marginalising over the neutrino error are not much larger than the case where the neutrino mass is fixed. \textbf{Right:} Degeneracy of axions with massive neutrinos. There is a significant degeneracy for $m_a=3\times 10^{-29}\text{ eV}$ and $\Sigma m_\nu=60\text{ meV}$. Figure derived from constraints presented in \cite{Hlozek:2016lzm}.
\label{fig:axions}}
\end{center}       
\end{figure}       

We show the forecasted constraints on the axion energy density from CMB-S4 including lensing in the left panel of Figure~\ref{fig:axions} (for fixed neutrino mass of $\Sigma m_\nu = 0.06~\rm{eV}$). Adding information from the lensing reconstruction using CMB-S4 will improve constraints on axion DM significantly. A percent-level measurement of the lensing deflection power at multipoles $\ell > 1000$ leads to an improvement in the error on the axion energy density of a factor of eight relative to the current \planck\ constraints, for an axion mass of $m_a=10^{-26}~{\rm eV}$.    This represents an ability to test the component nature of dark matter, and thus the CDM paradigm, at the percent level. Furthermore, since $\Omega_a\propto f_a^2$ this improves the expected constraint on the axion decay constant from $10^{17}\text{ GeV}$ with \planck\ to $10^{16}\text{ GeV}$ with CMB-S4, testing the predictions of the ``string axiverse’’ scenario~\cite{Arvanitaki:2009fg}. 

\planck\ is degenerate with CDM at $m_a=10^{-24}\,\mathrm{eV}$, and only has weak constraints at $m_a=10^{-25}\,\mathrm{eV}$. CMB-S4 could make a $>5\sigma$ detection of departures from CDM for masses as large as $m_a=10^{-25}\,\mathrm{eV}$, and improves the lower bound on DM particle mass to $m_a=10^{-23}\,\mathrm{eV}$ and fractions O(10\%). Realising this level of constraining power will, however, require improved understanding of the non-linear clustering of axions \cite{Marsh:2016vgj}.

\subsection{Axion Isocurvature \label{axion_iso}}
The axion decay constant, $f_a$, specifies the scale at which the underlying $U(1)$ symmetry is broken. If $H_I/2\pi<f_a$, then this symmetry is broken during inflation, and the axion acquires \emph{uncorrelated isocurvature perturbations} (e.g. Refs.~\cite{Axenides:1983hj,Fox:2004kb,Hertzberg:2008wr}).\footnote{We ignore the case where $H_I/2\pi>f_a$, since no isocurvature initial conditions are excited. The limit $r_{0.05}<0.12$ implies that isocurvature is produced if $f_a>1.8\times 10^{13}\text{ GeV}$. This accounts for the QCD axion in the ``anthropic'' window (roughly half of the allowed range of $f_a$ on a logarithmic scale), axions with GUT scale decay constants (such as string axions~\cite{Svrcek:2006yi,Arvanitaki:2009fg}) and axions with lower $f_a$ in models of low-scale inflation.} The uncorrelated CDM isocurvature amplitude is bounded by \planck\ to be $A_I/A_s<0.038$ at 95\% C.L.~\cite{Ade:2015lrj}. 

It is important to note, that while axion isocurvature gives a test of the energy scale of inflation (and as such has the same goals as those discussed in Section~\ref{sec:scale-of-inflation}) - this test is \textit{independent} of any other constraints on the tensor-to-scalar ratio, and uses the constraints on the axion energy density and its own signature isocurvature to probe the inflationary epoch.

The axion isocurvature amplitude is:
\begin{equation}
A_I = \left(\frac{\Omega_a}{\Omega_d}\right)^2\frac{(H_I/M_{\rm pl})^2}{\pi^2(\phi_i/M_{\rm pl})^2} \, .
\label{eqn:iso_amplitude}
\end{equation}
The initial axion displacement, $\phi_i$, fixes the axion relic abundance such that $\Omega_a=\Omega_a (\phi_i,m_a)$~\cite{Preskill:1982cy,Abbott:1982af,Dine:1982ah,Turner:1983he,Steinhardt:1983ia,Marsh:2010wq}. Thus, if the relic density and mass can be measured by independent means, \emph{a measurement of the axion isocurvature amplitude can be used to measure the energy scale of inflation, $H_I$}.

We forecast the errors on axion isocurvature for the base line CMB-S4 experiment with a 1 $\mu$K-arcmin noise level and a 1 arcminute beam: the isocurvature limit will be improved by a factor of approximately five compared to \planck, allowing for detection of axion-type isocurvature at 2$\sigma$ significance in the region $0.008<A_I/A_s<0.038$.

If the QCD axion is all of the DM, axion direct detection experiments can be used in conjunction with CMB-S4 to probe $H_I$ in the range
\begin{equation}
 2.5\times 10^6\lesssim H_I/\text{GeV}\lesssim 4\times 10^9\, 
\text{(QCD axion + direct detection)}\, \,
\end{equation}
This is demonstrated in Fig.~\ref{fig:qcd_isocurvature} (left panel) for the case of ADMX~\cite{Asztalos:2009yp} (in operation), and CASPEr~\cite{Budker:2013hfa} (proposed), where we have used the standard formulae relating the QCD axion mass and relic abundance to the decay constant (e.g. Ref.~\cite{Fox:2004kb}). \emph{Combining axion DM direct detection with CMB-S4 isocurvature measurements allows a unique probe of low-scale inflation, inaccessible to searches for tensor modes.}\footnote{In simple models of inflation, the high-$f_a$ QCD axion is incompatible with detection of tensor modes~\cite{Fox:2004kb,Hertzberg:2008wr,Visinelli:2014twa,Marsh:2014qoa,Visinelli:2014twa}, although non-standard cosmic thermal histories of PQ breaking mechanisms can lift constraints , e.g. \cite{Higaki:2014ooa,Fairbairn:2014zta,Nomura:2015xil}.}

\begin{figure*}[htbp!]
\begin{center}
\includegraphics[trim= 2cm 0 2cm 0,clip, width=0.45\textwidth]{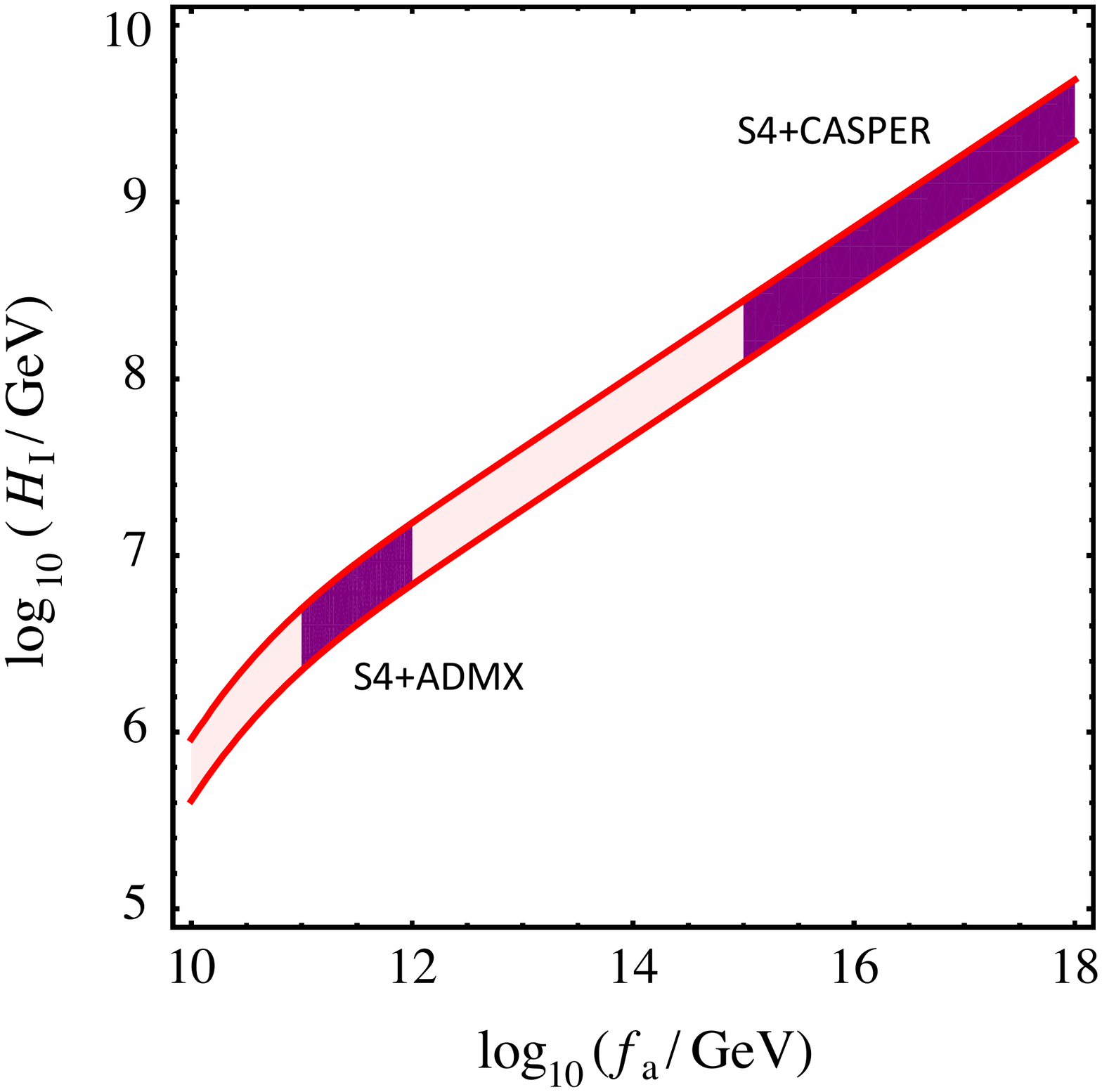} 
\includegraphics[trim= 2cm 0.45cm 2cm 0,clip, width=0.4585\textwidth]{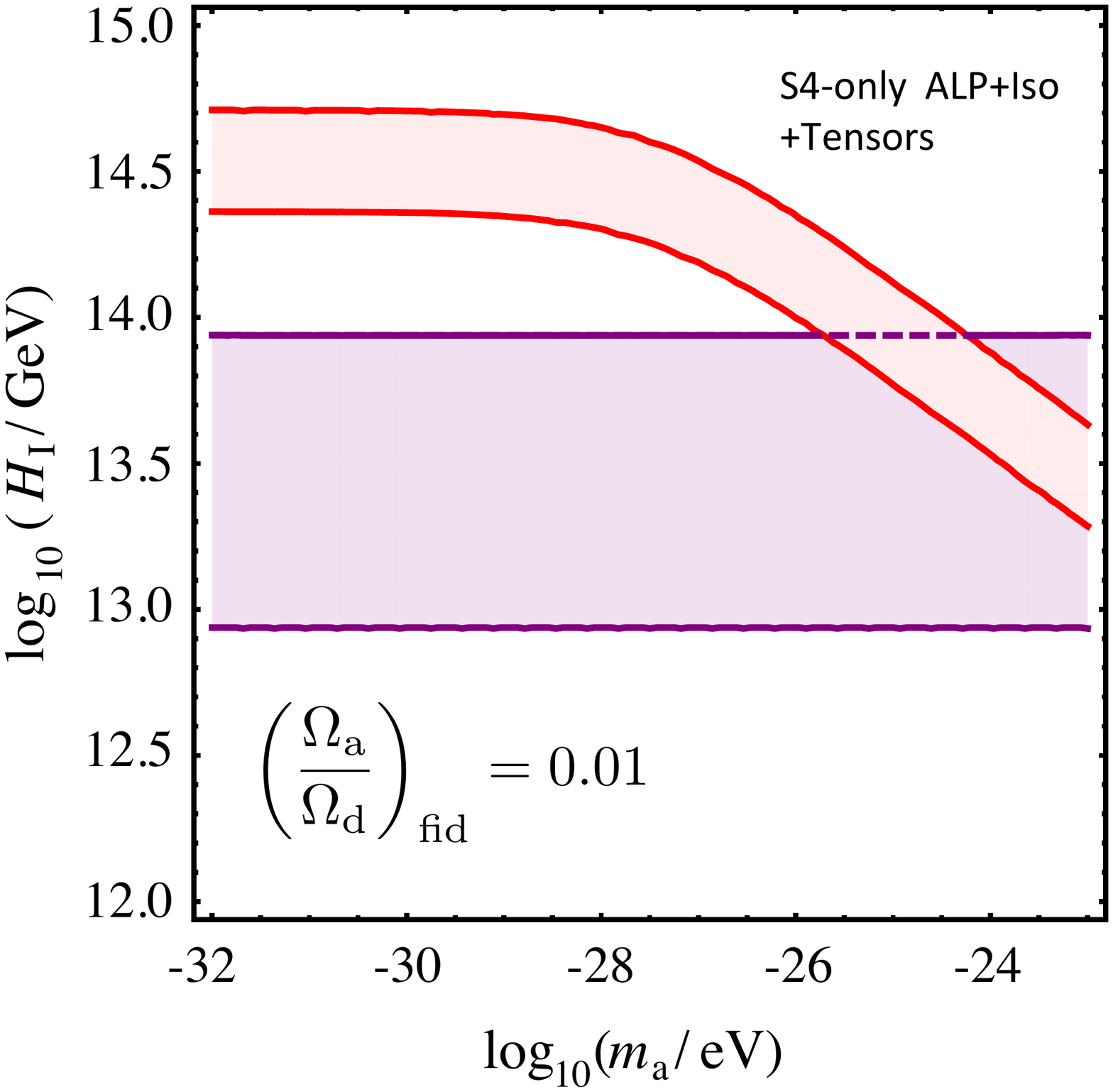}
 \end{center}
 \caption{Axion dark matter isocurvature. Red bands show the isocurvature amplitude consistent with \planck\ and detectable with CMB-S4. \textbf{Left Panel:} The QCD axion: measuring the energy scale of inflation with CMB-S4+axion direct detection. Here we restrict axions to be all of the DM. The purple regions show the range of $f_a$ accessible to axion direct detection experiments. Combining ADMX~\cite{Asztalos:2009yp} (in operation), CASPEr~\cite{Budker:2013hfa} (proposed), and CMB-S4 it is possible to measure $4\times 10^5\lesssim H_I/\text{GeV}\lesssim 4\times 10^9$. \textbf{Right Panel:} ALPs - a combination measurement using CMB-S4 alone. Assuming 1\% of the total DM resides in an ultralight axion, the mass and axion density can be determined to high significance using, for example, the lensing power. The isocurvature amplitude can also be determined, allowing for an independent determination of $H_I$ in the same regime as is accessible from tensor modes (purple band).}
\label{fig:qcd_isocurvature}
\end{figure*} 

\begin{figure}[htbp!]
\begin{center}
\includegraphics[width=0.6\textwidth]{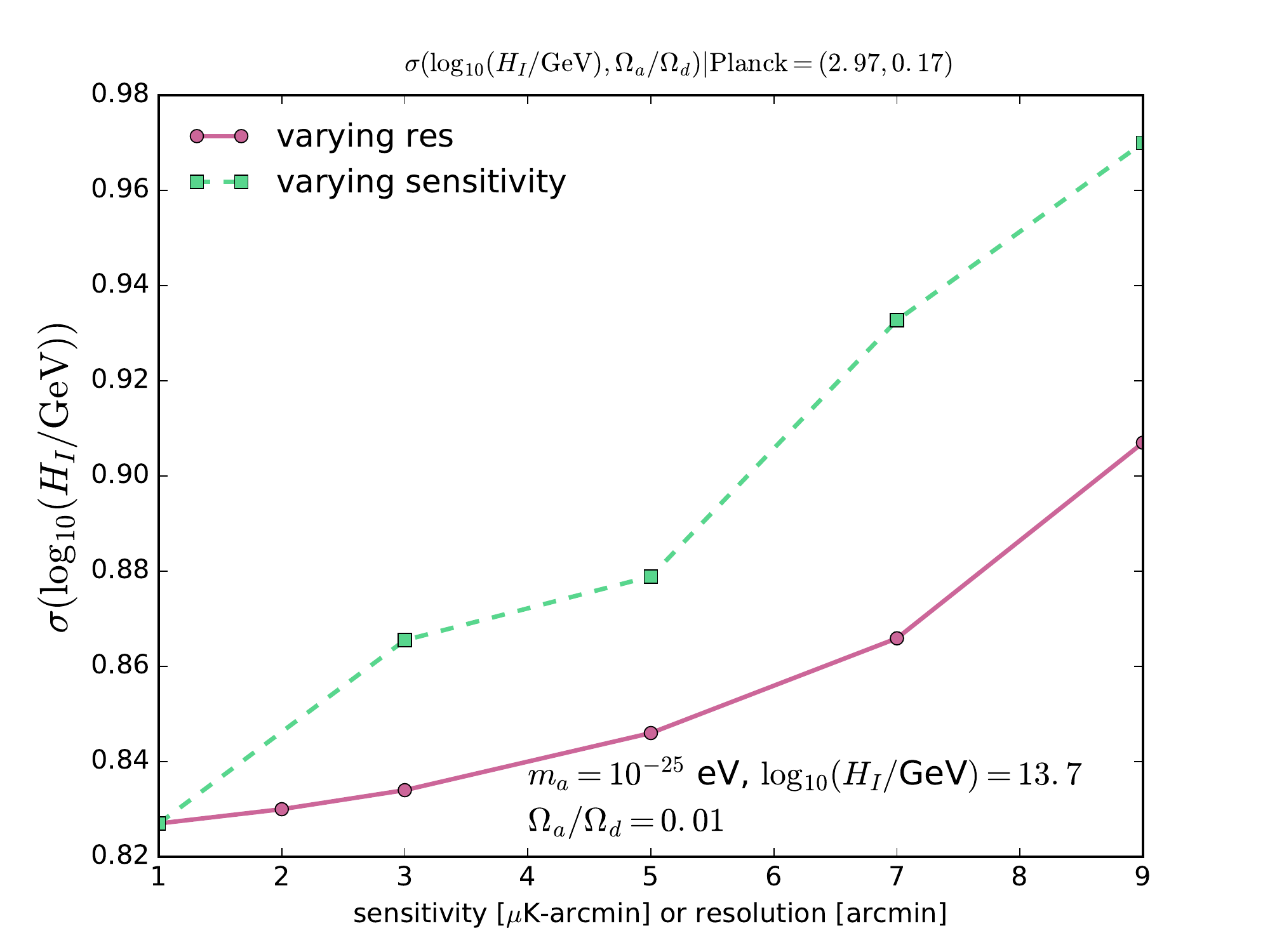}
\caption{\textbf{Optimizing constraints on the energy scale of inflation with CMB-S4.} We vary $H_I$ directly around a fiducial model of $H_I/\text{GeV} = 10^{13.7}$ for axions of mass $10^{-25} \mathrm{eV}$ making up 1\% of the total dark content, for a range of possible CMB-S4 survey parameters. While the error degrades as the resolution and sensitivity are worsened, this degradation is small compared to the factor of three improvemement in the error moving from \planck\ to CMB-S4. \label{fig:optimize_axioniso}}
\end{center}
\end{figure}

We now consider isocurvature in ULAs, see e.g. Refs.~\cite{Marsh:2013taa,Marsh:2014qoa}. ULA DM has a number of distinctive features in large scale structure and the CMB~\cite{Hlozek:2014lca,Marsh:2013ywa}. For ULAs with $10^{-32}\lesssim m_a/\text{eV}\lesssim 10^{-23}$ a DM fraction of $\Omega_a/\Omega_d$ in the range of 1\% is consistent with \planck\ \cite{Hlozek:2014lca} and high-$z$ galaxy formation~\cite{Bozek:2014uqa,Schive:2015kza}, yet can be distinguished from pure CDM using CMB-S4 lensing power at $>2\sigma$ (depending on the ULA mass, Sec.~\ref{sec:uladiabat}). Fig.~\ref{fig:qcd_isocurvature} (middle panel) shows isocurvature constraints possible with CMB-S4, compared to tensor constraints. We fix the fiducial ULA fraction to 1\%, such that $\Omega_a$ and $m_a$ can be separately measured using the CMB-S4 lensing power, and thus using Eq.~(\ref{eqn:iso_amplitude}) a measurement of $A_I$ is a measurement of $H_I$. 

In contrast to the QCD axion, there are masses, $m_a\lesssim 10^{-26}\text{ eV}$, for which tensor modes impose a stronger constraint on $H_I$ than isocurvature (such that isocurvature in these ALPs would be undetectably small). However, there are also regions of overlap between possible tensor and isocurvature measurements. Using CMB-S4 in these regions, it is possible to make a combination measurement of isocurvature and axion parameters, giving an independent measurement of $H_I$:
\begin{equation} 2.5\times 10^{13}\lesssim H_I/\text{GeV}\lesssim 10^{14}\,
\text{(ultralight ALPs, CMB-S4 alone)}\,\,.
\end{equation}
This applies to ALPs in the mass range $10^{-26}\lesssim m_a/\text{eV}\lesssim 10^{-23}$, where effects on lensing of a 1\% axion fraction can be distinguished from CDM. 

We show a range of possible CMB-S4 scenarios and their implications for constraints on $H_I$ in Figure~\ref{fig:optimize_axioniso}. The constraints on $H_I$ are only mildly sensitive to the beam size, until around 5 acrmin. The constraints degrade faster for a reducing in map sensitivity. In this figure we have kept one parameter (either beam sensitivity at 1$\mu$K-arcmin or beam size at 1 arcminute) fixed while varying the other parameters. We note that the \planck\ error bar is non-Gaussian - current data do not measure the energy scale of inflation in this scenario.

\emph{Detecting isocurvature and lensing effects from ULAs using CMB-S4 can provide a measurement of $H_I$ complementary to searches for tensor modes.}
 
\section{Summary}
Determining the nature of dark matter remains one of the main goals of the current cosmological epoch. CMB-S4 will provide not only a handle on distinguishing between different models of DM annihilation and other models of DM interaction, but will also place extremely tight constraints on axions and axion-like particles. CMB-S4 will tighten the bounds on DM annihilation by a factor of 2 to 3 with its improved sky coverage and sensitivity. Similarly, CMB-S4 will have the power to rule out many non-standard DM interactions (e.g. DM-dark radiation interactions and DM-baryon interactions). 
CMB-S4 could make a $>5\sigma$ detection of an axion that contributes only $3\%$ to the total energy budget of the dark sector at an axion mass of $m_a=10^{-25}\,\mathrm{eV},$ and even at masses as high as $m_a=10^{-23}\,\mathrm{eV},$ CMB-S4 will rule out axions with fractions greater than $64\%$ at $2\sigma$ confidence. Such axions are currently completely degenerate with a CDM component and are totally unconstrained with current data.  Moreover, detecting axion isocurvature with CMB-S4 will provide a probe of the energy scale of inflation that is complementary to the search for tensor modes. 
All of the above results will make CMB-S4 an excellent probe of the dark sector. CMB-S4 could constrain or detect departures from standard CDM at the sub-percent level. A detection would allow the particle- and component-nature of DM to be determined.


\def\gtrsim{\raise-.75ex\hbox{$\buildrel>\over\sim$}}
\def\lsim{\raise-.75ex\hbox{$\buildrel<\over\sim$}}

\chapter{Dark Energy}
\bigskip

\begin{quotation}

\end{quotation}

\section{Dark Energy and Modified Gravity}

The enigma of cosmic acceleration is among the most challenging problems in physics. Our most basic understanding about gravity---that objects fall towards one another under mutual gravitational attraction---simply does not apply on the largest distance scales. Instead, gravity is apparently repulsive at large distances and late times; the scale of spacetime itself is currently not only expanding but accelerating. The implication is either that our understanding of gravity is incomplete, or some other causative agent---dark energy---with exotic gravitational properties fills the Universe. In both cases, new physics is required beyond the four fundamental forces described by the Standard Model and general relativity.

The working hypothesis is that the cosmic acceleration is due to an exquisitely small cosmological constant, that Einstein's general relativity is valid from millimeter to beyond gigaparsec scales, and that dark matter consists of a single species of a cold, collisionless particle. Yet none of these offer insight or reflect the unity of physics demonstrated elsewhere as in the Standard Model of particle physics.

In particular, the cosmological constant suffers from a naturalness problem whose resolution may lie in a dynamical dark energy, quintessence. Theories of quintessence posit a new scalar field and predict a variety of testable phenomena.  They can also unveil new links to dark matter, neutrino physics, and cosmic parity violation. In their most general form, they represent
a scalar-tensor theory of gravity which can be described by  the effective field theory (EFT) of cosmic acceleration in the linear regime. 
CMB-S4 can provide the hard evidence  needed to pare down these possibilities and discover clues to the enigma of cosmic acceleration that will enable the development of  compelling theoretical 
alternatives to the cosmological constant.

In summary, the current observational evidence suggests a new frontier for physics at low energies and weak coupling, implied by the cosmological scales that characterize cosmic acceleration.  CMB lensing, thermal SZ cluster counts, and the kinematic SZ effect  all measure the influence of cosmic acceleration on the growth of 
structure while novel probes such as birefringence and the speed of gravitational waves test the physics of dark energy and modified gravity.   As such, CMB-S4 would be capable of helping to answer basic questions about dark energy and gravity in a manner complementary to ongoing precision measurements of the expansion history.    In Section \ref{sec:deparam}, we review the parameterizations and theories of dark energy.  We describe the 
dark energy observables that CMB-S4 will have the greatest impact on in Section \ref{sec:deobs} and forecast their impact on dark energy 
parameters in Section \ref{sec:deforecasts}.

\section{Models and parameters}
\label{sec:deparam}

In this section, we  briefly review the models and frameworks that have been proposed over the past years to test dark energy and modified gravity.   These fall into 
three families: ``trigger'', equations of motion, and theory parametrizations. The first ones are aimed at testing and falsifying the standard model
of $\Lambda$CDM, a cosmological constant with cold dark matter, and are agnostic as to its 
alternatives.  Given precise measurements from primary CMB anisotropy of the high redshift Universe, all low redshift observables related to the expansion history and growth of
structure are potential triggers.  Trigger parameters thus have the benefit that their relationship to the raw observables can be made as direct as desired.   The drawback is 
that deviant values for the trigger may not have any physical motivation.  Instead they help pare down the possibilities for the more model-dependent and theory-oriented tests.

  In the next section we discuss the cluster abundance,
 CMB lensing and pairwise kinematic SZ effects as the building blocks of triggers when
 combined with other measurements such as BAO and SNIa.   CMB-S4 will also enhance the precision and robustness
 of these other tests by measurements of the primary $E$-mode polarization.   For example
 the cold dark matter $\Omega_c h^2$ and effective relativistic degrees of freedom $N_{\rm eff}$ enter into the calibration of the BAO scale and inferences on $H_0$. 
 
In addition to triggers based on the expansion history, CMB-S4 provides triggers based
on the growth of structure.  The $\Lambda$CDM model predicts that the growth of structure
will slow in a precisely known manner as the expansion starts to accelerate.   
For example the rms amplitude of linear matter fluctuations at the $8 h^{-1}$Mpc
scale, $\sigma_8(z)$, is a trigger parameter that can be closely associated with the 
cluster abundance.   The linear growth rate index, $\gamma_L$, is another that is closely related
to peculiar velocities and the kSZ observables.   

The second way of parametrizing deviations from $\Lambda$CDM is by modifying the equations of motion for dark energy in a manner consistent with conservation laws.   These have the benefit of attempting to tie distance and growth
tests together in a physical, yet still phenomenological manner.    The next step up in
complexity from a cosmological constant is a model where the dark energy is dynamical
but spatially smooth relative to the dark matter.   In these models, the expansion history
can deviate from that of $\Lambda$CDM due to evolution in  the dark energy
equation of state $w(z)$, yet still predict the growth of structure.   A common parameterization
of this phenomenology is
\begin{equation}
w(z) = w_0 + w_a \frac{z}{1+z}.
\end{equation}
The figure of merit defined by the DETF is the inverse area of the
95\% CL region in the $w_0$--$w_a$ plane.  

There are generalizations of this type of parameterization that separate the expansion history from 
the growth of structure.   
A complete parametrization for observables for scalar-tensor theories in the linear regime would include in addition
the gravitational slip or effective anisotropic stress  (the ratio of the space curvature potential and Newtonian potential),  the effective Newton constant, and  $c_T$, the
speed of tensor perturbations. 
The last way of studying deviations from $\Lambda$CDM consists of directly testing theories beyond it.  Given the lack of a compelling specific theory to test, we can still make progress
by parameterizing all possible Lagrangians for fluctuations that are consistent with the
given symmetry.
 This approach maintains a strong connection with the underlying theory at the price of
complicating the relation to the raw data.  

More specifically, a systematic implementation of this approach is the effective field theory (EFT)
of cosmic acceleration \cite{Gubitosi:2012hu,Bloomfield:2012ff}, inspired by the EFT  of inflation described in the Inflation Chapter~\cite{Creminelli:2006xe,Cheung:2007st,Weinberg:2008hq,Creminelli:2008wc,Park:2010cw,Jimenez:2011nn}.    The EFT of cosmic acceleration describes the cosmological phenomenology of all universally coupled single scalar field dark energy and modified gravity models. Specifically, the EFT action is constructed in a unitary gauge to preserve isotropy and homogeneity of the cosmological background and reads:
\begin{eqnarray} \label{Eq:EFTaction}
\mathcal{S}_{\rm EFT} = \int d^4x \sqrt{-g}&& \bigg\{ \frac{m_0^2}{2} \left[1+\Omega(\tau)\right] R + \Lambda(\tau) - c(\tau)\,a^2\delta g^{00} + \frac{M_2^4 (\tau)}{2} \left( a^2\delta g^{00} \right)^2 \nonumber \\ 
&&   - \frac{\bar{M}_1^3 (\tau)}{2} \, a^2\delta g^{00}\,\delta K{^\mu_{\,\,\mu}}  - \frac{\bar{M}_2^2 (\tau)}{2} \left( \delta {K}{^\mu_{\,\,\mu}}\right)^2   - \frac{\bar{M}_3^2 (\tau)}{2} \,\delta {K}{^\mu_{\,\,\nu}}\,\delta {K}{^\nu_{\,\,\mu}} \nonumber \\
&& + m_2^2(\tau)\left(g^{\mu\nu}+n^{\mu} n^{\nu}\right)\partial_{\mu}(a^2g^{00})\partial_{\nu}(a^2g^{00}) +\frac{\hat{M}^2(\tau)}{2} \, a^2 \delta g^{00}\,\delta \mathcal{R}+	\ldots \bigg\}  \nonumber \\
&& + S_{m} [g_{\mu \nu}, \chi_m ]
\end{eqnarray}
where $R$ is the four-dimensional Ricci scalar, $\delta g^{00}$, $\delta K{^\mu_{\,\,\nu}}$, $\delta K{^\mu_{\,\,\mu}}$ and  $\delta \mathcal{R}$ are, respectively, the perturbations of the upper time-time component of the metric, the extrinsic curvature and its trace and the three dimensional spatial Ricci scalar of constant-time hypersurfaces. Finally,  $S_m$ denotes the action for all the matter fields conventionally considered in cosmology. 

In the action (\ref{Eq:EFTaction}), the  extra scalar degree of freedom is hidden inside metric perturbations. 
To study the dynamics of linear perturbations, however, it is convenient to make it explicit by means of  the St\"{u}ckelberg technique i.e.~performing an infinitesimal coordinate transformation such that $\tau\rightarrow \tau+\pi$, where the field $\pi$ describes the extra propagating degree of freedom. This approach allows us to maintain a direct link to the underlying theory so that we can keep under control its theoretical viability while exploring the cosmological implications of any of the models included in this language~\cite{Raveri:2014cka}. 

Since the choice of the unitary gauge breaks time diffeomorphism invariance, each operator allowed by the residual symmetry in action (\ref{Eq:EFTaction}) can be multiplied by a time-dependent coefficient that we shall call EFT function. To fully specify the phenomenology of linear perturbations only a restricted set of EFT functions are needed. These can be either parametrized to explore agnostically the space of dark energy and modified gravity models~\cite{Gleyzes:2013ooa,Bloomfield:2013efa,Piazza:2013coa,Gleyzes:2014rba} or can be fixed to reproduce exactly the phenomenology of some model of interest such as $f(R)$ gravity, quintessence and, more generally, the Horndeski class of theories and beyond~\cite{Gleyzes:2014dya,Frusciante:2015maa,Frusciante:2016xoj}.

\section{CMB  Dark Energy Observables}
\label{sec:deobs}

\subsection{Cluster abundance and mass}

Clusters of galaxies are the most massive ($\sim$10$^{14}$--10$^{15}$\,M$_{\odot}$) objects in the Universe to have undergone gravitational collapse, forming from regions $\sim$10--40\,Mpc in size.  This property makes clusters representative of the overall content of the Universe, and also makes them important tracers of the evolution of large-scale structure, sampling the most extreme peaks in the large-scale matter distribution. Cluster measurements have played a central role in establishing the modern standard model of cosmology, with key results including the discovery of dark matter in the Coma cluster \cite{Zwicky:1933gu}, early evidence that we live in a low-matter-density Universe ($\Omega_m < 1$) \cite{White:1993wm, Donahue:1997sp, Bahcall:1998ur}, constraints on the physical nature of dark matter \cite{Clowe:2006eq}, limits on neutrino masses (see the Neutrinos Chapter), and a broad swathe of constraints on properties of dark energy and modifications of gravity \cite{Vikhlinin:2008ym, Mantz:2009fw, Rapetti:2012bu, Benson:2011uta, Mantz:2014xba, Mantz:2014paa}.

In the report of the DETF, galaxy clusters were highlighted as having the highest sensitivity to dark energy parameters, but also potentially the largest systematic uncertainties, primarily relating to the perceived difficulty of estimating cluster masses. Since the time of the DETF report, however, significant progress has been made in quantifying and mitigating these uncertainties, and clusters currently provide, and are expected to continue to provide, constraints on cosmological parameters that are highly competitive with other leading probes. As described below, CMB-S4 is expected to play a key role in this effort, advancing cluster cosmology while simultaneously controlling systematic uncertainties.

Clusters are identified in CMB data through the inverse Compton scattering of CMB photons off of intra-cluster gas, otherwise known as the Sunyaev-Zel'dovich (SZ) effect \cite{Sunyaev:1972eq}. SZ cluster surveys have two important advantages: the SZ surface brightness is redshift independent, and the integrated SZ signal is expected to have a relatively small scatter at fixed mass \cite{Nagai:2005wx, Nagai:2007mt, Kravtsov:2012zs}. These properties enable SZ surveys to provide relatively clean, nearly mass-limited catalogs of clusters out to the highest redshifts where they exist; in particular, SZ surveys are easily the most efficient approach to finding massive clusters at $z > 1$. Since the first SZ-discovered clusters were reported in 2009 \cite{Staniszewski:2008ma}, catalogs of over 1000 SZ-selected clusters extending out to $z \sim 1.7$ have been produced \cite{Vanderlinde:2010eb, Reichardt:2012yj, Hasselfield:2013wf, Ade:2013skr, Bleem:2014iim, Ade:2015mva}. Going forward, the ability of SZ surveys to find the highest-redshift clusters offers a unique and powerful complement to optical surveys like DES and LSST, which will unveil the low-redshift Universe. These optical surveys will also provide key photometric redshifts and independent mass-calibration information (through galaxy-cluster lensing) to enable full utilization of the clusters found by CMB-S4.

Figure \ref{fig:cluster_counts} shows the projected mass thresholds and cluster counts for three possible CMB-S4 configurations. The figure illustrates the importance of high spatial resolution for cluster science: as the angular resolution of CMB-S4 is improved from 3 to 1 arc-minutes, the mass threshold for the sample drops from $\sim2$ to $\sim1\times 10^{14}\,M_{\odot}$. At a 99\% purity threshold, the 3, 2 and 1 arc-minute configurations would identify approximately 45,000, 70,000 and 140,000 clusters, respectively, providing a more than 100 fold increase in the number of SZ-identified clusters.  At the highest redshifts---the primary discovery space for CMB-S4 cluster science---the benefits of high resolution are even more apparent, with the predicted number of $z>1$ clusters for the 1 arc-minute configuration being $\sim$5 and 3 times larger than the 3 and 2 arc-minute configurations, respectively. 

\begin{figure}[t]
\begin{center}
\includegraphics[width=0.49\textwidth]{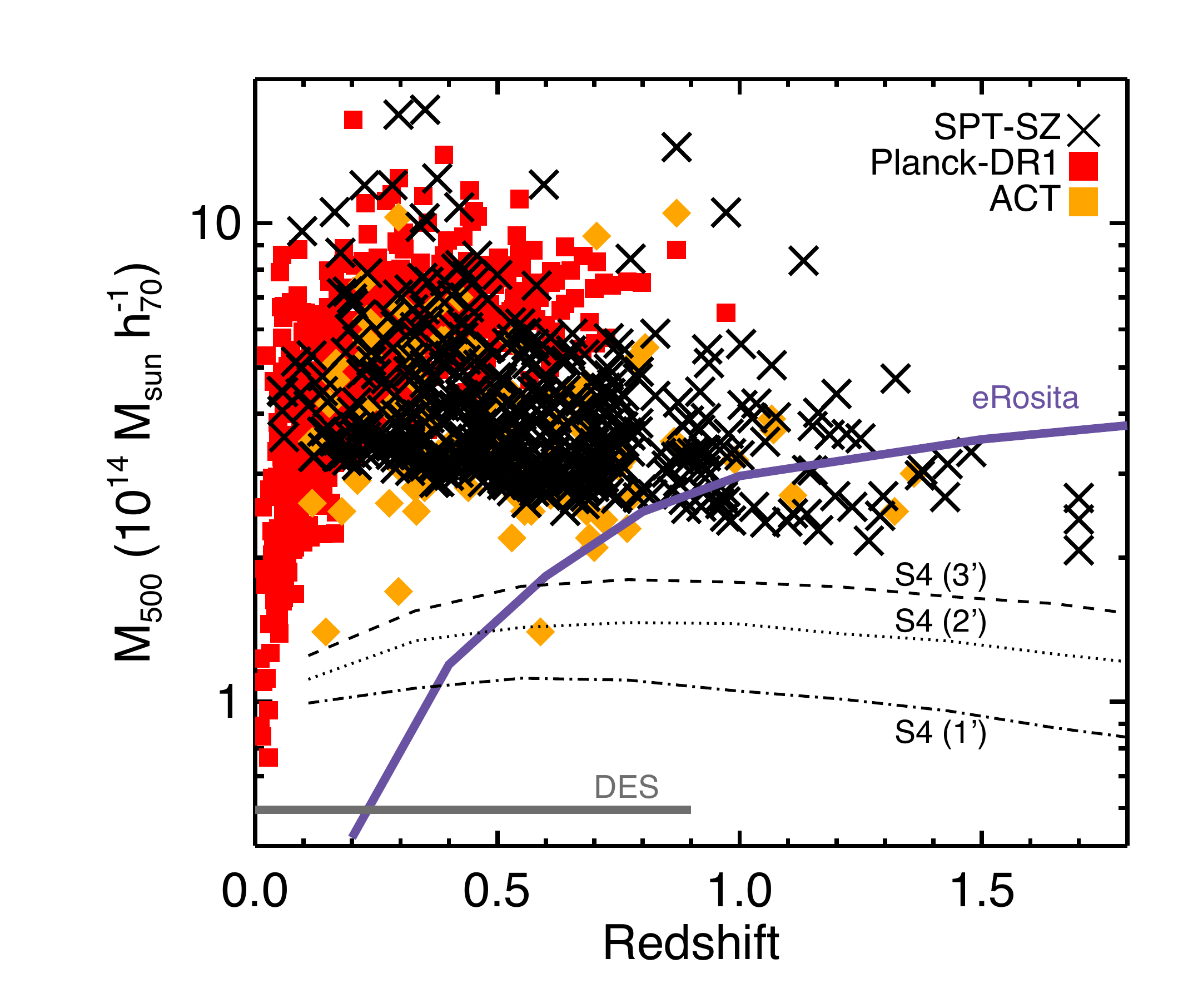}
\includegraphics[width=0.49\textwidth]{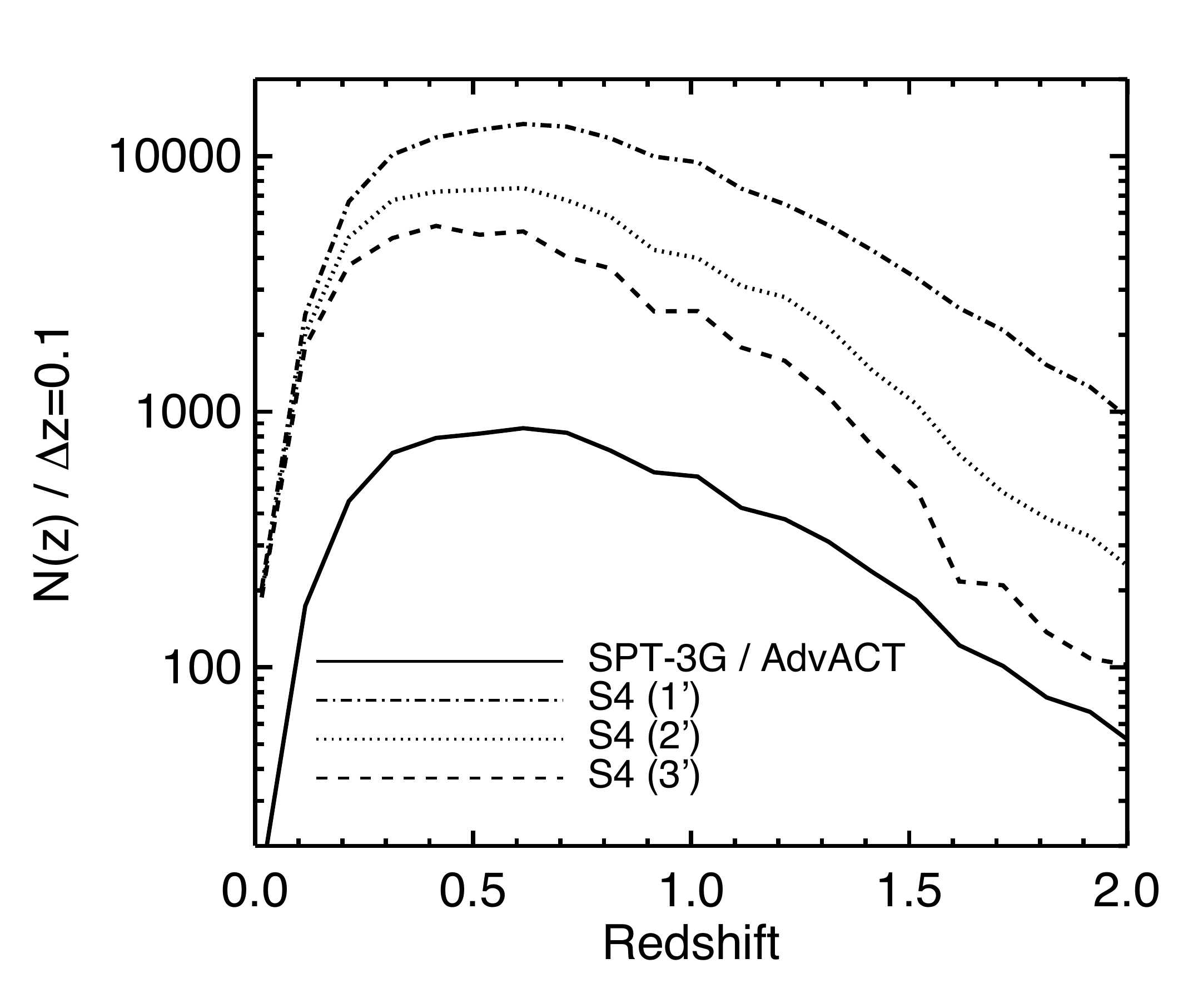}
\caption{(Left) The 50\% mass-completeness limits for three possible CMB-S4 instrumental configurations, with either 1, 2, or 3 arc minute angular resolution, are compared with existing SZ-selected cluster catalogs from \planck\ \cite{Ade:2015mva}, SPT-SZ \cite{Bleem:2014iim}, and ACT \cite{Hasselfield:2013wf}, and future thresholds expected for the optical Dark Energy Survey and the X-ray eRosita survey \cite{Pillepich:2011zz}.  (Right) The projected cluster counts for the three possible CMB-S4 configurations described above, compared with projections for the SPT-3G \cite{Benson:2014qhw} and AdvACT surveys.}
\label{fig:cluster_counts}
\end{center}
\end{figure} 

Beyond the cluster catalog itself, a key input to cluster dark energy and gravity tests is reliable cluster mass estimation. This can be usefully divided into two tasks: absolute mass calibration, i.e.\ our ability to measure cluster masses without bias on average, and relative mass calibration. The former is required for accurate inference of the power spectrum amplitude as a function of redshift, while the latter boosts cosmological constraints by providing more precise measurements of the shape and evolution of the mass function.  Precise relative mass information can be provided by X-ray observations of the intracluster medium using existing ({\it Chandra}, XMM-{\it Newton}) and future (eROSITA, ATHENA) facilities; for the most massive SZ-selected clusters found in existing CMB surveys, this work is already well advanced \cite{deHaan:2016qvy, Andersson:2010vy}. 

For absolute mass calibration, the most robust technique currently is galaxy-cluster weak lensing, which (with sufficient attention to detail) can provide unbiased results \cite{Corless:2009hi,Becker:2010xj}. At redshifts $z\lsim1$, residual systematic uncertainties in the lensing mass calibration are at the $\sim7\%$ level currently \cite{Applegate:2012kr}, and reducing these systematics further is the focus of significant effort in preparation for LSST and other Stage 4 data sets, with 1--2\% mass calibration at low redshifts seen as an achievable goal \cite{Abate:2012za}. Lensing of the CMB by clusters \cite{Madhavacheril:2014slf,Baxter:2014frs,Melin:2014uaa} provides an additional route to absolute mass calibration for CMB-S4 (see the CMB Lensing Chapter). CMB-cluster lensing is particularly well suited to calibrating clusters at high redshifts ($z\gtrsim1$) where ground-based galaxy-cluster lensing becomes inefficient. CMB data with sufficient resolution and depth (especially in polarization) can potentially provide a percent-level mass calibration at these redshifts~\cite{Hu:2007bt}, comparable to galaxy-cluster lensing at lower redshifts.

CMB-S4 can thus provide an exceptionally powerful cluster data set for dark energy and modified gravity science, by producing a large catalog of massive, cleanly selected clusters extending to high redshifts, and simultaneously constraining their masses. This data set would strongly complement contemporaneous dark energy programs, especially LSST and DESI. The impact of clusters as part of the CMB-S4 dark energy portfolio is strongly dependent on the angular resolution and depth of the survey, which affects the number of clusters discovered and the quality of the CMB-cluster lensing mass constraints; to achieve its full potential, a resolution of $\sim1$ arcminute and few-$\mu$K-arcmin depth will be required.

\subsection{Lensing}

As described in the Lensing Chapter, the CMB lensing deflection map measures the projected mass density all the way back to the decoupling epoch at $z \sim 1100$, with the majority of the contributions coming from $z > 1$.  CMB lensing is also dominated by structure on large scales in the linear regime.  Thus,  CMB lensing provides a clean probe of a particular integral over the linear growth of structure,
e.g.~$\sigma_8(z)$,
weighted by distances. The lensing power spectrum shape is predicted from the background cosmology; shape deviations indicate scale-dependent effects on the growth, including those caused by modified gravity or the 
gravitational effects of dark energy.  CMB lensing complements other dark energy probes by providing
a handle on effects at high redshift, e.g. in so-called early dark energy scenarios.

Cross-correlating the CMB lensing with other tracers of structure further permits extraction of information about the growth rate of structure in the Universe that is localized in redshift.  To the extent that other tracers have well-understood redshift distributions, cross-correlating a set of them to the CMB constitutes a tomographic study probing the evolution of the dark energy and its impact on the growth rate. The Lensing Chapter catalogs two broad categories of other tracers: galaxy density fields (and by extension the CIB) and galaxy shear maps. Combining lensing maps with maps of large scale flows from the kSZ will provide further constraints on the dark energy.

These cross-correlations with CMB lensing highlight the complementarity of CMB-S4 to other Stage IV experiments, including DESI, LSST, EUCLID, and WFIRST. 
For example, Figure \ref{LSSTdarkEnergy} shows the improvement on constraints of dark matter and dark energy from LSST cosmic shear after adding CMB lensing from CMB-S4.  The inclusion of CMB-S4 lensing improves the constraint on $w$ by roughly a factor of 2.  This forecast assumes LSST will have 30 galaixes per arcminute and cover 40\% of the sky overlapping CMB-S4.  This figure assumes just one wide redshift bin for the LSST shear analysis, and we note that the constraints indicated by both curves can improve with a tomographic LSST shear analysis.   In addition, CMB-S4 lensing can be used to calibrate the shear multiplicative bias for LSST, down to the level of the LSST requirements.  Details of this are given in Figures \ref{shear_calibration_cmbs4} and \ref{LSSTshearcalibration_vary_noise_beam_lmax} and the accompanying discussion in Section \ref{lensxlens} of the Lensing Chapter.

\begin{figure}[htbp]
\centering
\includegraphics[width=0.7\textwidth]{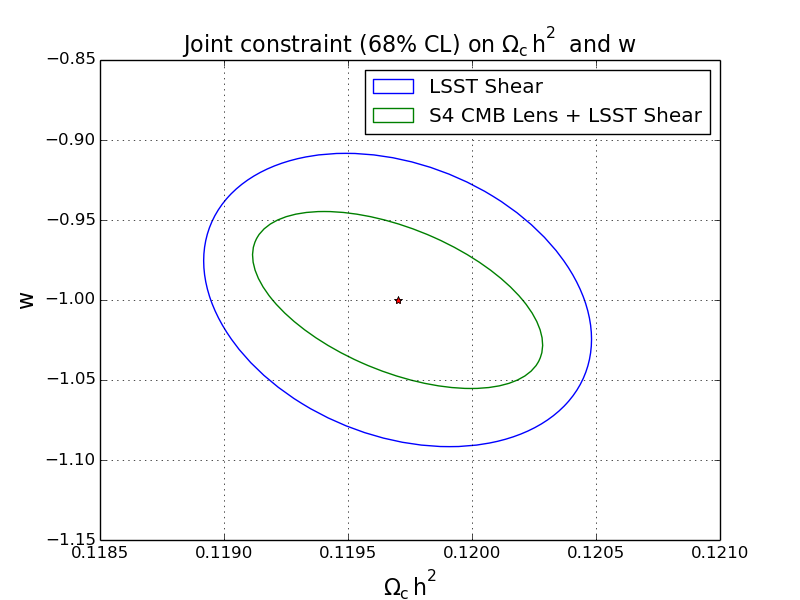}
\caption{The improvement in dark matter and dark energy parameters adding CMB-S4 lensing measurements to LSST cosmic shear.  The addition of CMB lensing results in a factor of 2 improvement in the constraint on the dark energy equation of state $w$.}
\label{LSSTdarkEnergy}
\end{figure}

\subsection{Kinematic SZ}

CMB-S4 will map with unprecedented precision the momentum field of the large scale structure via measurements of the kinematic Sunyaev Zel'dovich (kSZ) effect. Multi-frequency data can be used to remove other foregrounds and isolate the kSZ signal. CMB-S4 measurements with sufficient angular resolution can be used to reconstruct the diffuse kSZ anisotropy signal enabling sub-percent precision measurements of the amplitude of the matter density fluctuations $\sigma_8$ (see for example \cite{Calabrese:2014gwa}) which then acts as a trigger
parameter for testing $\Lambda$CDM. 
Measurements of the patchy kSZ can place strong constraints on the time and duration of reionization.

The combination of CMB-S4 with data from galaxy surveys will be able to measure the kSZ effect associated with galaxy clusters, which is proportional to their peculiar momentum. The large scale structure momentum field is an important cosmological observable that can place strong constraints on the cosmological parameters \cite{Bhattacharya:2007sk,Kosowsky:2009nc,Mueller:2014nsa,Mueller:2014dba} complementary to density fluctuation measurements. The mean pairwise velocity of galaxy clusters is sensitive to both the growth of structure and the expansion history of the Universe and it is an excellent probe for gravity on large scales. Being a differential measurement it is also particularly stable against residual foregrounds that might survive the frequency cleaning process. In \cite{Mueller:2014nsa,Mueller:2014dba} it has been shown that a S4 survey with high resolution can constrain the redshift dependent growth of structure at $\lesssim 5\%$ precision in generic models allowing also for a redshift dependent equation of state of the dark energy. These measurements will be able to distinguish dark energy from modified gravity and will provide complementary constraints to redshift space distortions and weak lensing measurements, probing larger physical scales.

Pairwise kSZ measurements can also constrain the sum of neutrino masses $M_{\nu}= \sum m_{\nu}$ with a $1\sigma$ uncertainty of $0.030$eV for a $1$ arcmin CMB-S4 overlapping 10000 deg$^2$ with a galaxy survey able to identify $M>10^{13}M_{\odot}$ clusters. With 5-arcmin resolution, separating the CMB background from the kSZ signal would be more difficult, providing  $\sigma_{M_{\nu}} = 0.076$eV. These forecasts include only priors on the 6 standard cosmological parameters from \planck\ temperature data and show the potential of the kSZ pairwise signal to provide constraints on the neutrino mass. 

Here we explore how the kSZ detection significance depends on the noise and aperture of the CMB experiment. 
To interpret the forecasts below, we note that the size of the temperature shift for a cluster with radial velocity $v_r$ and optical depth $\tau_{\rm cluster}$ is $(\Delta T/T)_{\rm kSZ} = - \tau_{\rm cluster} v_r \propto n_e v_r$, where $n_e$ is the free electron number density.  Therefore, the $S/N$ values quoted are on the product of optical depth and radial velocity, the latter being a sensitive probe of cosmology as discussed previously in this section.  If the optical depth is known externally (for example through the use of thermal Sunyaev-Zel'dovich or X-ray observations), then the quoted $S/N$ applies to the radial velocity, or equivalently to the growth factor of density perturbations.  If instead general relativity and a fiducial cosmology are assumed,
the measurement probes the total electron abundance associated with the halo, as well as the gas profile. Precision measurement of the gas profile through the kSZ effect will constrain galaxy cluster physics, feedback effects, and provide clues on galaxy formation. 
Since baryons amount to $\approx 20$\% of the total mass, they can have large effects on the matter power spectrum on small scales. Thus kSZ measurements with CMB-S4 will provide useful calibration information for weak lensing surveys aimed at measuring dark energy \cite{vanDaalen:2011xb, Mohammed:2014mba}.
There are two regimes that we will consider in the forecasts:

1) If spectroscopic or good photometric redshifts (e.g., $\sigma_z / (1+z) \lesssim 0.01$) are available, kSZ techniques such as ``velocity reconstruction'' \cite{Ho:2009iw} or ``pairwise momentum'' \cite{Ferreira:1998id} can be used.  To forecast the total detection significance we use the Fisher formalism in harmonic space, normalized to agree with current results \cite{Ade:2015lza,Schaan:2015uaa}. This forecasting technique has been validated on high-resolution simulations and shown to be accurate.
Figure \ref{fig:S4_kSZ_forecast} shows results when combining a CMB-S4-type experiment with 20 million spectroscopic galaxies from the upcoming DESI survey \cite{Levi:2013gra} on 14,000 square degrees.  \begin{figure}[ht]
\centering
\includegraphics[width=0.7\textwidth]{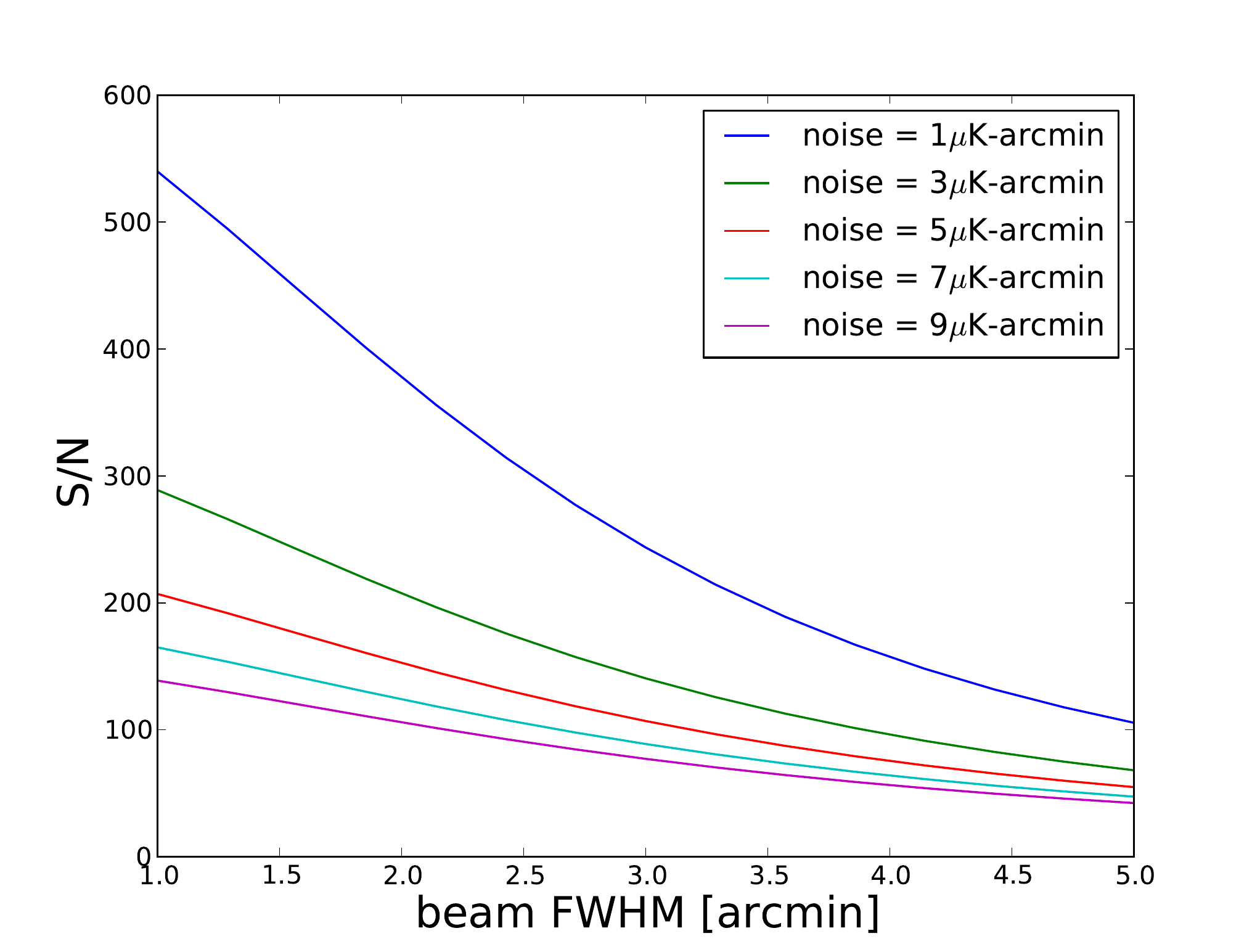}
\caption{kSZ $S/N$ for different beam FWHM and noise level between 1 $\mu$K-arcmin (top curve) to 9 $\mu$K-arcmin (bottom curve) for a DESI-like spectroscopic galaxy survey, using the ``velocity reconstruction'' or ``pairwise momentum'' techniques.}
\label{fig:S4_kSZ_forecast}
\end{figure}
We note that a very high-significance statistical detection can be achieved and that high resolution is beneficial. Because the estimators considered here are differential, other secondary anisotropy components (such as tSZ or the CIB) that are uncorrelated with the cluster velocities cancel on average, which makes the detection possible even on single-frequency maps.  Foregrounds are not expected to be a limiting factor in this kind of analysis.  Moreover, since the measurement is directly proportional to the local number of free electrons, it allows us to measure cluster profiles in an almost model-independent way.
Note that the values reported here are subject to significant astrophysical uncertainty on small scales, especially for small beam FWHM $\lesssim 2$ arcmin, with most of the uncertainty related to the concentration (i.e., compactness) of the gas profile.

2) If no redshift information is available or the photometric redshift uncertainty is large, a different technique can be used \cite{Dore:2003ex,Ferraro:2016ymw,Hill:2016dta}:  an appropriately filtered version of the CMB map can be squared in real space and then cross-correlated with tracers of the density field (e.g., galaxies, quasars, or lensing convergence).  Since the kSZ effect is expected to dominate the foreground-cleaned high-$\ell$ CMB anisotropy ($\ell \gtrsim 4000$), this correlation can yield very high $S/N$ in the high-resolution regime.  An advantage of this technique is that it does not require knowledge of redshifts for individual objects, but just a statistical redshift distribution $dn/dz$. The drawback is that very good foreground cleaning is required and residual foregrounds might limit the actual performance.  Here we follow \cite{Ferraro:2016ymw} and note that a CMB-S4-type experiment with (beam FWHM [arcmin], noise [$\mu$K-arcmin]) = (1, 1), (3, 1) and (3, 3) can achieve $S/N$ = 822, 702 and 296 respectively, when combined with a galaxy sample from the WISE survey \cite{Wright:2010qw}.  These are the statistical errors only, and the actual detection significance is likely to be dominated by systematics and imperfect foreground cleaning.  As previously noted, these numbers are also sensitive to assumptions about small-scale astrophysics, which are currently fairly uncertain. Moreover, these last forecasts have assumed that baryons trace the dark matter down to the scale of interest.  While this approximation is thought to be accurate for current analysis with data from the \planck\ satellite \cite{Hill:2016dta}, it is likely to be inaccurate for a high-resolution CMB-S4 experiment.

\subsection{Cosmic Birefringence}
\label{sec-biref}
CMB-S4 will measure parity violating two point correlations between B modes and  E or T modes.   Their detection would have paradigm changing implications
for cosmological physics.  In the dark energy context, they could arise from cosmic birefringence.

The simplest dynamical way to model the accelerated expansion of the Universe is to invoke a new slowly evolving scalar field that dominates its energy budget (the quintessence models for DE). Such a field generically couples to photons through the Chern-Simons term in the electromagnetic Lagrangian, causing linear polarization of photons propagating cosmological distances to rotate----the effect known as cosmic birefringence~\cite{Carroll:1998zi}. In the case of the CMB, such rotation converts the primordial E mode into B mode, producing characteristic TB and EB cross-correlations in the CMB maps \cite{Kamionkowski:2008fp,Gluscevic:2009mm}. Even though there is no firm theoretical prediction for the size of this effect, if observed, it would be a clear ``smoking-gun'' evidence for physics beyond the standard model in the form of a new scalar field. Previous studies have used quadratic estimator formalism to constrain this effect \cite{Gluscevic:2012me}, with the best current limit coming from sub-degree scale polarization measurements with POLARBEAR \cite{Ade:2015cao} ($<0.33$ deg$^2$ for the amplitude of a scale-invariant rotation-angle power spectrum). A promising way to pursue search for cosmic birefringence in the future is measurement of the off-diagonal EB cross correlations on small angular scales, and the measurement of polarization anisotropy on a wide range of scales is going to be essential for achieving this. 

Fig.~\ref{fig:CB-forecast} shows the current upper limit on the rotation-angle power spectrum from POLARBEAR and a projection for \planck, and a forecast for a Stage-IV experiment (with noise of $1.41$ $\mu$K-arcmin in polarization, and a resolution of $1'$). The improvement from the current constraint at all multipoles is about two orders of magnitude. We assumed access to polarization modes from $\ell=30$ to $\ell=5000$.
\begin{figure}[h!]
\centering \includegraphics[width=0.70\textwidth]{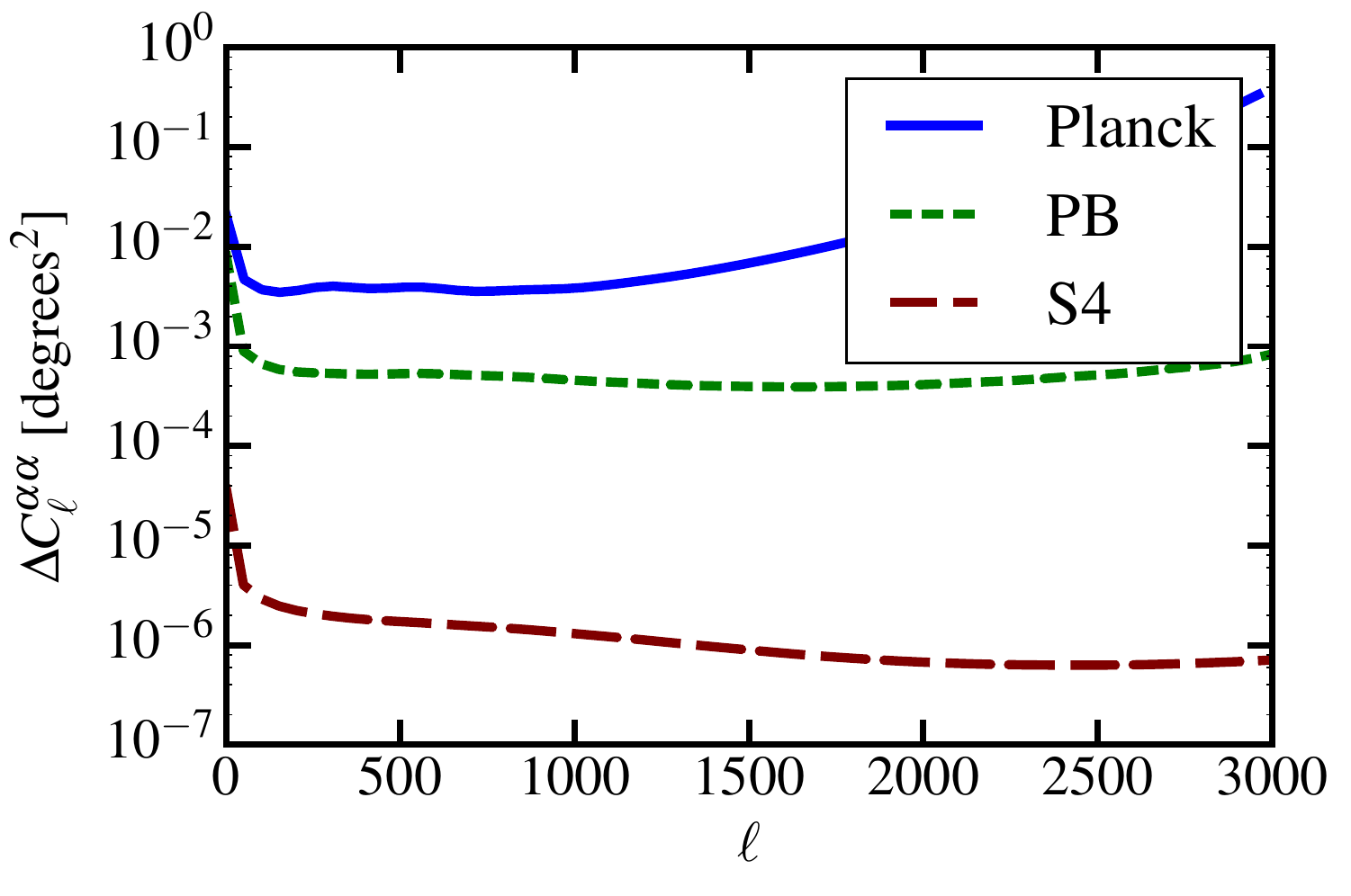}
\caption{The current (from POLARBEAR, labeled as PB) and projected (for \planck\ and Stage-IV experiment) 1$\sigma$ errobars on the birefringent rotation-angle power spectrum are shown on the vertical axis. A Stage-IV has the potential to improve the current best constraint on anisotropic birefringence by more than two orders of magnitude at all multipoles. For the Stage-IV forecast, we assumed noise of $1.41$ $\mu$K-arcmin (in polarization), a resolution of 1', and have considered polarization modes from $\ell=30$ to $\ell=5000$.}
\label{fig:CB-forecast}
\end{figure}

For a fixed integration time (and a varied noise level and sky coverage), large sky coverage optimizes sensitivity to low multipoles of the rotation angle and gives the best signal-to-noise ratio for rotation models that have power on large scales (such as, for example, a model with a scale-invariant power spectrum, which could result from fluctuations in a spectator scalar field present during inflation). Conversely, for models that have power on scales corresponding to multipoles above $\ell\sim 1000$, the best signal-to-noise is achieved with deeper integration on small sky patches.  For a measurement of the magnitude of the quadrupole of the rotation angle, reducing the resolution from 1' to 9' produces a factor of a few increase in the projected errorbar (for all other parameters fixed). Increasing the noise from $1.41$ to $12.7$ $\mu$K-arcmin produces a factor of about $20$ increase in the errorbar. Access to polarization modes down to $\ell=2$ does not significantly affect the forecasts.

\section{Dark Energy Forecasts}
\label{sec:deforecasts}

To forecast CMB-S4 performances on dark energy and modified gravity models we shall use the following specifications.
CMB-S4 is assumed to measure CMB fluctuations in temperature and polarization over $40 \%$ of the sky with a $1\,\mu {\rm K} \, {\rm arcmin}$ sensitivity in temperature and $1.4\,\mu {\rm K} \, {\rm arcmin}$ sensitivity in polarization, with a beam with $1'$ FWHM. This is added to \planck\ measurements of CMB fluctuations on the remaining part of the sky with specifications from \cite{Adam:2015rua}. To reproduce the noise levels of real \planck\ measurements at large angular scales in polarization the E and B mode polarization sky fraction is reduced to $0.01$.
 As such, 
these forecasts do not include the cluster abundance or kSZ but do include CMB-S4 lensing.  The former
can then be viewed as providing independent trigger tests for the consistency of $\Lambda$CDM or    on dark energy and modified gravity models 
using the projections described in the previous sections.

Along with CMB probes we shall use DESI to exploit the complementary sensitivity of LSS measurements and investigate the synergies with CMB-S4 in constraining DE/MG models.
We shall assume pessimistic specifications for the DESI survey as in \cite{Font-Ribera:2013rwa}.
When both CMB-S4 and DESI are considered we include in the forecast all the cross correlations between these two probes. 
When a result is presented it is always marginalized over all the other parameters of the model. 
In particular, when considering DESI, we marginalize over a constant scale independent bias, different in all the survey redshift bins.

We use the the CosmicFish code \cite{Raveri:2016xof,Raveri:2016leq} to perform the forecast presented in this section. The CosmicFish code uses CAMB sources \cite{Lewis:1999bs,Challinor:2011bk} for all the $\Lambda$CDM cosmological predictions, uses EFTCAMB sources \cite{Hu:2013twa,Raveri:2014cka} for all models enclosed in the EFT framework and MGCAMB sources \cite{Zhao:2008bn,Hojjati:2011ix} for the Growth Index forecast.

\subsection{Trigger}

\begin{figure}[!tb]
\begin{center}
\includegraphics[width=1.0\textwidth]{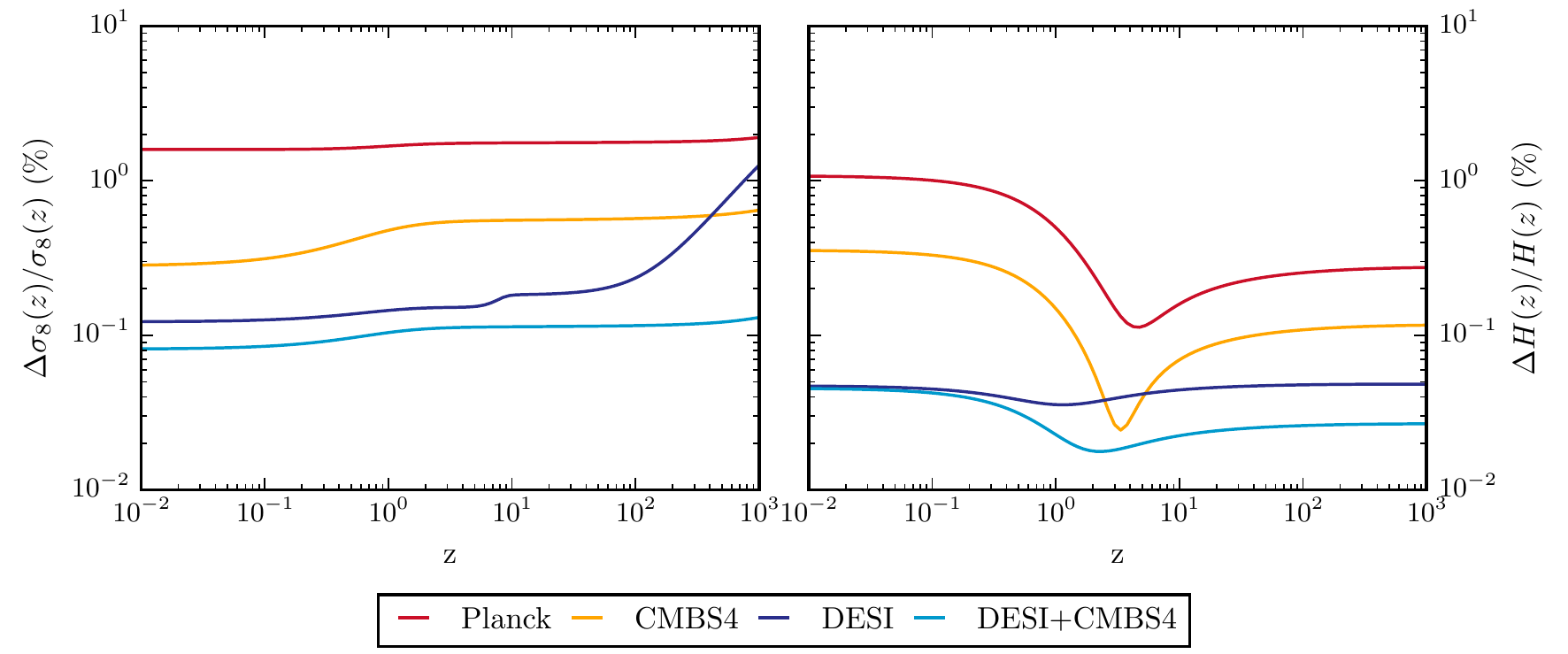}
\caption{Relative $68\%$ C.L. error on $\sigma_{8}$ and $H$ as a function of redshift. Different colors correspond to different experiments, as shown in legend.}\label{fig:GrowthExpansion}
\end{center}
\end{figure}

We consider $\sigma_8(z)$, $H(z)$ and the growth index $\gamma_{L}$ as trigger parameters. Figure \ref{fig:GrowthExpansion} shows the relative error on $\sigma_8(z)$ and $H(z)$, assuming an underlying $\Lambda$CDM model. The central panel of Figure \ref{fig:MixedDEG} shows the marginal forecast constraint on $\gamma_{L}$. 

As we can see from the left panel of Figure \ref{fig:GrowthExpansion} CMB-S4 will improve on \planck\ determination of $\sigma_{8}(z)$ substantially, pushing the sensitivity to this parameter to sub-percent accuracy, especially at late times. 
This level of accuracy is comparable with DESI measurements that are, in turn, just a factor two tighter.
At early times CMB-S4 sensitivity to $\sigma_{8}$ is slightly lower but significantly better than DESI, as soon as redshift increases.
Noticeably when CMB-S4 and DESI are joined $\sigma_{8}$ gets constrained to $\sim 0.1\%$ at all times. The gain in the joint constraint is higher than the gain in sensitivity in going from CMB-S4 to DESI, thanks to the cross correlation between the two surveys.
A similar picture emerges from the right panel of Figure \ref{fig:GrowthExpansion} with the noticeable difference that CMB measurements have a peak in sensitivity around $z\sim 3$ that makes CMB-S4 stronger than DESI. The joint DESI CMB-S4 constraints reflect this. The addition of CMB-S4 measurements improves the constraint on the expansion history significantly at redshifts higher than three.

We considered $\sigma_8(z)$, $H(z)$ as trigger parameters because these levels of sensitivity will be the key to resolve tensions between different experiments. 
CMB measurements and LSS surveys display a marginal disagreement on the determination of the growth of cosmic structures but these tensions are still in a low statistical significance phase.
In particular \planck\ data are in tension with measurements of the Canada-France-Hawaii Telescope Lensing Survey (CFHTLenS) \cite{Joudaki:2016mvz} and the Kilo Degree Survey (KiDS) \cite{Hildebrandt:2016iqg} when considering the parameter $S_8\equiv \sigma_8\sqrt{\Omega_m/0.3}$. 
On the other hand the disagreement with the Dark Energy Survey (DES) is only marginal \cite{Abbott:2015swa}.
Two experiments with similar, high, sensitivity can either confirm or falsify these tensions to high statistical significance making CMB-S4 and DESI instrumental to each other.

In Table \ref{table:ForecastTensionS8} we investigate the expected statistical significance of these two tensions, when assuming the \planck\ mean $S_8$ value and the KiDS and DES ones. As we can see if we replace the \planck\ error with the forecasted CMB-S4 one the statistical significance of these tensions is limited by the sensitivity of the weak lensing surveys. When considering \planck\ and DESI sensitivities the statistical significance improves becoming almost decisive but still being limited by \planck. Only when considering both CMB-S4 and DESI we will achieve definitive sensitivity and this will allow us to establish whether these discrepancies are due to new physical phenomena or just statistical fluctuations.

\begin{table}[t!]
\begin{center}
\begin{tabular}{|c|c|c|c|} 
\hline
    				  Datasets 			& $\sigma_{S_8(z=0)}$  & \planck-KiDS tension & \planck-DES tension  \\
				  \hline
\planck  		& 		0.025	& - & -	\\
\hline
KiDS-450 		& 			0.038  	& $2.3 \, \sigma$ & -	\\
\hline
DES             &			0.06	 & - & $0.6 \, \sigma$	\\
\hline
CMB-S4       &			0.003  & 	$2.7 \, \sigma$ & $0.6 \, \sigma$	\\
\hline
\planck\ and DESI            &   0.0009	& 	$4.1 \, \sigma$ & $1.5 \, \sigma$	 \\
\hline
CMB-S4 and DESI  &   0.0004	 & 	$33 \, \sigma$ & $12 \, \sigma$			\\
\hline
\end{tabular}
\caption{Forecasted constraints on $S_8\equiv \sigma_8\sqrt{\Omega_m/0.3}$ and statistical significance of the discrepancy between \planck\ and the DES and KiDS surveys.}
\label{table:ForecastTensionS8}
\end{center}
\end{table}

The power of CMB-S4 in constraining the growth of structures and its synergy with LSS surveys clearly shows when considering the Growth Index $\gamma_{L}$. As we can be see from both the central panel of Figure \ref{fig:MixedDEG} and Table \ref{table:ForecastGammaConstEFT} CMB-S4 will give stronger constraints with respect to \planck\ due to the additional leverage of CMB lensing. These constraints will be comparable with DESI ones and displaying a slightly different degeneracy with the amplitude of scalar perturbations.
Leveraging on the precision of both CMB and LSS measurements, the joint constraints with CMB-S4 and DESI are significantly stronger than the single probes considered alone.

\subsection{Equation of motion parametrization}

\begin{table}[t!]
\begin{center}
\begin{tabular}{|c|c|c|c|} 
\hline
Datasets 			&  $r$ fiducial & $\sigma ( r )$  & $\sigma ( c_{\rm GW}^2 )$ \\
\hline
\hline
CMB-S4               & 0.05 & 0.002 & 0.05 \\
\hline
CMB-S4               & 0.01 & 0.001 & 0.1 \\
\hline
CMB-S4               & 0.001 & 0.0008 & - \\
\hline
CMB-S4 + DESI     & 0.001 & 0.0007 & - \\
\hline
\end{tabular}
\caption{Forecast $68\%$ C.L. marginal constraints on the tensor to scalar ratio ($r$) and the speed of gravitational waves for different fiducial values of $r$.}
\label{table:ForecastCT}
\end{center}
\end{table}
Of the parameters for the equation of motion description of dark energy and modified gravity,  CMB-S4 is in a unique position to 
constrain deviations of the speed of gravitational waves from the speed of light, with the parameter $c^2_{\rm GW}$. If the effect of primordial GWs on the B-mode polarization of the CMB is detected then the same observations will be capable of constraining their propagation speed at the time of recombination.
In the left panel of Figure \ref{fig:MixedDEG} we show the marginalized joint forecast constraint on the tensor to scalar ratio and the speed of GWs for a fiducial value of $r=0.01$ and $c_{\rm GW}^2=1$. In Table \ref{table:ForecastCT} we show the expected marginal constraints when changing the fiducial value of $r$.

As we can see, if $r$ is detected in the $0.05$ range, CMB-S4 measurements will provide a $5\%$ bound on the speed of GWs at the time of recombination.
As soon as the GW induced component in the B-mode polarization spectrum, becomes weak the bound on the GW's speed gets looser. If the fiducial is $r=0.01$ then CMB-S4 measurements will provide a $10\%$ bound. If $r=0.001$ then the statistical significance of the tensor induced B-mode component detection weakens and correspondingly the speed of GWs gets unconstrained.

When $r=0.01$ we also notice a slight degeneracy between the speed of GWs at recombination and the tensor to scalar ratio. Correspondingly CMB-S4 measurements will be more sensitive to the sum of these two parameters. This degeneracy is alleviated as soon as the fiducial $r$ value is increased and becomes negligible for $r=0.05$.

We stress here that all the other experiment combinations considered in this section could not constrain the speed of GWs thus CMB-S4 will give us the unique opportunity to measure this quantity at the time of recombination.

\subsection{Theory parametrization}

\begin{table}[t!]
\begin{center}
\begin{tabular}{|c||c||c|c|c||c|c|c|} 
\hline
Datasets 			& $\sigma ( \gamma_{\rm L} )$  & $\sigma ( \Omega_0 )$ & $\sigma ( \gamma_0^{(2)} )$ & $\sigma ( \gamma_0^{(3)} ) $ & $\tilde{M}_0$ & $\alpha_0^{\rm B}$ & $\alpha_0^{\rm T}$    \\
\hline
\hline
\planck  		& 0.02 & 0.03 & 0.4 & 0.01 & 0.03 & 0.02 & 0.02 \\
\hline
CMB-S4       & 0.007 & 0.02 & 0.1 & 0.01 & 0.02 & 0.02 & 0.008 \\
\hline
DESI            & 0.007 & 0.2 & 0.4 & 0.1 & 0.02 & 0.07 & 0.03 \\
\hline
CMB-S4 + DESI  & 0.003 & 0.01 & 0.05 & 0.003 & 0.006 & 0.02 & 0.001 \\
\hline
\end{tabular}
\caption{Forecast $68\%$ C.L. marginal constraints on different models: the trigger parameter $\gamma_{\rm L}$; constant EFT couplings $\Omega_0$, $\gamma_0^{(2)}$ and $\gamma_0^{(3)}$; constant Horndeski couplings $\tilde{M}_0$, $\alpha_{\rm B}$ and $\alpha_{\rm T}$.}
\label{table:ForecastGammaConstEFT}
\end{center}
\end{table}

\begin{table}[t!]
\begin{center}
\begin{tabular}{|c||c|c||c|c|c|c|c|c|} 
\hline
Datasets 			& $\Omega_{\rm early}$ & $\Omega_{\rm late}$ & $\tilde{M}_{\rm early}$ &  $\tilde{M}_{\rm late}$ & $\alpha^{\rm B}_{\rm early}$ & $\alpha^{\rm B}_{\rm late}$ & $\alpha^{\rm T}_{\rm early}$  & $\alpha^{\rm T}_{\rm late}$   \\
\hline
\hline
\planck  		& 0.08 & 0.05 & 0.2 &  0.1 & 0.05 & 0.2 & 0.03 & 0.04 \\
\hline
CMB-S4       &  0.04 &  0.04 & 0.05 & 0.02 & 0.05 & 0.1 & 0.02 & 0.01 \\
\hline
DESI            & 0.3 & 0.2 & 1.0 & 0.04 & 0.4 & 0.4 & 0.08 & 0.03 \\
\hline
CMB-S4 + DESI  &  0.03 &  0.02 & 0.04 & 0.007 & 0.04 & 0.08 & 0.02 & 0.002\\
\hline
\end{tabular}
\caption{Forecast $68\%$ C.L. marginal constraints on early and late time values of different EFT couplings.}
\label{table:ForecastAlphaAtan}
\end{center}
\end{table}

We consider two parametrization bases for the functions describing the EFT of cosmic acceleration and, for the sake of simplicity, we focus on the Horndeski class of models \cite{Horndeski:1974wa}.
The first parametrization is obtained by making the couplings in action (\ref{Eq:EFTaction}) dimensionless, as in \cite{Hu:2014oga}.
The second one consists in re-parametrizing the couplings explicitly targeting the phenomenological features of Horndeski, as in \cite{Bellini:2014fua}.
In both cases we consider two functional forms: first we assume all the couplings are constant in time; next we allow all the EFT couplings to have different early and late time values, with a smooth transition in between, inspired by \cite{Linder:2015rcz}. Specifically this second parametrization is given by $f(a)= 1/2 (f_{\rm early} + f_{\rm late}) + ( f_{\rm late} -f_{\rm early} ) {\rm ArcTan}[(a - a_T)/\Delta a]/\pi$ where $f_{\rm early}$ and $f_{\rm late}$ are respectively the early and late time values of the considered EFT function, $a_T$ is the transition scale factor assumed to correspond to $z=10$ and $\Delta a=0.01$ is the transition sharpness. 

In Figure \ref{fig:ConstantEFT} and Table \ref{table:ForecastGammaConstEFT} we show the forecast constraints on constant EFT couplings. As we can see the sensitivity of CMB probes are unmatched when measuring the conformal coupling to gravity $\Omega_0$. 
CMB-S4 measurements, in addition, are found to be the most constraining measurements on the other two EFT higher order operators, $\gamma_0^{(2)}$ and $\gamma_0^{(3)}$. Confirming the picture previously presented, the synergy between CMB-S4 measurements and DESI, results in much tighter constraints on all the considered parameters.
For all the probes considered the kinetic operator, $\gamma_0^{(1)}$, is found to be unconstrained. 

A similar picture also emerges from the forecast constraints on Horndeski couplings, with CMB-S4 providing the tightest bounds, as we can see from Figure \ref{fig:ConstantAlpha} and Table \ref{table:ForecastGammaConstEFT}. 
When considering the effective Planck mass $\tilde{M}_0$ and the tensor speed excess, $\alpha_0^{\rm T}$, DESI and \planck\ sensitivities are comparable while CMB-S4 is a factor 1.5 and 2.5 stronger, respectively. 
CMB measurements, on the other hand, are the most powerful at constraining the braiding coefficient $\alpha_0^{\rm B}$ and we can notice that \planck\ measurements are slightly stronger than CMB-S4 ones leveraging on the constraining power of large angular scales. 
As expected, combining CMB-S4 to DESI, results in a significant improvement with respect to the single probes alone.
As in the constant EFT case the scalar field kineticity $\alpha_0^{\rm K}$ is unconstrained.

When considering all the EFT couplings having different values at early and late times we found that early times changes are constrained by physical viability requirements and late time values have comparable bounds with respect to the constant case considered before.
The only EFT coupling that does not display this behavior is the conformal one, $\Omega(a)$, and the corresponding forecast constraints are shown in the right panel of Figure \ref{fig:MixedDEG} and Table \ref{table:ForecastAlphaAtan}. For these parameters we find that marginalization slightly degrades the forecast bounds with respect to the constant case. Moreover we find that data are more sensitive to the sum of these two parameters rather than their difference. When CMB-S4 is combined to DESI this degeneracy in parameter space is relieved.

Horndeski couplings in turn display a qualitatively different picture, as we can see from Figure \ref{fig:VariationAlpha} and Table \ref{table:ForecastAlphaAtan}.
Not surprisingly CMB measurements are generally stronger than LSS surveys at constraining the early time values of these functions, i.e. $\tilde{M}_{\rm early}$, $\alpha^{\rm T}_{\rm early}$ and $\alpha^{\rm B}_{\rm early}$. However CMB-S4 measurements, leveraging on both the early and late time constraining power of CMB and CMB lensing, are sensitive to both early and late time values, to unmatched accuracy.

\begin{figure}[!tb]
\begin{center}
\includegraphics[width=1.0\textwidth]{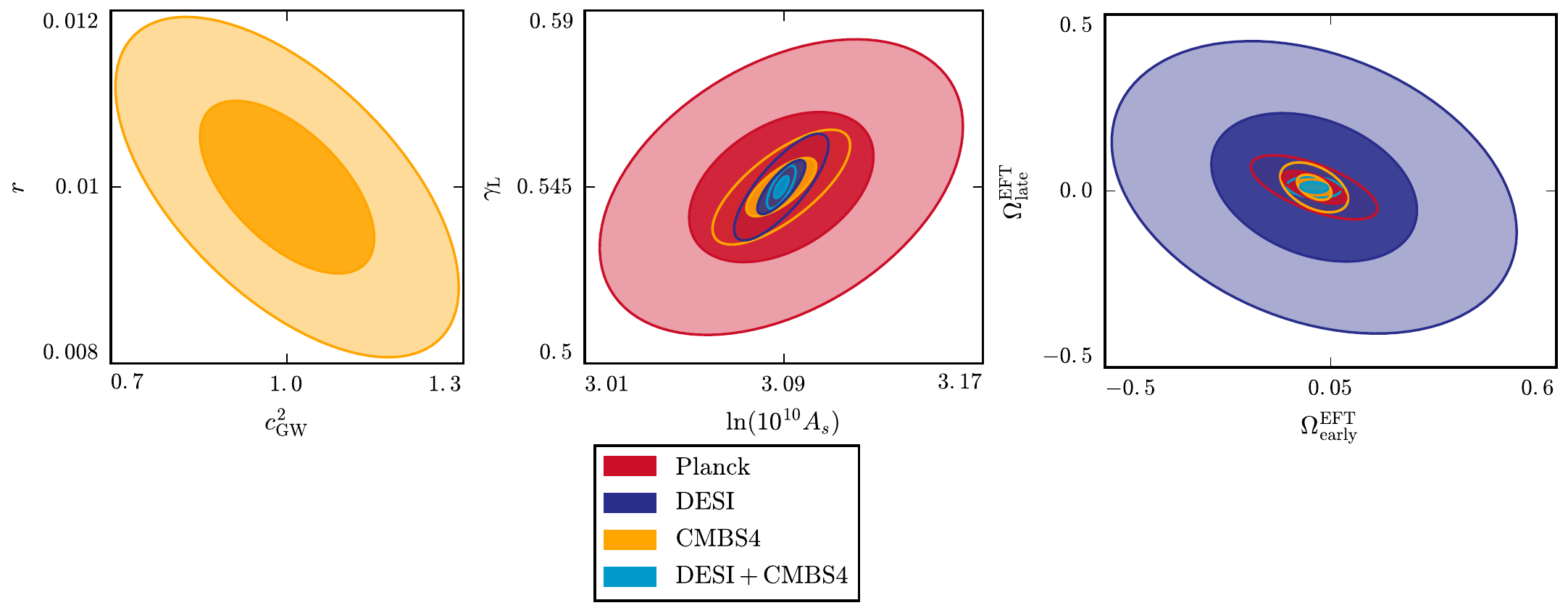}
\caption{Forecast marginalized constraints on different models. The left panel shows the joint constraints on the tensor to scalar ratio and the speed of gravitational waves. The central panel shows the joint constraints on the growth index and the amplitude of scalar perturbations. The right panel shows the joint constraints on relative variations of the gravitational constant at early times $\Omega_0^{\rm EFT}$ and late times $\Omega_1^{\rm EFT}$. Different colors correspond to different experiments, as shown in legend. The darker and lighter shades correspond respectively to the 68\% C.L. and the 95\% C.L. regions.}\label{fig:MixedDEG}
\end{center}
\end{figure}

\begin{figure}[!tb]
\begin{center}
\includegraphics[width=1.0\textwidth]{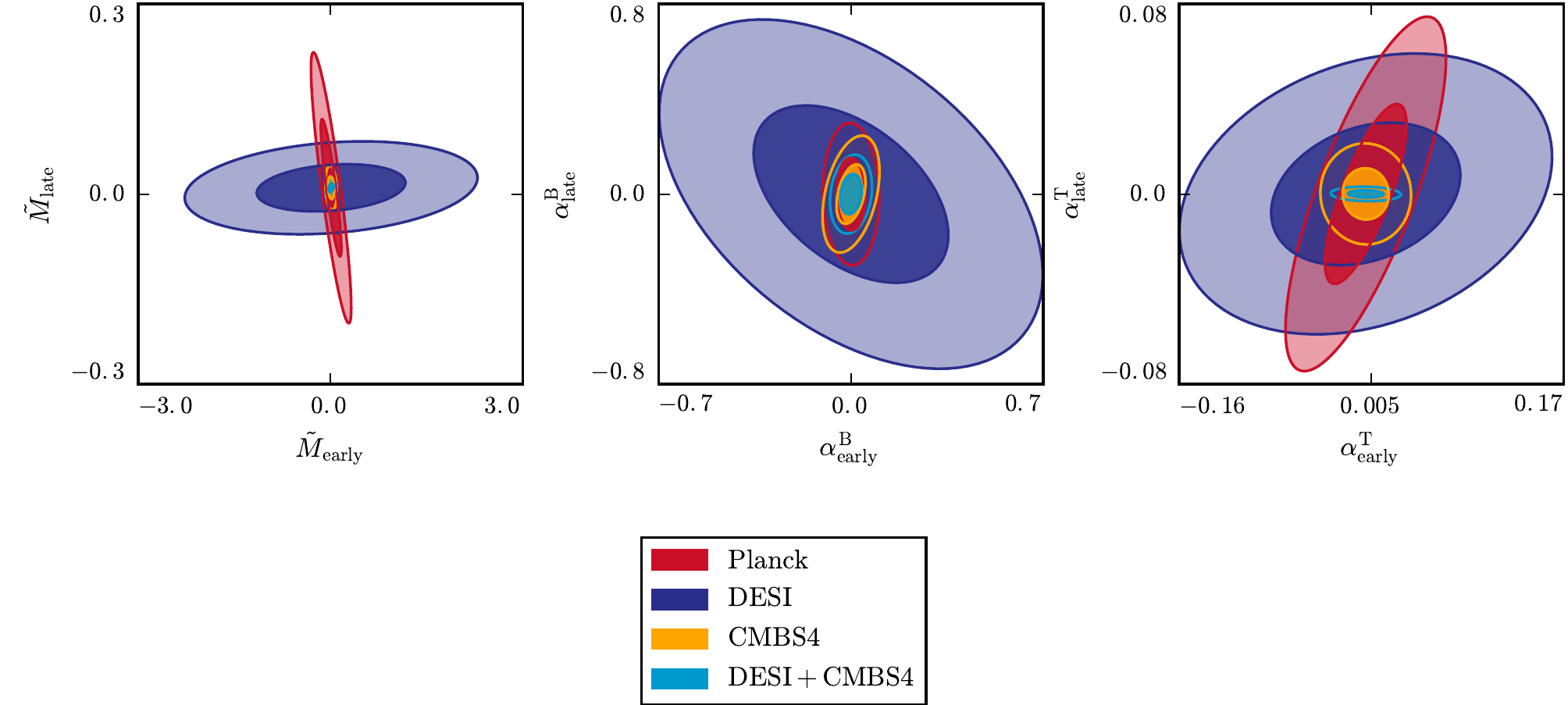}
\caption{Forecast marginalized constraints on Horndeski couplings. The left panel shows the joint constraints on the effective Planck mass at early times $\tilde{M}_{\rm early}$ and late times $\tilde{M}_{\rm late}$. The central panel shows the joint constraints on the Horndeski braiding coefficient at early times $\alpha^{B}_{\rm early}$ and late times  $\alpha^{B}_{\rm late}$. The right panel shows the joint constraints on the Horndeski tensor speed excess coefficient at early times $\alpha^{T}_{\rm early}$ and late times  $\alpha^{T}_{\rm late}$. The Horndeski kineticity coefficient is unconstrained by all the experimental combinations considered. Different colors correspond to different experiments, as shown in legend. The darker and lighter shades correspond respectively to the 68\% C.L. and the 95\% C.L. regions.}\label{fig:VariationAlpha}
\end{center}
\end{figure}

\begin{figure}[!tb]
\begin{center}
\includegraphics[width=1.0\textwidth]{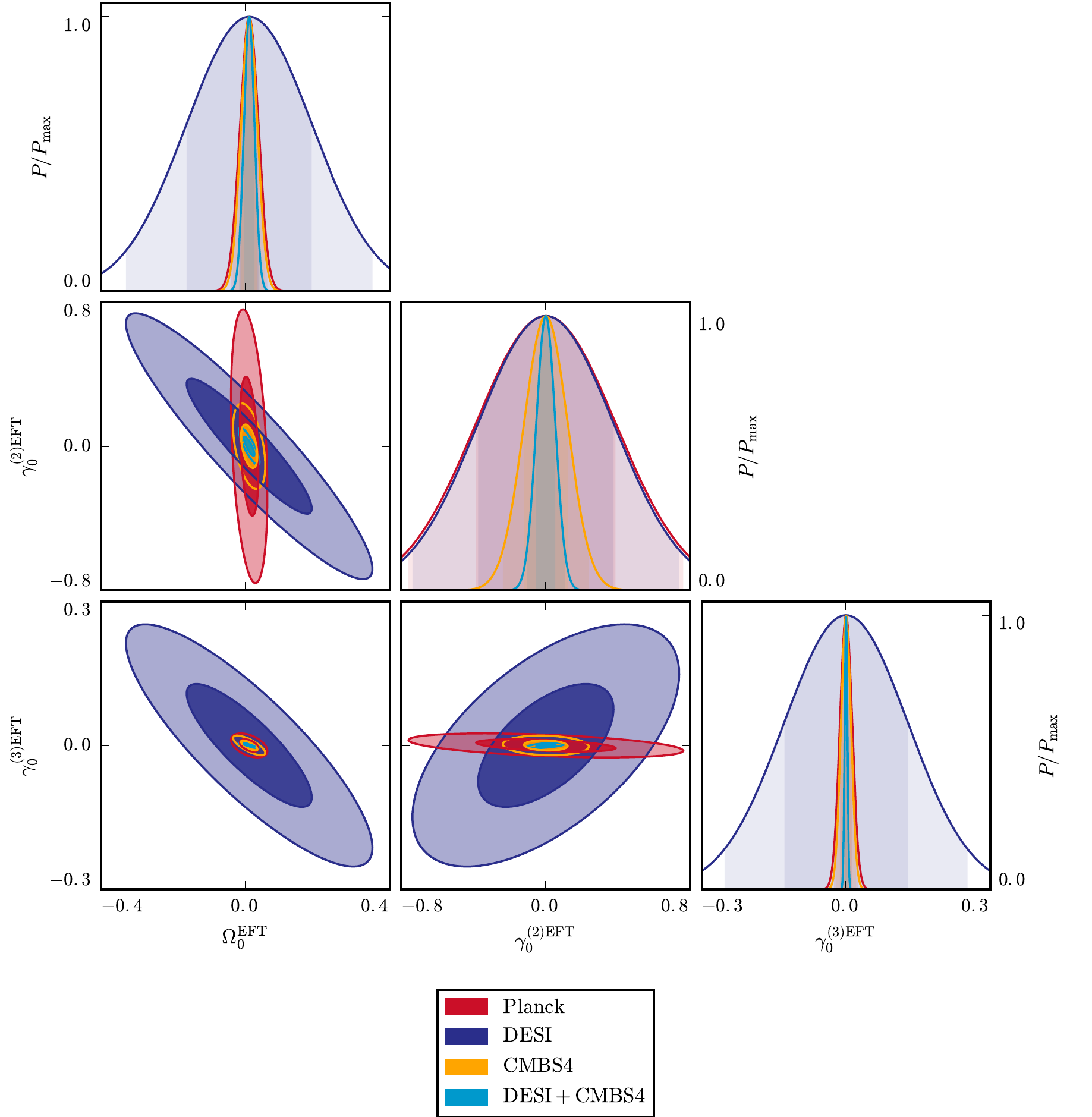}
\caption{Forecast marginalized constraints on constant EFT couplings: $\Omega_0^{\rm EFT}$, $\gamma_{0}^{(2){\rm EFT}}$ and $\gamma_{0}^{(3){\rm EFT}}$. Different colors correspond to different experiments, as shown in legend. The darker and lighter shades correspond respectively to the 68\% C.L. and the 95\% C.L. regions.}\label{fig:ConstantEFT}
\end{center}
\end{figure}

\begin{figure}[!tb]
\begin{center}
\includegraphics[width=1.0\textwidth]{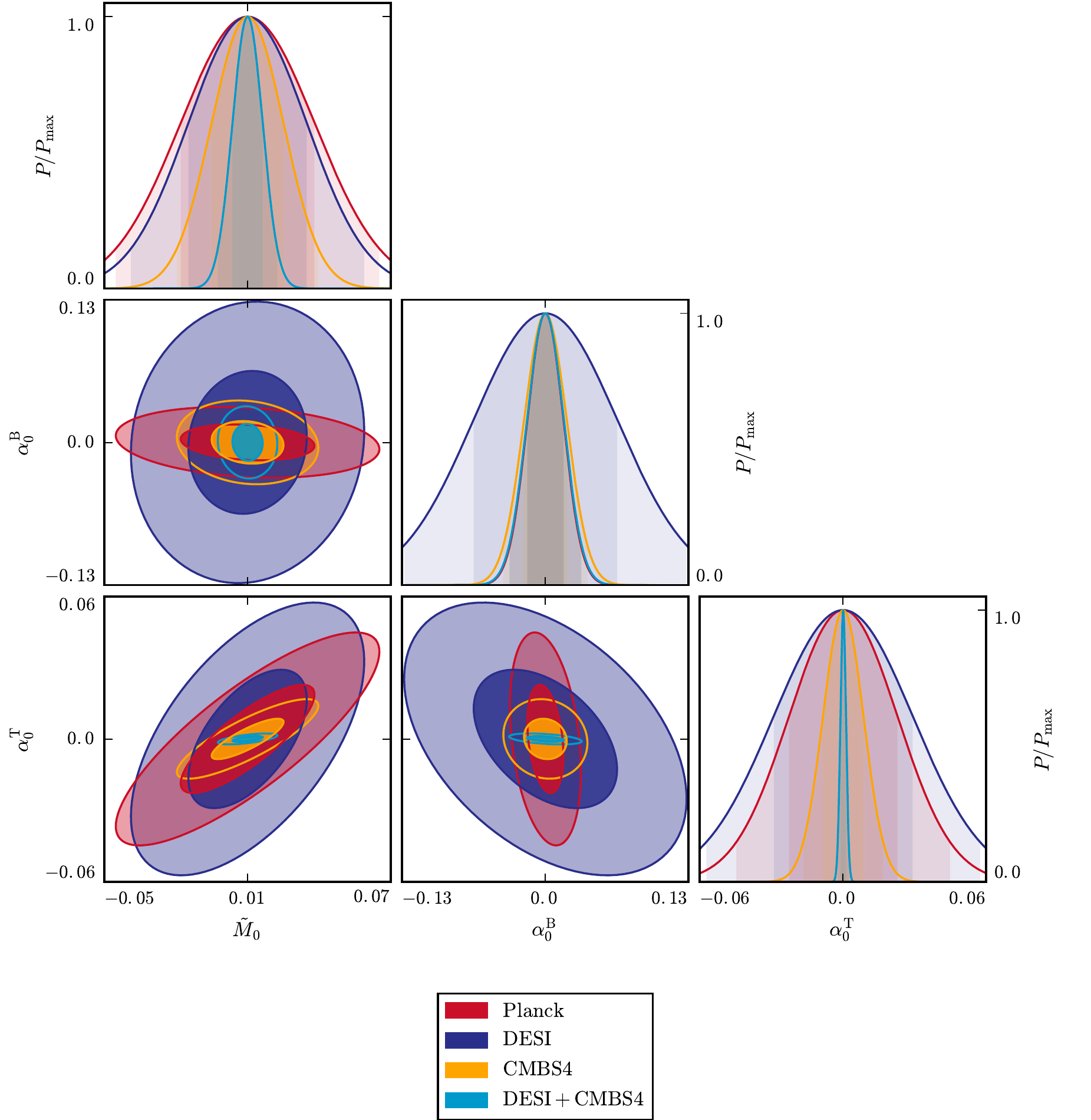}
\caption{Forecasted marginalized constraints on constant Horndeski couplings: $\tilde{M}_{0}$, $\alpha^{\rm B}_0$ and $\alpha_0^{\rm T}$. Different colors correspond to different experiments, as shown in legend. The darker and lighter shades correspond respectively to the 68\% C.L. and the 95\% C.L. regions.}\label{fig:ConstantAlpha}
\end{center}
\end{figure}

\section{Summary}
Dark energy is an elusive cosmological component which drives many current efforts in cosmology. The model parameter space is broad, and CMB-S4 provides a powerful probe to distinguish different models of dark energy through their effects on both the CMB power spectrum itself and the lensing of the CMB.
Through dark energy parameters and models, these measurements also make predictions for what should be seen in the cluster and kSZ observables 
that are independently measured by CMB-S4 as well as observables of and cross-correlation with other dark energy probes.   
Violations of these predictions would trigger a fundamental change in the paradigm that underlies
the parameters and models.  Verification of these predictions bringing us closer to the goal of 
robustly determining the physics of dark energy.

 
\chapter{CMB Lensing}

\bigskip

\def\nnu{N_{\mathrm eff}}
\def\gtrsim{\raise-.75ex\hbox{$\buildrel>\over\sim$}}

\begin{quotation}

The distribution of matter in the Universe contains a wealth of information about the primordial density perturbations and the forces that have shaped our cosmological evolution. Mapping this distribution is one of the central goals of modern cosmology. Gravitational lensing provides a unique method to map the matter between us and distant light sources, and lensing of the CMB, the most distant light source available, allows us to map the matter between us and the surface of last scattering.

Gravitational lensing of the CMB can be measured because the statistical properties of the primordial CMB are exquisitely well-known. As CMB photons travel to Earth from the last scattering surface, they are deflected by intervening matter which distorts the observed pattern of CMB anisotropies and modifies their statistical properties. These distortions can be used to create a map of the gravitational potential that altered the photons' paths. The gravitational potential encodes information about the formation of structure in the Universe and, indirectly, cosmological parameters like the sum of the neutrino masses.  CMB-S4 is expected to produce high-fidelity maps over large fractions of the sky, improving on the signal-to-noise of  the \planck\ lensing maps by more than an order of magnitude.  These maps will inform many of the science targets discussed throughout the book and can also be used to calibrate and enhance results of upcoming galaxy redshift surveys or any other maps of the matter distribution.  Unfortunately, lensing also obscures our view of the CMB.  By measuring and removing the effects of lensing from the CMB maps, we sharpen our view of primordial gravitational waves and our understanding of the very early Universe more generally.

\end{quotation}

\section{Introduction to CMB Lensing}
\label{sec:lensing_intro}

As CMB photons travel from the last scattering surface to Earth, their travel paths are bent by interactions with intervening matter in a process known as \textit{gravitational lensing}.
This process distorts the observed pattern of CMB anisotropies, which has two important consequences:\\
 
$\bullet$ CMB lensing encodes a wealth of statistical information about the entire large-scale structure (LSS) mass distribution, which is sensitive to the properties of neutrinos and dark energy.\\

$\bullet$ CMB lensing distortions obscure our view of the primordial Universe, limiting our power to constrain inflationary signals; removing this lensing noise more cleanly brings the early Universe and any inflationary signatures into sharper focus.\\

Gravitational lensing of the CMB can be measured by relying on the fact that the statistical properties of the primordial CMB are well known.
The primordial (un-lensed) CMB anisotropies are statistically isotropic.
Gravitational lensing shifts the apparent arrival direction of CMB photons, which breaks the primordial statistical isotropy;
lensing thus correlates previously independent Fourier modes of the CMB temperature and polarization fields.
These correlations can be used to make maps of the LSS projected along the line-of-sight; see the discussion in Section \ref{kappaMap}.

A CMB-S4 experiment will make radical improvements in CMB lensing science:
high sensitivity will enable lensing maps that have much higher signal to
noise; the high polarization sensitivity will allow
lensing maps that are much less sensitive to foreground contamination;
multi-frequency coverage will greatly reduce foreground 
contamination in the temperature-based lensing estimates, 
allowing lensing maps with higher resolution; and
large area coverage will provide maps for cross-correlation with maps of large
scale structure from next generation surveys, including WFIRST, Euclid, and LSST.

The information contained in lensing mass maps can be accessed and used in several ways.
First, the power spectrum of the lensing deflection map is sensitive to any physics that modifies how structure grows, such as dark energy, modified gravity, and the masses of neutrinos.
In Section \ref{measuringLensing}, we discuss how the lensing power spectrum is measured.
Second, lensing mass maps can be compared to other tracers of LSS at lower redshifts such as the distribution of galaxies and optical weak lensing shear maps.  By cross correlating, for example, CMB lensing and optical shear mass maps, which are each derived from lensed sources at widely differing redshifts, one can enhance dark energy constraints and improve the calibration of systematic effects.
Cross-correlation science with CMB lensing maps is discussed in Section \ref{cross}.
Finally, lensing distortions partially obscure potential signatures of cosmic inflation in the primordial B-mode polarization signal.
With precise measurements, this lensing-induced noise can be characterized and removed in a procedure known as ``delensing.''
Because B-mode polarization measurements from CMB-S4 are expected to be lensing-noise dominated, delensing will be critical to maximize the information we can infer about cosmic inflation; see the discussion in Section \ref{delens}.

We discuss systematics from astrophysical and instrumental effects that can impact the lensing signal, and ways to mitigate these effects, in Section \ref{syst}.  Section \ref{forecasts} discusses forecasted parameter constraints with and without CMB lensing and demonstrates the importance of CMB lensing measurements for all the key CMB-S4 science goals.

\section{Measuring CMB Lensing}\label{measuringLensing}

\subsection{Constructing a Lensing Map}\label{kappaMap}

A map of the CMB lensing deflection field is a direct probe of the projected matter distribution that exists in the observable Universe. This lensing map is a fundamental object for nearly all areas of CMB lensing science: it is used to measure the lensing power spectrum, measure cross correlations between CMB lensing and external data sets, and to delens maps of the B-mode polarization.  

To date, all maps of the lensing field have been constructed using the quadratic estimator by \cite{Hu:2001kj}.  This estimator uses information about the off-diagonal mode-coupling in spherical harmonic space that lensing induces in order to reconstruct the deflection field.  An estimate for the amount of lensing on a given scale is obtained by averaging over pairs of CMB modes in harmonic space separated by this scale. CMB-S4 will greatly improve over existing measurements by having high angular resolution and high sensitivity for both temperature and polarization CMB maps.

One way that the high angular resolution and sensitivity
of CMB-S4 improves upon the \planck\ measurement is simply
by increasing the number of CMB modes imaged on scales smaller than the \planck\ beam.  Imaging CMB modes between $l=2000$ and 4000, which can be achieved with CMB-S4 yields considerable gain in the accuracy of the lensing power spectrum measurement.

\begin{figure}[htbp]
\centering
\includegraphics[width=0.6\textwidth]{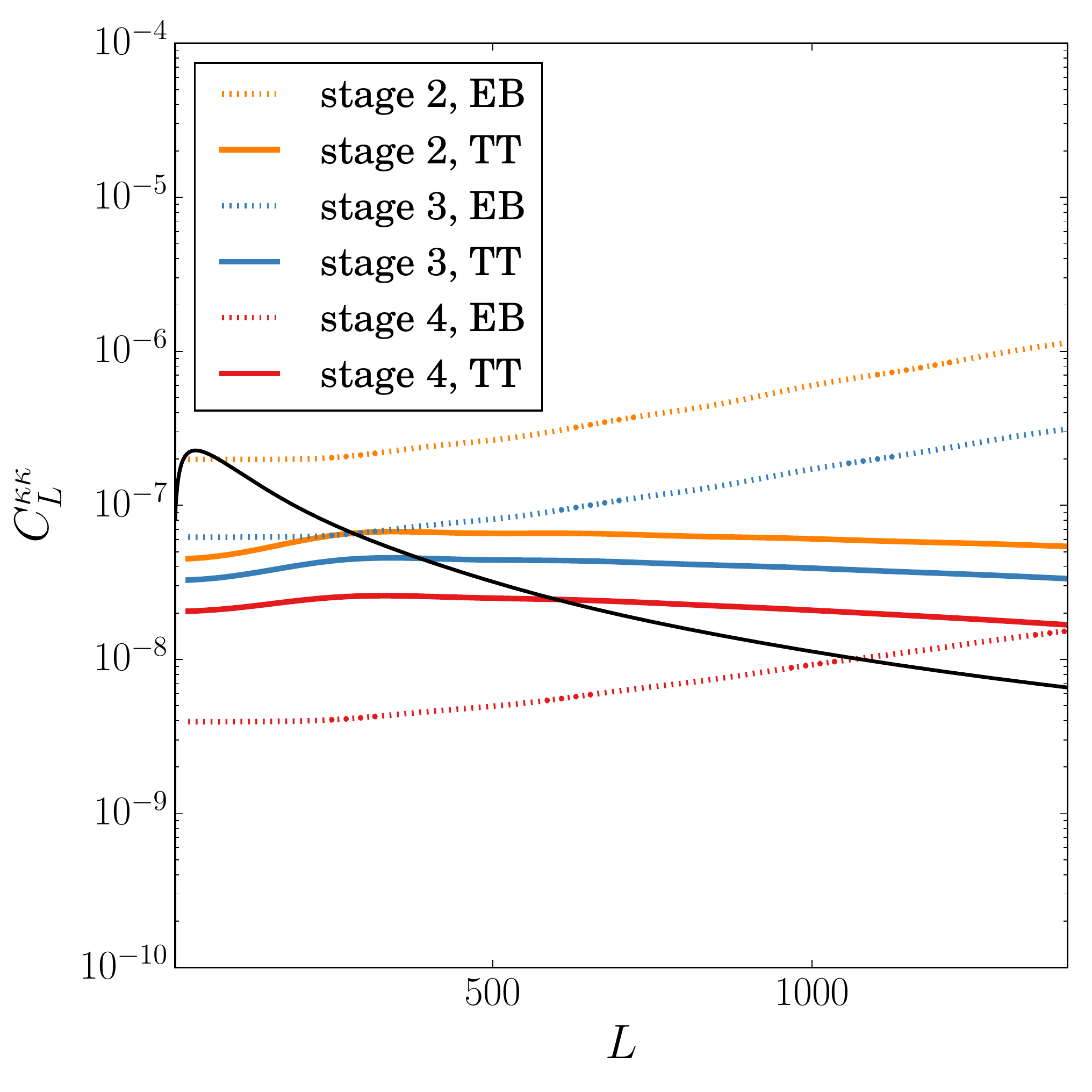}
\caption{Signal and noise-per-mode curves for three experiments. ``Stage 2'' is meant to represent a current-generation survey like SPTpol or ACTPol and has $\Delta_T = 9 \mu$K-arcmin; ``Stage 3'' is an imminent survey like SPT-3G or AdvACT, with $\Delta_T = 5 \mu$K-arcmin; and ``Stage 4'' has a nominal noise level of  $\Delta_T = 1 \mu$K-arcmin.   These noise-per-mode curves do not depend on the area of sky surveyed.  All experiments assume a 1.'0 beam, and a maximum $l$ of 5000.}  
\label{n0s_s4}
\end{figure}

\begin{figure}[htbp]
\centering
\includegraphics[width=0.45\textwidth]{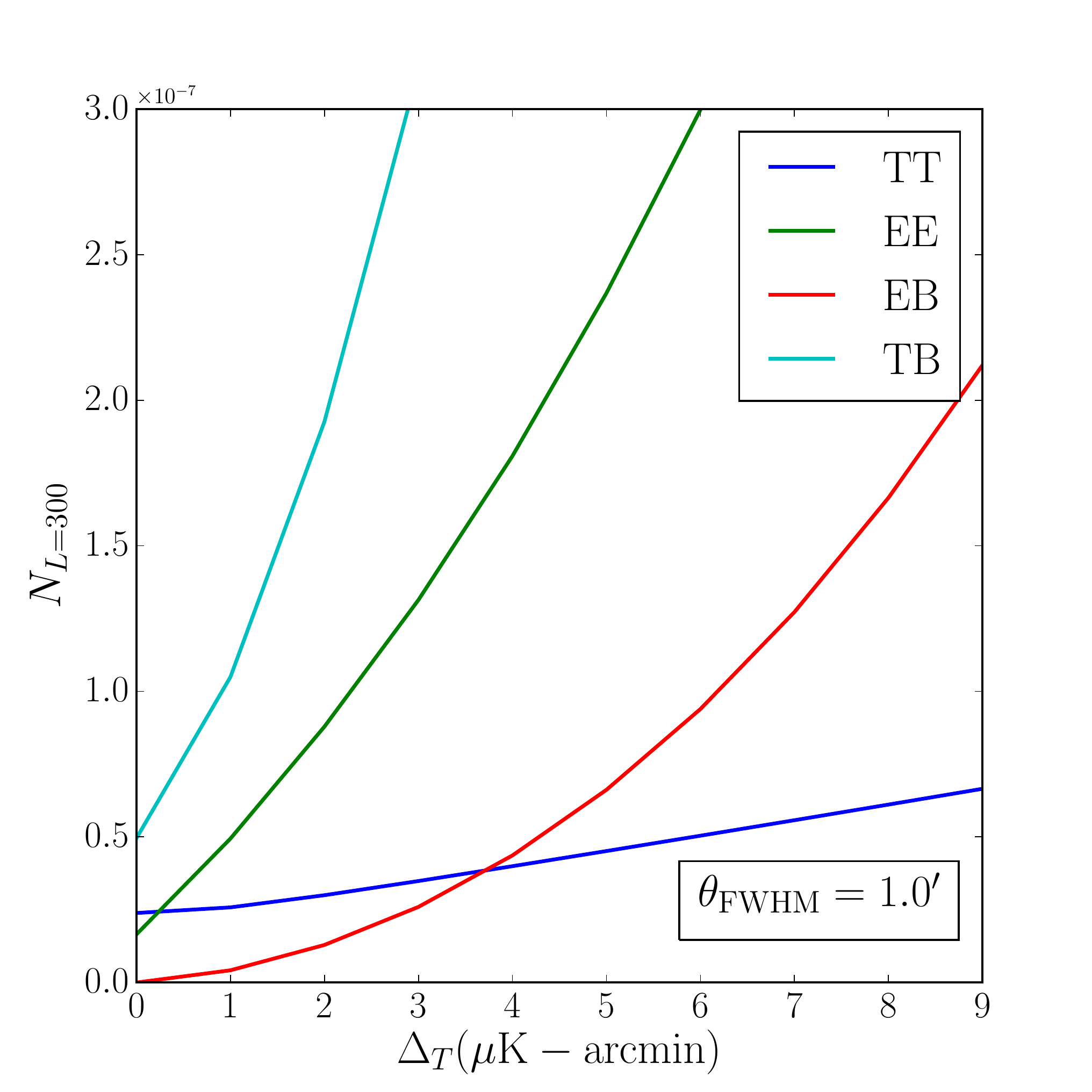}
\includegraphics[width=0.45\textwidth]{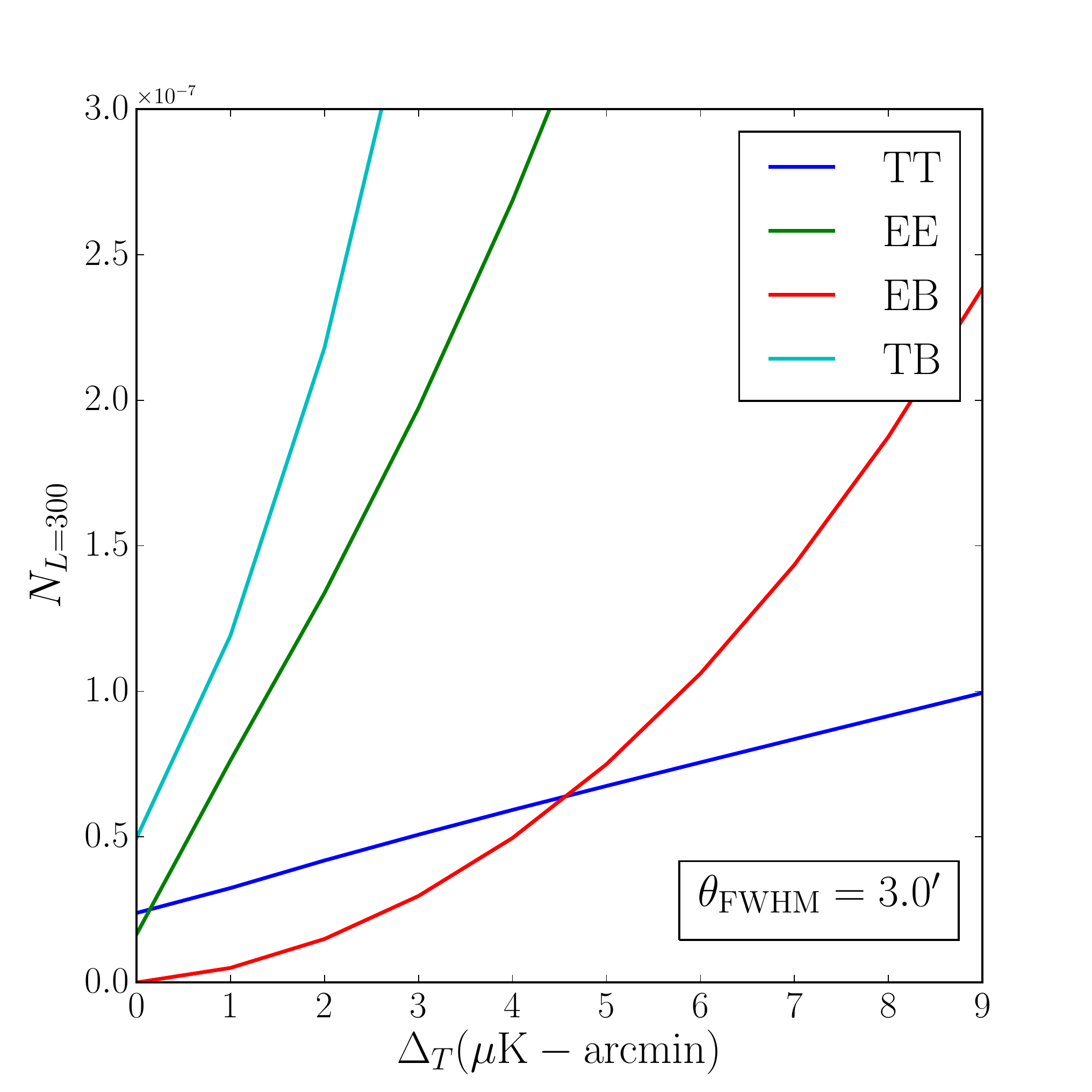}
\caption{Noise per mode in the lensing field for different lensing estimators at $L = 300$.  Left panel is for 1 arcmin resolution, and right panel is for 3 arcmin resolution.  For a 1 and 3 arcmin resolution experiment, the EB polarization estimator yields lower noise than the temperature estimator, below 4$\mu$K-arcmin and 5$\mu$K-arcmin noise in temperature respectively.}
\label{crossoverPlot}
\end{figure}

However, the primary reason for the increased power of CMB-S4 lensing measurements is this experiment's ability to measure CMB polarization with unprecedented sensitivity. To date, CMB lensing results have had their signal-to-noise dominated by lensing reconstructions based on CMB temperature data (see Figure \ref{n0s_s4}). Such lensing measurements in temperature are limited for two reasons. First, they are limited by systematic biases from astrophysical foregrounds and atmospheric noise. Second, the signal-to-noise on lensing measurements from temperature is intrinsically limited by the cosmic variance of the unlensed CMB temperature field. Due to the unprecedented sensitivity of CMB-S4, the bulk of the lensing signal-to-noise will now be derived from CMB polarization data (see Figures \ref{n0s_s4} and \ref{crossoverPlot}).  Polarization lensing reconstruction will allow CMB-S4 to overcome both of these limitations. For the former, the challenges of astrophysical emission and atmospheric noise are much reduced in polarization data. For the latter, low-noise polarization lensing measurements are not limited by primordial CMB cosmic variance, because they make use of measurements of the B-mode polarization, which contains no primordial signal on small scales. To fully exploit the lack of limiting primordial signal in the B-mode polarization, maximum likelihood lensing reconstruction algorithms can be used, which use iteration to surpass the quadratic estimator. This iterative lensing reconstruction procedure is discussed in more detail in Section \ref{delens}.

\subsection{Lensing Power Spectrum}\label{kappaPower}

The power spectrum of reconstructed CMB lensing maps is a measure of the matter power spectrum integrated over redshift.  The lensing power spectrum has a broad redshift response kernel, with most of the contribution coming from $z\sim 1-5$, with a peak at $z\sim 2$ (see Figure \ref{cmb-gal-kernels}).  
Most of the scales probed by the lensing power spectrum are on sufficiently
large scales that they are mainly in the linear regime.  As such, the lensing power spectrum is sensitive to physics which affects the growth of structure on large scales and at high redshift, such as the mass of the neutrinos.

\begin{figure}[htbp]
\centering
\includegraphics[width=0.95\textwidth]{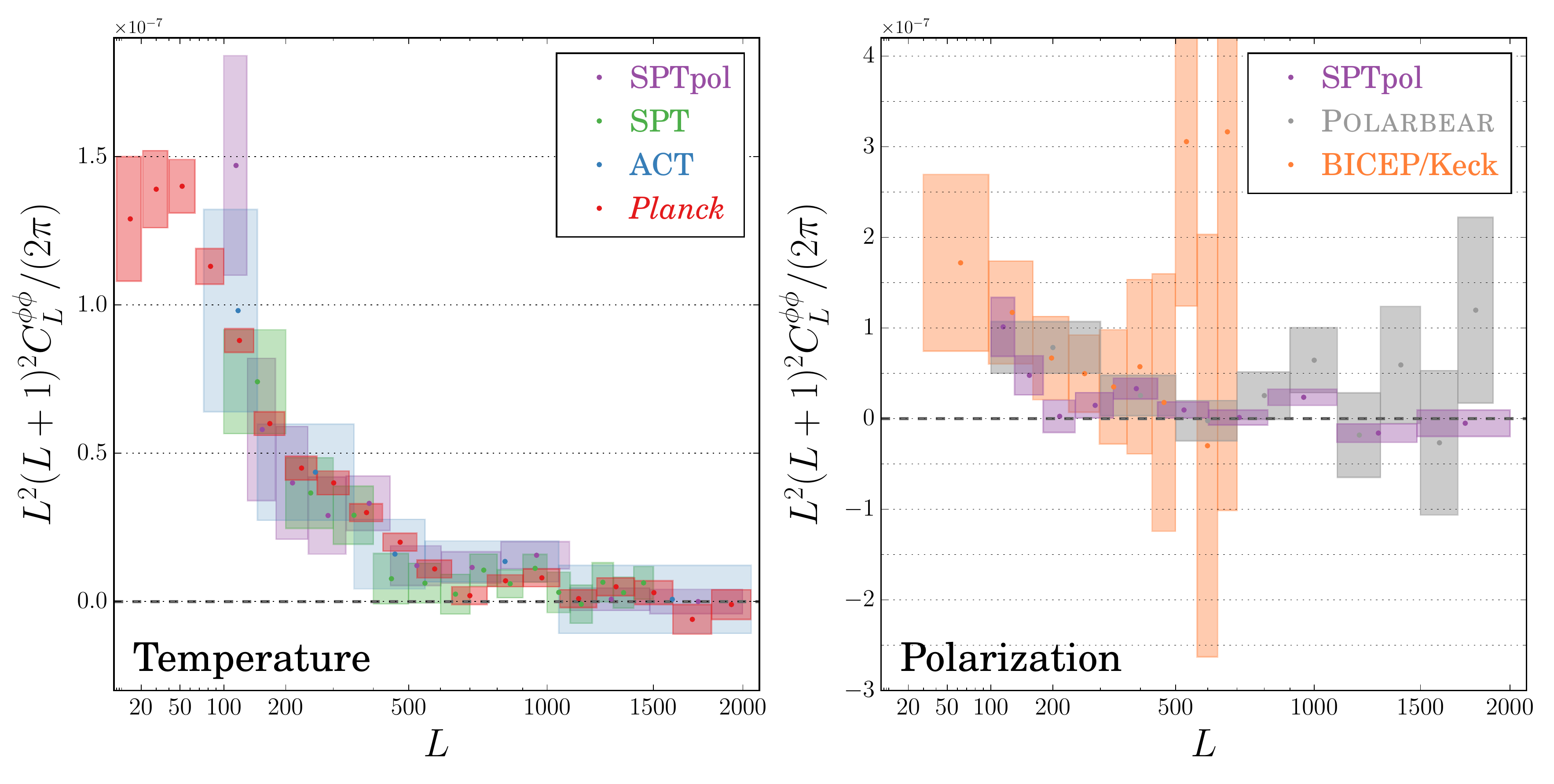}
\caption{Compendium of lensing power spectrum measurements since first measurements in 2011.} 
\label{CMBLensPower}
\end{figure}

The latest measurements of the CMB lensing autospectrum, as of early 2016, are shown in Figure \ref{CMBLensPower}. The first detections were obtained by the Atacama Cosmology Telescope (ACT; \cite{Das:2011ak}) and South Pole Telescope  (SPT; \cite{vanEngelen:2012va}) teams, who analyzed maps of several hundreds of square degrees yielding precisions on the lensing power spectrum of approximately 25\% and 18\% respectively.  The \planck\ collaboration has since provided all-sky lensing maps whose precision on the power spectrum amplitude is approximately 4\% in the 2013 data release and 2.5\% in the 2015 data release.  The first detections of the lensing autospectrum using CMB polarization, which is ultimately a more sensitive measure of lensing for low-noise maps,  have also been obtained \cite{Ade:2013gez,Story:2014hni,Array:2016afx}.

There has been rapid improvement in these measurements over the period of just a few years. 
Early detections of the CMB lensing autospectrum were not sample variance limited over a broad range in $L$ and were only covering a relatively small sky area;  
the  power spectrum of the noise in the CMB lensing reconstruction in the 2015 \planck\ data release is approximately equal to the lensing power spectrum only at its peak of $L \sim 40$, but smaller scales are noise-dominated. Lensing reconstructions from current ground-based surveys (like SPTpol, ACTPol, POLARBEAR) 
are strongly signal-dominated below $L \sim 200$ and noise-dominated on smaller scales.  However, they have been obtained over relatively small sky areas of several hundreds of square degrees. A ground-based survey such as CMB-S4, with wide sky coverage, low-noise, and high resolution, will provide a sample-variance-limited measurement to scales below $L \sim 1000$ (see Figure \ref{n0s_s4}) over a wide area.

\begin{figure}[htbp]
\centering
\includegraphics[width=0.85\textwidth]{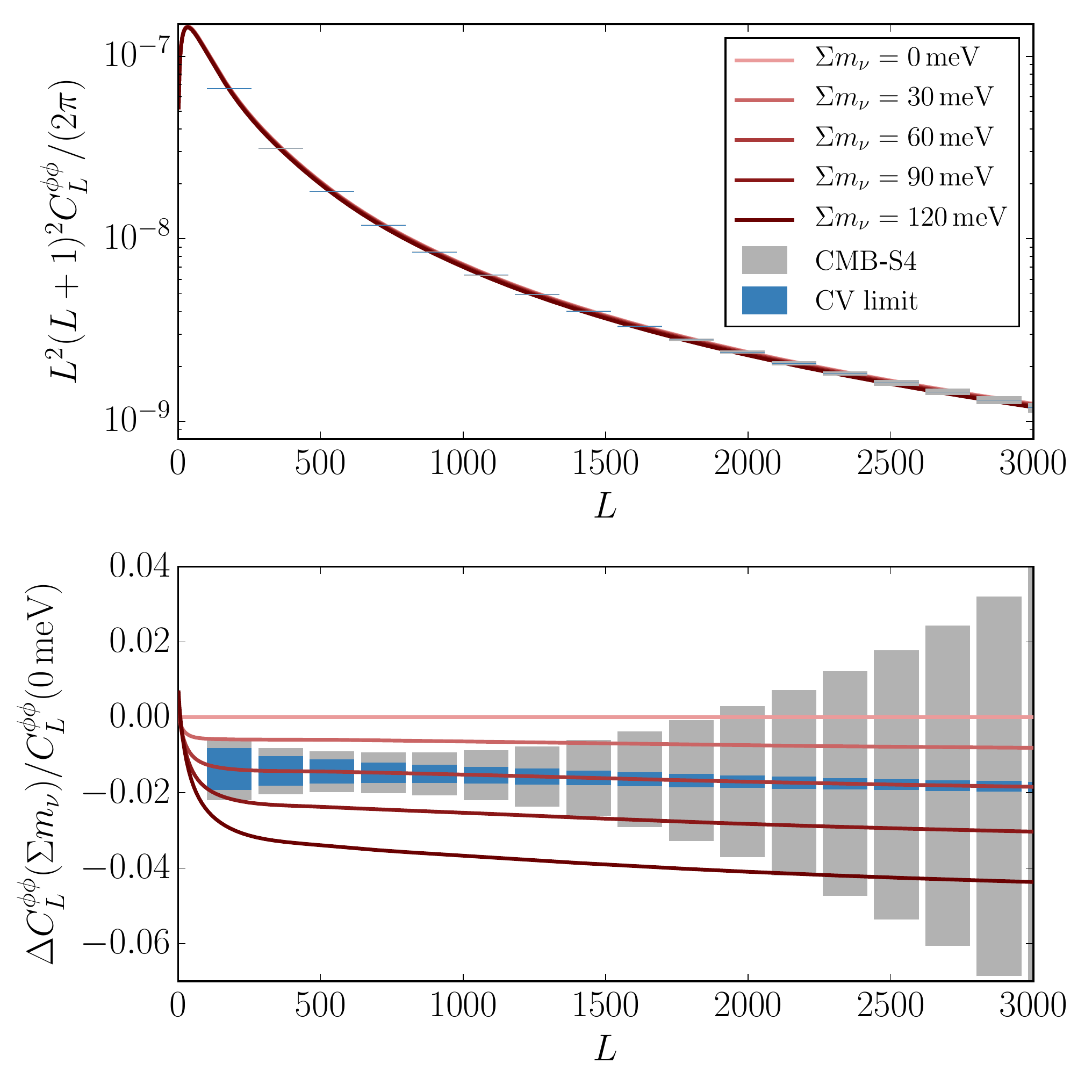}
\caption{Constraining neutrino mass with CMB-S4.  Top: lensing power spectra for multiple neutrino masses (curves) together with forecasted errors for S4.  Bottom: residual from curve at zero neutrino mass.  Error boxes are shown centered at the minimal value of $60$\,meV.  S4 will be targeted to resolve differences in neutrino mass of 20\,meV. }
\label{mnuS4errors}
\end{figure}

Such a measurement holds the promise to qualitatively improve our understanding of cosmology.  While the cosmological parameters describing the standard Lambda-Cold Dark Matter model have been precisely measured, extensions to this model can be constrained by including growth or geometrical information at a new redshift.  From the redshifts probed by CMB lensing, extensions to the standard model such as a non-minimal mass for the sum of the neutrinos, a dark energy equation of state deviating from the vacuum expectation, and a non-zero curvature of the Universe can all be probed to much higher precision than with the primordial CMB alone. Figure \ref{mnuS4errors} shows the expected precision of a CMB-S4 lensing power spectrum measurement and demonstrates its potential, for example, to discriminate between different neutrino mass scenarios (see the Neutrino Chapter for additional details).

\section{Cross Correlations with CMB Lensing}\label{cross}

Cross-correlating CMB lensing maps with other probes of large-scale structure provides a powerful source of information inaccessible to either measurement alone. Because the CMB last-scattering surface is extremely distant, the CMB lensing potential includes contributions from a wide range of 
intervening distances extending to high redshift. As a result, many other cosmic observables trace some of the same large-scale structure that lenses the CMB. These cross-correlations can yield high-significance detections, are generally less prone to systematic effects, and given the generally lower redshift distribution of the other tracers, are probing large-scale structure in exactly the redshift range relevant for dark energy studies (see Figure \ref{cmb-gal-kernels}). 
With CMB-S4, cross-correlations will transition from detections to powerful cosmological probes. 

\begin{figure}[htbp]
\centering
\includegraphics[width=0.85\textwidth]{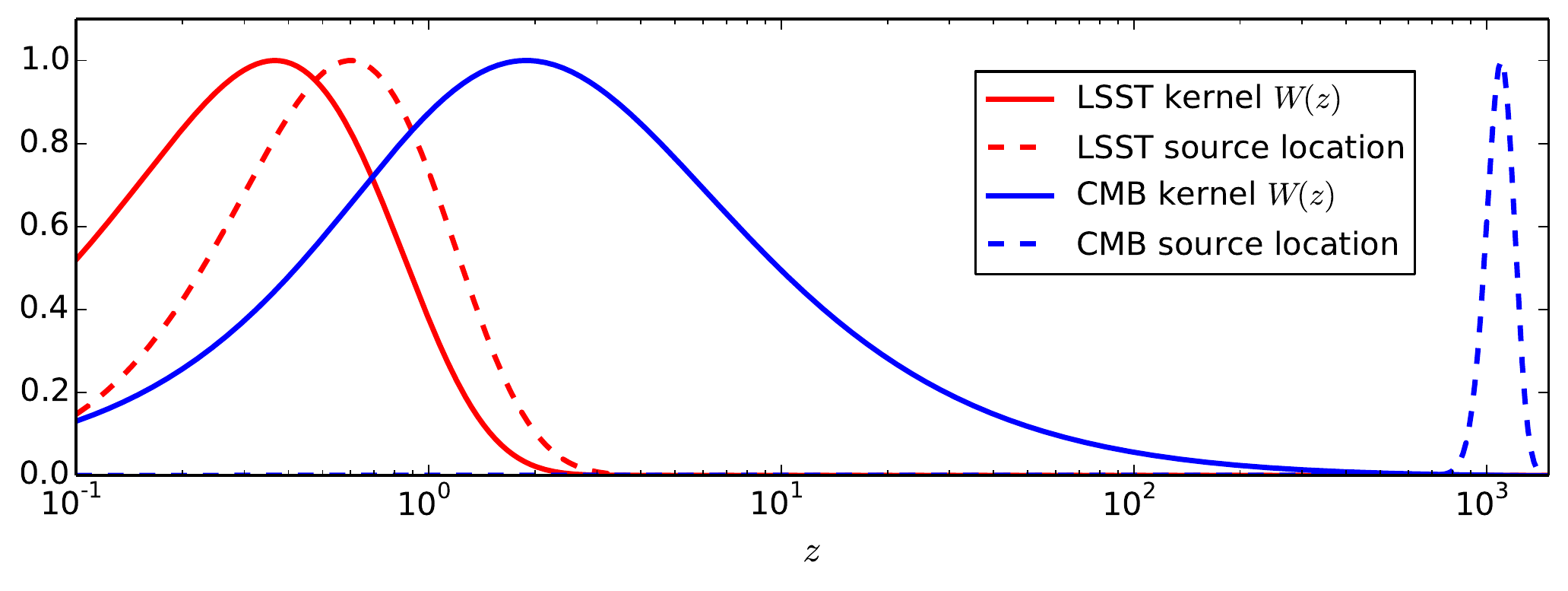}
\caption{Redshift kernel for CMB lensing (blue solid) and for cosmic shear with LSST (red solid), together with the expected redshift distribution of LSST galaxies (red dashed) and the CMB source redshift (blue dashed).}
\label{cmb-gal-kernels}
\end{figure}

\subsection{CMB Lensing Cross Galaxy Density}
Galaxies form in the peaks of the cosmic density field; thus the distribution of galaxies traces the underlying dark matter structure.  This same dark matter structure also contributes to the CMB lensing potential.
Cross-correlating galaxy density distributions with CMB lensing is thus a powerful probe of structure and is highly complementary to galaxy clustering measurements.
Galaxy surveys measure luminous matter while CMB lensing maps directly probe the underlying dark matter structure. Thus these cross-correlations provide a clean measurement of the relation between luminous matter and dark matter.
Cross-correlations between independent surveys are also more robust against details of selection functions or spatially inhomogeneous noise that could add spurious power to auto-correlations.
Additionally, while CMB lensing maps are projected along the line-of-sight, galaxy redshift surveys provide information about the line-of-sight distance; thus cross-correlating redshift slices of galaxy populations allows for tomographic analysis of the CMB lensing signal (see, e.g., \cite{Baxter:2016ziy}, \cite{Miyatake:2016gdc}).
These benefits can lead to improved constraints on cosmology: for example, with LSST galaxies, it has been shown that including cross-correlation with CMB lensing can substantially improve constraints on neutrino masses \cite{Pearson:2013iha}.

CMB lensing was first detected using such a cross-correlation \cite{Smith:2007rg, Hirata:2008cb}.  Since these first detections, cross-correlation analyses have been performed with tracers at many wavelengths, including optically-selected sources \cite{Bleem:2012gm, Sherwin:2012mr, Ade:2013tyw, Baxter:2016ziy, Pullen:2015vtb}, infrared-selected sources \cite{Bleem:2012gm, Geach:2013zwa, DiPompeo:2014yea}, sub-mm-selected galaxies \cite{Bianchini:2014dla}, and maps of flux from unresolved dusty star-forming galaxies \cite{Holder:2013hqu, Hanson:2013hsb, Ade:2013aro, vanEngelen:2014zlh}. 

These cross-correlations between CMB lensing and galaxy clustering have already been used to test key predictions of general relativity, such as the growth of structure \cite{Baxter:2016ziy} as a function of cosmic time, and the relation between curvature fluctuations and velocity perturbations \cite{Pullen:2015vtb}. Cross-correlations using CMB-S4 lensing data will enable percent level tests of general relativity on cosmological scales (see the Dark Energy Chapter for futher details).

On the timescale of the CMB-S4 experiment, a number of large surveys are expected be concurrent or completed, including DESI, WFIRST, Euclid, and LSST.  Due to the high number density of objects detected, wide area coverage, and accurate redshifts, the precision of cross-correlation measurements with these surveys will be much higher than those performed to date.  For example, the amplitude of cross-correlation between the CMB-S4 convergence map and the galaxy distribution from LSST is expected to be measured to sub-percent levels.

\subsection{CMB Lensing Cross Galaxy Shear}\label{lensxlens}

There have been several recent detections of the cross-correlation between lensing of the CMB and galaxy shear \cite{Hand:2013xua, Liu:2015xfa, Kirk:2015dpw}, demonstrating the emergence of a new cosmological tool. In addition, CMB and galaxy lensing can be combined with galaxy surveys for lensing tomography measurements, providing the ability to reconstruct the 3D mass distribution. CMB lensing offers similar signal-to-noise as galaxy shear surveys but provides the most distant source possible, allowing this 3D reconstruction to extend to the edge of the observable Universe and providing a high-redshift anchor for dark energy studies.  The combination of CMB and galaxy lensing with galaxy surveys can also be used to measure cosmographic distance ratios \cite{Miyatake:2016gdc, Singh:2016xey}, which provides a clean, complementary probe of the geometry of our Universe and dark energy. 

CMB lensing can also be used as an external calibration for galaxy shear studies. It has been shown \cite{Vallinotto:2011ge, Vallinotto:2013eva, Das:2013aia} that CMB lensing, galaxy clustering, and galaxy shear data taken together can in principle cross-calibrate each other while still providing precise constraints on cosmological parameters. This has been successfully applied to existing surveys \cite{Liu:2015xfa, Baxter:2016ziy, Miyatake:2016gdc, Singh:2016xey} as a proof of principle. 

In particular, CMB lensing from CMB-S4 can calibrate the shear multiplicative bias for LSST, down to the level of the LSST requirements of $\sim 0.5\%$, as shown in Figure \ref{shear_calibration_cmbs4} \cite{Schaan:2016ois}. This calibration is possible while simultaneously varying cosmological parameters and nuisance parameters (photometric redshift uncertainties and galaxy bias). It is robust to a reasonable amount of intrinsic alignment, to uncertainties in the non-linearities and baryonic effects, and to changes in the photo-z accuracies. This shear calibration is weakly dependent on the sensitivity of CMB-S4, and mostly independent of the beam and maximum multipole included in the CMB lensing reconstruction (see Figure \ref{LSSTshearcalibration_vary_noise_beam_lmax}). A similar shear calibration occurs with CMB-S4 lensing combined with WFIRST or Euclid. 

\begin{figure}[htbp]
\centering
\includegraphics[width=0.7\columnwidth]{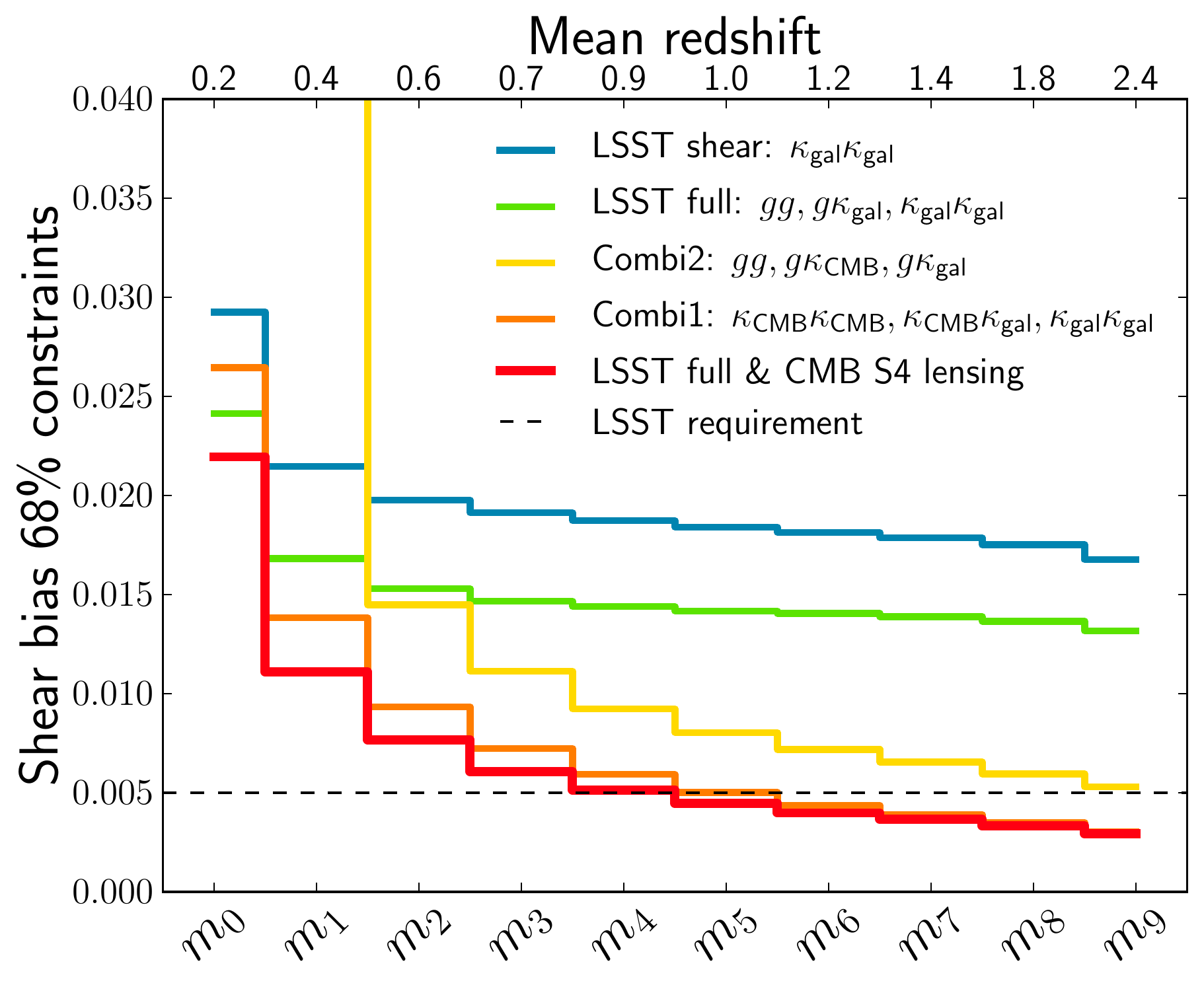}
\caption{$68\%$ confidence constraints on the shear biases $m_i$ for the 10 tomographic bins of LSST, when self-calibrating them from LSST cosmic shear alone (blue), LSST full (i.e. clustering, galaxy-galaxy lensing and cosmic shear; green), combination 1 (orange), combination 2 (yellow) and the full LSST \& CMB-S4 lensing (red). The self-calibration works down to the level of LSST requirements (dashed lines) for the highest redshift bins, where shear calibration is otherwise most difficult. We stress that all the solid lines correspond to self-calibration from the data alone, without relying on image simulations. Calibration from image simulations is expected to meet the LSST requirements, and CMB lensing will thus provide a valuable consistency check for building confidence in the results from LSST.}
\label{shear_calibration_cmbs4}
\end{figure}

\begin{figure}[htbp]
\centering
\includegraphics[width=0.32\columnwidth]{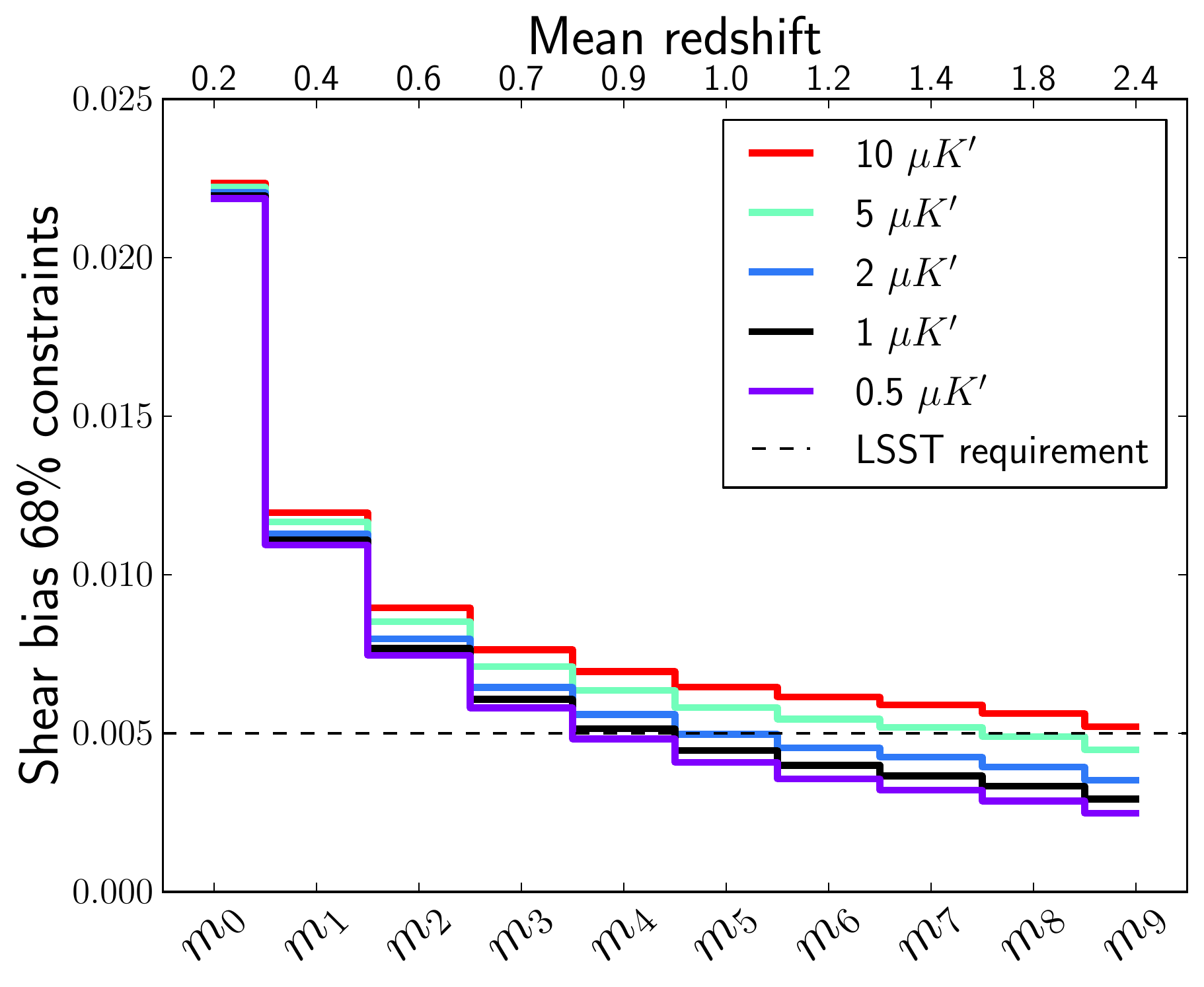}
\includegraphics[width=0.32\columnwidth]{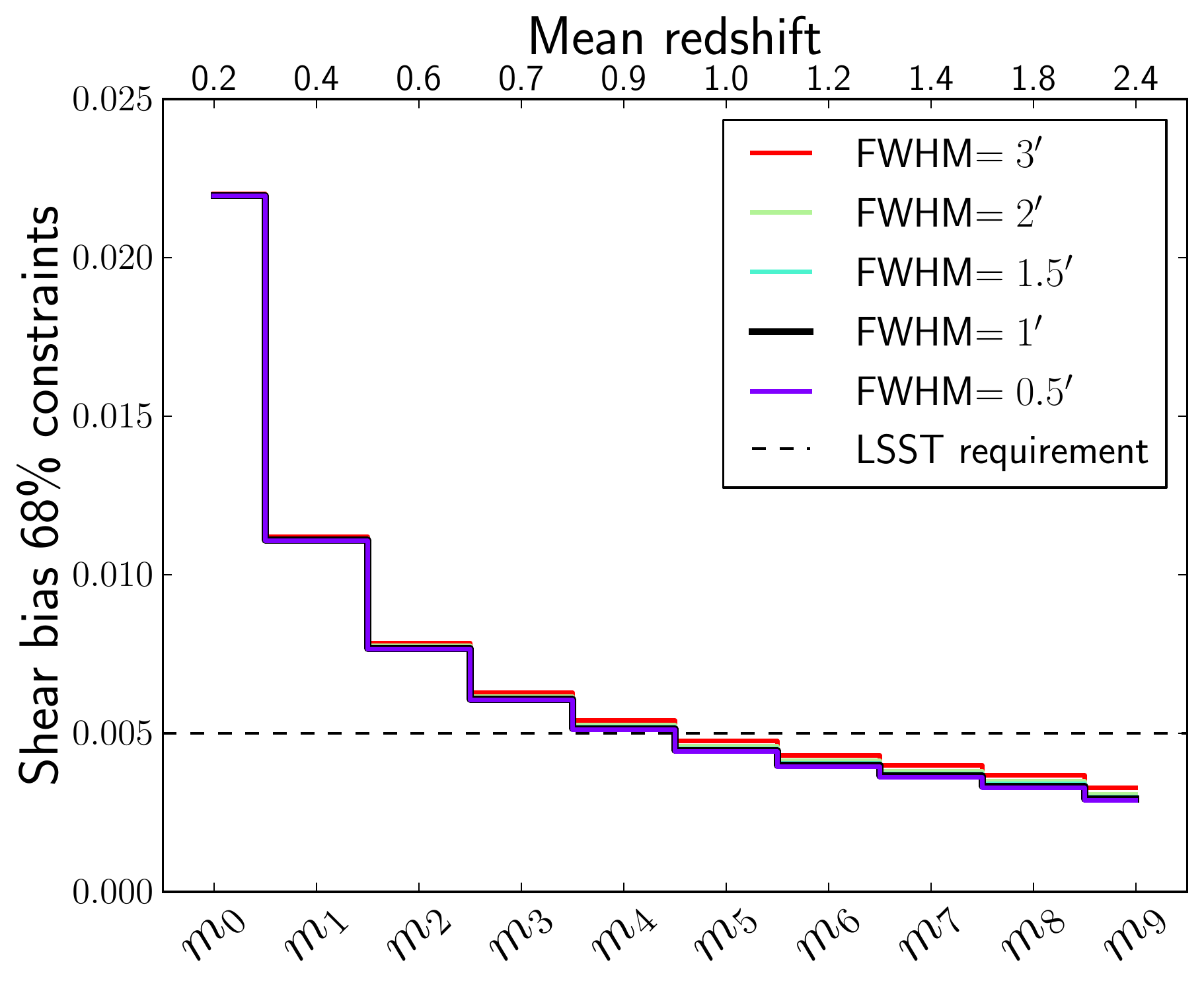}
\includegraphics[width=0.32\columnwidth]{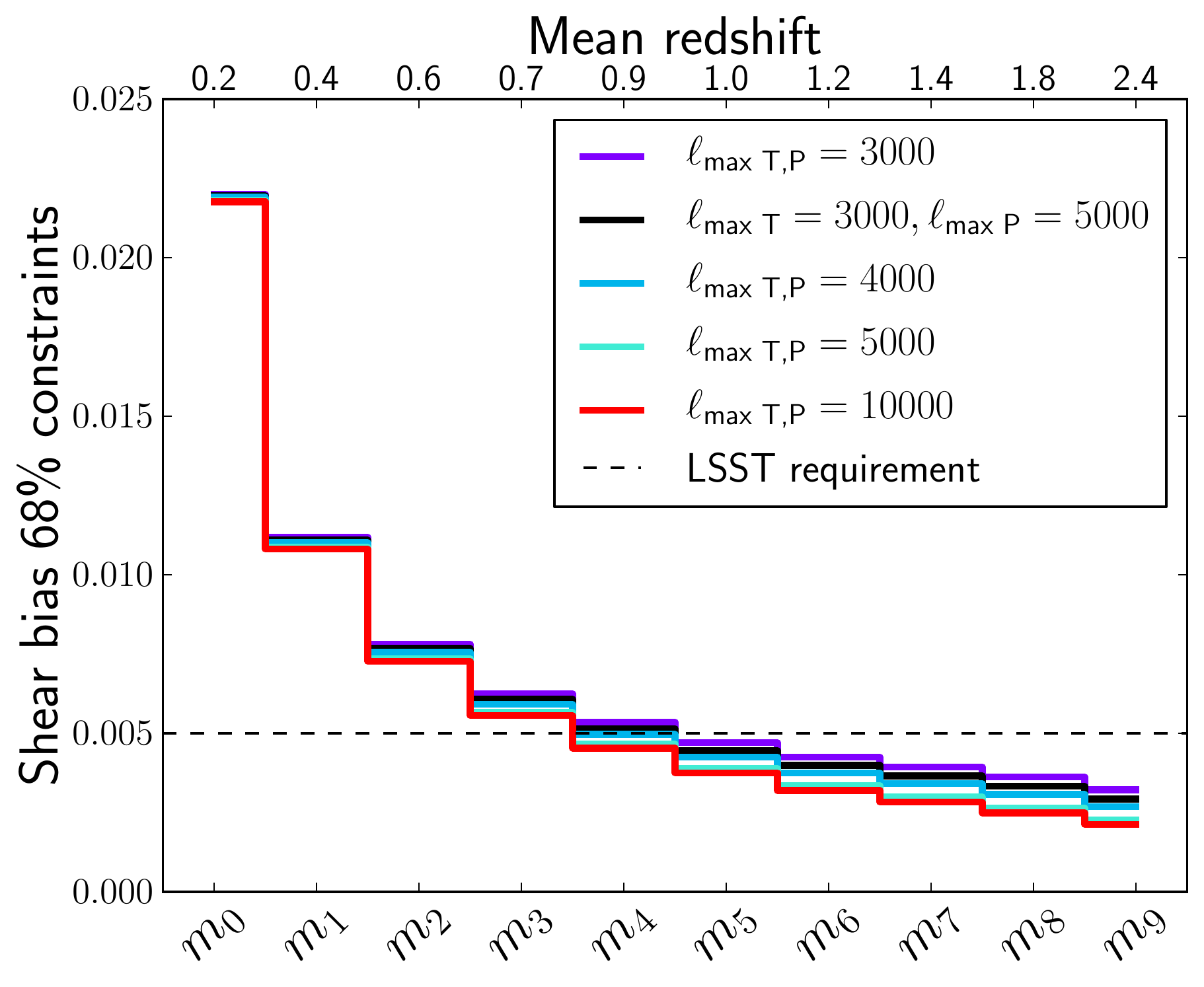}
\caption{Impact of varying CMB-S4 specifications on the shear calibration for LSST. The shear calibration level for the fiducial CMB-S4 (1 arcmin resolution, $1\mu$K-arcmin white noise, f$_{\rm{sky}}$ = 44\%; black solid line) is compared to the one obtained for each variation (solid colored lines). The LSST requirement is shown as the black dashed lines. The shear calibration is weakly dependent on the sensitivity of CMB-S4, and mostly independent of the resolution and maximum multipole included in the CMB lensing reconstruction.}
\label{LSSTshearcalibration_vary_noise_beam_lmax}
\end{figure}

\subsection{CMB Halo Lensing}\label{haloLensing}

In addition to constructing CMB lensing maps of matter fluctuations on relatively large scales ($> \sim 5$ arcmin) as discussed in the preceding sections, one can also make CMB lensing maps capturing arcminute-scale matter distributions. Such small-scale measurements capture lensing of the CMB by individual dark matter halos, as opposed to lensing by larger scale structure represented by the clustering of halos.  This small-scale lensing signature, called CMB halo lensing, allows one to obtain measurements of the mass of these halos.  

Using CMB halo lensing, CMB-S4 will be sensitive to halo masses in the range of $10^{13} M_{\odot}$ to $10^{15} M_{\odot}$.  This corresponds to halos belonging to galaxy groups and galaxy clusters.  Measuring the abundance of galaxy clusters as a function of mass and redshift provides a direct handle on the growth of matter perturbations and consequently, on the equation of state of dark energy (see the Dark Energy Chapter for details).  Galaxy clusters can be identified internally in CMB maps through their wavelength-dependent imprint caused by the thermal Sunyaev-Zeldovich (tSZ) effect.  This technique provides a powerful redshift-independent way of detecting clusters. However, the scaling between the tSZ observable, which is sensitive to baryonic physics, and the cluster mass, which is dominated by dark matter is not precisely constrained.  Calibration of this mass scaling and scatter is currently the dominant systematic for extracting dark energy constraints from cluster abundance measurements. 

Weak lensing of galaxies behind the galaxy cluster is a promising method for mass calibration since it is directly sensitive to the total matter content of the cluster.  Reconstructing the mass profiles of clusters using measurements of the shapes of distant galaxies in deep photometric surveys is an active research program; however, it is often limited by the poor accuracy of source redshifts and the availability of sufficient galaxies behind the cluster, especially for very high-redshift clusters. CMB halo lensing has an advantage here, because the CMB is a source of light which is behind every cluster, has a well defined source redshift, and has well understood statistical properties.  

A general approach for obtaining the average mass of a sample of clusters using CMB halo lensing is to reconstruct the lensing deflection field using a variation of the standard quadratic estimator, stack the reconstructed lens maps at the positions of the clusters, and fit the resulting signal to a cluster profile (e.g NFW). A modified quadratic estimator is used to reconstruct small-scale lensing signals since the standard estimator tends to underestimate the signal from massive clusters \cite{Hu:2007bt}.  This modified estimator makes use of the fact that halo lensing induces a dipole pattern in the CMB that is aligned with the background gradient of the primordial CMB.  The halo lensing signal can be measured with both temperature and polarization estimators, which can be used to cross check each other and reduce systematics. 

\begin{figure}[htbp]
\centering 
\includegraphics[width=0.65\textwidth]{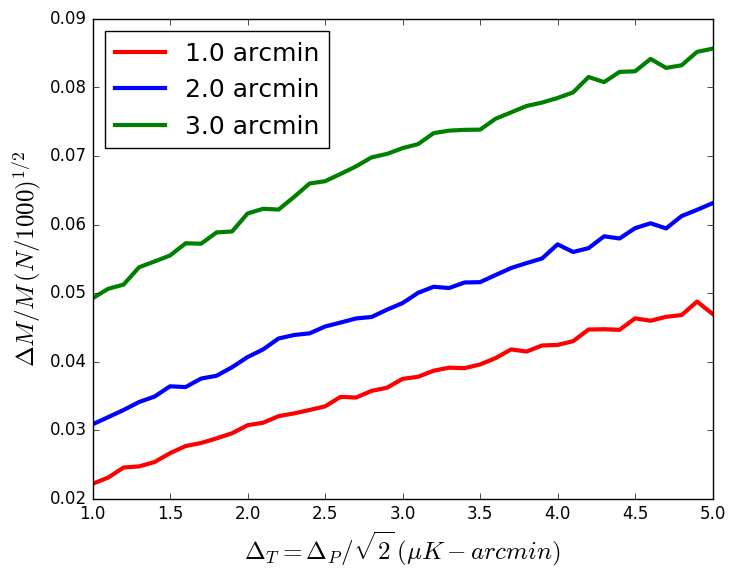}
\caption{Mass uncertainty from CMB halo lensing measurements stacking $10^3$ halos of mass $M_{180\rho_{m_0}} \approx 5\times 10^{14} M_{\odot}$, as a function of instrumental noise and varying instrumental resolution.}
\label{haloLens}
\end{figure}

CMB experiments have only very recently reached the sensitivity required to detect the lensing signal on scales of dark matter halos.  The first detections were reported in 2015 by ACTPol \cite{Madhavacheril:2014slf}, SPT \cite{Baxter:2014frs}, and \planck\ \cite{Ade:2015fva}.  CMB-S4 will be capable of providing precision mass calibration for thousands of clusters which will be an independent cross check of galaxy shear mass estimates and will be indispensable for high-redshift clusters. Figure \ref{haloLens} shows that an arcminute resolution experiment with a sensitivity of around 1$\mu$K-arcmin can determine the mass of 1000 stacked clusters to $\sim 2\%$ precision, combining temperature and polarization maps. The primary systematic in temperature maps is contamination from the thermal SZ effect and radio and infrared galaxies coincident with the halos. This systematic can be mitigated using multi-frequency information due to the spectral dependence of the thermal SZ effect and galaxy emission, a procedure that requires the high sensitivity at multiple frequencies allowed by CMB-S4.  Halo lensing from polarization maps is relatively free of these systematics, and ultimately may be the cleanest way to measure halo masses.  This requires the high polarization sensitivity provided by CMB-S4.

\section{Delensing}\label{delens}

To probe an inflationary gravitational wave signal it is important to have low-noise B-mode polarization maps as discused in the Inflation Chapter. However, for instrumental noise levels below $\Delta_P \simeq 5 \mu$K-arcmin in polarization, the dominant source of noise is no longer instrumental, but instead is from the generation of B-mode polarization by lensing of E-mode polarization from recombination (see Figure \ref{snowmssDelens}).  This B-mode lensing signal has a well-understood amplitude, but the sample variance in these modes in the CMB maps leads to increased noise in estimates of the inflationary B modes. Unlike other sources of astrophysical B-mode fluctuations in the map, it cannot be removed with multifrequency data.  Fortunately, this signal can be removed using map-level estimates of both the primordial E-mode map and the CMB lensing potential $\phi$ with a technique called delensing. However, this procedure requires
precise maps of both the E modes and of the gravitational lensing potential
(which can be obtained from the CMB-S4 data itself).

\begin{figure}[htbp]
\centering
\includegraphics[width=0.7\textwidth]{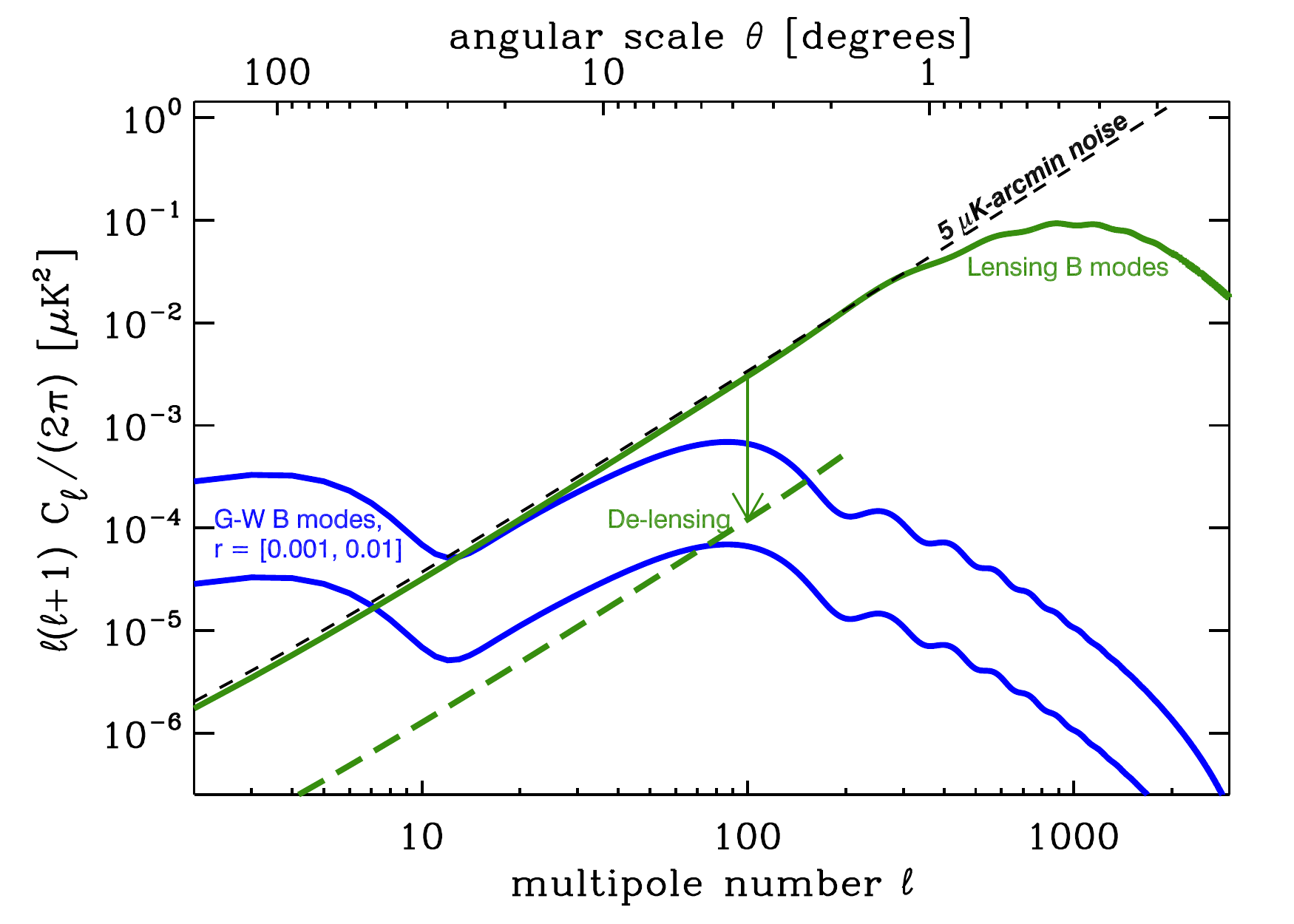}
\caption{The green curve is the power spectrum of lens-induced E-to-B mixing.  Delensing can reduce the amplitude of this effect by large factors (green dashed curve) yielding lower effective noise in B-mode maps.}
\label{snowmssDelens}
\end{figure}

Moreover, delensing will be a crucial part of the reconstruction of the CMB lensing field for CMB-S4, even for science goals like measuring the neutrino mass.  This is because at low noise levels the standard quadratic reconstruction of lensing using the EB estimator \cite{Hu:2001kj} can be improved upon by cleaning the B-mode CMB maps of the lens-induced B-mode fluctuations and then performing lens reconstruction again.  This procedure can be repeated until CMB maps cleaned of the lensing signals are produced (see Figure \ref{iterative}).  

Delensing in principle can be a nearly-perfect procedure: in the limit of no instrumental noise or primordial B modes, the lensing potential and the primordial E-mode map can be nearly-perfectly imaged \cite{Hirata:2003ka}.  However, the finite noise in a CMB-S4 survey will lead to residual lensing B modes which cannot be removed and will act as a noise floor for studying primordial B modes from tensors.  In addition, higher order effects may ultimately limit the reconstruction.

It is important to have relatively high-angular resolution maps in order to obtain the small-scale E and B fluctuations needed for the EB quadratic lensing estimator.  As shown in Figure \ref{sigCon}, quadratic EB lens reconstruction requires high-fidelity measurements of the E and B polarization fields on a variety of angular scales.  For large-scale lenses, such as those at degree scales of $L=300$, the E and B fields contain information to scales of several arcminutes ($l \sim 2000$).  For arcminute-scale lenses at $L = 2000$, the B field must be measured to even smaller scales, $l > 3000$. 

\begin{figure}[htbp]
\centering
\includegraphics[width=0.55\textwidth]{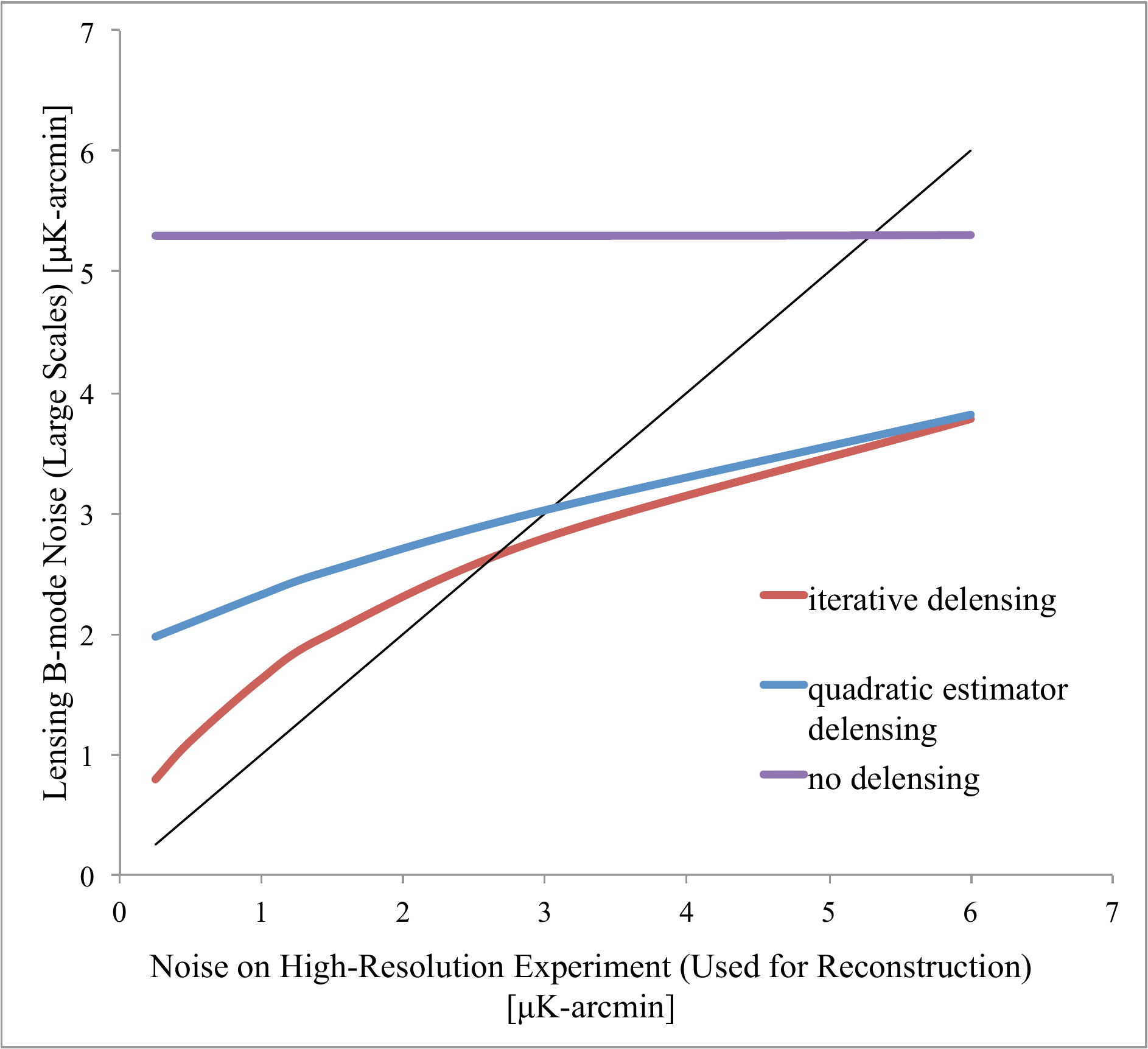}
\vspace{0.3cm}
\caption{The B-mode noise on large scales as a function of the noise level used in EB-based lens reconstruction.  The purple line is for no delensing and shows that lens-induced  E to B mixing manifests as an effective 5 uK-arcmin white noise level.  The blue curve shows the improvement possible when using a lens reconstruction to remove this source of effective noise.  The red curve shows further improvement when the delensing is performed in an iterative fashion.}
\label{iterative}
\end{figure}

\begin{figure}[htbp]
\centering
\includegraphics[width=0.60\textwidth]{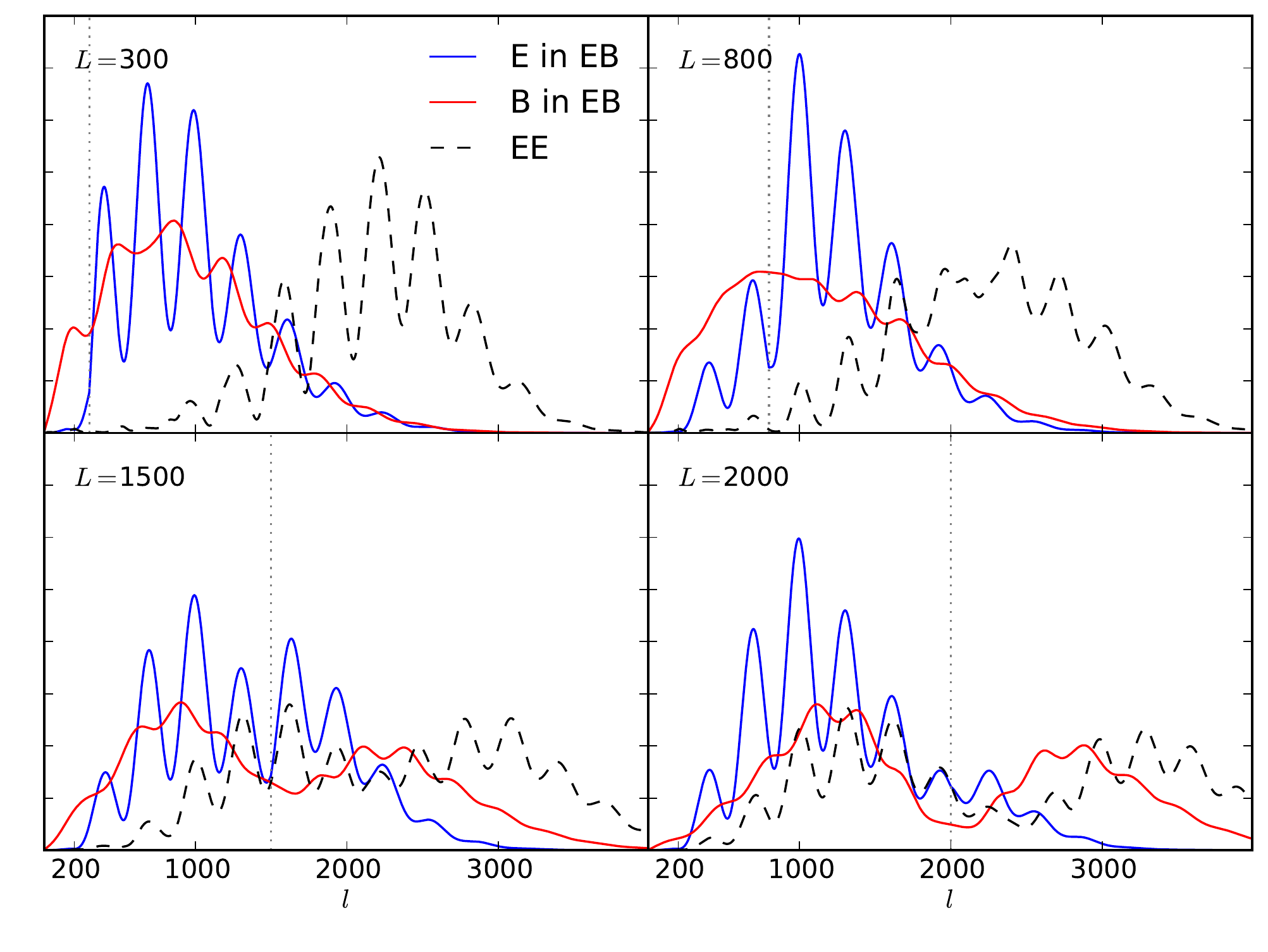}
\caption{Contributions from CMB scales ($\ell$) to lensing reconstruction on four lensing scales ($L$).  The EB estimator is expected to be the main channel for lensing science with CMB-S4.  On degree and sub-degree scales, $L = 300$ and $800$, the estimator uses E and B modes at $\ell \sim 1000$.  On scales of several arcmin, $L = 1500$ and $2000$, the estimator uses B modes on significantly smaller scales.  Figure taken from \cite{Pearson:2014qna}.}
\label{sigCon}
\end{figure}

Potential systematic biases with the delensing procedure are similar to those for measuring the lensing power spectrum. The impact on the reconstructed lensing field of polarized dust and synchrotron emission from the Galaxy and from extragalactic sources are discussed in Section \ref{systAst}, as are ways to mitigate these effects.

Additionally, rather than using an estimate of the CMB lensing field obtained internally from the CMB itself, it is also possible to use other tracers of large-scale structure which are correlated with  CMB lensing \cite{Smith:2010gu}.  In particular the dusty, star-forming galaxies that comprise the cosmic infrared background (CIB) are strongly correlated with CMB lensing due to their redshift distribution which peaks near $z \sim 2$ \cite{Sherwin:2015baa, Simard:2014aqa}.  The level of correlation can be as high as $80\%$ \cite{Ade:2013aro} and can in principle be improved using multifrequency maps of the CIB which select different emission redshifts \cite{Sherwin:2015baa}. However, as shown in \cite{Smith:2010gu}, the gain from delensing with external galaxy tracers is modest, and delensing internally with CMB maps holds far more promise.

\section{Systematic Effects and Mitigation}\label{syst}
The quadratic estimators used for lens reconstruction search for departures from statistical isotropy.  The lens effect locally changes the CMB power spectrum via shear and dilation effects (e.g. \cite{Bucher:2010iv}).  Other sources of deviation from statistical isotropy can thus be confused with lensing effects; these can be of both instrumental and astrophysical origin.

\subsection{Astrophysical Systematics}\label{systAst}
	
Extragalactic sources and tSZ clusters in temperature maps can be troublesome for lensing estimates in two ways: they tend to cluster more strongly in overdense regions (i.e, are non-Gaussian), an effect which
lensing estimators can mistakenly attribute to lensing, while individual sources show up as strong local deviations from statistical isotropy.  

\planck\ \cite{Ade:2013tyw} had to remove the effect of Poisson sources biasing the CMB lensing power spectrum, which left untreated would have shifted their measured lensing power spectrum amplitude by $4\%$, a  1$\sigma$ shift.  For an experiment with lower map noise level and smaller beam, such as CMB-S4, sources can be found and removed to much fainter flux thresholds, making this a much smaller effect.  The largest sources of bias thus come from the three-point and four-point correlation functions of the non-Gaussian clustering of sources and non-Gaussian clustering between the sources and the lensing field.  These biases can be as large as several percent \cite{vanEngelen:2013rla, Osborne:2013nna} and their amplitude is highly model-dependent in temperature-based CMB lensing estimates. However, the extragalactic sources and tSZ clusters that can cause large sources of bias in temperature-based CMB lensing estimates are expected to be nearly unpolarized and therefore not a concern for polarization-based lensing estimates. In
addition, sensitive multi-frequency temperature measurements should be able to spectrally remove these foregrounds through their unique frequency
signatures. In addition, a robust campaign to measure these non-Gaussianities in the CMB data should allow a careful empirical understanding of these
effects, an approach known as ``bias-hardening'' \cite{Osborne:2013nna}.  

Observed levels of the polarization fraction of the diffuse Galactic emission at intermediate and high latitudes, reaching $10\%$ or more, 
have been shown to impact non-negligibly on quadratic estimators for lensing extraction \cite{Fantaye:2012ha}. 
This is due to leakage of the dominating long wavelength modes of the foreground signal onto the scales at which the lensing pattern is reconstructed. 
Therefore, as was the case for the \planck\ data analysis \cite{Ade:2015zua}, lensing extraction has to be validated on foreground cleaned maps output from a component separation process. 

\subsection{Instrumental and Modeling Systematics}\label{systInst}
 	
Given the unprecedented precision targeted by CMB-S4 lensing measurements, the effects of instrumental systematic errors must be investigated and well-controlled. Since lensing results in a remapping or distortion of the sky, beam systematics are a particular concern. 

The main beam systematics that affect CMB measurements are commonly described by differential gain, differential beamwidth, differential ellipticity, and differential pointing and rotation. In \cite{Smith:2008an}, the impact of all these beam systematics on lensing measurements and hence on $r$ and $\sum m_\nu$ was investigated using a Fisher matrix formalism. It was found that for a CMB-S4-type experiment, with $1 \mu $K-arcmin noise and a 3 arcmin beam, the beam characterization from planets or other point sources will be sufficiently accurate that the biases arising from differential gain, differential beamwidth and differential ellipticity are less than one tenth of the one-sigma error on key parameters. Differential pointing and rotation must be controlled to within 0.02 arcmin and 0.02 degrees respectively in order to be similarly negligible.

While ideally the instrument can be designed or shown using measurements to have systematic errors that are negligible, one can also estimate residual beam systematics directly from the data, in a manner analogous to bias-hardening. Many beam systematics result in a known mode-coupling \cite{Yadav:2009eb}.  Their levels can hence be estimated by quadratic estimators and projected out, though complications due to the scan strategy must be accounted for. This method of beam-hardening was first demonstrated in \cite{Ade:2013tyw}.

Another challenge in making high precision lensing power spectrum measurements is improving the theoretical modeling. First, improvements must be made to the modeling of the true lensing power spectrum given cosmological parameters. Second, the sophistication of lensing power spectrum estimators must be improved in order to remain unbiased to high precision, as the presence of higher-order corrections, which have been previously neglected in calculations, can cause simple power spectrum estimators to be biased. For example, often the Gaussianity of the lensing potential is assumed in deriving the lensing power spectrum estimator.  However, when not taking into account the full large-scale structure non-linearity, biases in the lensing measurement can result that can be at the one-sigma level over many bandpowers.  The use of accurate N-body simulations which make use of ray tracing through N-body simulations \cite{Calabrese:2014gla} will be valuable both for testing that the lensing power spectrum is theoretically well modeled and for verifying that the estimator is unbiased to the required precision.

In addition, the lensing power spectrum itself may not be exactly known due to baryonic effects which modify the mass distribution. While this is a challenge for optical weak lensing measurements, investigations with simulations have found that such baryonic effects can be neglected for CMB lensing, at least at the precision achievable by CMB-S4 \cite{Natarajan:2014xba}.

\section{Impact of CMB Lensing/Delensing on Parameters}\label{forecasts}

Measurements of CMB lensing are essential to all the key science goals of CMB-S4.  Since CMB lensing is a sensitive probe of the matter power spectrum, CMB lensing measurements added to measurements of the primordial CMB power spectrum serve to significantly tighten parameter constraints.  In particular, measurements of the CMB lensing power spectrum (4-point signal) and the peak-smearing lensing induces in the CMB primordial power spectrum (2-point signal) yield tight constraints on the sum of the masses of the neutrinos ($\sum {m_\nu}$) (see also Chapter 3).  Cross-correlations of CMB lensing maps with maps of galaxy density and galaxy shear can provide tight constraints on curvature, the dark energy equation of state ($w$), and modified gravity (see also Chapter 6).  Delensing B-mode and E-mode polarization maps will be crucial for maximizing constraints on the tensor-to-scalar ratio ($r$) from inflationary primordial gravity waves and will also tighten constraints on the number of neutrino species ($N_{\rm{eff}}$) (see also Chapters 2 and 4).  

\vspace{0.3cm}
\begin{figure}[htbp]
\centering
\includegraphics[width=0.7\textwidth]{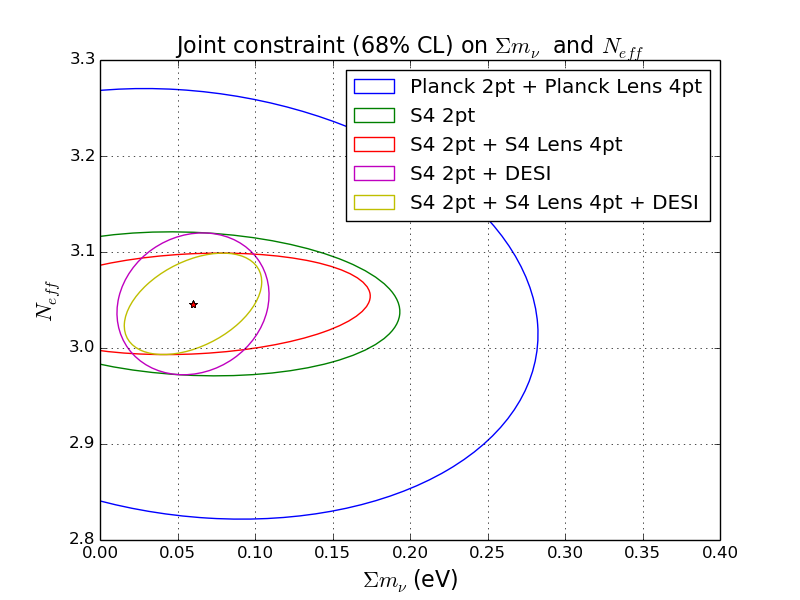}
\caption{CMB lensing signals measured both from lensing-induced mode coupling in the CMB power spectrum (2-point function) and from the CMB lensing power spectrum (4-point function) drive the constraint on the sum of the neutrino masses.  The constraint on the number of relativistic species is improved by about 30\% after delensing the E-mode power spectrum using 4-point lensing information.}
\label{neutrinos}
\end{figure}

Figures \ref{neutrinos}, \ref{fig_rforecast1},   \ref{fig_rforecast2},  and \ref{LSSTdarkEnergy} show the importance of CMB lensing for measuring key cosmological parameters.  Figure \ref{neutrinos} shows constraints on $\sum {m_\nu}$ and $N_{\rm{eff}}$ including the lensing signals from the 2-point and 4-point functions.  In particular for $\sum {m_\nu}$, there is almost no constraining power on $\sum {m_\nu}$ from the primordial CMB power spectrum, so the constraint on $\sum {m_\nu}$ is driven by the lensing-induced mode coupling in the 2-point function and from the 4-point lensing signal.  

For Figure \ref{neutrinos}, we use the CMB-S4 survey and instrument specifications for the wide survey given in Chapter 8.  In particular, no foregrounds are assumed and only white instrumental noise.  Since most of the lensing signal-to-noise is coming from the EB lensing estimator (see Figure \ref{crossoverPlot}), minimal foregrounds and white noise may be reasonable assumptions as foregounds and atmospheric noise are greatly reduced in polarization maps.  This figure also assumes that CMB-S4 will observe $40\%$ of the sky, and \planck\ primordial CMB data is included in the non-overlapping region of the sky ($65\% - 40\% = 25\%$ of sky).  For both CMB-S4 and \planck, temperature modes between $l=50-3000$ and polarization modes between $l=50-5000$ are used.  \planck\ low-ell data between modes $l=2-50$, and lensing modes between $L=40-3000$ are also included.  Here the 2-point CMB power spectrum is iterativley delensed to get tighter parameter constraints, and the covariance between the residual lensing in the 2-point function and the lensing in the 4-point signal is taken into account (see Chapter 8 for details).     

In Figures  \ref{fig_rforecast1} and  \ref{fig_rforecast2} in the Inflation Chapter, the impact of delensing on measuring the tensor-to-scalar ratio $r$ is shown.  Figure \ref{fig_rforecast1} shows how the importance of delensing for measuring $r$ increases as the sky area of the survey gets smaller and the effective map depth gets deeper, given a fixed number of detectors.   For a survey targeting 1\% of the sky, for example, delensing improves $\sigma(r)$ by a factor of 5 to almost an order of magnitude, depending on the actual value of $r$.   Figure \ref{fig_rforecast2} shows that for fixed sky area, delensing is more critical as the number of detectors is increased (or equivalently the map sensitivity is improved).   From these figures, it is clear that to reach the CMB-S4 target of $\sigma(r) = 0.001$, delensing will play a critical role in all survey configurations considered.
 
Figure \ref{LSSTdarkEnergy} in the Dark Energy Chapter shows the constraints on dark matter and dark energy from CMB-S4 obtained from LSST cosmic shear alone and with CMB lensing from CMB-S4 included.  The inclusion of CMB-S4 lensing improves the constraint on $w$ by roughly a factor of 2.  Here it is assumed that LSST will have 30 galaixes per arcminute and cover 40\% of the sky overlapping CMB-S4.  Also just one wide redshift bin for LSST shear analysis is assumed, and we note that the constraints indicated by both blue and green curves can improve with a tomographic analysis.  This is just one example of a number of cross-correlations with CMB-S4 lensing that can constrain the geometry of the Universe, matter abundance, and dark energy properties.

 
\chapter{Data Analysis, Simulations \& Forecasting}


\bigskip

\begin{quotation}

Extracting science from a CMB dataset is a complex, iterative process requiring expertise in both physical and computational sciences. An integral part of the analysis process is played by high-fidelity simulations of the millimeter-wave sky and the experiment's response to the various sources of emission. Fast-turnaround versions of these sky and instrument simulations play a key role at the instrument design stage, allowing exploration of instrument configuration parameter space and projections for science yield. In all three of these areas (analysis, simulations, forecasting), the large leap in detector count and complexity of CMB-S4 over fielded experiments presents challenges to current methods. Some of these challenges are purely computational---for example, performing full time-ordered-data simulations for CMB-S4 will require computing resources and distributed computing tools significantly beyond what was required for \planck. Other challenges are algorithmic, including finding the optimal way to separate the CMB signal of interest from foregrounds and how to optimally combine data from different experimental platforms. To meet these challenges, we will bring the full intellectual and technical resources of the CMB community to bear, in an effort analogous to the unified effort among hardware groups to build the CMB-S4 instrument. A wide cross-section of the CMB theory, phenomenology, and analysis communities has already come together to produce the forecasts shown elsewhere in this document, including detailed code comparisons and agreement on unified frameworks for forecasting.

\end{quotation}

\section{Introduction}

In this chapter we start with an overview of the data analysis pipeline before diving more deeply into its subsets - time-ordered data processing, map-domain processing, and the estimation of statistics and parameters. We then discuss the drivers for the simulation pipeline, and describe in detail its sky modeling and data simulation subsets. From these pieces we then assemble the full production simulation and data analysis pipeline, including its various feedback loops, and consider the computational challenges posed by its implementation and execution. Finally we detail some approaches to mission forecasting, bypassing the most computationally challenging steps in the production pipeline in order to be able to explore the full instument and observation parameter space. Throughout our goal is to describe the current state of the art, note the particular challenges posed by \cmbexp\, and describe how these challenges might be addressed. 

\section{Data Analysis Overview}

The central challenge of CMB data analysis is to reduce both the systematic and statistical uncertainties in the data to a sufficient level to enable well-constrained estimates of the parameters of cosmology and fundamental physics. Typically this is achieved through an iterative process, first mitigating the systematic effects exposed in the particular data domain, and then performing a data-compressing domain transformation in order to reduce the statistical uncertainty by increasing the signal-to-noise.

\begin{figure}[htbp]
\hspace{0.75in}\includegraphics[width=0.875\textwidth]{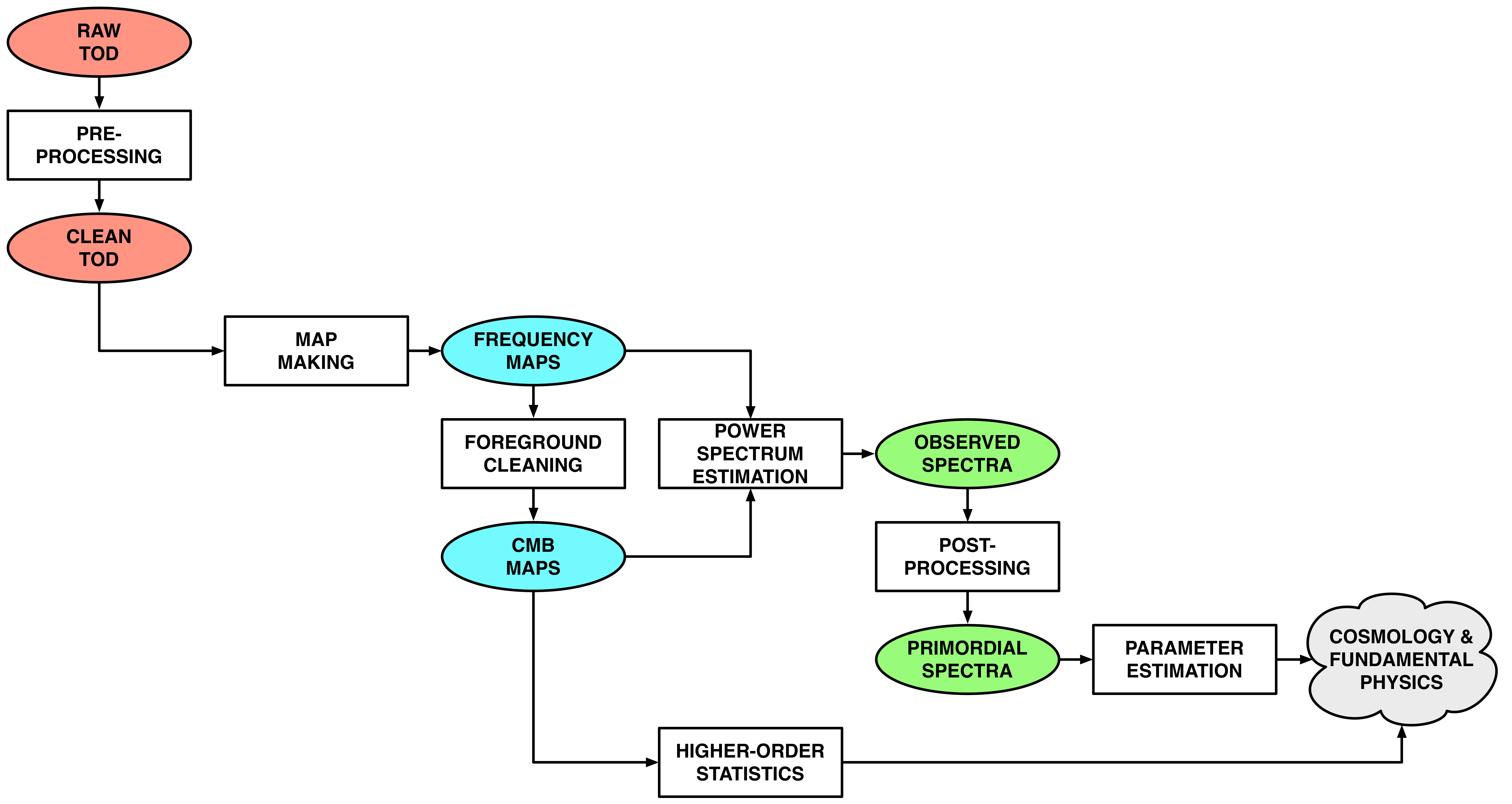}\\
\caption{The schematic CMB data analysis pipeline, showing the reduction from time-samples (red) to map-pixels (blue) to spectral-multipoles (green), including processes to reduce both systematic and statistical uncertainties.}
\label{fig_da}
\end{figure}

As illustrated in Figure \ref{fig_da}, this analysis proceeds in a series of steps:
\begin{description}
\item[ Pre-processing:] The raw time-ordered detector data are calibrated, and gross time-domain systematics are either removed (typically by template subtraction, filtering, or marginalization) or flagged. The goal here is to make the real data match a model that will underpin all subsequent analyses.
\item[Map-making:] At each observing frequency, estimates of the intensity I and the Stokes Q- and U-polarizations of the sky signal are extracted from the cleaned time-ordered data based on their spatial stationarity, typically using some degree of knowledge of the instrument's noise properties.
\item[Foreground Cleaning:] If a sufficient number of frequency maps are available, the CMB can be separated from foreground emission based on its unique spectral signature; if insufficient frequency maps are available then we must use a combination of masking and marginalizing over foreground templates from other sources.
\item[Power spectrum estimation \& higher-order statistics:] The observed two-point correlation functions (power spectra) of the CMB temperature T and E- and B-mode polarizations are estimated from the CMB and/or frequency maps; various higher-order (typically three- and four-point) correlation functions may also estimated from the CMB maps.
\item[Post-processing]: The primordial power spectra are derived from the observed spectra, with the details of this step depending on both the input maps and the algorithms used in the spectral estimation; examples of this step include debiasing of pseudo-spectra, delensing, and spectral-domain point-source removal.
\item[Parameter estimation:] The best-fit parameters for any cosmological model are derived by comparing the theoretical correlation function(s) predicted by the model with the data.
\end{description}

Note however that the data can only remain a sufficient statistic at each step in the reduction if we also propagate its full covariance matrix. This is an ${\cal N}_b \times {\cal N}_b$ matrix in the dimension of the basis, so its construction, manipulation and reduction typically poses the greatest computational challenge to this analysis. In particular the full pixel-domain data covariance matrix is generally dense and unstructured, requiring O(${\cal N}_p^3$) operations to build and O(${\cal N}_p^2$) bytes to store. A mission that covers a fraction of the sky \fsky\ with a beam of $b$ arcminutes generates O($10^9 \, \fsky/b^2$) pixels per CMB component per observing frequency, so that increases in sky fraction, resolution, polarization sensitivity or frequency coverage (as have been the science drivers to date) necessarily increases ${\cal N}_p$. For the last decade or more the computational intractability of the resulting pixel-domain matrices has forced us to replace explicit covariance propagation with Monte Carlo methods in all but a limited set of small sky fraction/low resolution cases---although it should be noted that, as outlined previously, this case may be of great interest for determining the tensor-to-scalar ratio $r$.

 
\section{Time-Ordered Data Processing}

This section discusses the first stage of the CMB data analysis pipeline, in which the raw time-ordered data from every detector are first pre-processed and then combined into an estimate of the temperature and polarization on the sky. We discuss those steps---and some of the challenges we expect to face in implementing those steps in the \cmbexp\ era---below.

\subsection{Pre-Processing and Mission Characterization}

The first stage of analysis in typical CMB experiments involves pre-processing the raw 
time-ordered data in an attempt to clean the data of time-domain systematics and 
make the real data match a model that will underpin all subsequent analyses. Typical 
steps in this pre-processing are finding and removing cosmic-ray hits (``glitches'') on individual detectors
and narrow-band filtering of spectral-line-like contamination to the time-ordered data (often 
from detector sensitivity to a mechanical apparatus such as the cryocooler). 
A challenge in the \cmbexp\ era will be to properly account for these steps---which can involve
the data from the entire set of detectors over a long observing period---in the time-ordered data
simulation pipeline, in order to characterize their effects on the final science results. The data 
volume is sufficiently large at this step that multiple full simulations may be unfeasible.

The pre-processing phase is also where many of the inputs to the mission model (a key piece of the data simulation pipeline) are measured. These inputs include the overall observation pointing reconstruction, and the individual detector beam profiles, bandpasses, and noise charateristics, complementing dedicated instrument-characterizing observations and other laboratory or field measurements.

\subsection{Map-Making}
\label{sec_mapmaking}

Map-making is the stage of the analysis when the major compression of the time-ordered data happens and some estimate of the sky signal is produced at each observing frequency. It is usually a linear operation, characterized by some operator, $\mathbf{L}$, which transforms the input time-ordered data, $\mathbf{d}$, into a pixel domain map, $\mathbf{m}$, 
e.g., \cite{Tegmark:1996qs},
\begin{eqnarray}
\mathbf{m} = \mathbf{L}\mathbf{d},
\end{eqnarray}
typically under the condition that the estimator is unbiased over the statistical ensemble of instrumental noise realizations, i.e.,
\begin{eqnarray}
\langle \mathbf{m} - \mathbf{s}\rangle = 0,
\label{eq:condMaps}
\end{eqnarray}
where $\mathbf{s}$ is the underlying pixelized sky signal. Given the usual model for the time-ordered data as the sum of sky-synchronous signal and time-varying noise, 
\begin{eqnarray}
\mathbf{d} = \mathbf{A}\mathbf{s} + \mathbf{n},
\end{eqnarray}
for a pointing matrix $\mathbf{A}$, this condition leads to,
\begin{eqnarray}
\langle \mathbf{m} - \mathbf{s}\rangle =  (\mathbf{L}\mathbf{A}-\mathbf{1})\mathbf{s} 
+ \langle \mathbf{n} \rangle = (\mathbf{L}\mathbf{A}-\mathbf{1})\mathbf{s},
\end{eqnarray}
as the average noise is assumed to vanish. Hence,
\begin{eqnarray}
\mathbf{L}\mathbf{A} = \mathbf{1},
\end{eqnarray}
which is solved by,
\begin{eqnarray}
\mathbf{L} = (\mathbf{A}^{\rm T} \mathbf{W} \mathbf{A})^{-1} \mathbf{A}^{\rm T} \mathbf{W}.
\end{eqnarray}
Here the matrix $\mathbf{W}$ is an arbitrary positive definite weight matrix, and different choices of $\mathbf{W}$ lead to different estimates of the sky signal.
\begin{itemize}
\item If $\mathbf{W}$ is taken to be the inverse of the time-domain noise covariance, i.e., $\mathbf{W} = \mathbf{N}^{-1}$, then the sky signal estimate, $\mathbf{m}$, will correspond to the {\bf maximum likelihood} and {\bf minimum variance} solution. 
\item If $\mathbf{W}$ is taken to be proportional to some diagonal matrix minus some low-rank correction, i.e. $\mathbf{W} \propto \mathbf{1} - \mathbf{T}\mathbf{T}^{\rm T}$,  with $\mathbf{T}$ assumed to be column-orthogonal, then the modes defined by its columns are marginalized over, effectively removing them from the solution. This approach includes as a special case so-called {\bf destriping} map-making, e.g.,~\cite{Poutanen:2004hy, Keihanen:2003pu}, which has gained recognition thanks to its successful applications to the \planck\ data, e.g.,~\cite{Keihanen:2009tj, Tristram:2011gq, Ade:2015uua, Adam:2015vua}, and is therefore of potential interest to any experiments aiming to cover a large fraction of the sky. More generally, however $\mathbf{T}$ can be constructed to remove any unwanted modes present in the time domain data, e.g.,~\cite{Stompor:2002jy, Cantalupo:2009if, Dunner:2012vp}.
\item If $\mathbf{W}$ is taken to be diagonal, then the map-making solution corresponds to {\bf binning}, i.e. the weighted co-addition of the samples falling within each pixel.
\end{itemize}
If the instrument beams display complex, non-axially symmetric structure, the proper estimation of the sky signal may require correcting for their effects at the map level, leading to the so-called {\bf deconvolution} map-making ~\cite{Armitage:2004pk, Harrison:2011xt, Keihanen:2012rm}.  However, further work is needed to demonstrate the effectiveness of such an approach in general.

If map-making is used primarily as a data compression operation on the way to deriving constraints on the statistical properties of the sky signal (such as its power spectra), one may choose to relax the condition in Eq.~(\ref{eq:condMaps}) in favor of the more computationally tractable, albeit potentially biased, sky estimate,
\begin{eqnarray}
\mathbf{m} = (\mathbf{A}^{\rm T} {\rm diag}( \mathbf{W}) \mathbf{A})^{-1} \mathbf{A}^{\rm T}\mathbf{W} \mathbf{d},
\end{eqnarray}
where ${\rm diag}( \mathbf{W})$ denotes the diagonal part of $\mathbf{W}$. In this approach any bias is then corrected at the next level of the data processing, e.g.,~\cite{Hivon:2001jp}. This approach has been proven to be very effective, at least in the context of experiments with small sky coverage, e.g., \cite{Culverhouse:2010ya, Schaffer:2011mz, Ade:2014afa, Ade:2014xna}.

Formally the linearity of the mapmaking operation permits the propagation of the uncertainty due to the instrumental noise from time- to pixel-domain as
\begin{eqnarray}
{\hat{\mathbf{N}}} = \mathbf{L} \mathbf{N} \mathbf{L}^{\rm T},
\end{eqnarray}
which leads to a particularly simple expression for maximum likelihood estimators 
\begin{eqnarray}
{\hat{\mathbf{N}}} = (\mathbf{A}^{\rm T} \mathbf{N}^{-1} \mathbf{A})^{-1}.
\end{eqnarray}
However, as noted above, the computational cost of computating such pixel-domain noise correlation matrices make it impractical for all but special cases today, and the uncertainty is carried over to the next stages of the data processing in implicit form and ultimately estimated using Monte Carlo simulations.

          
\section{Component Separation}
\label{sec:compsep}

This section discusses the algorithms and methods for disentangling different sources of sky emission in multi-frequency maps, whether these come exclusively from \cmbexp\ or include well-characterized external maps. Under this assumption, we are able to go beyond simple foreground cleaning to full component separation, which provides both important consistency checks on our results and critical inputs to sky modeling (see below). We first present the motivations and the general ideas of existing approaches. We then give some specifics of parametric and blind methods. Finally, we summarize several questions which might be answered by follow-up studies.

Recent measurements by BICEP2/Keck/\planck\ \cite{Ade:2015tva} confirm that on degree scales, where 
\cmbexp\ is expected to search for 
the imprint of B modes from primordial gravitational waves, the contamination from polarized foreground emission is  comparable to or higher than the cosmological signal at 150~GHz even in one of the cleaner patches of the sky.
Given that 150~GHz is expected to be close to the minimum of foreground contamination vs.~CMB signal, this 
is likely to be the case at all frequencies and all but the smallest fractions of the sky.
Given the power law behavior in $\ell$ found on larger scales by \planck\ and \wmap\ \cite{Adam:2015tpy,Page:2006hz}, foregrounds are expected to be even more relevant at larger angular scales. Foregrounds are expected to be subdominant  with respect to the B-mode lensing signal on the scale of a few arcminutes (see Figure~\ref{fig:power_spectrum_fgs}); nevertheless, dust polarization fractions around $10\%$ (comparable to observed levels) have been shown to have non-negligible impact on the 4-point function used for achieving lensing extraction \cite{Fantaye:2012ha}. Therefore, component separation is a necessary and important step in gaining insight into the amplitude of primordial gravitational waves, the neutrino masses, and the abundance of dark energy through CMB lensing studies.

\begin{figure*}[htbp]
\centering
\includegraphics[width=0.5\textwidth]{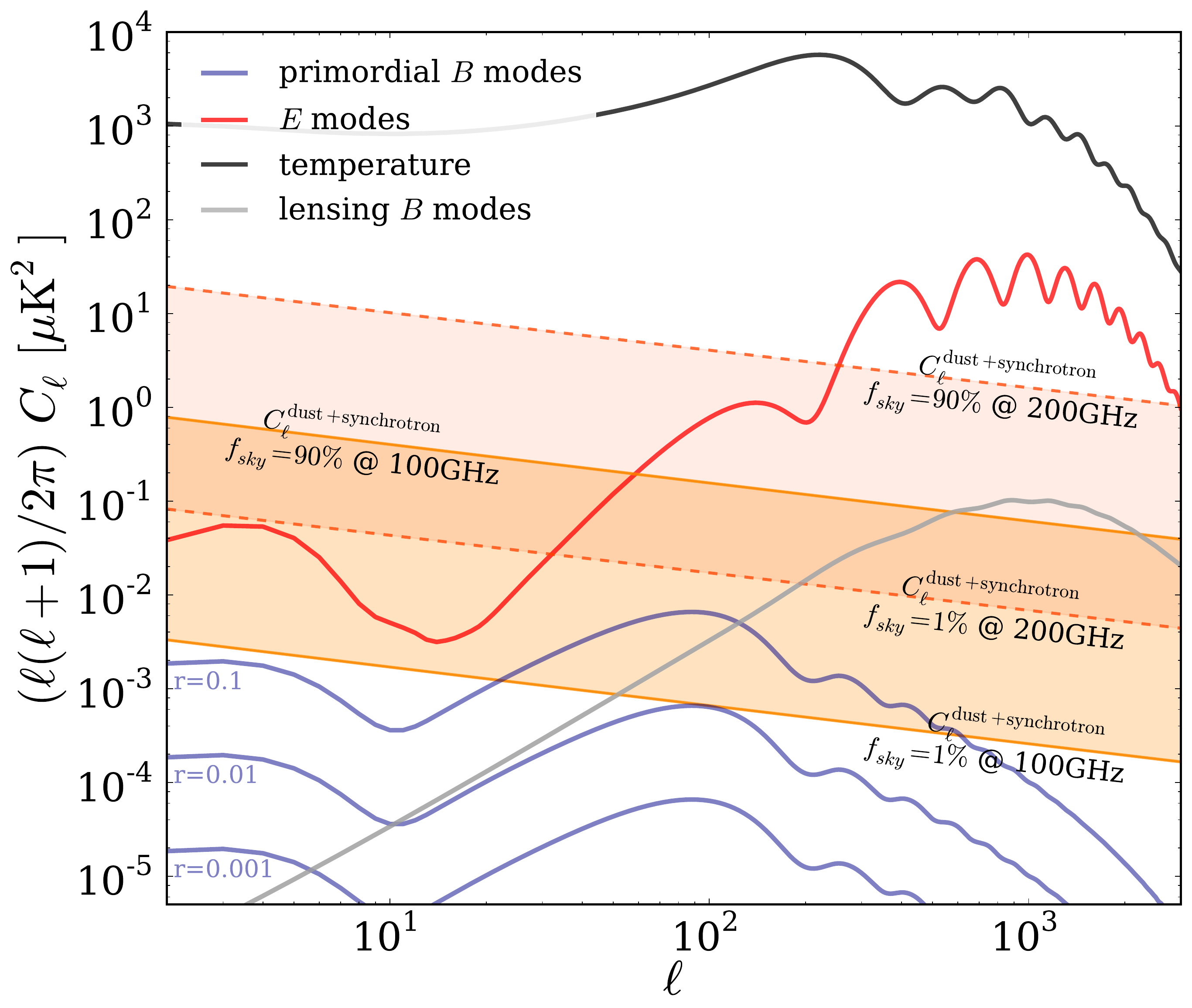}
\caption{Angular power spectra showing primordial B modes, lensing B modes, total intensity, and E modes, as well as the total contribution of polarized B-mode foregrounds (dust plus synchrotron), expected on the cleanest $1-90\%$ of the sky, at $100$ and $200\;$GHz. Note that these results are derived from \planck\ data on large patches of the sky, and the estimates for 1\% patches of the sky are extrapolations; there may in fact be individual 1\% patches that are cleaner than the levels shown here.
From~\cite{Errard:2015cxa}.}
\label{fig:power_spectrum_fgs}
\end{figure*}

Broadly defined, the process of component separation would generally
\begin{itemize}
	\item include any data processing that characterizes and exploits correlations between observations at multiple frequencies
	\item use external constraints and physical modeling
	\item aim at distinguishing between different physical sources of emission.
\end{itemize}

The general data modeling reads
\begin{eqnarray}
	\centering	
		d_p &=& \sum_{\rm comp, p} a_p^{\rm comp} s^{\rm comp}_p + n_p \equiv \mathbf{A}\,s_p + n_p
	\label{eq:comp_sep_data_modeling}
\end{eqnarray}
where the vector $d_p$ contains the measured signal in each observing band, $\mathbf{A}$ is the so-called mixing matrix which encapsulates the emission law $a_p^{\rm comp}$ of each component, $s_p$ is a vector containing the unknown CMB and foregrounds amplitude, and $n_p$ is a vector containing the noise level at each observing band. The index $p$ refers to sky pixels $\left( \theta, \phi \right)$, or modes of a spherical harmonic decomposition $\left( \ell, m\right)$, or a set of Fourier modes $\left(k_x,k_y\right)$, etc. Note that this modeling assumes spatial templates $s_p$ that are the same in all observing bands.

Component separation aims at inverting Eq.~\ref{eq:comp_sep_data_modeling}, thus estimating the foreground-cleaned CMB signal encapsulated in $s_p$, as well as the foreground maps which are relevant for testing and updating our knowledge of astrophysical processes (and hence improving the sky model).
The estimate $\tilde{s}_p$ of the true sky templates $s_p$---given $\mathbf{A}$, $d_p$ and the statistical properties of the noise --- minimizes the following $\chi^2$:
\begin{eqnarray}
	\centering
		\chi^2 \equiv \sum_p \left| s_p - \tilde{s}_p\right|^2
	\label{eq:chi2_compsep}
\end{eqnarray}
and can be taken to have the following general form
\begin{eqnarray}
	\centering
		\tilde{s}_p = \mathbf{W}\,d_p
	\label{eq:sp_solution}
\end{eqnarray}
where the weighting operator $\mathbf{W}$ is chosen to optimize some criterion regarding $\tilde{s}_p$ and $s_p$ (variance of the cleaned map, unbiasedness, etc.) while keeping statistical consistency and robustness. In particular, a common requirement for all component separation algorithms is the ability of propagating errors due to foreground subtraction, while having the flexibility of including foreground modeling and external constraints in a transparent way. 
Component separation is then defined as a method of estimating the mixing matrix $\mathbf{A}$ and finding the weighting $\mathbf{W}$ that provides closest possible estimate $\tilde{s}_p$ to the true sky signal.

For example, a solution to Eqs.~\ref{eq:chi2_compsep} and~\ref{eq:sp_solution} is obtained by taking $\mathbf{W} \equiv \left( \mathbf{A}^T\mathbf{N}^{-1}\mathbf{A} \right)^{-1}\mathbf{A}^T\mathbf{N}^{-1}$ with $\mathbf{N} \equiv \langle n_p^T n_p\rangle$, leading to an unbiased estimate of the sky. As mentioned below, this expression can be changed (see, e.g., \cite{Delabrouille:2009aa}), depending on the desired level of generality and complexity and on the level of prior knowledge of the sky signal.

Studies have demonstrated the applicability of classes of component separation 
algorithms to certain simulated multi-frequency datasets, either balloon-borne or ground-based, and targeting limited frequency ranges and sky areas \cite{Stivoli:2010rs,Fantaye:2011zq,Fantaye:2012ha}. Results indicate that generally, for a frequency range extending from $90$ to $250$ GHz, polarized foregrounds may be removed effectively through a multi-frequency combination, at the price of enhancing the white noise contribution due to channel mixing; moreover, a possible bias may be introduced if, at the lowest frequency interval edge, the synchrotron component is not negligible: lower frequency templates/data are required to avoid such a contribution \cite{Essinger-Hileman:2014pja}. 
The most comprehensive application of component separation to data, in terms of completeness of algorithms and frequency range, is represented by \planck\ \cite{Adam:2015tpy}, 
although the targeted CMB components in that analysis (total intensity and E-mode polarization) are
not the same as in the \cmbexp\ case.

\subsection{Component Separation Methods}

The CMB extraction may be achieved essentially through two basic concepts: the fitting of foreground unknowns along with CMB, or the minimization of the variance of a linear combination of the data, constrained to have the frequency scaling of a blackbody.
The first class of algorithms, known as ``parametric,'' makes the maximum use of prior knowledge of foreground emission. By contrast, the second class, known as ``blind,'' makes the minimum set of assumptions. 
These two broad classes and other possibilities are discussed in turn below.
Both approaches are used widely in the field and have been tested with multiple implementations applied
to multiple data sets. Blind techniques have included those using internal template subtraction (e.g. \cite{Bennett:1992aa,Hansen:2006rj,Katayama:2011eh}) and those exploiting statistical independence of sky components (e.g. \cite{Delabrouille:2002kz,Maino:2006pq,Bonaldi:2006qw,Stolyarov:2004xp}). Implementations of parametric fitting have included the studies in \cite{Brandt:1994aa,Eriksen:2005dr, Stompor:2008sf}.

\begin{itemize}
	\item \textbf{Parametric} -- The overall idea of these methods boils down to two steps: 1) the estimation of the mixing matrix, $\mathbf{A}$; and 2) the inversion of Eq.~\ref{eq:comp_sep_data_modeling} to recover an estimate of the sky signal, $s_p$.
	Parametric methods assumes that the mixing matrix, used in Eq.~\ref{eq:comp_sep_data_modeling}, has a functional form which is known and which can be parametrized by so-called ``spectral'' parameters $\beta$, i.e., $\mathbf{A} = \mathbf{A}(\beta)$. The functional form of $\mathbf{A}$ being fixed, the estimation of the mixing matrix is therefore equivalent to an estimation of the parameters $\beta$. The parameters of the model are determined via a fitting procedure, often performed over sky pixels. This can be achieved by maximizing the following so-called ``spectral" likelihood \cite{Brandt:1994aa,Eriksen:2005dr}:
	\begin{eqnarray}
		\centering
			-2\log \mathcal{L}(\beta) = - \sum_p \left( \mathbf{A}^T\mathbf{N}^{-1} d\right)^T\left( \mathbf{A}^T\mathbf{N}^{-1} \mathbf{A}\right)^{-1}\left( \mathbf{A}^T\mathbf{N}^{-1} d\right).
	\end{eqnarray}
Any deviation between the true mixing matrix $\mathbf{A}$ and the estimated $\mathbf{\tilde A} \equiv A(\tilde \beta)$ leads to the presence of foreground residuals in the reconstructed component maps.
	\item \textbf{Blind} -- Under the  assumption that sky components are statistically independent, blind methods aim to recover these components with an a priori unknown mixing matrix. Blind methods make minimal assumptions about the foregrounds and focus on reconstructing the CMB from its well known blackbody spectral energy distribution. The Internal Linear Combination (ILC, \cite{Tegmark:2003ud}) belongs to this class of methods. It only uses the CMB column of the mixing matrix elements (noted $a$ hereafter) to perform the minimum variance reconstruction, cf. Eq.~\ref{eq:sp_solution}:
\begin{equation}
  \tilde s_p = \sum_{i=0}^{i=m} w_i d_{p,i}
\end{equation}
with $\sum_i  w_i a_i = 1$, leading to the following solution:
\begin{equation}
  w_i = a^T N^{-1} (a^TN^{-1} a)^{-1}
\end{equation}

In this scheme, no attempt is made to design a foreground model. The decorrelation property between CMB and foregrounds alone is used to project out the contamination into a $m$-1 subspace (with $m$ being the number of frequency maps).

The main caveat in this method is its well known bias (\cite{Hinshaw:2006ia, Delabrouille:2009aa}, etc) which comes from empirical correlation between the CMB and the foregrounds. The ILC bias is proportional to the number of detectors $m$ and inversely proportional to the number of pixels used to compute $N$. In order to reduce this effect, one could think of reducing the foreground subspace size by adding further constraints. The SEVEM template fitting method (\cite{MartinezGonzalez:2003dy}, etc) follows this idea, by building some foreground templates with a combination of a subset of the input frequency maps.

The semi-blind SMICA method~\cite{Cardoso:2008qt} also works at containing the foreground in a smaller dimension space, but in a more general way. The idea of Independent Component Analysis (ICA) is to blindly recover the full mixing matrix $\mathbf{A}$ by using the independence property of the different components. As we know that they are spatial correlations between the foregrounds, the ICA principle is used to disentangle the CMB from the noise and the foregrounds taken as a whole.

The main advantage of such blind or semi-blind methods is their ability to face any unknown and/or complex foreground contamination, to reconstruct a clean CMB signal. This is a big advantage when real data comes, one can then focus on instrumental effects, or data set combination issues at first, and leave the complex task of the foreground modeling and reconstruction for a future analysis step.

Moreover, in a framework like SMICA, the level of blindness can be adjusted via the plugin of any parametric component to its flexible engine as described in~\cite{Cardoso:2008qt}, allowing for a step by step fine grain design of the foreground model.\\
	\item \textbf{Template fitting} -- 	In this variant, emission laws are not modeled, and the analysis is reduced to the maximisation of a likelihood over 
the CMB contribution and the amplitudes of each foreground component (see, e.g., \cite{Katayama:2011eh}).
\end{itemize}

For all of the approaches discussed above, Eq.~\ref{eq:sp_solution} can be implemented equivalently with any representations of the map---i.e. pixel, harmonic, wavelet, etc. The resulting component separation is independent of this choice as long as the linear data modeling (Eq.~\ref{eq:comp_sep_data_modeling}) holds. 
This complementarity, and the internal comparison of results through these pipelines has been proven to be relevant in actual analysis of \planck\ data \cite{Adam:2015tpy}. That said, the difference between domain of application will lie in the computational needs: for high number of sky pixels, the implementation of Eq.~\ref{eq:sp_solution} might be significantly more efficient in harmonic space. 

One practical difference of particular interest between these approaches is how uncertainties in the 
foreground model get propagated to uncertainties in data products derived from the foreground-cleaned
maps, such as angular power spectra and cosmological parameters. In parametric models, any statistical 
covariance between foreground parameters and the cleaned CMB is properly captured (at the 
Gaussian-approximation level) in the resulting covariance matrix. For non-parametric methods, this 
covariance can be approximated using Monte Carlo methods, which can be problematic for foregrounds
on the largest scales, the distribution of which is quite clearly non-Gaussian. This issue is discussed further
in Section~\ref{se:challenges}.

\subsection{Open Questions}
\begin{itemize}
	\item \textbf{E/B or Q/U basis of analysis} -- Component separation between CMB radiation and its foregrounds can be performed either dealing with Stokes parameters $Q$ and $U$ maps of the sky in real space or Fourier space, and either before or after the separation between the E and B modes. Several approaches have been followed by CMB experiments so far \cite{Gold:2010fm,Aghanim:2015xee, Ade:2015tva},
and each of them has some advantages and some caveats. For example, processing $Q/U$ data in the map domain 
simplifies the treatment of foreground components that have non-Gaussian and/or non-stationary
spatial distributions.
However, in the $Q$ and $U$ basis, the CMB E and B modes are mixed and the CMB E modes will be the dominant contribution to the variance at intermediate and small scales in the CMB observing frequencies, limiting the accuracy of the separation.
To overcome this limitation, E and B observables can be constructed in Fourier space. 
The separation of the B-mode components (primordial CMB, Galactic foregrounds, lensing, etc.) can then be done in the angular power spectrum domain (where the final accuracy might be limited by the cosmic variance associated to foregrounds),
in the two-dimensional, phase-full Fourier domain (where the treatment of non-stationary components
will be complicated) or in the map domain (where the final accuracy might be limited by ringing of the foregrounds due to the non-local transformation). 
Although these different approaches are currently giving satisfactory results on simulated data, these effects will become crucial at the sensitivity of \cmbexp\ and merit a dedicated study. 
	\item \textbf{Combining data from multiple instruments} -- Ground-based instruments heavily filter time streams because of atmosphere contamination, ground emission, etc. 
	In particular, large angular scales are usually suppressed anisotropically, and this suppression is corrected in the power spectrum estimate. 
Component separation using observations from different platforms will be made more straightforward if all maps are derived from common filters. 
As stressed already, the first attempt at component separation or foreground cleaning for  B modes on multi-platform data was recently implemented in \cite{Ade:2015tva}, using data from BICEP2, Keck Array and \planck. A template fitting analysis was implemented with the primary objective of minimizing the variance in the CMB solution.
The simultaneous analysis of combined data sets 
required an additional layer in the analysis, namely the simulated scans of the \planck\ data through the filtering by the ground observatories, along with validation through simulations of the whole procedure. A simpler approach for \cmbexp\ would be to have a single pipeline reducing and combining different datasets. With a common filtering implemented from scratch in a multi-site experiment, the 
combination would be built-in, thus avoiding the extra layer and increasing confidence and robustness of results.
	\item \textbf{Various resolutions} -- Under the approximation that the mixing matrix does not significantly vary as a function of resolution, the impact of different beam sizes can be propagated to the noise level of the final CMB map by incorporating the beam for each frequency channel in the expression of the noise covariance matrix 
	\begin{eqnarray}
		\centering
			\mathbf{N}(i) \equiv \mathbf{N}(i)_\ell = \left( \sigma_i\right)^2\exp\left[ \frac{\ell(\ell+1)\theta_{\rm FWHM}^2}{8\log(2)} \right]
	\end{eqnarray}
	where $i$ is a frequency channel and $\sigma_i$ is the noise level in the corresponding map. The noise variance in the reconstructed CMB map, i.e. after component separation, would then be given by
	\begin{eqnarray}
		\centering
			N_\ell^{\rm post\ comp\ sep} = \left[\left(\mathbf{A}^T\left(\mathbf{N}_\ell\right)^{-1}\mathbf{A}\right)^{-1}\right]_{\rm CMBxCMB}.
	\end{eqnarray}
	\item \textbf{Atmosphere residuals} -- Atmosphere residuals appear at large scales in ground-based CMB observations, and they scale with frequency in a similar way as dust, $\propto \nu^\beta$ \cite{Errard:2015twg}. Having redundant frequencies among the different observatories could help mitigate the atmospheric and astrophysical foregrounds. Furthermore, the small intrinsic polarization of the atmosphere \cite{Battistelli:2012we,Errard:2015twg} will
limit the contamination to component separation of the polarized signals.
Still, this effect will have to be investigated quantitatively with realistic simulations.
\end{itemize}

\section{Statistics \& Parameters}

In this section, we discuss the process of going from sky maps at different frequencies---or, in light
of the previous section, foreground-cleaned CMB maps and an estimate of foreground residuals---to
post-map products such as angular power spectra, estimates of lensing potential $\phi$, and finally
cosmological parameters, as well as covariance estimates for all of these quantities. We briefly describe
the current practice for this process, then we address specific challenges anticipated in the \cmbexp\ era.

\subsection{Current practice}
\label{se:current}
Early measurements of CMB temperature anisotropy, with comparatively few map pixels or angular modes
measured, often used maximum-likelihood methods to produce maps of the sky (e.g., \cite{Wright:1996dk}) and
either a direct evaluation of the full likelihood or a quadratic approximation to that likelihood (e.g., \cite{Bond:1998zw}) to go from 
maps to angular power spectra. With the advent of the \wmap\ and \planck\ space 
missions, which would map the entire sky at sub-degree resolution, it became apparent that computing
resources could not compete with the $\mathcal{O}({\cal N}^3)$ scaling of the full-likelihood approach 
(e.g., \cite{Borrill:1998tn}). The solution for power spectrum analysis
that has been adopted by most current CMB experiments is a
Monte-Carlo-based approach advocated in \cite{Hivon:2001jp}. In this approach, a biased estimate of
the angular power spectrum of the data is obtained by simply binning and averaging the square 
of the spherical harmonic transform of the sky map. That estimate (known as the 
``pseudo-$C_\ell$ spectrum'') is related to the unbiased 
estimate that would be obtained in a maximum-likelihood procedure through the combined effect
of noise bias, sky windowing, and any filtering applied to the data before or after mapmaking
(including the effects of instrument beam and pixelization). These effects are estimated by ``observing''
and analyzing simulated data and constructing a matrix describing their net influence on simulated data. 
This matrix is inverted, and the inverse matrix is applied to the pseudo-$C_\ell$s to produce the 
final data product. Some version of this Monte-Carlo treatment is likely to be 
adopted for \cmbexp. 

Pseudo-$C_\ell$ methods are also now commonly used in analysis of CMB polarization anisotropy
\cite{Aghanim:2015xee,Naess:2014wtr,Crites:2014prc}. An added complication in polarization analyses is that 
pseudo-$C_\ell$ methods do not cleanly separate E and B modes (e.g., \cite{Challinor:2005jy}).
``Pure'' B-mode estimators can be constructed that suppress the spurious B-mode contribution
from estimating E and B on a cut sky with pseudo-$C_\ell$ methods \cite{Smith:2005gi}), but 
other analysis steps (such as particular choices of filtering) can produce spurious B modes that
are immune to the pure estimators \cite{Keisler:2015hfa}. These can also be dealt with using Monte-Carlo
methods, either by estimating the statistical bias to the final B-mode spectrum or by constructing
a matrix representing the effect of any analysis steps on the true sky \cite{Ade:2014xna}. The latter
approach involves constructing an ${\cal N}_\mathrm{pixel}$-by-${\cal N}_\mathrm{pixel}$ matrix, equal in size to the 
full pixel-pixel covariance, and will not be feasible for high-resolution \cmbexp\ data but could be 
used in analyzing lower-resolution data.

In addition to the two-point function of CMB maps, higher-order statistics of the maps have recently 
been of great interest to the community. In particular, the four-point function encodes the effect of 
gravitational lensing, and estimators can be constructed to go from CMB temperature and polarization
maps to estimates of CMB lensing $\phi$ and the associated covariance (e.g., \cite{Hu:2001kj,Okamoto:2003zw}).
These quadratic estimators are the first step in an iterative estimation of the true likelihood, and in
the weak-lensing limit they are nearly optimal; as a result, they remain the state of the art for estimating
the large-scale $\phi$ from CMB lensing (e.g., \cite{Ade:2013zuv}). For \cmbexp\ sensitivity levels, 
it is possible that further gains can be made with more iterations (see Section \ref{delens}).
Even with multiple steps, the computational burden involved
in this step of the analysis is unlikely to be significantly greater for \cmbexp\ than for \planck.

An additional post-map product of interest for \cmbexp\ is the location and properties of compact
sources, in particular clusters of galaxies identified through the thermal SZ effect. The standard 
practice for extracting SZ clusters from multifrequency millimeter-wave maps is through the application
of a Fourier-domain spatial-spectral filter \cite{Melin:2006qq}.
The computational effort involved in this step is small compared to the estimation of power spectra 
and higher-order correlations, and the algorithms are well-developed and fully implemented for 
multi-frequency data sets (e.g., \cite{Ade:2013skr,Bleem:2014iim})---however, the cluster density
could be high enough in \cmbexp\ data that approaches more sophisticated than the simple matched
filter (e.g., \cite{Pierpaoli:2004bp}) could be required to maximize cluster yield.

The final step in the analysis of a CMB data set is the estimation of cosmological parameters from
the various post-map statistics discussed above.
This involves estimating the likelihood of the data
given a model parameterized by the standard six $\Lambda$CDM parameters, possible extensions
of the cosmological model, and any nuisance parameters involving the instrument, foregrounds, and
other sources of systematic uncertainty. The current industry standard for this part of the analysis are
Monte-Carlo Markoalv-Chain (MCMC) methods, in particular the implementation in CosmoMC
\cite{Lewis:2002ah}, and it is expected that \cmbexp\ will use similar methods. 

\subsection{Challenges}
\label{se:challenges}
There will be several 
aspects of the \cmbexp\ dataset that will necessitate going beyond what past analyses
have done at the post-map step. First of all, the data from several different telescopes and cameras will need
to be combined in as lossless a fashion as possible---such that combining at the parameters stage
may be sub-optimal.
Further, as shown by \cite{Ade:2015tva}, foregrounds cannot be ignored in the 
estimation of the B-mode power spectrum, even in the cleanest parts of the sky and in the 
least contaminated observing bands. Foreground modeling will be used to mitigate the contamination,
but there will be foreground residuals (both from noise and imperfect modeling), and these need
to be properly characterized and accounted for in parameter extraction. 
Similarly, algorithms to separate the contributions to the B-mode power spectrum from a background of gravitational
waves and from lensing of E modes (so-called ``de-lensing'', see the Section \ref{delens} for 
details) will leave an uncertain level of lensing residuals in the primordial B-mode spectrum, and
this residual will need to be treated properly. Finally, for information from angular power spectra
and lensing potential $\phi$ to be properly combined, the covariance between the two-point and
four-point functions of the CMB needs to be taken into account.

We treat each of the following challenges individually in the sections below:
\begin{itemize}
\item{The combination of data from different telescopes and cameras (with different heritage 
and observation/analysis techniques) without significant loss of information.}
\item{The impact of uncertainties in foreground modeling on cosmological parameters, particularly the tensor-to-scalar ratio $r$.}
\item{The covariance between different observables (for example the lensed CMB power spectrum and the reconstructed lensing potential power spectrum).}
\item{The impact of delensing---the separation of the gravitational lensing signal and the primordial B-mode signal, lowering the effective lensing background---and lensing residuals on cosmological parameters.} 
\end{itemize}

\subsubsection{Combining different data sets}
\label{se:combine}
At what stage in the analysis does it make the most sense to combine data from different experimental platforms? 
One possibility is to estimate
angular power spectra or even cosmological parameters from every data set individually and combine them at that stage. This would
be computationally efficient but sub-optimal from a sensitivity standpoint unless every experiment
covered fully independent patches of the sky. For any overlap between data sets, combining at
the map or time-ordered data stage (adding before squaring) will lead to lower final uncertainties
than combining at the power spectrum stage (squaring before adding). Of course, the earlier in the analysis
we choose to combine data, the more work it will be to standardize the data between experimental platforms---the
time-ordered data is generally quite instrument-specific, the maps less so, etc. The trade-off between
maximizing constraining power and possibly placing undue burdens on the individual  
pipelines will need to be balanced in answering this question.
Furthermore, the frequency coverage may not be identical across the individual experimental platforms.
In this case, combining data at the stage of independent frequency-band maps will result in
different sky coverage at different frequencies; combining data at the stage of foreground-cleaned
CMB maps will result in different foreground residuals and noise levels in different parts of the sky.
These factors must also be balanced in the decision of when to combine data.

\subsubsection{Foreground-related uncertainty on cosmological parameters}
\label{se:paramforeg}

To separate the CMB signal from the contaminating signals of Galactic and extragalactic foregrounds, 
data from multiple bands will be combined, either 
in a cross-spectrum analysis or, as detailed in Section \ref{sec:compsep}, by making linear 
combinations of maps in different bands to produce a ``pure-CMB'' map for power spectrum estimation.
In either case, an underlying model of foreground behavior is assumed---even if that model is simply
an assumption regarding the level to which the spectral behavior of foregrounds varies over the sky.
There are two challenges related to uncertainties in foreground modeling: one statistical and one
systematic. The statistical issue is simply how to propagate the statistical uncertainty on the foreground 
model to uncertainties on cosmological parameters. In explicitly parameterized foreground models, 
this happens automatically through the covariance resulting from the fit. For non-parametric models,
this covariance can be assessed through Monte Carlo methods, but making many independent 
realizations of large-scale Galactic foregrounds is problematic because of the strongly non-Gaussian
behavior of these foregreounds.

Perhaps more importantly, 
any model of foreground behavior is by definition imperfect, and the resulting component separation
or frequency-cross-spectrum fit will have systematic leakage between the foreground and CMB components.
At the sensitivity levels attainable by \cmbexp, these residuals have the potential to dominate the
error budget on cosmological parameters and, more troublingly, to significantly bias the best-fit 
parameter values if they are not properly taken into account.

Section~\ref{sec:skymodel} discusses the baseline plan for, and challenges involved in, modeling
Galactic and extragalactic foregrounds. It is possible that more information will be needed---from 
Stage-3 experiments, or from a possible dedicated, balloon-borne CMB foreground mission---before
we can confidently assess the level to which foregrounds will limit the final parameter constraints
from \cmbexp\ and how flexible we will need to make the underlying foreground models that 
inform component separation and parameter extraction. 

\subsubsection{CMB lensing covariances for \cmbexp }
\label{se:covs}

The measured lensing power spectrum is given by a four-point function of the lensed CMB. 
This is not statistically independent from the lensed CMB two-point function, because both depend on the same observed, lensed CMB maps. 
As a consequence, measured lensing power spectra and lensed CMB power spectra may be correlated. 
This correlation should be taken into account when combining these measurements to avoid spurious double counting of information. 
For the specific case of \planck\ this correlation is negligible \cite{Schmittfull:2013uea}. 
However, the level of correlation depends on experiment specifications and the multipole range where power spectra have high signal-to-noise. 
The correlation should thus be included in analyses that combine two- and four-point measurements unless it is known to be negligible for a specific experiment.
\\

The forecasted noise level for CMB-S4 is much lower than the noise level for \planck, and the reconstruction of the lensing potential power spectrum will come from a mixture of temperature and polarization data.
In this context, modelling the correlations only in the case of temperature is not accurate enough.
Instead, a minimum variance lensing estimator out of all the measured quadratic pairs needs to be considered, and contributions arising from all couplings of the CMB six-point function would need to be modelled.
\\

In \cite{Schmittfull:2013uea} three main contributions to the cross-covariance are identified: The noise contribution arising because fluctuations of the unlensed CMB and instrumental noise change both the Gaussian reconstruction noise $N^{(0)}$ and the CMB power spectra;
the signal contribution from the cosmic variance fluctuations of the true lensing potential (i.e.~fluctuations of matter along the line of sight); and contributions coming from the connected trispectrum part of the CMB six-point function.
Those contributions have been modelled with high precision and agree well with simulations.
The noise contribution is not present if the Gaussian reconstruction bias $N^{(0)}$ is subtracted in a realization-dependent way using the measured CMB power spectrum in our Universe (see e.g.~\cite{Hanson:2010rp,Schmittfull:2013uea}).
However for an experiment such as CMB-S4, the last two contributions contribute the most to the correlation between lensing power spectra and lensed CMB power spectra, reaching few tens of percent in some cases.
The signal covariance could in principle also be avoided by delensing CMB power spectra with the estimated lensing reconstruction by mimicking the realization-dependent technique as shown in \cite{Schmittfull:2013uea}, or by applying more advanced delensing methods as described in Section \ref{delens}.
In addition to the cross-covariance between two-point and four-point functions of the CMB, these quantities can each have non-trivial auto-covariances which can again be modeled as shown in e.g.~\cite{Smith:2005ue,Smith:2006nk,Li:2006pu,BenoitLevy:2012va,Schmittfull:2013uea}. 
However we want to emphasize that the discussion above applies to the standard quadratic lensing reconstruction estimators, and the situation may be different for iterative or maximum-likelihood lensing estimators \cite{Hirata:2002jy}.

\subsubsection{Delensing}
For noise levels below $\Delta_P \simeq 5 \mu$K-arcmin (after foreground cleaning),  the dominant source of effective noise in the search for primordial B modes is the fluctuation induced by the lensing of E modes from recombination.  This signal has a well-understood amplitude, and unlike many other sources of astrophysical fluctuation in the map, it cannot be removed with multifrequency data.  Instead it must be removed either using map-level estimates of both the primordial E-mode maps and the CMB lensing potential $\phi$ or using a prediction for the lensing B-mode spectrum. The latter approach necessarily leaves some cosmic variance residual of the lensing signal after cleaning, while the former can in principle result in nearly perfect cleaning, so we will concentrate on that approach here.

Even in the map-level approach, the finite noise in the \cmbexp\ survey will lead to residual lensing B modes which cannot be removed and will act as a noise floor for studying B modes from tensors.  The amplitude of these residual lensed B modes are discussed in Section \ref{delens} as a function of the angular resolution and the noise level of the S4 survey; in particular, it is crucial to have high-angular-resolution maps in order to measure the small-scale E- and B-mode fluctuations needed for the EB quadratic lensing estimator.

The concerns with the delensing procedure are similar to those for measuring the lensing power spectrum. The impact of polarized dust and synchrotron emission from the Galaxy, and the impact of polarized point sources on small scales on the lensing reconstruction are addressed in Section~\ref{systAst}. Left untreated the effects may be large; however the use of  multi-frequency data together with the application of dedicated point-source estimators can mitigate these effects.

Another question to be answered for delensing in \cmbexp\ is what to use as the estimate of lensing $\phi$.
Rather than using an estimate of the CMB lensing field obtained from the CMB itself, it is also possible to use other tracers of large-scale structure which are correlated with  CMB lensing \cite{Smith:2010gu}.  In particular the dusty, star-forming galaxies that comprise the cosmic infrared background (CIB) are strongly correlated with CMB lensing, due to their redshift distribution which peaks near $z \sim 2$ \cite{Sherwin:2015baa,Simard:2014aqa}.  The level of correlation is approximately $80\%$ \cite{Ade:2013zuv} and can in principle be improved using multifrequency maps of the CIB which select different emission redshifts \cite{Sherwin:2015baa}.  

Lensing of the CMB can also impact the measurement of features of the CMB power spectrum on small scales, in particular the CMB damping scale and the precise location of the acoustic peaks in harmonic space.  Delensing can therefore improve not just our measurement of primordial B modes but also constraints on parameters that affect the damping tail and peak location.
This includes parameters such as the effective number of neutrino species,  the primordial helium fraction, and running of the spectral index of fluctuations.  Using completely unlensed CMB spectra, rather than lensed spectra, can improve constraints on these parameters. (For example,~\cite{Green:2016} find a 20\% improvement on $\sigma(\neff)$ and $\sigma(Y_p)$ relative to lensed spectra.)  While the delensing procedure will not completely recover the unlensed CMB fluctuations for the S4 experiment, the low noise levels will enable the primordial CMB fluctuations to be measured with good enough fidelity that delensing should have a non-negligible impact on these parameter constraints~\cite{Green:2016}.

\section{Simulation Overview}
Simulations of a CMB mission's data play a number of critical roles; specifically they are required for
\begin{itemize}
\item Mission design and development: ensuring that the mission is capable of meeting its science goals.
\item Validation and verification: ensuring that all of our data analysis tools meet their requirements and specifications.
\item Uncertainty quantification and debiasing: providing an alternative to the full data covariance matrix when this is computationally intractable.
\end{itemize}

As shown in Figure \ref{fig_sim}, given a mission model (both instrument and observation) and a sky model (both CMB and extragalactic and Galactic foregrounds) we can generate a simulation of the mission data in any of its domains. However, there is an inevitable trade-off between how representative the simulation is of real data and the complexity of the input models and computational cost of generating the simulation. The choice of the simulation data domain will then be determined by the balance between the realism requirements and the complexity/cost constraints for the particular task at hand.

\begin{figure}[htbp]
\includegraphics[width=0.75\textwidth]{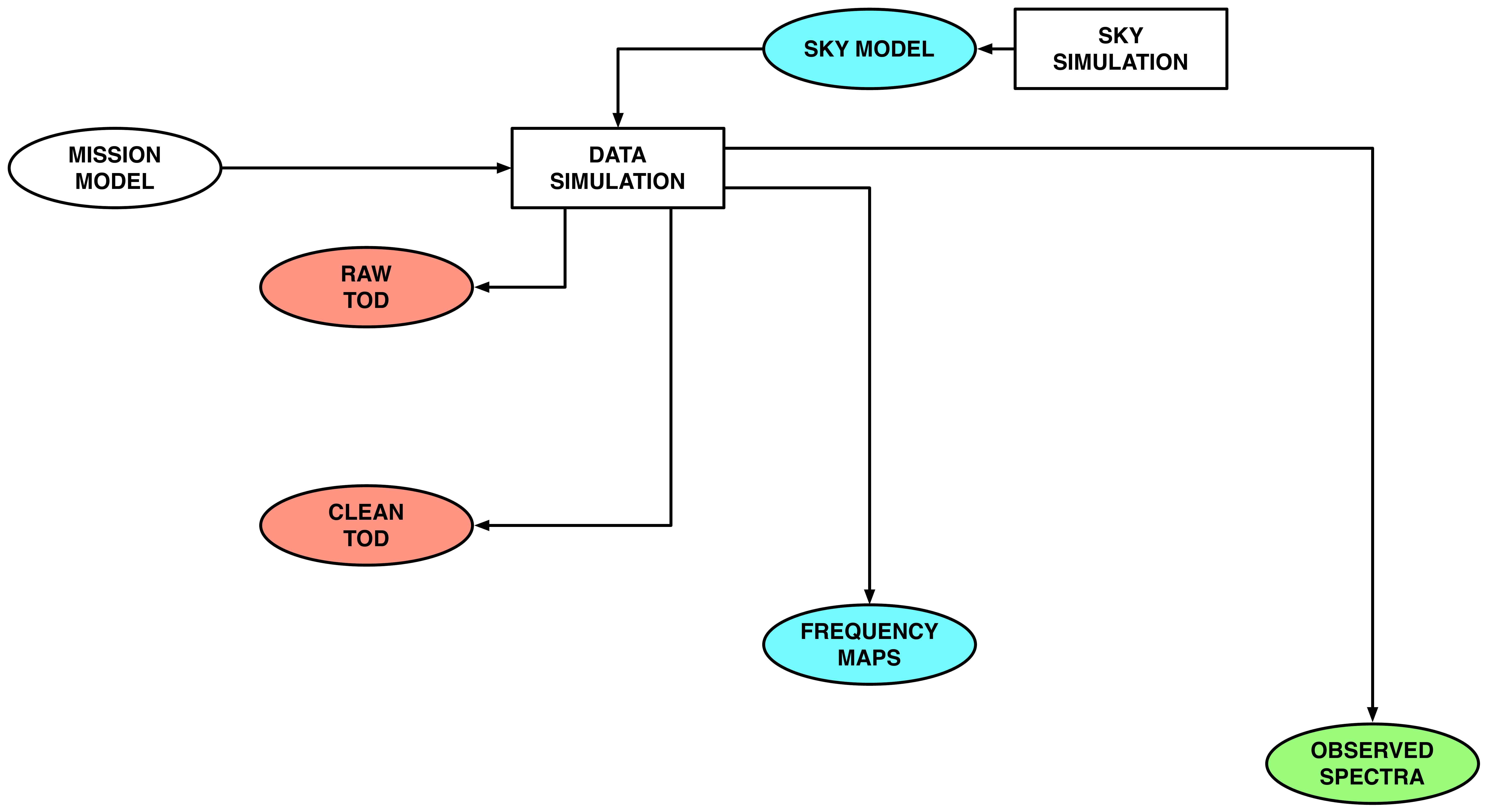}\\
\caption{The CMB simulation pipeline, including both mission-independent sky modeling and mission-specific data simulation in the time (red), pixel (blue) and spectral (green) domains.}
\label{fig_sim}
\end{figure}

The generation of the input mission and sky models are themselves far from trivial tasks. The mission model is typically derived from pre-deployment measurements of the instrument properties refined by characterization from the data themselves, together with ancilliary telescope and environmental data characterizing the observation; the sky model requires its own dedicated simulation capability which - since it is independent of the details of any single mission - can be a community-wide endeavor.


\section{Sky Modeling}
\label{sec:skymodel}

The capability of \cmbexp\ to address its science program depends crucially on the ability to separate the signals of interest from other sources of astrophysical emission. Furthermore, the final accuracy of \cmbexp\ parameter constraints will be limited by the accuracy of the characterization of foreground residuals after such cleaning is performed. 

At the degree scales targeted by ground-based B-mode searches, the polarized CMB is mostly contaminated by diffuse emission from the interstellar medium of our own Milky Way. Both synchrotron and thermal dust are polarized---up to the level of tens of percents, depending on the observed region). Their integrated emission dominates over both the CMB E modes and the CMB B modes on large angular scales, and cannot be safely neglected at scales where B modes from gravitational lensing dominate without robust analyses of their impact on lensing science. Other components, such as spinning dust, free-free emission, emission of molecular lines such as CO, could in principle be polarized at a lower level, of order 1-2 per cent or less, but measurements or upper limits are scarce, and not sufficient at this stage for robust predictions of the polarized amplitude of their emission over large patches of sky. For science on smaller angular scales, the presence of polarized extragalactic radio and infrared sources constitutes an additional source of contamination, which must be removed with a combination of masking or subtracting individual sources, and modeling residuals at the power-spectrum level.

Estimating more precisely the impact of foreground emission on the main science targets of \cmbexp\ will require realistic simulations of the sky emission that can be used to test the effectiveness of component separation techniques and to assess any degradation of the error bars or possible biases due to residual foreground contamination. 

The sky emission is naturally modeled as the sum of emission from different sources. These sources may identified by their emission process (e.g. Galactic synchrotron, due to electrons spiralling in the Galactic magnetic field), or by their place of origin (e.g. emission of a particular extragalactic source). These emissions, as a function of sky pixel and frequency, must then be integrated over the instrument observing band and angular response function (beam) to produce individual frequency maps as observed by the instrument. 
This latter part of the simulation pipeline is treated in Section \ref{sec_datasim}; we concentrate in this section on the sky model.

A sky model is useful only as far as it captures the characteristics of the real sky emission sufficiently for testing the performance of cleaning techniques, in particular the amplitude and statistical properties of residual contamination.
The key characteristics of sky emission for foreground cleaning are:
\begin{itemize}
\item The level of coherence of diffuse emission across observing frequencies, as any decoherence will limit the efficacy of cleaning a foreground from one observing band using the measurement in a different band;
\item The existence or not of a simple parametric emission law for each component emission, such as power laws (for synchrotron) or modified blackbody emission (for dust components);
\item The absolute level of foreground emission (in particular for those components that do not scale simply as a function of frequency, such as the superposition of many individual sources with a specific emission law each);
\item Whether or not emissions for which the level of polarization is unknown or unclear 
must be modeled and treated for \cmbexp\ or can be safely neglected;
\item The level at which foregrounds can be treated as Gaussian random fields, which is an assumption of certain foreground cleaning approaches.
\end{itemize}

The key challenges for constructing a sky model are hence:
\begin{itemize} 
\item The reliability of models based on observations at angular resolution lower than that of \cmbexp\, integrated in broad frequency bands, and with a sensitivity limit at least an order of magnitude worse than what will be achieved with \cmbexp\;
\item The self-consistency of CMB secondary anisotropies (lensing, SZ emission from hot intra-cluster gaz and filaments, late ISW) and extragalactic foregrounds (CIB, radio and infrared sources) is crucial to both de-lensing, and to extragalactic science; generating reliable models over the entire Hubble volume is challenging, the evaluation of errors of such models even more so;
\item The practical usability of the modeling software, in particular the ability to generate many independent simulations quickly.
\end{itemize} 

\subsection{The Galactic interstellar medium}

Strong evidence exists for variability of the physical properties of the interstellar medium of the Milky Way as a function of the line of sight. This variability implies that the properties vary across different regions of the Milky Way, with the total ISM emission in each line of sight being a superposition of emission from various regions. Even assuming that each such region has a simple parameteric emission law, such as a power law or a modified blackbody, the superposition of such emission cannot be modeled with a single simple emission law. Modeling the Galactic ISM for future sensitive surveys such as \cmbexp\ requires  modeling  this complexity at the appropriate level. One possibility is to use a multi-layer approach, in which each ISM component is modeled as a superposition of several optically thin layers, each with a simple (though pixel- and polarization-dependent) emission law.

\subsubsection{Synchrotron}

The baseline Galactic synchrotron model we use here has a power-law scaling with a modestly spatially varying spectral index.  The emission templates are the Haslam 408 MHz data reprocessed by \cite{Remazeilles:2015hpa}, 
and the \wmap\ 7-year 23 GHz Q/U maps \cite{Jarosik:2010iu}
smoothed to 3 degree FWHM and with smaller scales added using the PSM code \cite{Delabrouille:2012ye}.
The spectral index map is derived using a combination of the Haslam 
408 MHz data \cite{Haslam:1981aa} and \wmap\ 23 GHz 7-year data \cite{MivilleDeschenes:2008hn}.
The same scaling is used for intensity and polarization.  This is the same prescription as used in the \planck\ Sky Model's v1.7.8 `power law' option, but with the Haslam map updated to the version in \cite{Remazeilles:2015hpa}.

Extensions to this model that we are exploring include a curved power 
law model with a single isotropic curvature index, and a polarization spectral index that steepens with Galactic latitude by $\delta \beta \sim 0.2$ from low to high latitude, as this is currently consistent with \wmap\ and \planck\ data. 

\subsubsection{Thermal dust}
The baseline model we consider has thermal dust modelled as a single-component modified  blackbody. We use dust templates for emission at 545 GHz in intensity and 
 353 GHz in polarization from the \planck-2015 analysis, and scale these to different  frequencies with a modified black-body spectrum using the spatially varying dust temperature and emissivity obtained from the \planck\ data using the Commander code \cite{Adam:2015wua}. This therefore assumes the same spectral index for  polarization as for intensity.  These templates are smoothed to degree-scale resolution.

Variations on this model that appear consistent with current data include a model with more strongly varying emissivity, e.g. up to $\sigma \sim 0.2$ dispersion on degree scales, and a model with different prescriptions for small-scale behavior, accounting for turbulence in the magnetic field. A two (or more) component model for the dust, composed of the spatially varying sum of silicon and carbonaceous dust, each with a different emissivity, is also physically motivated.

\subsubsection{Spinning dust}
Spinning dust, or anomalous microwave emission, is nominally unpolarized. However, a fractional polarization of a few percent is physically possible and not excluded by current data. We construct a possible model for this polarization using the intensity templates for spinning dust from the \planck-2015 Commander fits
\cite{Adam:2015wua}, combined with the thermal dust polarization angles and an overall polarization fraction.

\subsubsection{Other components}
Other contributions to the intensity and polarization of the Milky Way at \cmbexp\ frequencies, such as 
free-free emission and molecular line emission, are not expected to be at the same amplitude and degree
of polarization as the components treated individually above. However, the full sky model will need to include
these components in the most pessimistic scenarios, unless further data is obtained that 
conclusively demonstrates they can be fully neglected.

\subsection{CMB Secondary Anisotropies and Extragalactic Sources}

The key goal for the extragalactic sky models of \cmbexp\ is to provide fast and self-consistent simulations of CMB secondary anisotropies and extragalactic sources. These models will allow us to make more realistic forecasts. In our cosmological analyses they will allow us to Monte Carlo over the underlying astrophysical uncertainties of these secondaries and sources. Our plan to meet these challenges is modular and can be broken down as follows: 

\begin{itemize}
\item We will use full hydrodynamical simulations of cosmological volume as the basis to parametrically model the complicated {\it gastrophysical} processes associated with extragalactic foregrounds.
\item As the backbone of our model we will require fast simulations of growth of structure that generate halo catalogs for a large set of cosmological parameters.
\item To have self-consistent maps we will have a flexible pipeline that generates simulated all sky maps which applies the parametric models from the hydrodynamical simulations to our backbone large-scale structure simulations and halo catalogs.
\end{itemize}

Hydrodynamical simulations of cosmological volumes are currently available which we can already use to model extragalactic foregrounds. These simulations will be used for the development and testing phases of the simulation pipeline. However, they are limited in their size and sub-grid modeling accuracy, and thus will not meet our accuracy requirements of \cmbexp. We will develop new full hydrodynamical simulations of cosmological volumes that include a variety of physical processes. An essential requirement of these simulations will be to capture growth and evolution of galaxies to cluster-size halos throughout cosmic time at a sufficient spatial resolution. Hydrodynamic simulations of this size and scale are already computationally feasible, the challenges will be the appropriate modeling of radiative cooling, star formation, and feedback processes in order to capture the global stellar and gas contents of these halos.

There are many different approaches already developed to provide us with the underlying large-scale structure simulations that will we build our extragalactic model upon. They vary in speed which tend to inversely scale with accuracy. A benefit of our modular and flexible approach is that we do not need to limit ourselves to one approach. In fact we will compare the various approaches to see how they bias our answers. It is in these simulations where we will vary cosmological parameters assuming that they only affect the growth of structure and not the {\it gastrophysical} properties of extragalactic foregrounds.

Our final product will be all sky maps. They will be in HEALPix \cite{Gorski:2004by} format to seamlessly interface with galactic and CMB simulated maps. The map products will include:

\begin{itemize}
\item Optical galaxies that correspond to the various overlapping surveys including LSST.
\item Radio and dusty star-forming galaxy point sources.
\item Unresolved CIB.
\item Projected density maps (both total and gas) of the large-scale structure.
\item Thermal and kinetic SZ maps.
\end{itemize}

\noindent We will explore the parameter space for each of the maps listed above and provide a sufficient number of realizations that we can marginalize over the many model uncertainties. For example, the lensing field can be constructed through a proper ray-tracing method from the projected density maps or via the Born approximation. Our self-consistent extragalactic sky model allows us to test various sources of contamination and systematic biases in our estimators. Additionally, any cross-correlation analyses can easily be checked and evaluated using these maps. {\bf All the simulation products we create will become public.}

 
\section{Data Simulation}
\label{sec_datasim}

The data simulation subset of the CMB simulation pipeline (Figure \ref{fig_sim}) takes the sky model and applies the mission model to it to generate a simulated data set for that mission. The mission model consists of two parts: the instrument model defines the data acquisition system (telescope, detectors, read-out), while the observation model defines its deployment (scanning strategy, environment). Depending on the degree of detail of the sky, instrument and observation models that we include, the resulting data set can be in any of the data domains - time-ordered (raw or clean), map, or spectral. Inevitably there is a trade-off between the realism of the simulation and the complexity and cost both of generating the model inputs and of performing the simulation, with the choice reflecting both the requirements of the subsequent analyses of the data set and the availability of computational resources.

At the most detailed level, the observation model includes the telescope pointing (typically sampled more sparsely that the detectors) and its environment (comprising the atmosphere and surroundings for a ground-based telescope). Correspondingly, the instrument model includes each detector's polarized $4 \pi$-beam and bandpass (defining the optical power incident on the detector for a given pointing), and a model of its electronics and readout (defining the recorded output data resulting from that optical power).

\subsection{Time Domain}

TOD simulations are necessarily the most expensive to perform, but provide the most precise representation of the mission data. In particular they enable the injection of the full range of systematic effects into the data to assess strategies for their mitigation and to quantify any residuals. As such they are critical for the quantification of uncertainties due to inherently temporal data components such as noise. The TOD simulation is separated into signal and noise components, which are then added prior to the reduction of the total TOD.

For the signal simulation for a given detector, we first apply the detector's bandpass to the sky model, component by component, to build up the total sky for that detector. We then reconstruct the detector pointing from the overall telescope pointing model and generate the astrophysical sky signal for each pointing by convolving the sky model map with the $4 \pi$ beam. The astrophysical sky signal is added to additional simulated signals from atmospheric signal fluctuations and ground pickup (both of which will obviously induce correlated signals across the detectors), and the total signal is propagated through a simple model of the optics to include the polarization angle rotations and optical efficiencies of the optical stages. This results in the total millimeter-wave power incident on the detector. For simulating the clean TOD this is sufficient. However, for the raw TOD we now need to apply a physical model of the detector system and associated readout to convert the optical power into detector output. The details of the physical model depend on the detector technology, but as an example we consider a transition-edge superconducting (TES) bolometer read out with a multiplexed SQUID amplifier. The simulation would then need to model the flow of heat in the TES absorber and the flow of current and magnetic flux through the SQUID readout. Variations in ambient magnetic field could also be added at this stage. Such a simulation would also need to incorporate detector-detector correlations induced by crosstalk or thermal fluctuations.  Additional filters applied by the readout electronics would also be included, including digitization with an analog to digital converter. For MKID or coherent receivers, the physical model would be different in detail, but would include a similarly detailed model.

For the noise simulation we can simply generate a white noise timestream and convolve it with the detector's noise power spectral density (PSD), given in either analytic or numerical form. Cross-correlated noise can be included by simulating multiple noise timestreams each with their own PSD, with some being common to multiple detectors, while piecewise stationary noise simply requires us to use the appropriate PSD for each stationary interval.

\subsection{Map Domain}

The next level of abstraction from full TOD simulations is simulating the sky map that would be made
from the TOD. The signal part of such simulations is straightforward: the various components of the sky
model are bandpass-integrated and convolved with a beam and any filtering kernel that is applied to the 
actual data. Both of these operations can be performed on a per-detector basis (using the measured 
individual-detector beams and bandpasses), but this reduces much of the computational gain in going from
TOD to map-level simulations, so it is more likely that maps would be simulated for large groups of
detectors---possibly all detectors at a given observing frequency---at once. For experimental platforms that 
apply TOD filtering before mapmaking, it will be necessary to create a map-level (or two-dimensional 
Fourier-space) representation of the TOD filtering that results in simulated maps with the same modes
missing or altered as the real map (or a map that has been constructed from full TOD simulations).

The simplest implementation for the noise part of map-level simulations is adding constant-amplitude
white noise to every simulated map pixel. This ignores pixel-pixel correlations and incomplete coverage, 
both of which are naturally accounted for in full-TOD simulations. The exact nature of the 
correlated pixel-domain noise 
(or non-white Fourier-space noise) arises from a combination of non-white noise in the TOD 
and the scan strategy, and for some scan patterns can be analytically projected from the time domain
to the pixel or 2d Fourier domain \cite{Wandelt:2001fz,Crawford:2007mh,Bucher:2016ysw}. 

The effect of non-uniform coverage on the noise properties of the map is simple to simulate
in the white-noise case: the uniform-coverage white-noise map is simply multiplied by the square
root of the inverse of the ``hit-count'' map. The combined effect of non-white noise and non-uniform
coverage---particularly if the coverage map is not smooth on scales of the noise correlation---will
be more difficult to simulate purely in the map domain.

\subsection{Spectral Domain}

Simulations at the one-dimensional power spectrum level are fast, computationally light, and can 
be used to explore large experimental and observing-strategy parameter spaces quickly and efficiently.
As such, they will constitute the bulk of simulations used in forecasting for \cmbexp, particularly in
the era in which the experimental design is not final, and reasonably fast communication and iteration
between the experiment design and forecasting teams is crucial. Section 
\ref{sec_specforecast} contains a detailed discussion of plans for
implementing maximally realistic spectral-domain simulations.

 
\section{The Production Data Pipeline}

With real data in hand, the production data pipeline (Figure \ref{fig_prod}) will include data analysis, time-domain simulation for uncertainty quantification, and feedback loops to improve both the mission model and the sky simulation. Time domain data are extensively used to refine our model of the mission; for the instrument model this can include such steps as determining beam profiles and estimating noise properties (including cross-correlations); for the observation model, it includes reconstructing the detector pointing and polarization orientation from telescope sensor data, and incorporating atmosphere records in the data-flagging. The foreground component maps, observed spectra, and parameters of cosmology and fundamental physics are used to refine our sky model. In both cases the feedback loops both improve the consistency of the simulations with the data and enable us to refine the validation and verification of the analysis methods themselves.

\begin{figure}[htbp]
\includegraphics[width=1.0\textwidth]{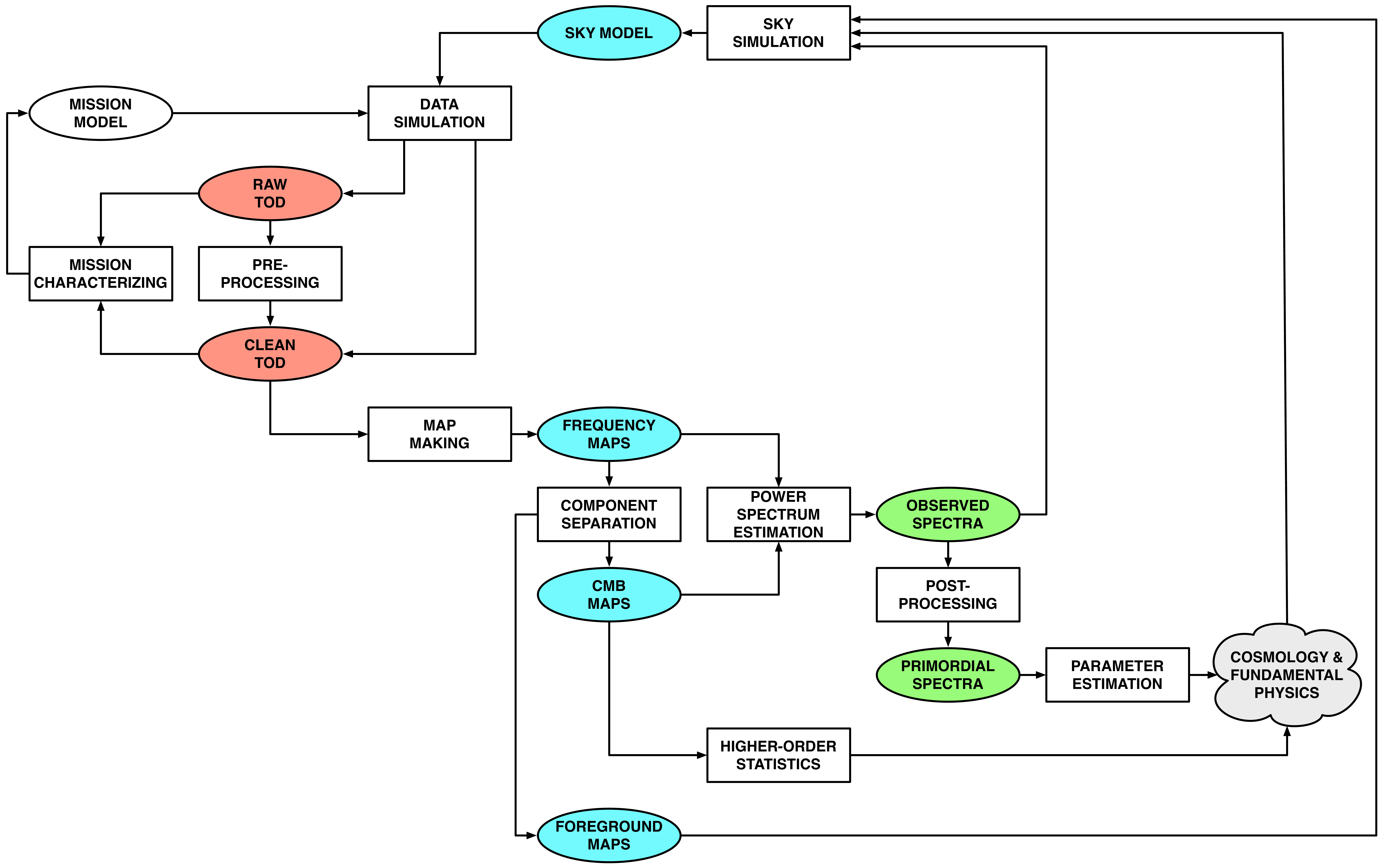}\\
\caption{The production data pipeline, including data analysis, simulation, and feedback.}
\label{fig_prod}
\end{figure}

\subsection{Implementation Issues}

The quest for ever-fainter signals in the CMB drives us to gather ever-larger time-ordered data (TOD) sets to obtain the necessary signal-to-noise to uncover them. As Figure \ref{fig_cmb_hpc_scaling} shows, the volumes of ground-based, balloon-borne and satellite CMB data sets have exhibited exponential growth over the last two decades, and are anticipated to do so again over the coming two. Moreover, for suborbital experiments the exponent exactly matches that of Moore's Law for the growth of computing capability, where we use as a proxy here the peak performance of the flagship high performance computing (HPC) system at the DOE's National Energy Research Scientific Computing (NERSC) Center at any epoch (reflecting the widespread use of NERSC for CMB data analyses over the last 20 years). 

\begin{figure}[htbp]
\centering
\includegraphics[width=0.75\textwidth]{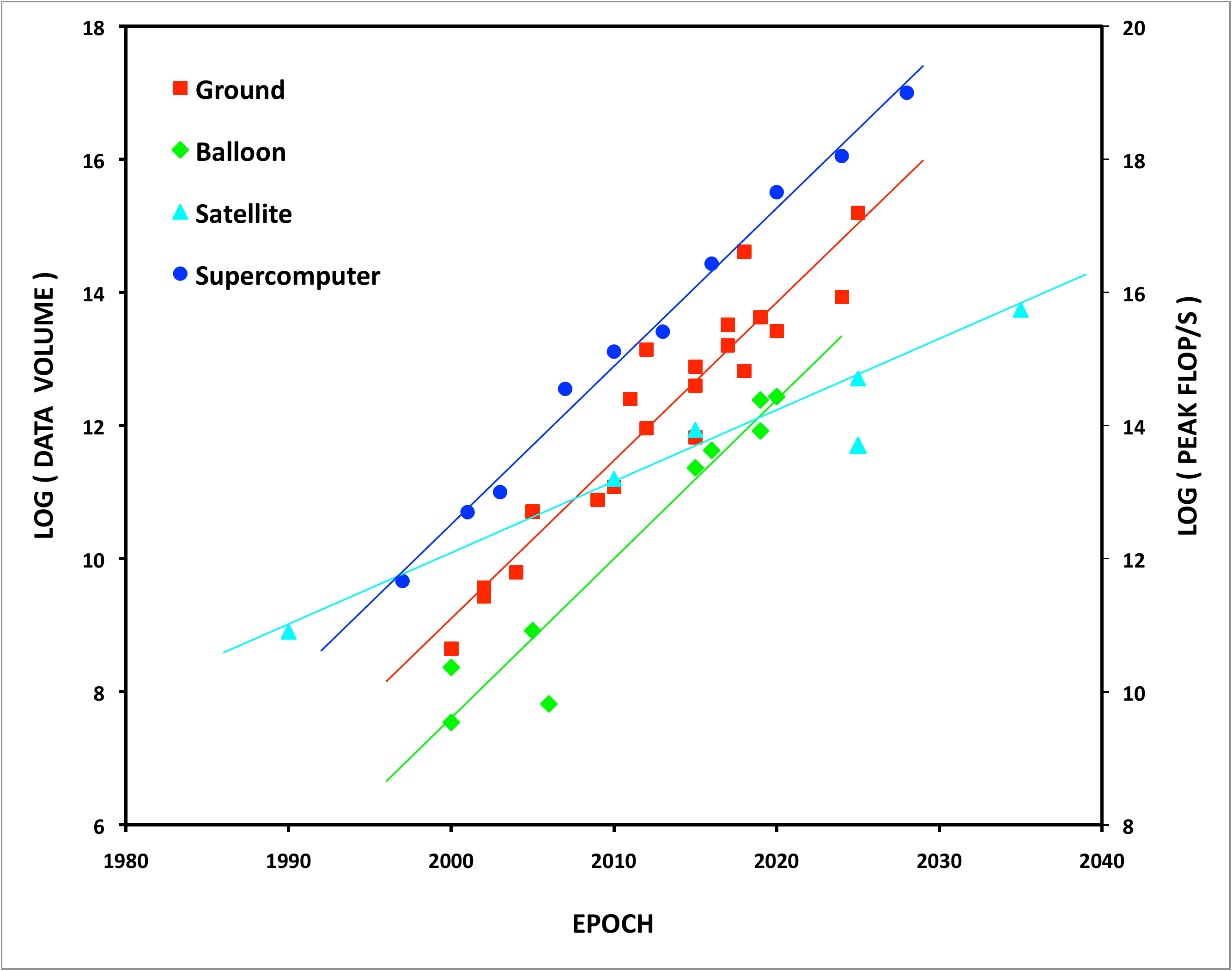}
\caption{Exponential growth of CMB time-ordered data volume and HPC capability: 1990 -- 2030.}
\label{fig_cmb_hpc_scaling}
\end{figure}

As noted above, in the absence of a full data covariance matrix we rely on Monte Carlo methods for uncertainty quantification and debiasing, and achieving the desired percent-level statistical uncertainty requires us to simulate and reduce $10^4$ realizations of the data. This implies that all TOD-processing steps (in simulation or analysis) must employ algorithms that scale no worse than linearly in the number of samples, and that these algorithms must {\em collectively} be implemented efficiently on the largest high performance computing (HPC) platforms available to us. 

The most massive Monte Carlo sets generated to date have been the Full Focal Plane (FFP) sets in support of the analysis of the \planck\ satellite data \cite{Ade:2015via}, with FFP8 comprising $10^4$ realizations of the mission reduced to O($10^6$) maps. Key to achieving this scale has been an aggressive optimization of the software stack, coupled with system-specific tuning over 6 generations of NERSC supercomputer. In particular wherever possible TOD input/output (IO) is removed from the pipeline so that, for example, instead of pre-computing the TOD and then pre-processing/mapping it, each realization is generated on demand and passed to the analysis pipeline in memory. While this necessitates the re-simulation of a realization should further analysis be required, it is still very substantially faster than writing it to disk and reading it back in. Similarly, inter-process communication is optimized by using a hybridized MPI/OpenMP implementation that employs explicit message passsing for inter-node, and threading for intra-node, communication.

The critical challenge for CMB-S4 will be to develop these capabilities for a dataset 1000x the size of \planck's and 100x the size of those from existing S2 ground-based experiments. This scale of computing will require substantial development effort from the CMB community, but is still much smaller than some existing experiments (e.g. ATLAS, CMS) and, with appropriate tooling, should be possible on existing or forthcoming computing facilities. The S3 experimental efforts are currently exploring a number of computational tools to reach the required level, including investigating use of the US grid where feasible and code optimization for the upcoming generations of energy-constrained HPC architectures, with their increased heterogeneity and deeper memory hierarchies and based, in the short term, on either graphical programming unit (GPU) or many integrated core (MIC) technologies.


\section{Forecasting}\label{sec:Forecasting}

Given the computational constraints on time-domain simulations and analyses, it is currently not possible to use the production pipeline to explore the full instrument and observation parameter space. However, such exploration is critical in the mission design and development phase, and so we employ a forecasting pipeline (Figure \ref{fig_fcast}) to bypass this bottleneck. This section outlines the forecasting methodology; specific results obtained from this appraoch are given in the preceeding science chapters.

\begin{figure}[htbp]
\includegraphics[width=0.9\textwidth]{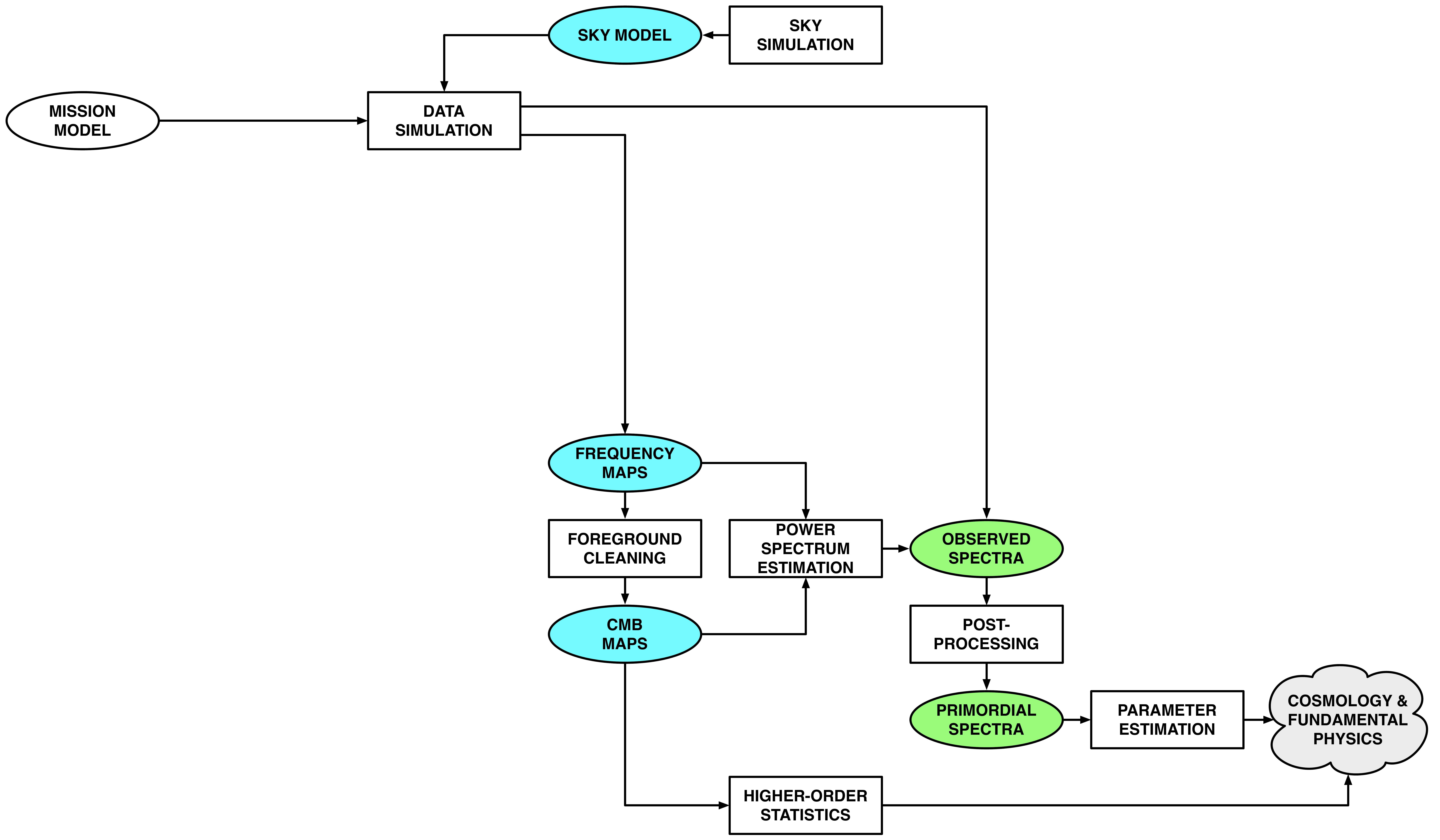}\\
\caption{The forecasting pipeline, including pixel- and spectral-domain simulations and analyses}
\label{fig_fcast}
\end{figure}

Forecasting efforts for Stage 1 and 2 experiments were hampered by lack of experience with previous deep polarization maps and little knowledge of high latitude Galactic foregrounds; forecasting for \cmbexp\ will therefore be built on the solid foundation of map-derived evaluations of instrumental noise performance and astrophysical foreground levels from the Stage 2 experiments and the \planck\ satellite.

The forecasting approach will combine Fisher matrix-derived estimates of power spectrum errors with detailed map-level simulations.
The spectral-domain projections are computationally easy, making them useful to explore the large parameter space of instrument and survey configurations.
Map-domain simulations are used to ground the spectral-domain projections in reality and to challenge them with cases of real world astrophysical complexity.
To gain the benefit of this complementarity, it is important that we maintain compatibility between these two forecasting approaches and establish agreement between them for simple questions before proceeding to more difficult tests.

A key input to the forecasting process are full-season noise maps from existing Stage 2 experiments, which encode actual noise performance and have been verified by null tests on real datasets.
Performance of \cmbexp\ can be estimated by rescaling these noise maps, which already contain reality factors such as detector yield, weather, observing efficiency, and filtering of sky modes.
Systematic errors should be included in the projections, with unknown systematics allowed at a level that scales with the map noise used for jackknife null tests.
Forecasting should also include our best knowledge of the astrophysical foregrounds and account for the impact of component separation on \cmbexp\ science goals.
The forecasting inputs will improve as we acquire data from Stage-3 experiments and possible complementary balloon-borne experiments.

Here we describe the main approaches used by our community for forecasting the expected performance of \cmbexp. The central considerations for assessing the expected performance for large-scale B modes are Galactic foregrounds, ability to delens the data, and a realistic assessment of instrument noise at large scales. 
For the smaller-scale polarization two-point functions (TE, EE) and the lensing four-point function ($\kappa \kappa$), extragalactic foregrounds and instrumental noise are the key considerations.
To forecast the return of the thermal Sunyaev-Zel'dovich effects, an estimate of the expected cluster counts as a function of mass and redshift is the core statistic, combined with an estimate of how well the masses can be calibrated using optical or CMB weak lensing. For the kinetic Sunyaev-Zeldovich effect, extragalactic foregrounds and overlap with spectroscopic surveys must all be considered. 

\subsection{Forecasting \cmbexp\ constraints on the tensor-to-scalar ratio}

\subsubsection{Spectrum-based domain forecasting}
\label{sec_specforecast}

Power spectra are the primary tool used for CMB analysis.
Forecasting the power spectrum uncertainty and resulting parameter constraints for \cmbexp\ is an efficient and powerful tool to explore trade-offs in experiment design.

The bandpower covariance matrix describes the raw sensitivity of all auto and cross-spectra obtained between maps of T, E, and/or B modes at multiple observing frequencies, as well as the signal and noise correlations that exist between these spectra.
This covariance matrix includes contributions from the sample variance of signal fields (CMB and foreground) and instrumental noise, including signal$\times$noise terms.
The signal variance depends on the assumed sky model, which can be modified to explore optimistic or pessimistic scenarios.
As discussed above, estimates of the noise variance should be obtained by rescaling of noise levels that have actually been obtained by Stage 2 experiments (or Stage 3, when available).
Only these scaled noise levels will include all the ``reality factors'' that are incurred in operating a CMB experiment.

Once we have a projection for the bandpower covariance matrix of \cmbexp\, we can derive constraints on a parametrized model of cosmological and foreground signals via the Fisher information matrix.
While we are most interested in parameter $r$, it is necessary to also consider the amplitude, spectrum, and spatial distribution of the dust and synchrotron foregrounds (see \cite{Ade:2015tva} for an example).
The Fisher matrix formalism allows us to calculate the marginalized error on each parameter, with priors if desired, or to explore the degeneracies between parameters.

By compressing the data down to power spectra, it is feasible to use this technique to evaluate a wide range of survey designs.
The parametrized signal model is also quite flexible and can include complications such as dust--synchrotron correlation or spatially varying foreground spectral indices.
The limitation is that by considering the power spectrum only we are treating all signals as Gaussian, an approximation which must break down at some point for foregrounds.
For this reason, it is important to have the ability to spot check the spectrum-based forecasts against map-based forecasts at specific choices of signal model.

As described in Section~\ref{sec:needs}, the specific implementation of Fisher forecasting for CMB-S4 constraints on $r$ in this document assume the following baseline configuration parameters:
\begin{itemize}
\item{250,000 detectors operating for four years, dedicated solely to the combination of measuring degree-scale B modes and measuring lensing B modes on the same patch of sky (for delensing). 
The split in effort between degree-scale observations and lensing observations is
optimized for, and depends on the fraction of sky observed. The optimziation sets the level of delensing necessary.}
\item{Eight frequency bands (30, 40, 85, 95, 145, 155, 215, 270), two each in the four atmospheric windows. The framework optimally distributes effort among each of the 
eight frequencies, assuming comparable focal plane area. To mitigate against 
potential unknown foreground complexities, we also force an equal effort 
split among each pair of frequencies in each atmospheric window.}
\item{For the degree-scale effort, bandpower covariance matrices scaled directly from achieved BICEP2/Keck performance.}
\item{Foregrounds as measured in the BICEP2/Keck patch of sky, including decorrelation of the dust signal between high frequencies and the frequencies of interest for CMB. The decorrelation parameter is a function of both frequency and multipole number.}
\item{Delensing efficiency based solely on noise level in the high-resolution map used for delensing---i.e., no degradation to delensing from foregrounds or systematics.}
\end{itemize}
In Section~\ref{sec:needs}, these assumptions are used to search sky fraction and detector effort parameter space, with 
$\sigma(r)$ as the figure of merit. We verify these results using a second Fisher code (described below) and a map-based
forecasting method (see next section).

The second Fisher code---detailed in \cite{Errard:2015cxa}---parameterizes the CMB-S4 instrument by its sky and multipole coverage, along with the central frequency, bandwidth, resolution and white-noise level of each channel. The ability of the instrument to remove diffuse foregrounds is estimated using a parametric maximum-likelihood forecast \cite{Errard:2011vi,Errard:2012qx} based on \planck\ foreground measurements \cite{Adam:2015tpy,Adam:2015wua}. The impacts of foreground subtraction---residual foregrounds and increased noise relative to the raw combination of all channels---are propagated to a delensing forecast (based on \cite{Smith:2010gu}), which also estimates the sensitivity to the lensing convergence power spectrum. Constraints on $r$ are produced with a standard Fisher forecast (see Section~\ref{sec:ttee} for a complete description) using temperature, E-mode, delensed-B-mode and lensing convergence information, marginalizing over the amplitude and multipole-dependence of the foreground residuals. 

The first code folds in a number of realism factors that results in more conservative
constraints, typically by a factor of 2.5: including dust decorrelation, using a 
fully descriptive BPCM including correlation between adjacent bins, 
using realistic $N_l$'s that are grounded in achieved performance and include
an appropriate level of low ell excess noise, mode filtering which affect sky coverage
and S/N per mode, and noise contributions to the BPCM that take into account the 
non-uniformity of the survey which effectively yields wider/shallower maps than the 
``Knox formula''-derived \cite{Knox:1995dq} equivalent products used in the second code. When these realism factors
are stripped away, the two codes agree. It is worth noting that it is possible that in the future we might do better on 
some of the points above, in which case the factor of 2.5 brackets the list of 
possible outcomes.  

\subsubsection{Map-based domain forecasting}

Foregrounds are intrinsically non-Gaussian and anisotropic, so we also consider approaches directly in map space to explore the robustness of spectrum-based approaches, in particular in the case of pessimistic foregrounds where the spectral indices or dust emissivities have non-trivial spatial variation. The map-based method used in this Science Book is a Bayesian model fitting method, where the foregrounds are described parametrically using a physical model for each component. It is described in \cite{Alonso:2016xft} and follows similar implementations in e.g., \cite{Eriksen:2005dr}.

Using this method, we simulate maps of the CMB, Galactic foregrounds and expected noise at each of the \cmbexp\ frequencies and integrate them across the expected bandpass width for each channel. We use the PySM numerical code \cite{Thorne:2016ifb} which generates Galactic models as described for example in Section \ref{sec:skymodel}. Other similar codes exist in our community \cite{Delabrouille:2012ye}. In this framework it is straightforward to include simulations at ancillary frequencies that might be provided by other experiments, for example the \planck\ data. We fit a parameterized model to the simulated maps, fitting the CMB, thermal dust, and synchrotron in small pixels, and the synchrotron spectral index and dust emissivity and temperature in larger pixels of order degree-scale or larger. We estimate the BB power spectrum of the foreground-marginalized CMB map using the MASTER \cite{Hivon:2001jp} algorithm, and convert this into an estimate of $r$ and its uncertainty. In this Science Book we use the BFoRe code \cite{Alonso:2016xft}; our community also has access to the Commander code which can be used to perform similar analyses.

This method provides an assessment of the expected bias on $r$ if the model does not match the simulation, and shows how much the expected uncertainty on $r$ would increase if more complicated foreground models are explored e.g. \cite{ArmitageCaplan:2011sn,Remazeilles:2015hpa}. It is more computationally expensive than spectral-domain forecasts though, so we limit this approach to a smaller subset of explorations. 

We compare this map-based forecasting to the results of the spectrum-based Fisher-matrix 
forecasting described above. For discrete points in sky fraction and detector effort parameter space, 
we recover the Fisher results with the map-based code, both finding for example $\sigma(r)=0.001$ for $f_{\rm sky}=0.1$ and $r=0$. In this map-based approach we approximate the scaled achieved noise properties by modeling the noise as white noise plus a power-law noise component that starts dominating at a scale below $\ell_{\rm knee}$. The power law was determined by fitting to the achieved BICEP2/Keck noise curves.  This consistency of the different approaches indicates that foreground residuals due to assumptions of Gaussianity and isotropy are not significantly biasing the spectrum-based Fisher forecasts, but this will be the subject of further study as the S4 design evolves.

\subsection{Forecasting \cmbexp\ constraints on parameters from TT/TE/EE/$\kappa\kappa$}
\label{sec:ttee}

Throughout this Science Book we forecast the expected constraints on cosmological parameters from TT/TE/EE and $\kappa\kappa$ (the CMB lensing convergence power spectrum) using Fisher-matrix methods. This assumes that the resulting parameter distributions are close to Gaussian, which is sufficient for the majority of parameters we consider.

We assume that S4 data will be combined with existing \planck\ satellite data. We also assume that other non-CMB data will be available. In particular we consider measurements of Baryon Acoustic Oscillations from the DESI spectroscopic galaxy redshift survey. In some places we consider measurements of cosmic shear from the Large Synoptic Survey Telesope.

In most cases, the codes we use either consider the unlensed maps and the lensing convergence map as the basic statistics, or the lensed power spectra of those maps together with the reconstructed $\kappa \kappa$ spectrum. For the power spectrum approach, to compute the Fisher matrix for the CMB we use the lensed power spectrum between each pair of fields $X, Y$:
\begin{equation}
\label{eqEstimator}
\hat{C}^{XY}_\ell = \frac{1}{2\ell+1}\sum_{m=-\ell}^{m=\ell} x^{*}_{\ell m} y_{\ell m}.
\end{equation}
The estimated power spectrum is Gaussian-distributed to good approximation at small scales. In this case a full-sky survey has
\begin{equation}
-2\ln\mathcal{L}(\boldsymbol{\theta}) = -2\sum_\ell \ln p( \hat{C}_\ell | \boldsymbol{\theta}) \\
=  \sum_\ell  \Big[ (\hat{C}_\ell - C_\ell(\boldsymbol{\theta}) )^\top  \mathbb{C}^{-1}_\ell(\boldsymbol{\theta}) \big(\hat{C}_\ell - C_\ell(\boldsymbol{\theta})) + \ln \det(2 \pi \mathbb{C}_\ell(\boldsymbol{\theta})) \Big]
\end{equation}
where $ \hat{C}_\ell = (\hat{C}_\ell^{TT}, \hat{C}_\ell^{TE}, ...) $ contains auto- and cross-spectra and $\mathbb{C}_\ell$ is their covariance matrix. Discarding any parameter dependence in the power spectrum covariance matrix gives
\begin{equation}
F_{ij} = \sum_\ell \frac{\partial C^\top_l}{\partial \theta_i} \mathbb{C}^{-1}_\ell \frac{\partial C_l}{\partial \theta_j}.
\end{equation}
Here the covariance matrix for the power spectra has elements
\begin{equation}
\mathbb{C}(\hat{C}_l^{\alpha \beta}, \hat{C}_l^{\gamma \delta}) = \frac{1}{(2l+1)f_{\rm sky}} \big[ (C_l^{\alpha \gamma} + N_l^{\alpha \gamma}) (C_l^{\beta \delta} + N_l^{\beta \delta})  \\
+ (C_l^{\alpha \delta} + N_l^{\alpha \delta}) (C_l^{\beta \gamma} + N_l^{\beta \gamma}) \big],
\end{equation}
where $\alpha, \beta, \gamma, \delta \in \{T, E, B, \kappa_c\}$ and $f_{\rm sky}$ is the effective fractional area of sky used.  Other codes construct the Fisher matrix using the unlensed temperature and polarization fields, and the lensing convergence field, rather than the suite of lensed two-point spectra and the lensing four-point function. Both approaches give broadly consistent estimates.  

For certain cosmological parameters, unlensed spectra show moderately stronger constraints compared to lensed spectra (e.g. 20-30 percent difference $N_{\rm eff}$~\cite{Baumann:2015rya}).  This difference is largely attributable to the lens-induced peak smearing which reduces the sensitivity to the acoustic peak locations in both T and E.  In reality one does not have access to the unlensed spectra and we must delens T and E.  Of course, delensing is not a perfect procedure and should be modeled including the noise in the lensing reconstruction.  In cases where the modest improvement from delensing are important (e.g. $\neff$, $Y_p$), we model the delensed spectra in our forecasts.  Delensing reverses the deflections of the CMB due to lensing using an observed map of the lensing potential, $\phi^{\rm obs}$.  Following~\cite{Green:2016}, the delensed temperature can be written schematically as
\begin{equation}
T^{\rm delensed}(x) = \bar h \star T^{\rm obs}\big(x \big) + h \star T^{\rm obs}\big(x-g\star \nabla \phi^{\rm obs}(x)\big)
\end{equation}
where $\bar h$, $h$ and $g$ are filters and $^{\rm obs}$ indicates the observed maps.  The filters are chosen to maximize the Fisher information in the delensed spectra given the noise in the maps.  A similar procedure is applied to the E modes.  In addition, the lens-induced covariances discussed in Section~\ref{se:covs} are included to correctly account for the residual lensing left in the maps after delensing.  Details of the methods used to generate the relevant forecasts can be found in~\cite{Green:2016}.

The CMB lensing reconstruction noise is calculated using the \cite{Hu:2001kj} quadratic-estimator formalism.  We also avoid including information from both lensed BB and the four-point $\kappa \kappa$, as they are covariant. The BB spectrum will not contribute as significantly to S4 constraints, compared to $\kappa \kappa$, and has a highly non-Gaussian covariance \cite{BenoitLevy:2012va}. 

\subsubsection{CMB-S4 specifications}
For CMB-S4, we approximate the noise part of the covariance as
\begin{equation}
N^{\alpha \alpha}_\ell = (\Delta T)^2 \exp \left( \frac{\ell(\ell + 1) \theta^2_{\rm FWHM}}{8 \ln 2} \right)
\end{equation}
for $\alpha \in \{T, E, B\}$, where $\Delta T$ ($\Delta P$ for polarization) is the map sensitivity in $\mu$K-arcmin and $\theta_{\rm FWHM}$ is the beam width. 

We approximate the wide-field part of the S4 experiment as a 4-year survey using approximately 250,000 detectors covering 40\% of the sky in the lowest Galactic foreground region. We consider beam widths of both 1' and 2', and in some cases consider the effect of greater variation in the beam width. By scaling the map depths achieved by current CMB experiments, we estimate a white noise level of 1 $\mu$K-arcmin in intensity, and $\sqrt{2}$ higher in polarization, if the detectors were all concentrated at 150~GHz. In practice we will distribute detectors among a set of bands and a component separation method will be used to estimate the CMB, but in this phase of our study we assume that maps from these bands are optimally combined together, and do not model the removal of Galactic foregrounds for these `non-r' parameters. Including these multiple frequencies will be the focus of future work; since Galactic foregrounds have a smaller effect on lensing and the CMB damping tail, we expect them to impact forecast constraints much less than for gravitational wave limits.

For these smaller-scale forecasts we do not account for any possible mode filtering due to the mapping. For polarization our nominal estimate is white noise, assuming that the tiny intrinsic polarization of the atmosphere, potentially combined with the use of polarization modulators, minimizes atmospheric contamination. In the longer term, these forecasts may be refined using scaled versions of noise spectra achieved in the field by experiments at the appropriate site. Eventually, full bandpower covariance matrices scaled from fielded experiments can also be used.

To address the issue of extragalactic foregrounds, we set as the default a maximum multipole for the recoverable information of $\ell^T_{\rm max} = 3000$ and $\ell^P_{\rm max} = 5000$ for \cmbexp, as foregrounds are expected to be limiting at smaller scales. We also set a minimum multipole due to the challenge of recovering large scales from the ground, and consider in general $\ell=30$. 

\subsubsection{Non-S4 data specifications}

We include \planck\ data at the scales $\ell<\ell_{\rm min}$, nominally with $\ell_{min}=30$, and we also add \planck\ data at all scales over the part of the sky not measured by S4 from Chile or the South Pole, approximated as covering an additional $f_\mathrm{sky}=0.2$.

For the noise levels of \planck, we assume that a data release including reliable polarization data will have happened before \cmbexp\ data is taken and forecast results that include TE and EE data and also large-scale temperature and polarization from HFI. This follows approaches in e.g. \cite{Allison:2015qca}. For the optical depth to reionization, we assume that \planck\ has reached currently published results, so impose a prior of $\tau=0.06\pm0.01$.

In some cases we consider the addition of a cosmic-variance limited large-scale polarization measurement, as we might expect to get from a PIXIE or LiteBIRD satellite or potentially a high-altitude balloon.

To add information from Baryon Acoustic Oscillation (BAO) experiments, some of our codes add the BAO Fisher matrix
\begin{equation}
F_{ij}^{\rm BAO} = \sum_{k} \frac{1}{\sigma_{f,k}^2}\frac{\partial f_k}{\partial \theta_i}\frac{\partial f_k}{\partial \theta_j}
\end{equation}
where $f_k = r_s/d_V(z_k)$ is the sound horizon at photon-baryon decoupling $r_s$ over the volume distance $d_V$ to the source galaxies at redshift $z_k$. Other codes include the forecasted power spectra directly. We also follow standard approaches to including other low redshift probes.

\subsubsection{Fisher code validation}

We use six different Fisher matrix codes in the Science Book, but set up to use the same settings. We check that they all give consistent results for the $\Lambda$CDM model. These are shown in Table \ref{tab:fisher}, which indicates the expected improvement of S4 over \planck\ for these parameters. 

For forecasts quoted in this Science Book, we take the approach of adding just the individual parameters of interest to the basic LCDM set, unless stated.

\begin{table}
  \centering
\caption{\small Forecasted LCDM parameters}
\begin{tabular}{c  c  c  c  }
\hline
\hline
  & fiducial & \planck &  S4+\planck \\
 \hline
$100\Omega_bh^2$   & $2.22$ & $\pm 0.017$ & $\pm 0.003$ \\
$\Omega_ch^2$      & $0.120$ &$\pm 0.0014$  & $\pm 0.0006$  \\
$H_0$              & $69.0$ &  $\pm 0.7$     & $\pm 0.24$    \\
$10^{9}A_s$        & $2.2$  &$\pm 0.039$   & $\pm0.021$  \\
$n_s$             & $0.966$&  $\pm 0.004$   & $\pm0.002$ \\
$\tau$            & $0.06$ &  $\pm 0.01$   & $\pm0.006$ \\
\hline
\end{tabular}
\label{tab:fisher}
  \end{table}

\subsection{Forecasting \cmbexp\ constraints on parameters from tSZ/kSZ}

As discussed in Section 4.1, some of the most important constraints on dark energy and tests of 
general relativity will come from the thermal and kinematic Sunyaev-Zel'dovich effects. The information
from the tSZ will mostly be in the form of the abundance and clustering of galaxy clusters, while the 
exact way in which kSZ information will be extracted is not fully determined, as this is a fairly new probe
with rapidly developing analysis methods.

Forecasting constraints from cluster abundance is complicated by the fact that even current CMB 
experiments are not limited in their cluster-based constraints by raw sensitivity but rather by systematic
uncertainties in the scaling relation between the tSZ observable and the cluster mass 
\cite{Reichardt:2012yj,Ade:2015fva}. Thus the cluster-based forecasting for \cmbexp\ will likely be more
focused on constraints on the observable-mass relation such as those that come from CMB-cluster
lensing (see Section 4.1.2.1).

Forecasting constraints from kSZ will be an ongoing avenue of development. As early results become
more mature, and the community explores new ways of measuring this signal 
\cite{Hand:2012ui,Keisler:2012eg,Ade:2015lza,Schaan:2015uaa,Hill:2016dta,Soergel:2016mce}, the exact methods for forecasting
will become more clear.

\bibliography{cmbs4}



\end{document}